\documentclass[twocolumn,aps,rmp,amsmath,amssymb,longbibliography,10pt]{revtex4-1}

\usepackage{graphicx}
\usepackage{xcolor}
\usepackage{longtable}
\usepackage{bm}
\usepackage{lineno}
\usepackage{hyperref}
\usepackage{enumitem}
\hypersetup{backref=true,pagebackref=true,hyperindex=true,colorlinks=true,breaklinks=true,urlcolor=
red,linkcolor=blue,bookmarks=true,bookmarksopen=true}

\begin{document}

\title{Current-induced spin-orbit torques in ferromagnetic and antiferromagnetic systems}

\author{A. Manchon}
\email{aurelien.manchon@kaust.edu.sa}
\affiliation{King Abdullah University of Science and Technology (KAUST), Physical Science and Engineering Division (PSE), and Computer, Electrical, and Mathematical Science and Engineering (CEMSE), Thuwal, 23955-6900, Saudi Arabia}
\author{J. \v Zelezn\'y}
\affiliation{Institute of Physics, Academy of Sciences of the Czech Republic, 162 00 Praha, Czech Republic}
\author{I. M. Miron}
\affiliation{University of Grenoble Alpes, CNRS, CEA, INAC-SPINTEC, F-38000 Grenoble, France}
\author{T. Jungwirth}
\email{jungw@fzu.cz}
\affiliation{Institute of Physics, Academy of Sciences of the Czech Republic, 162 00 Praha, Czech Republic}
\affiliation{School of Physics and Astronomy, University of Nottingham, Nottingham NG7 2RD, United Kingdom}
\author{J. Sinova}
\affiliation{Institut f\"ur Physik, Johannes Gutenberg Universit\"at Mainz, 55128 Mainz Germany}
\affiliation{Institute of Physics, Academy of Sciences of the Czech Republic, 162 00 Praha, Czech Republic}
\author{A. Thiaville}
\email{andre.thiaville@u-psud.fr}
\affiliation{Laboratoire de Physique des Solides, Univ. Paris-Sud, CNRS UMR 8502 - 91405 Orsay Cedex, France}
\author{K. Garello}
\affiliation{IMEC, Kapeeldreef 75, 3001 Leuven, Belgium}
\author{P. Gambardella}
\email{pietro.gambardella@mat.ethz.ch}
\affiliation{Department of Materials, ETH Z\"urich, H\"onggerbergring 64, CH-8093 Z\"urich, Switzerland}

\date{\today}

\begin{abstract}
Spin-orbit coupling in inversion-asymmetric magnetic crystals and structures has emerged as a powerful tool to generate complex magnetic textures, interconvert charge and spin under applied current, and control magnetization dynamics. Current-induced spin-orbit torques mediate the transfer of angular momentum from the lattice to the spin system, leading to sustained magnetic oscillations or switching of ferromagnetic as well as antiferromagnetic structures. The manipulation of magnetic order, domain walls and skyrmions by spin-orbit torques provides evidence of the microscopic interactions between charge and spin in a variety of materials and opens novel strategies to design spintronic devices with potentially high impact in data storage, nonvolatile logic, and magnonic applications. This paper reviews recent progress in the field of spin-orbitronics, focusing on theoretical models, material properties, and experimental results obtained on bulk noncentrosymmetric conductors and multilayer heterostructures, including metals, semiconductors, and topological insulator systems. Relevant aspects for improving the understanding and optimizing the efficiency of nonequilibrium spin-orbit phenomena in future nanoscale devices are also discussed.
\end{abstract}

\maketitle
\tableofcontents

\section{Introduction\label{s:0}}

The interplay between spin and orbital degrees of freedom in condensed matter physics has been intensively studied for more than a century, starting from the seminal experiments of Barnett \cite{Barnett1915} and Einstein and de Haas \cite{Einstein1915}. At the time of these pioneering experiments on the transfer between magnetic and lattice angular momenta, the electron's spin was unknown and the phenomena could only be explained on the level of macroscopic angular momentum conservation principles. A microscopic insight into spin-orbit coupling emerged later from the relativistic quantum-mechanical Dirac equation. In magnetic materials, the relativistic spin-orbit coupling is now understood to play a fundamental role in a number of phenomena, including magnetocrystalline anisotropy, magnetization precession damping \cite{Stohr}, anomalous Hall effect \cite{Nagaosa2010}, anisotropic magnetoresistance \cite{McGuire1975}, and spin relaxation \cite{Dyakonov2008a}. In nonmagnetic semiconductors, the correlation between nonequilibrium charge and spin currents has been extensively studied since the 1970s \cite{Dyakonov1971,Ivchenko1978,Aronov1989}, allowing for the manipulation of spin states using both electrical and optical techniques \cite{Rashba1991,Ganichev2002,Kato2008}. In the past fifteen years, the interest in materials with strong spin-orbit coupling has substantially intensified. Heterostructures, surfaces and interfaces displaying unprecedentedly large spin-momentum locking have been recognized as powerful platforms for investigating the relativistic motion of electrons in condensed matter systems \cite{Hasan2010,Manchon2015} as well as the formation of chiral magnetic textures \cite{Nagaosa2013,Soumyanarayanan2016}. In this context, recent predictions \cite{Bernevig2005c,Tan2007,Obata2008,Manchon2008b,Manchon2009b,Garate2009,Zelezny2014} and observations of current-induced magnetization dynamics mediated by spin-orbit coupling in ferromagnets and antiferromagnets \cite{Ando2008b,Chernyshov2009,Miron2010,Miron2011,Liu2011,Wadley2016} have revolutionized the field of spintronics, leading to new opportunities to integrate electronic and magnetic functionalities in a wide variety of materials and devices.
 

 \onecolumngrid

 \begin{figure}[h!]
\centering
\includegraphics[width=12cm]{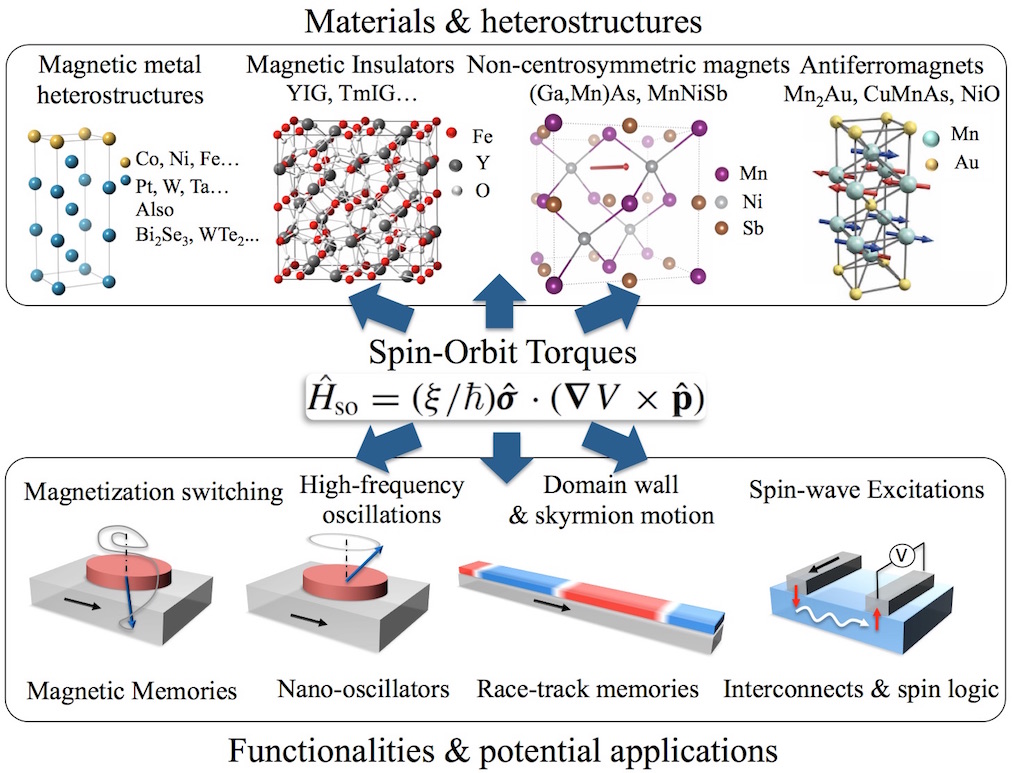}
\caption{\label{fig0}(Color Online) Materials in which spin-orbit torques have been observed range from metallic heterostructures involving transition metals, topological insulators and other heavy element substrates, to bulk non-centrosymmetric ferro- and antiferromagnets. Spin-orbit torque is a promising candidate mechanism to drive disruptive spintronics devices such as magnetic memories, nano-oscillators, race-track storage devices, as well as interconnects and spin logic gates.}
\end{figure}
 
 \twocolumngrid

Research in spintronics explores the possibilities to add the spin degree of freedom to conventional charge-based microelectronic devices or to completely replace charge with spin functionalities \cite{Wolf2001,Zutic2004}. Over the past three decades of research and development, spintronics has offered means to replace magnetic fields for reading and writing information in nanomagnets by more scalable current-induced spin-torques \cite{Chappert2007,Brataas2012}. Spin transfer torques (STT), which mediate the transfer of spin angular momentum between two magnetic layers having noncollinear magnetizations \cite{Slonczewski1996,Berger1996,Ralph2008}, are currently the method of choice for controlling the bit states in magnetic random access memories (MRAMs) \cite{Kent2015,Apalkov2016}. In STT, spin-orbit coupling already plays an important, but passive role: by inducing spin relaxation and magnetic damping, it enables the spin-polarization of the charge current passing through the reference layer and permits the magnetization switching. This review focuses on a new family of spin torques, whose physical origin is the transfer of orbital angular momentum from the lattice to the spin system. These torques rely on the conversion of electrical current to spin \cite{Gambardella2011,Sinova2015} and are called \emph{spin-orbit torques} (SOTs) in order to underline their direct link to the spin-orbit interaction.

Because of the ubiquity of spin-orbit coupling, SOTs provide efficient and versatile ways to control the magnetic state and dynamics in different classes of materials, as schematically shown in Fig. \ref{fig0}. Several microscopic mechanisms can give rise to SOT. In one picture, a charge current flowing parallel to an interface with broken inversion-symmetry generates a spin density due to spin-orbit coupling, which in turn exerts a torque on the magnetization of an adjacent magnetic layer via the exchange coupling \cite{Manchon2008b}. Several names have appeared in the literature for this model mechanism, such as the Rashba-Edelstein effect \cite{Edelstein1990} or the inverse spin galvanic effect (iSGE) \cite{Belkov2008}. In this review we will use the term iSGE-SOT. 

In the other model scenario, spin-orbit coupling generates a spin current in the nonmagnetic metal layer due to the spin Hall effect (SHE) \cite{Dyakonov1971b,Hirsch1999,Sinova2015}. The spin current propagates towards the interface, where it is absorbed in the form of a magnetization torque in the adjacent ferromagnet \cite{Ando2008b,Liu2011}. The SHE-SOT and iSGE-SOT can act in parallel. This is reminiscent of the early observations in semiconductors of the SHE and iSGE as companion phenomena, both allowing for electrical alignment of spins in the same structure \cite{Kato2004d,Kato2004b,Wunderlich2004,Wunderlich2005}.

Considering the SOT as originating from either the iSGE or SHE model scenarios can provide a useful physical and materials guidance. The necessary condition for the iSGE-induced non-equilibrium spin polarization is the broken inversion symmetry, which is automatically fulfilled in the above mentioned interfaces. However, also uniform crystals can have unit cells that lack a center of symmetry. The initial discovery of the iSGE-SOT was made in such a crystal, namely in the zinc-blende diluted magnetic semiconductor (Ga,Mn)As \cite{Chernyshov2009} and later also reported in asymmetric metal multilayers \cite{Miron2010}. This line of research was subsequently extended to crystals whose individual atomic positions in the unit cell are locally non-centrosymmetric, leading to the discovery of a staggered iSGE polarization and current-induced switching in an antiferromagnet \cite{Zelezny2014,Wadley2016}.
 
 \onecolumngrid

\begin{figure}[h!]
\centering
\includegraphics[width=16cm]{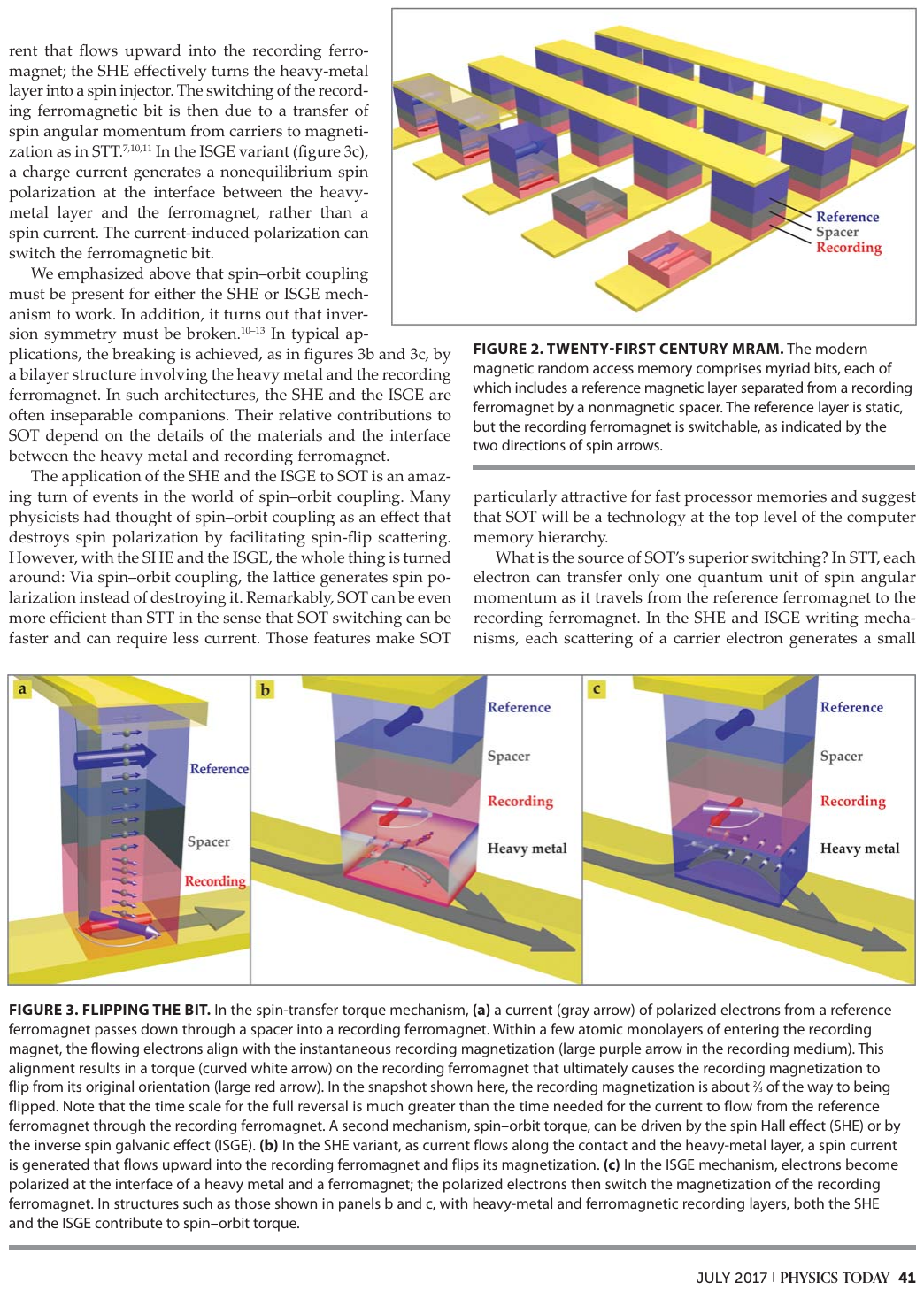}
\caption{\label{fig-SOT-STT}(Color Online) STT versus SOT switching of a magnetic tunnel junction. (a) In the STT case, a current of spin-polarized electrons (gray arrow) flows from a reference ferromagnet through a spacer layer into the recording ferromagnetic layer. Within a few atomic monolayers of entering the recording magnet, the flowing electrons align with the instantaneous magnetization due to the exchange interaction (large purple arrow in the recording medium). This alignment results in a torque (curved white arrow) on the recording ferromagnet that ultimately causes the magnetization to switch from its original orientation (large red arrow). In the snapshot shown here, the magnetization is about $2/3$ of the way to being
switched. Note that the time scale for the full reversal is much greater than the time needed for the current to flow from the reference ferromagnet through the recording ferromagnet. A second mechanism, SOT, can be driven by the SHE or by
the iSGE. (b) In the SHE variant, as current flows along the contact and the nonmagnetic metal layer, a spin current
is generated that flows upward into the recording ferromagnet and switches its magnetization. (c) In the iSGE mechanism, electrons become spin-polarized at the interface of a heavy metal and a ferromagnet; the polarized electrons then switch the magnetization of the recording ferromagnet. From \citet{Sinova2017}.}
\end{figure}

\twocolumngrid
 
The notion of the SHE-induced SOT, on the other hand, led to systematic studies correlating trends in the magnitude and sign of the SOT in ferromagnetic/nonmagnetic metal bilayers \cite{Ando2008b,Liu2012,Pai2012} with the magnitude and sign of the SHE in the nonmagnetic material calculated by ab-initio methods \cite{Tanaka2008,Freimuth2010} or measured by spin absorption in nonlocal spin valves \cite{Morota2011,Idzuchi2015}. However, in the commonly used bilayers with a nm-scale spin-diffusion length and nm-thick magnetic film, the distinction between SOTs generated by "bulk"-SHE or "interface"-iSGE remains principally blurred. Moreover, the experimentally observed complex SOT phenomenology in the bilayer structures is often not captured by either of the two idealized model scenarios \cite{Garello2013,Kim2013}. Other studies have pointed out further contributions to the SOT due to interface oxidation \cite{Miron2011b,Qiu2015,Demasius2016,An2018} and spin-dependent scattering of the spin-polarized current flowing in the ferromagnet \cite{Saidaoui2016,Amin2018}, which can add to the SHE-induced SOT.

Independently on their origin, SOT allow for new device architectures and efficient control of the magnetization. Figure~\ref{fig-SOT-STT} compares the out-of-plane current geometry employed in STT-MRAMs for both the write and read operations of a magnetic tunnel junction (MTJ), shown in (a), with the in-plane writing geometry enabled by SOT based on either iSGE (b) or SHE (c). SOT-induced magnetization switching, first demonstrated by \citet{Miron2011b} and \citet{Liu2012}, allows for decoupling the write and read current paths, with great advantages in terms of endurance of the junction and switching speed relative to STT \cite{Prenat2016,Fukami2017,Cubukcu2018}. Differently from STT, SOT allows also for the switching of magnetic insulators \cite{Avci2017} and antiferromagnets \cite{Wadley2016}, as well as for the generation of coherent spin waves \cite{Demidov2012,Collet2016} and interconversion of electric and magnon currents \cite{Kajiwara2010,Cornelissen2015,Goennenwein2015} in single-layer ferromagnets and ferrimagnets.

In this article we review the present theoretical understanding of the SOT in various types of material systems and summarize the experimentally established SOT phenomenology. We also discuss links of SOT to other currently highly active research fields, such as the topological phenomena in condensed matter, and outline foreseen technological applications of the SOT. A brief overview is given in Section~\ref{overview}. Readers interested in a more detailed discussion of theoretical and experimental aspects of SOT are referred to the subsequent sections.


\section{Overview}
\label{overview}

\subsection{Magnetization dynamics induced by spin-orbit torques}

\begin{figure}[h!]
\centering
\includegraphics[width=5cm]{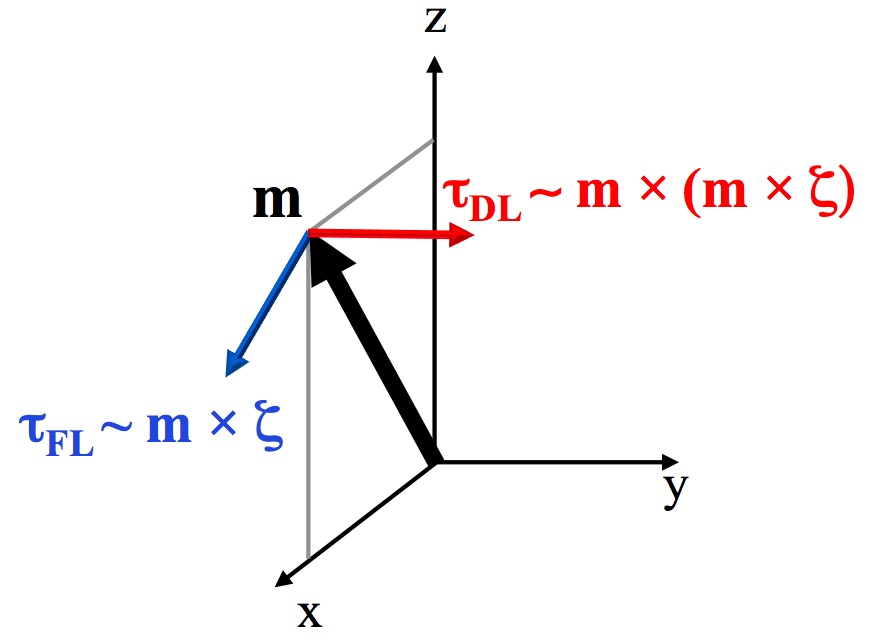}
\caption{\label{Fig3bis}(Color Online) Directions of the field-like and damping-like SOT components. In the present configuration and for clarity, we take ${\bm\zeta}={\bf y}$.}
\end{figure}

The dynamics of the recording layer subject to STT or SOT is governed by the Landau-Lifshitz-Gilbert (LLG) equation,
\begin{equation}
\label{eq:LLG}
\frac{d{\bf m}}{dt} =-\gamma {\bf m}\times {\bf B}_{\bf M}
+\alpha {\bf m} \times \frac{d{\bf m}}{dt} + \frac{\gamma}{M_{\rm s}}{\bf T},
\end{equation}
where $\gamma >0 $ is the (absolute value of the) gyromagnetic ratio (1.76$\times10^{11}$~s$^{-1}~$T$^{-1}$
for free electrons), $\alpha$ is the Gilbert damping parameter (dimensionless), $M_s$ is the saturation magnetization and ${\bf m}={\bf M}/M_s$ is the magnetization unit vector. The first term accounts for the precession of the magnetization ${\bf m}$ about the effective field, ${\bf B}_{\bf M}$, defined as the functional
derivative of the magnetic energy density $\mathcal{E}$, ${\bf B}_{\bf M}=- \delta \mathcal{E} / \delta {\bf M}$. The second term accounts for the relaxation of the magnetization towards its equilibrium position, and the third term represents the other torques ${\bf T}$ that may not derive from an energy density, notably the torques induced by
an electrical current. Such torques are by definition orthogonal to the magnetization ${\bf m}$ and adopt the most general form
\begin{equation}
\label{eq:torquedef}
{\bf T} =\tau_{\rm FL}{\bf m}\times{\bm\zeta}+\tau_{\rm DL}{\bf m}\times({\bf m}\times{\bm\zeta}).
\end{equation}
Here ${\bm\zeta}$ is a unit vector that depends on the microscopic mechanism at the origin of the torques, and the coefficients $\tau_{\rm FL, DL}$ may depend on the magnetization angle. In the STT configuration depicted on Fig. \ref{fig-SOT-STT}(a), ${\bm\zeta}$ is the polarization vector and is oriented along the magnetization direction of the reference layer. In the case of SOT, ${\bm \zeta}$ is determined by the charge-spin conversion process induced by spin-orbit coupling. In the literature, $\tau_{\rm DL}$ is usually referred to as the longitudinal (Slonczewski-like) component, which lies in the ({\bf m},\;${\bm\zeta}$) plane. In contrast, $\tau_{\rm FL}$ is normally referred to as the perpendicular (or transverse) component, which lies normal to the ({\bf m},\;${\bm\zeta}$) plane. The directions of these two torque components are represented in Fig. \ref{Fig3bis}. To understand the impact of these torques on the magnetization dynamics, it is instructive to remark that the perpendicular torque, $\tau_{\rm FL}$, acts on the magnetization like an effective magnetic field [first term in Eq. \eqref{eq:LLG}], while the longitudinal torque, $\tau_{\rm DL}$, acts like an effective magnetic damping [second term in Eq. \eqref{eq:LLG}, to the lowest order in $\alpha$]. Because of these similarities, these two torque components, $\tau_{\rm FL}$ and $\tau_{\rm DL}$, are also commonly called {\em field-like} and {\em damping-like} terms, respectively, a denomination we adopt in this review.

Initially, the damping-like component of the SOT was identified by measuring the damping of the ferromagnetic resonance (FMR) in nonmagnetic metal/ferromagnetic metal bilayers (NM/FM) \cite{Ando2008b}. A change of the damping factor was induced in the experiment by an in-plane dc current and interpreted as a consequence of the SHE-SOT. This was a new concept in which a dc electrical current driven through a conductor adjacent to the ferromagnet controls the FMR damping, in contrast to traditional means of controlling the FMR frequency by the dc current-induced Oersted field. Since the SHE in nonmagnetic metals was an emerging topic at the time of these pioneering SOT experiments, one of the key perceived merits of the SOT then was in providing an experimental measure of the spin Hall angle in the nonmagnetic material \cite{Ando2008b}. From our present perspective, however, we emphasize that such measurements have to be taken with great caution as the "spin Hall angles" inferred from these experiments are only effective parameters capturing, besides the bulk SHE, also the iSGE and other potential spin-orbit coupling and spin current contributions originating from the spin-orbit coupling from the interface \cite{Sinova2015,Lifshits2009,Saidaoui2016,Amin2016a,Amin2016b,Kim2017d,Miron2011b}.

While in experiments pioneered by \citet{Ando2008b} the FMR is generated externally and the SOT only modifies the dynamics, \citet{Liu2011} demonstrated that the SOT itself can drive the FMR when an alternating in-plane current is applied to the NM/FM bilayer. The method was again conceived to provide additional means to utilize ferromagnet dynamics for measuring the SHE in the adjacent nonmagnetic metal layer \cite{Liu2011}. A remarkable turn of events appeared, however. \citet{Miron2011b} and, subsequently, \citet{Liu2012} observed that SOTs can not only trigger small angle FMR precession but, for large enough electrical currents, it can fully and reversibly switch the ferromagnetic moments. The roles of the ferromagnetic and nonmagnetic metallic layers got reversed: In the original experiment by \citet{Ando2008b}, the ferromagnet provided the tool and the relativistic effects in the nonmagnetic metal layer were the object of interest. From now on, the new means to manipulate the magnetization took central stage.

Phenomenologically, SHE-SOT may appear as a mere counterpart of the STT \cite{Ando2008b,Liu2012}. At first sight, the spin current injected from the nonmagnetic metal layer due to the SHE just replaces the spin-injection from the reference to the recording ferromagnet in the STT stack (see Fig.~\ref{fig-SOT-STT}). However, the change in the writing electrical current geometry from out-of-plane in the STT to in-plane in the SOT has major consequences for the operation of memory devices as well as for the transport properties of layered structures. 

In the STT, each electron injected perpendicular to the plane of the heterostructure transfers one quantum unit of spin angular momentum as it travels from the reference to the recording ferromagnet. This transfer can be enhanced by using resonant tunneling \cite{Vedyayev2006,Theodonis2007}, which is difficult to realize experimentally. In the relativistic SOT utilizing no reference spin polarizer and where electrons are injected in the plane of the heterostructure, the spin angular momentum generated from the linear momentum in between collisions gives a little kick in every collision or acceleration that the electron feels, all along the plane. This configuration inherently enables the effective transfer of more than one spin unit per electron and allows for exerting spin torque on large sample areas. This fact has opened an entirely new space for material and device optimization of the switching process in SOT MRAMs. 


Present MRAM bit cells utilize the tunneling magnetoresistance (TMR) for readout \cite{Chappert2007}. The TMR effect is maximized when the recording and reference magnetizations switch between parallel and antiparallel configurations. For STT writing in such a device, however, the injected spin from the reference ferromagnet with a precisely aligned or anti-aligned orientation to the recording magnetization exerts no torque. This implies that the STT mechanism relies on thermal fluctuations of magnetization and the associated incubation time for initializing the magnetization dynamics slows down the switching process. A means to limit the incubation time is to engineer the polarizing and recording layers with orthogonal magnetizations. In this configuration, however, the polarizing layer cannot be used as a TMR sensor to probe the magnetic state of the recording layer and a third reference layer needs to be inserted in the device for this purpose \cite{Kent2004,Ye2014}.

In the SOT approach, the orientation of the current-induced spin polarization that exerts the torque in the recording ferromagnet is independent of the magnetization in the reference ferromagnet of the TMR stack, and can be engineered to be misaligned with the recording magnetization. Therefore, the in-plane writing current geometry can make the SOT more efficient and faster than STT \cite{Garello2014,Prenat2016,Fukami2017,Aradhya2016,Baumgartner2017}.


Sharing the read and write current paths in STT-MRAMs is also problematic \cite{Kent2015,Apalkov2016}. The distributions of read and write current values need to be well separated to avoid undesired writing while reading the memory. However, high writing currents go against energy efficiency. They also require thin tunnel barrier separating the recording and reference ferromagnetic layers, resulting in reliability issues due to barrier damage at high writing currents. Moreover, optimizing the tunnel barrier (and other components of the STT-MRAM stack) for writing can have detrimental effect on the magnitude of the readout TMR. In contrast, the three-terminal SOT-MRAM bit cell with separate write and read paths allows for optimizing separately these two basic memory functionalities and to remove the endurance issue by not exposing the tunnel barrier to the writing current. These advantages come at the expense of a larger area of the three-terminal SOT-MRAM cell (that is, a lower memory density) compared to the two-terminal STT-MRAM. Overall, SOT-MRAMs can find a broad utility and appear to be particularly well suited for the top of the memory hierarchy, namely for the embedded processor caches \cite{Prenat2016,Fukami2017,Hanyu2016}.

\subsection{Non-uniform magnetic textures\label{s:overdw}}

\begin{figure}[h!]
\centering
\includegraphics[width=8cm]{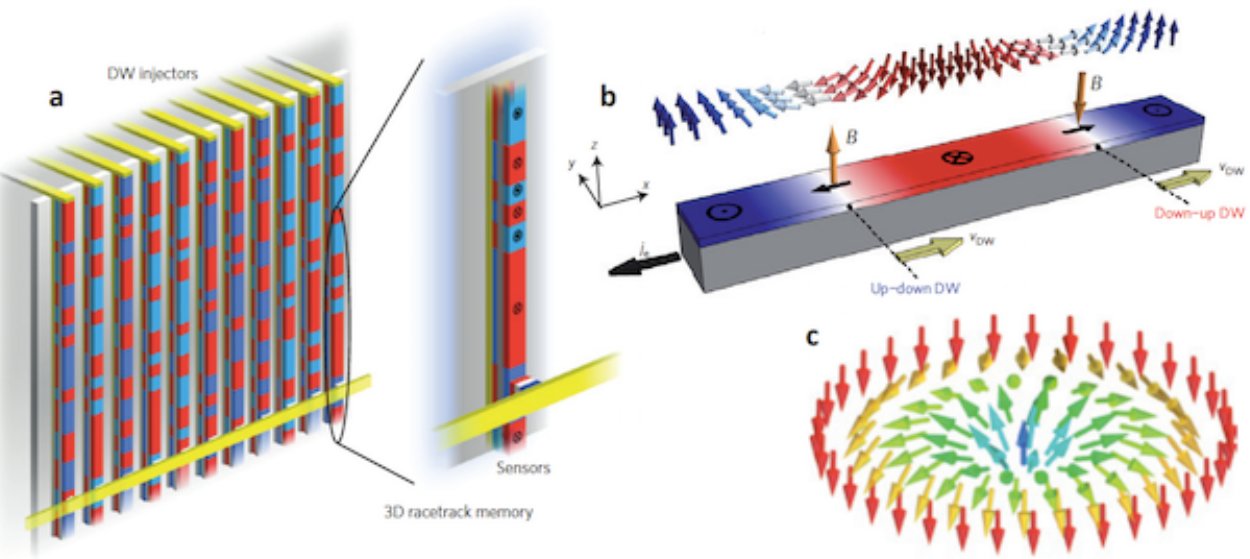}
\caption{\label{fig-DW_Skyrmion} (Color online) (a) Domain wall racetrack memory with red and blue regions representing areas that are oppositely magnetized. Adapted from \citet{Parkin2015}. Illustration of left-handed chiral N\'eel domain walls in a NM/FM bilayer. The effective field $B$ of the damping-like SOT moves adjacent up-down and down-up domains (with velocity $v_{\rm DW}$) in the same direction. Adapted from \citet{Emori2013}. (c) Skyrmions in a 2D ferromagnet with uniaxial magnetic anisotropy along the vertical axis. Magnetization is pointing down on the edges and pointing up in the center. Moving along a diameter, the magnetization rotates by $2\pi$ around an axis perpendicular to the diameter due to the DMI. Adapted from \citet{Fert2013}.}\end{figure}

Non-uniform magnetic textures are the basis of the racetrack memory concept illustrated in Fig.~\ref{fig-DW_Skyrmion} \cite{Parkin2008,Parkin2015}. A domain-wall racetrack memory consists of a series of alternating up and down magnetization domains that can be synchronously shifted along the corresponding multi-bit track and by this sequentially read by a single magnetoresistive sensor [see Fig.~\ref{fig-DW_Skyrmion}(a)].

An applied uniform (easy-axis) magnetic field cannot be used to operate the racetrack since it favors one of the two types of domains and thus pushes neighboring domain walls in opposite directions. Initially, this problem was resolved by replacing magnetic field with the STT that is induced by an in-plane current driven along the racetrack \cite{Parkin2008}. The physics is analogous to STT switching by a vertical current in an MRAM stack where the preferred magnetization direction is controlled by the direction of the applied spin current. In the racetrack, one direction of the applied electrical current moves electrons from, say, the up-domain to the down-domain at one domain wall, and from down-domain to the up-domain at the neighboring domain wall. As a result, the sense of the spin current is opposite at the two domain walls. It implies that, say, the up-domain is preferred at the first domain wall while the down domain is preferred at the second domain wall and the two domain walls then move in the same direction. 

At first sight, SOT in a racetrack fabricated from a NM/FM bilayer cannot be used to synchronously move multiple domain walls along the track. For example, a field-like SOT due to the iSGE would act as a uniform magnetic field. Also the damping-like SOT due to the SHE seems unfavorable as it is driven by a uniform vertical spin current. This makes the SHE-SOT fundamentally distinct in the domain wall racetrack geometry from the STT mechanism that exploits the repolarization of the in-plane spin current when carriers enter successive domains. 

Remarkably, theory and experiment have shown that the damping-like SOT can also move neighboring domain walls in the same direction, provided that the walls are of N\'eel type and have the same spin chirality \cite{Thiaville2012,Emori2013,Ryu2013}. Chiral N\'eel domain walls are stabilized by the Dzyaloshinskii-Moriya interaction (DMI) which relies on the interfacial spin-orbit coupling and broken inversion symmetry, similarly to the SOT. In this chiral case, the effective field driving the damping-like SHE-SOT in the domain wall is oriented along the easy axis in a direction that alternates from one domain wall to the next so that current drives them in the same direction [see Fig.~\ref{fig-DW_Skyrmion}(b)]. Moreover, in analogy to switching in MRAMs, the racetrack SOT can be more efficient than STT, resulting in higher current-induced domain wall velocities \cite{Miron2011,Yang2015a}.

In an alternative racetrack memory concept, the one-dimensional (1D) chiral domain walls are replaced with the skyrmion topological 2D chiral textures [see Fig.~\ref{fig-DW_Skyrmion}(c)] \cite{Fert2013}. While current-driven depinning can be achieved at substantially lower current density in skyrmion lattices \cite{Jonietz2010}, individual metastable skyrmions are expected to behave as point-like particles and are in principle less sensitive to the boundaries and pinning to boundary defects as compared to domain walls \cite{Sampaio2013}. Research is currently focusing on current-driven motion of individual skyrmions \cite{Woo2016,Jiang2017b,Legrand2017,Litzius2017}.

\subsection{Microscopic origin of spin-orbit torques}

We mentioned in the introduction that two main model mechanisms have been proposed to generate SOT. SHE originates from asymmetric spin deflection in the bulk of, e.g., a heavy metal induced by spin-orbit coupling. Such a deflection induces a pure spin current, transverse to the direction of the applied electrical current, that is subsequently absorbed in the adjacent magnetic layer, as depicted in Fig.~\ref{fig-SOT-STT}(b). The SHE-SOT model mechanism shares with the STT the basic concept of the angular momentum transfer from a carrier spin current to magnetization torque. As a consequence, the dominant component of the SHE-SOT in this picture is damping-like and takes the form \cite{Ando2008b}, 
\begin{equation}
{\bf T}=(j_{\rm s}^{\rm SHE}/t_{\rm F}){\bf m}\times({\bf m}\times{\boldsymbol\zeta}),\label{eq:sht}
\end{equation}
in units of eV/m$^3$. Here $j_{\rm s}^{\rm SHE}$ is the SHE spin current density absorbed by the recording magnet of thickness $t_{\rm F}$, and ${\boldsymbol\zeta}$ is a unit vector of the in-plane spin-polarization of the out-of-plane SHE spin current. The magnitude of the injected SHE spin current density into the magnet is modelled as $j_{\rm s}^{\rm SHE}=(\hbar/2e)\eta\theta_{\rm sh}\sigma_{\rm N} E$, where $\eta$ is the spin-injection efficiency across the NM/FM interface, also called transparency, $\theta_{\rm sh}=\sigma_{\rm sh}/\sigma_{\rm N}$ is the spin-Hall angle in the nonmagnetic material of spin-Hall conductivity $\sigma_{\rm sh}$ [expressed in units of $\Omega^{-1}$~m$^{-1}$] and electrical conductivity $\sigma_{\rm N}$, and ${\bf E}\perp{\boldsymbol\zeta}$ is the applied in-plane electric field. The SHE-SOT, being damping-like, directly competes with the damping term in the LLG equation of magnetization dynamics. This situation favors the current-induced switching of in-plane magnetized layers, as, for a damping-like torque, the critical current has to overcome the magnetic anisotropy barrier multiplied by the Gilbert damping factor $\alpha$, the latter being typically $\ll 1$ \cite{Ralph2008}.

Another common favorable feature of both SOTs and STT is that the switching condition is given by the applied current density and not the absolute current, which makes the mechanism scalable and, therefore, suitable for high-density memories. In contrast, for the traditional writing method by the current-induced Oersted magnetic field, the switching condition is determined by the absolute current. 

\begin{figure}[h!]
\centering
\includegraphics[width=8cm]{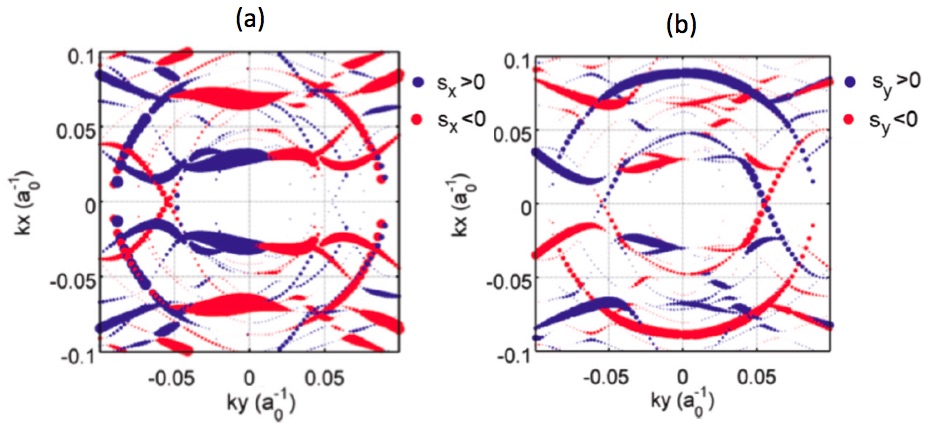}
\caption{\label{fig-Haney} (Color online) Spin texture in momentum space calculated for the interfacial Co layer of Pt(8ML)/Co(2ML) slab using density functional theory. ML stands for monolayer. Both in-plane spin components, (a) $S_x$ and (b) $S_y$, are {\em odd} in momentum space, enabling iSGE. Adapted from \citet{Haney2013a}.}
\end{figure}

The iSGE-SOT, as depicted in Fig.~\ref{fig-SOT-STT}(c), arises from spin-orbit coupling in non-centrosymmetric systems such as interfaces [see Fig. \ref{fig-SOT-STT}(c)], or zinc-blende (Ga,Mn)As crystals (see Subsection \ref{s:overaf}). In such systems, the band structure acquires a spin texture that is {\em odd} in momentum $k$. An example of such a spin texture is given in Fig. \ref{fig-Haney} for the prototypical case of the Pt/Co interface. This interfacial spin texture exhibits several features similar to the ideal case of Rashba spin-orbit coupling \cite{Manchon2015} and promotes iSGE. The iSGE-SOT resembles at first glance a mechanism in which the applied current generates a field rather than a damping-like torque. 

\begin{figure}[h!]
\centering
\includegraphics[width=8cm]{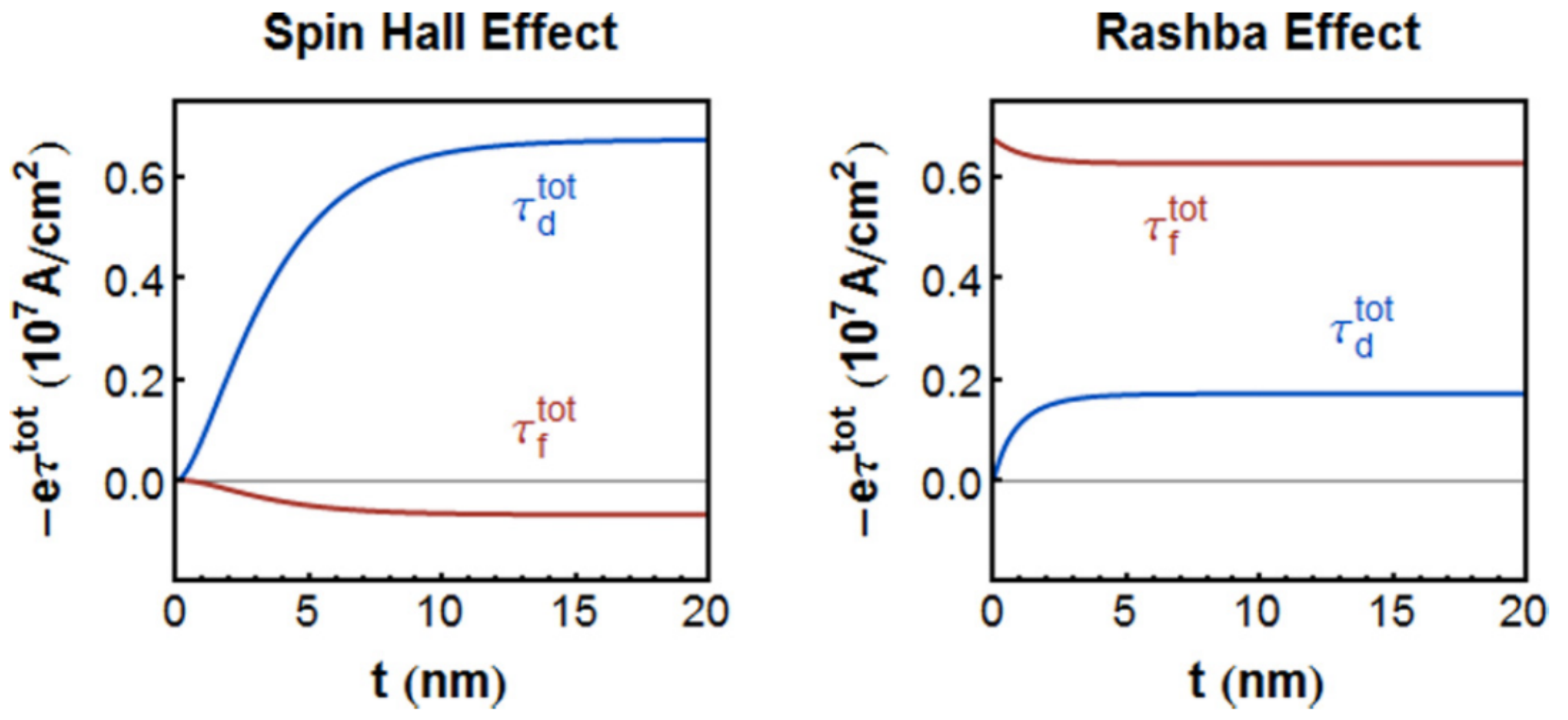}
\caption{\label{fig:she_bolt} (Color online) Torque components as a function of the nonmagnetic metal thickness, in the case of SHE (left) and iSGE (right). Both mechanisms produce field-like and damping-like components. From \citet{Amin2016b}.}
\end{figure}

In the iSGE mechanism in a NM/FM bilayer, the carrier spin density and the corresponding non-equilibrium effective magnetic field acting on the magnetization form directly at the inversion-asymmetric interface. The damping-like SHE-SOT, on the other hand, has been primarily viewed as a consequence of the spin current pumped from the bulk of the nonmagnetic material (which can be centrosymmetric) to the ferromagnet where it transfers its angular momentum to the magnetization. In the SHE, however, the spin current also yields a non-equilibrium spin density at the edges of the nonmagnetic material where the inversion-symmetry is broken. This implies an alternative picture of the SHE-SOT caused by the non-equilibrium spin density at the NM/FM interface. Correspondingly, the SHE can be also expected to contribute to the field-like SOT. {\em Vice versa}, as further discussed in Section \ref{s:6}, the iSGE mechanism can yield not only field-like but also damping-like SOT terms \cite{Miron2011b,Kurebayashi2014}. While the original iSGE models consider the effect of a uniform spin density on the magnetization dynamics, additional torques arise in models where the spin density generated at the interface is allowed to diffuse away from the interface \cite{Manchon2012,Haney2013b,Amin2016a,Amin2016b}. An example of numerical results is shown on Fig. \ref{fig:she_bolt}, where the torque magnitude is plotted against the nonmagnetic metal thickness in the case of pure SHE and pure iSGE \cite{Amin2016b}. For the reasons mentioned above, the decomposition into ${\bf T}_{\rm FL}$ and ${\bf T}_{\rm DL}$ does not allow to disentangle the microscopic iSGE and SHE mechanisms of the SOT. Moreover, the factors $\tau_{\rm FL}$ and $\tau_{\rm DL}$ can depend on the angle of ${\bf{m}}$ \cite{Garello2013}. This makes not only the microscopic analysis but also the phenomenological LLG description of the SOT more complex.

In general, SOT can be directly linked to the applied electric field ${\bf E}$ by a linear-response expression, ${\bf T}=\chi_T{\bf E}$ \cite{Freimuth2014a}. Alternatively, SOT can be written as ${\bf T}={\bf M}\times{\bf B}_{\rm T}$, where ${\bf B}_{\rm T}\approx-\Delta{\bf S}/M_{\rm s}$ is an effective current-induced spin-orbit field, $\Delta$ is the exchange coupling between carrier spins and magnetic moments, and ${\bf S}=\chi_S{\bf E}$ is the current-induced carrier spin density expressed again in the linear response. The different torque terms in the LLG equation are obtained from the expansion, $\chi_{S,ij}=\chi_{S,ij}^{(0)}+\chi_{S,ij,k}^{(1)}m_k+\chi_{S,ij,kl}^{(2)}m_km_l+\cdot\cdot\cdot$, where $m_i$ are the components of the magnetization unit vector. Here the response function coefficients for each order in ${\bf m}$ are independent of ${\bf m}$ and their matrix form reflects the underlying crystal symmetry of the considered material or structure \cite{Hals2013,Hals2013b,Wimmer2016,Zelezny2017}. For example, the field-like SOT corresponds to the zeroth order term while the damping-like SOT term appears in the first order of the expansion of $\chi_S$. 

Note that an analogous expansion can be written for $\chi_T$ and that the approaches using $\chi_T$ or $\chi_S$ expansion are in principle equivalent. Using $\chi_S$ appeals to the two-step physical picture of the SOT in which, first, the applied current polarizes the carriers (which can also appear in nonmagnetic metals) and, second, the non-equilibrium carrier spins generate the torque on magnetic moments via exchange coupling $\Delta$. When considering $\chi_T$, the physical intuition based on SHE, iSGE or other non-equilibrium spin density phenomena may be less apparent but the experimentally measured quantity, which is the SOT, is accessed directly. In microscopic theories, SOT has been calculated either from $\chi_S$ or $\chi_T$. In the former case, one obtains a spin-density averaged over the unit cell which is then multiplied by an averaged exchange field to obtain the net torque. In the latter case, the cross product of the spin density and exchange field is calculated locally and then averaged over the unit cell to get the torque. Hence, using $\chi_T$ always represents the more rigorous approach.

In the Kubo linear response formalism, the microscopic expression for $\chi_S$ (or $\chi_T$) can be split into the intra-band contribution (Boltzmann theory) and the inter-band term \cite{Garate2009}. The former one scales with conductivity, i.e., diverges in the absence of disorder, and contributes to the field-like SOT \cite{Manchon2008b}. The latter one is finite in the disorder-free intrinsic limit \cite{Freimuth2014,Kurebayashi2014} and contributes to the damping-like SOT. As a result, the field-like SOT tends to dominate the damping-like SOT in clean systems while the trend reverses in more disordered structures \cite{Freimuth2014a,Li2015b}. This is an example of basic guidelines that theory can provide when analyzing SOT experiments. We emphasize, however, that other terms beyond the lowest order field-like and damping-like torques can also significantly contribute to the total SOT, as seen in experiments \cite{Garello2013,Fan2014a}.

Finally, we note that unlike the rigorous and systematic methods based on the response functions $\chi_S$ or $\chi_T$, considering the SHE spin current as an intermediate step between the applied electrical current and the resulting SOT is an intuitive but not rigorous approach. This is because other mechanisms beyond the bulk-like SHE can contribute, and because in spin-orbit coupled systems the spin current is not uniquely defined, in contrast to the well-defined and directly measurable spin density or torque. 
As a result, the "Hall angle" $\theta_{\rm sh}$ inferred from Eq. \eqref{eq:sht}, relating the measured torque to a hypothetical SHE spin current, should not be understood in the original sense of the term "Hall angle" but rather as an effective experimental parameter providing a simple, and therefore rather vague, characterization of the charge-to-spin conversion efficiency in a given structure. For similar reasons, the spin current approach has not been applied for the systematic crystal and magnetization symmetry analysis of the series of SOT terms identified in experiment. From now on, to avoid unnecessary confusion we use $\xi$ to designate the charge-to-spin conversion efficiency (see Section~\ref{MML:phen}) and $\theta_{\rm sh}$ in the specific context of SHE. 

\subsection{Spin-orbit torques in antiferromagnets\label{s:overaf}}

For antiferromagnets, the STT or SOT phenomenology is modified by considering a current-induced spin density at a particular atomic site that tends to produce a torque which acts locally on the magnetic moment centered on that site \cite{MacDonald2011,Gomonay2014,Zelezny2014,Jungwirth2016}. In analogy to ferromagnets, the local torques acting on the $a$-th antiferromagnetic sublattice magnetization, ${\bf M}_a$, have a field-like component of the form ${\bf T}_{a}={\bf M}_{a}\times{\bf B}_{a}$, with ${\bf B}_{a}\sim{\boldsymbol\zeta}_a$, and a damping-like component ${\bf T}_{a}={\bf M}_{a}\times{\bf B}'_{a}$, with ${\bf B}'_{a}\sim{\bf M}_{a}\times{\boldsymbol\zeta}_a$, respectively. Note that in a rigorous systematic theory, these and all other torque terms acting in an antiferromagnet can be again obtained from the linear response expressions in which the coefficients of the magnetization-expansion of $\chi_{T,a}$ (or $\chi_{S,a}$) reflect local crystal symmetries of the $a$-th antiferromagnetic sublattice \cite{Zelezny2017}. Assuming a collinear antiferromagnet, two model scenarios can be considered for the field-like and damping-like SOTs: One with ${\boldsymbol\zeta}_1={\boldsymbol\zeta}_2={\boldsymbol\zeta}$ and the other one with ${\boldsymbol\zeta}_1=-{\boldsymbol\zeta}_2$. 

The former case corresponds, e.g., to injection of uniformly polarized carriers from an external reference ferromagnet, from a nonmagnetic SHE material, or to the generation of a uniform spin density at a nonmagnetic metal/antiferromagnetic metal (NM/AF) interface by iSGE [see e.g. \cite{Manchon2017}]. The field-like torque in the antiferromagnet would then be driven by a uniform non-staggered effective field ${\bf B}_{1}={\bf B}_{2}\sim{\boldsymbol\zeta}$, i.e., would be equally inefficient in switching an antiferromagnet as a uniform external magnetic field acting on an antiferromagnet. On the other hand, the local non-equilibrium effective field, ${\bf B}'_{a}\sim{\bf M}_{a}\times{\boldsymbol\zeta}$, driving the damping-like torque has an opposite sign on the two spin sublattices since ${\bf M}_1=-{\bf M}_2$. This staggered effective field cants the magnetizations of the two sublattices and triggers the dynamics of the antiferromagnetic order resulting in current-driven switching and excitations, somewhat similar to what is obtained in ferromagnets subject to damping-like torque \cite{Gomonay2010,Cheng2016,Khymyn2017}.

\begin{figure}[h!]
\centering
\includegraphics[width=8cm]{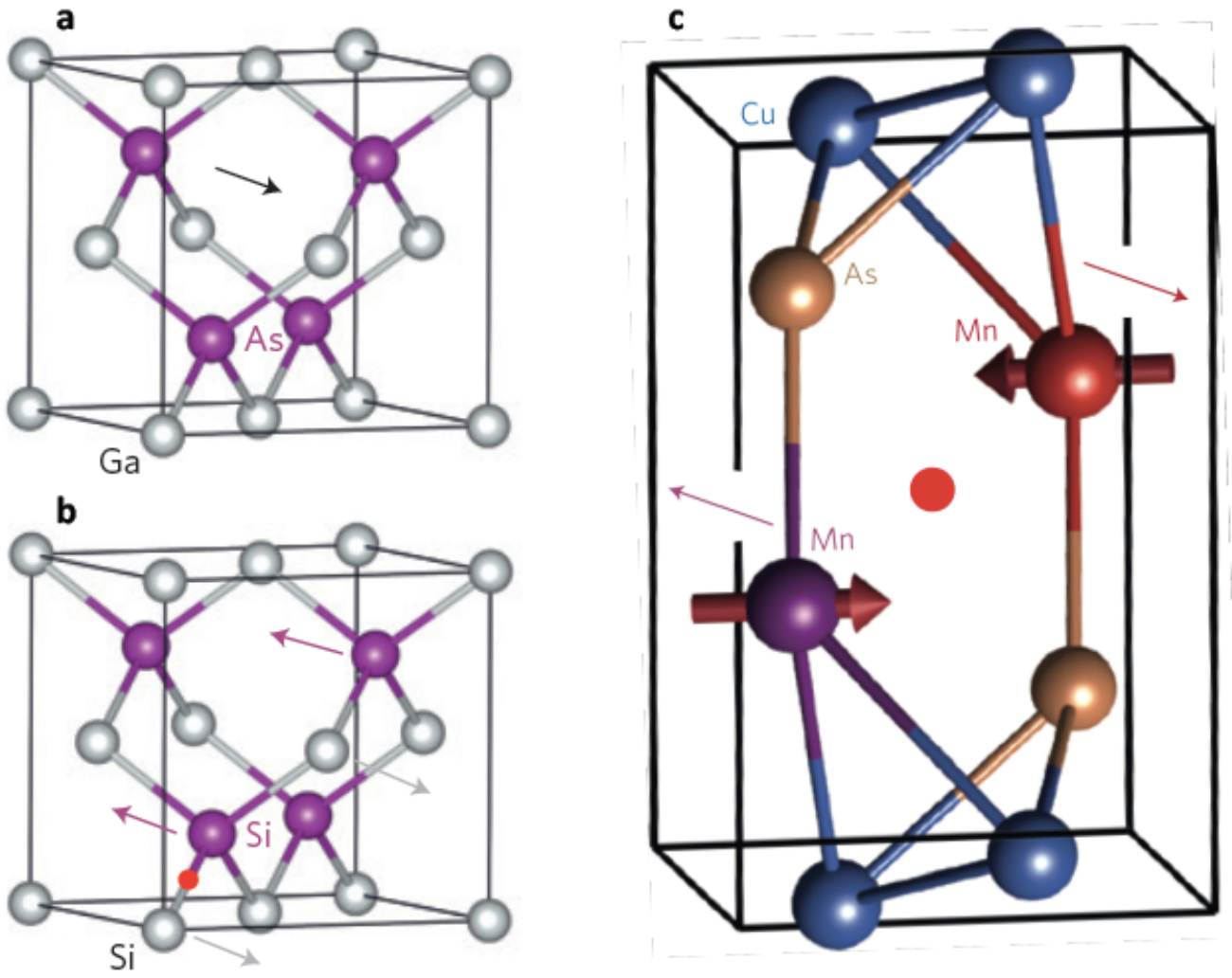}
\caption{\label{fig-GaAs_Si_CuMnAs} (Color online) (a) Global uniform non-equilibrium spin density generated by electrical current in a nonmagnetic lattice with global inversion-asymmetry (e.g. GaAs) due to the iSGE. (b) Local staggered, antiferromagnetic-like non-equilibrium spin density in a nonmagnetic lattice with local inversion-asymmetry (e.g. Si) due to the iSGE. Red dot shows the inversion-symmetry center of the Si lattice. The two Si atoms on either side of the center occupy inversion-partner lattice sites with locally asymmetric environments. In GaAs lattice, the inversion-symmetry center is absent since the two inversion-partner sites in the unit cell are occupied by different atoms. (c) Local staggered non-equilibrium spin density inducing a local staggered effective field in an antiferromagnetic lattice with local inversion-asymmetry (e.g. CuMnAs). Thin arrows represent the current-induced staggered effective field and thick arrows the antiferromagnetic moments. Adapted from \citet{Jungwirth2016}.}\end{figure}

The microscopic realization of the second scenario in which ${\boldsymbol\zeta}_1=-{\boldsymbol\zeta}_2$ is illustrated in Fig.~\ref{fig-GaAs_Si_CuMnAs} \cite{Zelezny2014,Ciccarelli2016,Jungwirth2016}. It is the staggered counterpart of the uniform iSGE spin density discussed above. As mentioned in the previous section, iSGE only exists in non-centrosymmetric systems. For instance, the unit cell of zinc-blende GaAs [Fig.~\ref{fig-GaAs_Si_CuMnAs}(a)] lacks a center of inversion, enabling an electrical current to induce a non-equilibrium uniform spin density in the bulk crystal. In contrast, the related diamond lattice of, e.g., Si [Fig.~\ref{fig-GaAs_Si_CuMnAs}(b)] has global inversion symmetry and therefore cannot promote a net iSGE spin-density when integrated over the unit cell. However, the two identical atoms in the unit cell sitting on the inversion partner sites have locally non-centrosymmetric environments. As a result, the diamond lattice is an example where the iSGE can generate {\em local} non-equilibrium spin density with opposite sign and equal magnitude on the two inversion-partner atoms while the global spin density integrated over the whole unit cell vanishes. Here a uniform electrical current induces a non-equilibrium {\em staggered} spin density in the bulk crystal. 

In Si there is no equilibrium antiferromagnetic order that could be manipulated by these local staggered non-equilibrium spin densities. However, antiferromagnets like CuMnAs shown in Fig.~\ref{fig-GaAs_Si_CuMnAs}(c), share the crystal symmetry allowing for the current-induced staggered spin density whose sign alternates between the inversion-partner atoms. Moreover, one inversion-partner lattice site is occupied by the magnetic atom belonging to the first antiferromagnetic spin sublattice and the other inversion partner is occupied by the magnetic atom belonging to the second spin sublattice. As a result, the corresponding field-like N\'eel SOT can reorient antiferromagnetic moments with an efficiency similar to the reorientation of ferromagnetic moments by an applied uniform field. This scenario has been confirmed experimentally in CuMnAs and Mn$_2$Au memory devices \cite{Wadley2016,Bodnar2018,Meinert2018,Zhou2018b}.

\subsection{Spin-orbit torques in topological materials}


The distribution of spin texture in momentum space is a crucial ingredient to understand SOT. In semiconductor materials where the iSGE and SHE were initially discovered, and even more in metal structures, multiple bands cross the Fermi level and their respective contributions to the current-induced spin density tend to compensate each other. Also the spin-textures are more complex, which can further reduce the net effect.

\begin{figure}[h!]
\centering
\includegraphics[width=8cm]{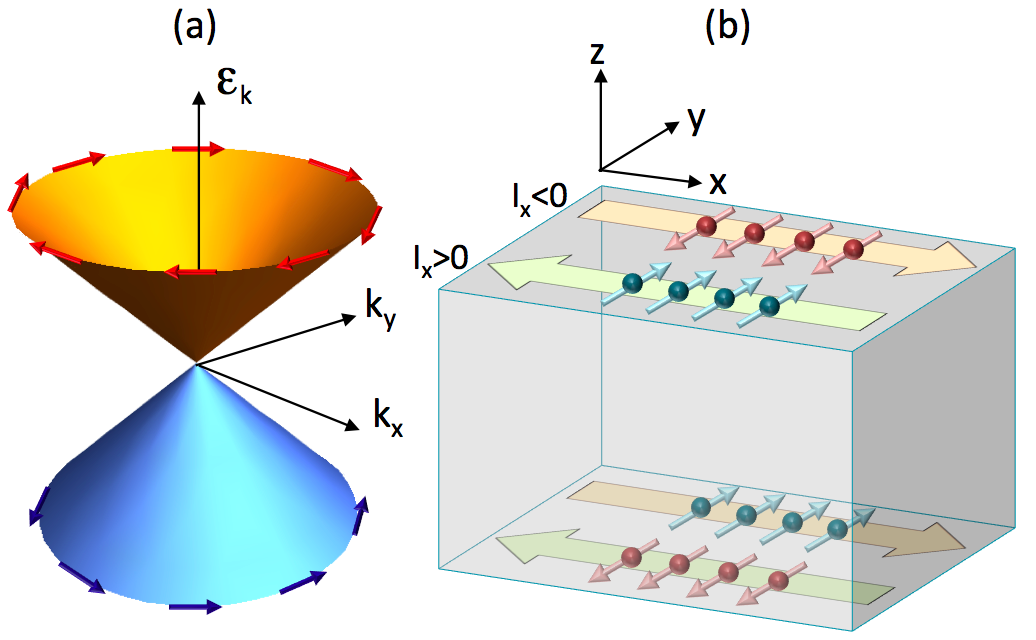}
\caption{\label{fig-TI}(Color online) Charge current-induced surface spin density in topological insulator. (a) Schematic illustration of the spin-momentum locked helical spin texture of the surface states in topological insulator: clockwise spin texture above the Dirac point while anticlockwise spin texture below the Dirac point. (b) Schematic of surface spin density on two opposite surfaces for a charge current flowing along $-x$ direction (i.e., $I_x < 0$, yellow thick arrow) and for a charge current flowing along $+x$ direction (i.e., $I_x > 0$, green thick arrow).}
\end{figure}

From this perspective, topological insulators \cite{Hasan2010,Pesin2012b} are regarded as optimal materials for the SOT. The surface states of a three-dimensional (3D) topological insulator form a Dirac cone with a single Fermi surface and a helical locking of the relative orientations of the spin and the momentum [see Fig.~\ref{fig-TI}(a)]. Indeed, SOT-FMR measurements in a metallic ferromagnet interfaced with a topological insulator showed an exceptionally large spin conversion efficiency $\xi$ \cite{Mellnik2014}. However, compared to common nonmagnetic metals, the increase of $\xi$ in the studied topological insulators turned out to be primarily due to its decreased electrical conductivity while the inferred effective spin-Hall conductivity was similar to the nonmagnetic metals (see Table \ref{tableSOT}). 

Interfacing a highly resistive topological insulator with a low resistive metal FM has also a practical disadvantage that most of the applied electrical current is shunted through the metallic magnet and does not contribute to the generation of the spin density at the topological insulator surface. A possible remedy is in using an insulating magnet. An example is a study of highly efficient magnetization switching at cryogenic temperatures in a topological insulator/magnetic topological insulator bilayer, in which the inferred spin conversion efficiency $\xi$ was three orders of magnitude larger than in nonmagnetic metals \cite{Fan2014a,Fan2016b}.

In the above studies, Dirac quasiparticles exhibiting strong spin-momentum locking are considered to enhance the efficiency of the SOT control of magnetic moments. {\em Vice versa}, a scheme has been recently proposed for the electric control of Dirac band crossings by reorienting magnetic moments via SOT \cite{Smejkal2017}. Instead of 2D surface states of a topological insulator, these predictions consider Dirac bands in the bulk of a topological 3D semimetal. Since Dirac bands can only exist in systems with a combined space-inversion and time-inversion (${\cal PT}$) symmetry, ferromagnets are excluded. On the other hand, the combined ${\cal PT}$-symmetry in an antiferromagnet is equivalent to a magnetic crystal symmetry in which antiferromagnetic spin sublattices occupy inversion-partner lattice sites. This in turn allows for an efficient SOT, as discussed in the previous section. 

\subsection{Inverse effect of the spin-orbit torque}

\begin{figure}[h!]
\centering
\includegraphics[width=8cm]{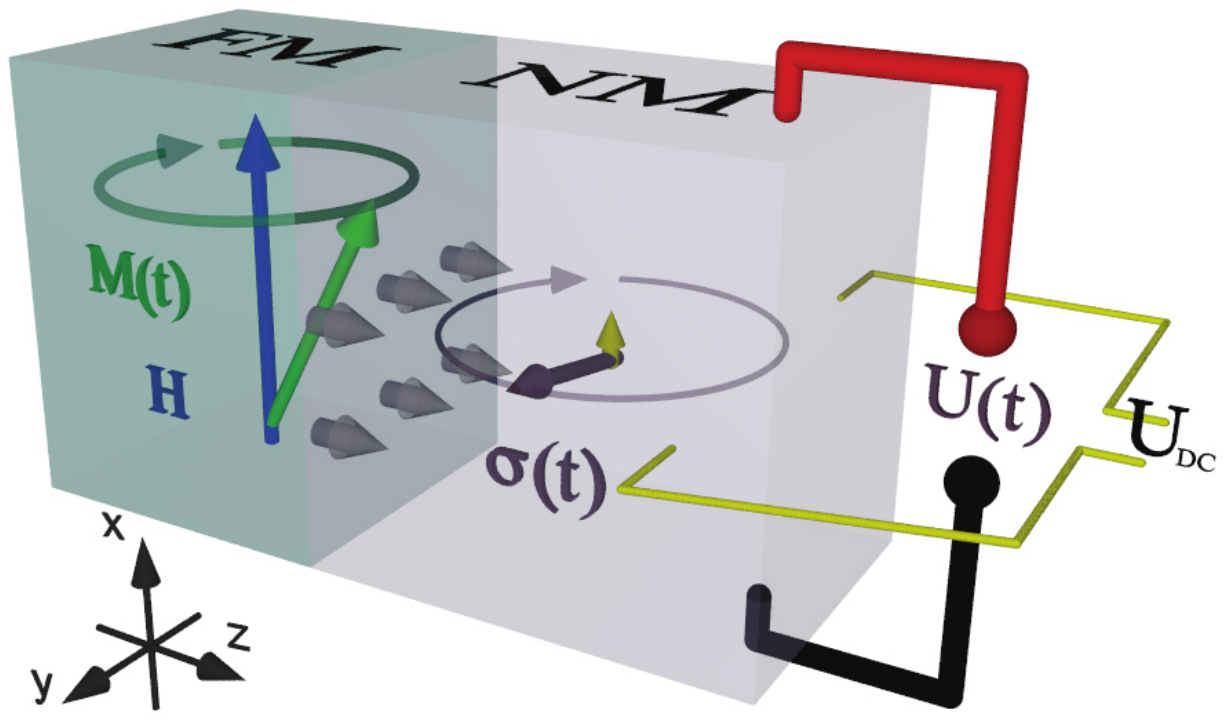}
\caption{\label{fig-inv-SOT}(Color online) A spin current is generated by spin-pumping at the NM/FM interface (grey arrows). The time dependent spin density $\sigma(t)$ of this current (indicated as a dark grey arrow) rotates almost entirely in the $y-z$ plane. The small time averaged dc component (yellow arrow) appears along the $x$ axis. Both components lead to charge currents in the nonmagnetic metal and can be converted into ac and dc voltages, $U(t)$ and $U_{\rm DC}$, by placing probes along the $x$ and $y$ direction, respectively. From \citet{Wei2014}.}
\end{figure}

The Onsager reciprocity relations imply that there is an inverse phenomenon to the SOT, which we call the spin-orbit charge pumping \cite{Hals2010,Kim2012,Tolle2017}. The underlying physics of the spin-orbit charge pumping generated from magnetization dynamics is the direct conversion of magnons into charge currents via spin-orbit coupling, as illustrated on Fig. \ref{fig-inv-SOT}. This effect evolves from the spin pumping predicted by \citet{Tserkovnyak2002b,Brataas2002} when SOC is included, either in the bulk of the nonmagnetic metal or at the interface. Thus, any external force that drives magnetization precession can generate spin-orbit charge pumping. Similarly to the SOT, two model microscopic mechanisms can be considered for the spin-orbit charge pumping: one due to the inverse effect of the iSGE \cite{Rojas-Sanchez2013b,Ciccarelli2014}, called the spin galvanic effect (SGE), and the other one due to the inverse SHE \cite{Saitoh2006}. Together with the non-local detection in a lateral structure \cite{Valenzuela2006}, the spin-orbit charge pumping across the NM/FM interface provided the first experimental demonstration of the inverse SHE \cite{Saitoh2006}. Since then it has evolved into one of the most common tools for electrical detection of magnetization dynamics.

\section{Theory of spin-orbit torques\label{s:6}}

In this section we review the progress that has been made towards the theoretical understanding of SOTs in both layered heterostructures and bulk materials. The most general treatment of the SOT that has been considered so far is based on the (spin-)density functional theory, in which the system is described by a Hamiltonian for non-interacting particles

\begin{align}
 \hat{H} = \hat{\cal K} + \hat V_\text{eff}(\mathbf{r}) + \hat{\bm\sigma} \cdot {\boldsymbol\Omega}_\text{xc}(\mathbf{r}) + \hat H_\text{so},\label{eq:KS_hamiltonian}
\end{align}
where $\hat{\cal K}$ is the kinetic energy, $\hat V_\text{eff}$ is the effective crystal potential, ${\boldsymbol\Omega}_\text{xc}$ is the exchange-correlation field, and $\hat H_\text{so}$ is the spin-orbit coupling. Assuming this form of the Hamiltonian, the torque on magnetization at point $\mathbf{r}$ is given by \cite{Haney2008,Manchon2011c}
\begin{align}
 \mathbf{T}(\mathbf{r}) = - \mathbf{S}(\mathbf{r}) \times {\boldsymbol\Omega}_\text{xc}(\mathbf{r}),\label{eq:torque}
\end{align}
where $\mathbf{S}=(1/V)\langle\hat{\bm\sigma}\rangle$ is the current-induced spin density, and $V$ is the volume of the unit cell. This equation is valid for the STT as well as for the SOT. When spin-orbit coupling is neglected, the torque can be equivalently expressed as a divergence of a spin current \cite{Ralph2008}
\begin{align}
 T_i(\mathbf{r}) = - {\bm\nabla} \cdot \bm{\mathcal J}_{\rm s}^i,\label{eq:torque_spincurrent}
\end{align}
where $\bm{\mathcal J}_{\rm s}^i=(\hbar/4)\langle\{\hat{\sigma}_i,\hat{\bf v}\}\rangle$ is a vector representing the $i$-th spin component of the spin current tensor. The $j$-th component of the vector $\bm{\mathcal J}_{\rm s}^i$ is noted $j_{{\rm s},i}^j$ and denotes a spin current polarized along the $i$-th direction and propagating along the $j$-th direction. In the absence of spin-orbit coupling, the torque is thus directly given by the absorption of the spin current. However, when spin-orbit coupling is not neglected, the spin angular momentum is not a conserved quantity and the spin current in Eq. \eqref{eq:torque_spincurrent} is then not uniquely defined, while Eq. \eqref{eq:torque} remains valid.

The total torque is obtained by integrating Eq. \eqref{eq:torque} over the whole unit cell, and a local torque is obtained by integrating over a particular magnetic atom. This torque can then be inserted into LLG equation to evaluate the magnetic dynamics induced by the SOT. When using this approach it is necessary to ensure that the dynamics of the non-equilibrium carrier spins is much faster than the dynamics of magnetic moments arising from equilibrium electrons; otherwise the dynamics of the two could not be separated. This is well justified in ferromagnets whose magnetization dynamics lies in the GHz range but it could become an issue when discussing antiferromagnets whose dynamics can reach several THz. We also note that Eq. \eqref{eq:torque} assumes that Eq. \eqref{eq:KS_hamiltonian} accurately describes the electronic system. This is reasonable for most materials of interest, namely metals, but fails in strongly correlated systems. In these systems, more sophisticated many-body approaches are necessary. So far SOTs have been studied only using non-interacting model (free electron or {\em k.p}) Hamiltonians or Kohn-Sham Hamiltonians originating from density functional theory.

At weak applied electric fields, the SOT is well described by linear response theory, ${\bf T}=\chi_T{\bf E}$, where the response tensor $\chi_T$ can be calculated using Eq. \eqref{eq:torque}. Equivalently, the torque can be rewritten as, ${\bf T}={\bf M}\times{\bf B}_{\rm T}$, with the effective field obtained from the linear response expression, ${\bf B}_{\rm T}=\chi_B {\bf E}$. In many calculations of the SOT, especially those based on model Hamiltonians, an approximation is used in which the effective magnetic field is made directly proportional to the current-induced spin density, ${\bf B}_{\rm T}\approx-\Delta{\bf S}/M_{\rm s}$. Here $\Delta$ is an exchange coupling energy corresponding to exchange between the carrier spins and magnetic moments, and ${\bf S}$ is again evaluated using linear response, ${\bf S}=\chi_S{\bf E}$. 

\begin{figure}[h!]
\centering
\includegraphics[width=8cm]{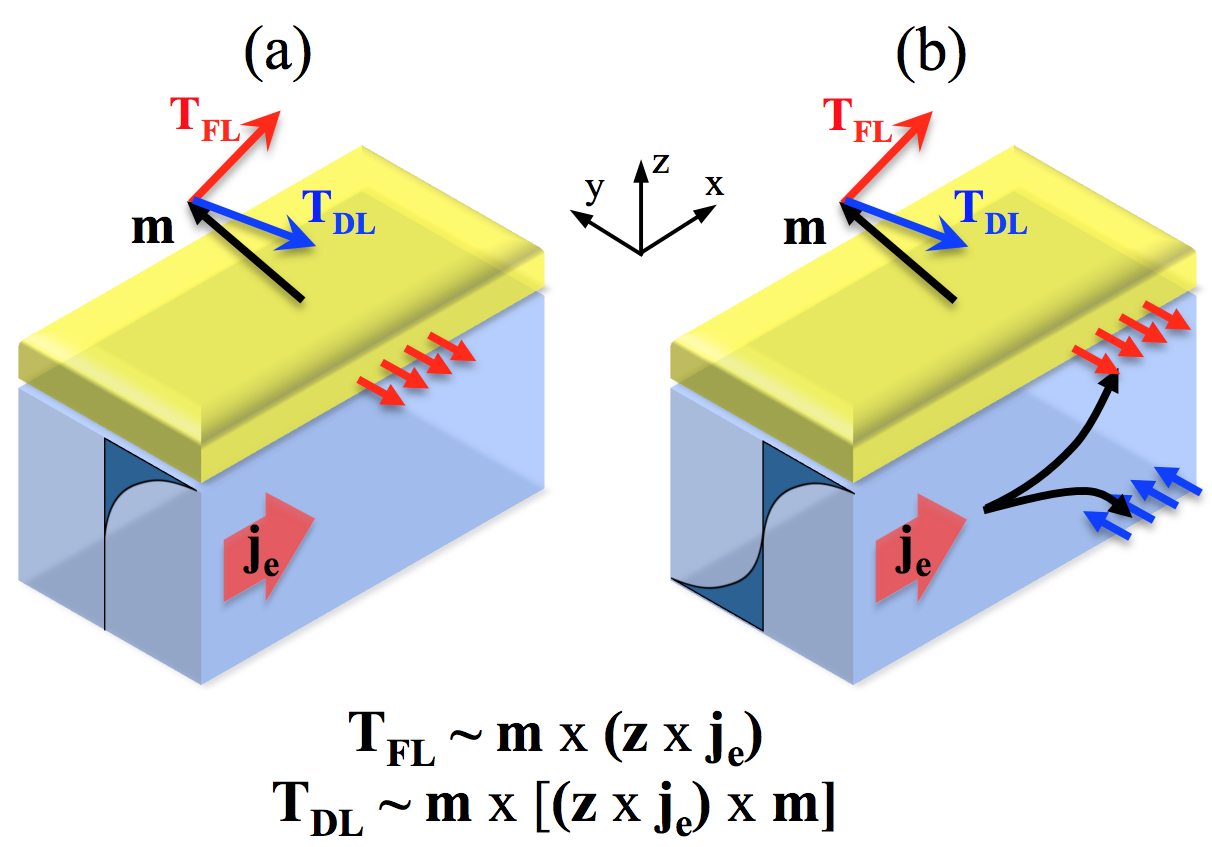}
\caption{\label{fig:Torques} (Color online) Two main model spin-charge conversion mechanisms at NM/FM interface: (a) iSGE and (b) SHE. Both mechanisms produce damping-like and field-like torques. The small red and blue arrows denote the non-equilibrium spin density accumulating at the interfaces, and their corresponding spatial distribution is sketched as a shaded area on the structure's side. The large red and blue arrows represent the field-like and damping-like torques, respectively.}
\end{figure}

As discussed in Section \ref{overview}, the origin of the SOT in the bilayer systems is often attributed to two different effects, the SHE and the iSGE, where the SHE-SOT is assumed to originate from the absorption of a spin current generated in the nonmagnetic metal [see Fig. \ref{fig:Torques}(a)] and the iSGE-SOT is due to spin density generated locally in the ferromagnet or at the interface [see Fig. \ref{fig:Torques}(b)]. Equation \eqref{eq:torque} shows however that the torque always originates from a current-induced spin density. Thus the SHE-SOT can be more fundamentally understood not in terms of the absorption of a spin current but in terms of a spin density induced by the spin Hall current. Consequently, both contributions can be treated on the same footing and there is no clear way how to theoretically separate them.

Still, it is intuitively appealing to separate the total torque into a contribution associated with the absorption of a spin current, as given by Eq. \eqref{eq:torque_spincurrent}, and a contribution due to a locally generated spin density, described by Eq. \eqref{eq:torque}. One could then attribute the former contribution to SHE and the latter one to iSGE. However, such an approach has several drawbacks. First, spin currents are not necessarily due to the bulk SHE alone and substantial contributions can also come from the interface with the ferromagnet \cite{Wang2016b,Amin2016a,Amin2016b,Kim2017d,Ghosh2018}. Second, even in bulk non-centrosymmetric materials where SOT is considered of purely iSGE origin, local spin currents within the unit cell can contribute to the torque. Third, in a slab geometry, interface and bulk are not well defined notions, and the terminology of what should be referred to as iSGE or SHE becomes unclear \cite{Freimuth2014a}. Conventionally, iSGE refers to spin density generated internally in the material. However, even the spin density induced by SHE in the bilayers could be referred to as iSGE, since it is also a spin density induced by a charge current. In conclusion, although models based on bulk SHE or iSGE due to interfacial Rashba spin-orbit coupling can be useful to explain some aspects of the experiments, in real systems there is not much point in trying to rigorously parse the torque into these two contributions.

In many experimental studies, the origin of the SOT is analyzed in terms of its symmetries. The damping-like torque ${\bf T}_{\rm DL}$ is often referred to as \emph{spin Hall} or \emph{Slonczewski torque}, and the field-like torque ${\bf T}_{\rm FL}$ as \emph{spin-orbit} or \emph{Rashba torque}. This is based primarily on the assumption that any torques associated with transfer and absorption of spin-angular momentum would have dissipative-like character and the one arising from the iSGE would be primarily of field-like character. This is however not the case, as interband transitions, spin-dependent scattering, spin relaxation, spin precession and size effects significantly complicate the SOT scenario. Hence, symmetry considerations alone cannot disentangle directly the two contributions. However, symmetry analysis remains a powerful tool. SOTs obey Neumann's principle and must be invariant under the symmetry operations of the material system. This can restrict significantly the forms of the response coefficients, and aids the formulation of the proper phenomenological description of the SOTs, reflecting the underlying crystal symmetry of the considered material or structure \cite{Hals2013,Wimmer2016,Zelezny2017}.

This section is organized as follows. We review the linear response formalism commonly used for microscopic calculations of the SOT in Subsection \ref{s:intrainter}, and the general symmetry properties of SOT are then discussed in Subsection \ref{s:sots-nut}. Because of the great challenge of incorporating the full complexity of the bilayer systems at once, almost all theoretical studies have been focused either on iSGE in model systems or on the SHE mechanism only, with a handful of them attempting a comprehensive modeling. In Subsection \ref{s:sheTorque} we review calculations of the SOT in bilayer systems based on the SHE mechanism. In Subsection \ref{s:rashbatorque} we review calculations of the SOT in bulk systems which includes the 2D Rashba model and 3D non-centrosymmetric materials. Microscopic calculations carried out using density functional theory calculations for bilayer structures are presented in Subsection \ref{s:dftsot}. In Subsection \ref{s:antiferromagnets} we review calculations of the SOT in bulk antiferromagnets, and a discussion of SOTs in topological insulators and other systems is presented in Subsections \ref{s:sot-ti} and \ref{s:th-others}, respectively.

\subsection{Kubo linear response: intraband versus interband transitions\label{s:intrainter}}

From a microscopic linear response perspective, basic quantum mechanics states that the statistical average of an operator $\hat{\cal O}$ reads ${\cal O}=\sum_{n,{\bf k}}\langle n,{\bf k}|\hat{\cal O}|n,{\bf k}\rangle f_{n,{\bf k}}$, where $f_{n,{\bf k}}$ is the carrier distribution function and $|n,{\bf k}\rangle$ is the quantum eigenstate of the system. Under a small perturbation, such as an external electric field, both the distribution function $f_{n,{\bf k}}$ and the eigenstates $|n,{\bf k}\rangle$ are modified, giving rise to different nonequilibrium contributions to the observable ${\cal O}$ as consistently modeled by quantum field theory \cite{Rammer1986,Mahan2000}. Within the constant relaxation time approximation, the distribution function and eigenstates become
\begin{eqnarray}
&&f_{n,{\bf k}}\rightarrow f_{n,{\bf k}}^0-\tau\langle n,{\bf k}|e{\bf E}\cdot\hat{\bf{v}}|n,{\bf k}\rangle\frac{\partial}{\partial \epsilon} f_{n,{\bf k}}^0,\\
&&|n,{\bf k}\rangle\rightarrow |n,{\bf k}\rangle_0 - \sum_{n'}\frac{\langle n',{\bf k}|e{\bf E}\cdot\hat{\bf{r}}|n,{\bf k}\rangle_0}{\epsilon_{n,{\bf k}}-\epsilon_{n',{\bf k}}}|n',{\bf k}\rangle_0,
\end{eqnarray}
where $f_{n,{\bf k}}^0$ is the Fermi-Dirac distribution, ${\bf E}$ is the electric field, $\hat{\bf{v}}$ and $\hat{\bf r}$ are the velocity and position operators, $\epsilon_{n,{\bf k}}$ is the eigenenergy associated with the unperturbed eigenstate $|n,{\bf k}\rangle_0$, and $e>0$ is the absolute value of the electron charge. As a result, within the linear response approximation and to the lowest order in relaxation time, ${\cal O}={\cal O}^{\rm Intra}+{\cal O}^{\rm Inter}$, where
\begin{eqnarray}\label{eq:intra}
{\cal O}^{\rm Intra}&=&-\tau\sum_{n,{\bf k}}{\rm Re}\langle n,{\bf k}|e{\bf E}\cdot\hat{\bf{v}}|n,{\bf k}\rangle\langle n,{\bf k}|\hat{\cal O}|n,{\bf k}\rangle \frac{\partial}{\partial \epsilon} f_{n,{\bf k}}^0,\nonumber\\\\
{\cal O}^{\rm Inter}&=&-\hbar\sum_{n,n',{\bf k}}{\rm Im}\langle n,{\bf k}|e{\bf E}\cdot\hat{\bf{v}}|n',{\bf k}\rangle\langle n',{\bf k}|\hat{\cal O}|n,{\bf k}\rangle\nonumber\\
&&\times\frac{(f_{n,{\bf k}}-f_{n',{\bf k}})}{(\epsilon_{n,{\bf k}}-\epsilon_{n',{\bf k}})^2}.\label{eq:inter}
\end{eqnarray}
The first contribution, Eq. (\ref{eq:intra}), is proportional to the relaxation time $\sim\tau$ and only involves intraband transitions, $|n,{\bf k}\rangle\rightarrow|n,{\bf k}\rangle$. The second one, Eq. (\ref{eq:inter}), is weakly dependent on disorder and sometimes called {\em intrinsic}. It only involves interband transitions $|n,{\bf k}\rangle\rightarrow|n',{\bf k}\rangle$. The intrinsic contribution can be related to the Berry curvature of the material that connects intrinsic transport properties to the topology of the phase space \cite{Xiao2010,Sinova2015}. Equations \eqref{eq:intra}, \eqref{eq:inter} are valid only under the assumption of a constant and large relaxation time. More generally, the linear response can be expressed in terms of the Kubo-Bastin formula \cite{Freimuth2014a,Wimmer2016}

\begin{align}
 {\cal O } &= {\cal O}^{I(a)} + {\cal O}^{I(b)} + {\cal O}^{II},\\
 {\cal O}^{I(a)} &= \frac{e}{h} \int_{-\infty}^{\infty} d\varepsilon \frac{\partial}{\partial \varepsilon} f^0_\varepsilon\text{Tr} \langle\hat{\cal O} \hat G^R_\varepsilon(\mathbf{E}\cdot \hat{\mathbf{v}})\hat G^A_\varepsilon\rangle_c,\\
 {\cal O}^{I(b)} &= -\frac{e}{h} \int_{-\infty}^{\infty} d \varepsilon \frac{\partial}{\partial \varepsilon} f^0_\varepsilon \text{Re}\text{Tr} \langle\hat{\cal O} \hat G^R_\varepsilon(\mathbf{E}\cdot \hat{\mathbf{v}}) \hat G^R_\varepsilon\rangle_c,\\
 {\cal O}^{II} &= \frac{e}{h} \int_{-\infty}^{\infty} d \varepsilon f^0_\varepsilon \text{Re}\text{Tr} \nonumber\\
 &\langle\hat{\cal O}\hat G^R_\varepsilon(\mathbf{E}\cdot \hat{\mathbf{v}}) \frac{\partial}{\partial \varepsilon} G^R_\varepsilon - 
\hat{\cal O}\frac{\partial}{\partial \varepsilon}\hat G^R_\varepsilon(\mathbf{E}\cdot \hat{\mathbf{v}})\hat G^R_\varepsilon\rangle_c,
\end{align}
where $\hat G^{R(A)}_\varepsilon$ denotes the retarded (advanced) Green's function respectively and $\langle...\rangle_c$ denotes an average over disorder configurations. \text{Tr} is the trace over spin, momentum and orbital spaces. For concreteness, the operator $\hat{\cal O}$ is simply the spin operator $\hat{\bm\sigma}$ or the spin current operator $\bm{\mathcal J}_{\rm s}^i$ defined above. This formula is often simplified by assuming that the only effect of disorder is to induce a constant energy broadening $\Gamma=\hbar/2\tau$, such that $\hat G^{R(A)}_\varepsilon= \hbar (\varepsilon - \hat H \pm i\Gamma)^{-1} $. In the limit of large relaxation time, $\Gamma\rightarrow0$ and the Kubo-Bastin formula reduces to Eqs. \eqref{eq:intra} and \eqref{eq:inter}. Notice that extrinsic contributions to SHE (side-jump and skew scattering) are overlooked within this approximation \cite{Sinova2015}. Thus for a complete treatment, more sophisticated approaches are necessary.

\subsection{Symmetry of spin-orbit torques \label{s:sots-nut}}

As mentioned above, the torque can always be rewritten in terms of an effective magnetic field ${\bf B}_{\rm T}$, ${\bf T} = \mathbf{M} \times {\bf B}_{\rm T}$. The symmetry of the SOT can be studied either in terms of the linear response tensor $\chi_T$ or, equivalently, in terms of $\chi_B$. Here we focus on the effective field since its symmetry relations are simpler. In terms of symmetry, the effective magnetic field is equivalent to the iSGE, i.e., the tensors $\chi_B$ and $\chi_S$ have the same form (although they are not necessarily proportional, as often assumed in model calculations). To understand the symmetry properties of the SOT, it is convenient to parse the effective field into two parts, even and odd under time-reversal (or, equivalently, under the reversal of all magnetic moments). This is similar to the case of conductivity in magnetic systems \cite{Grimmer1993}. However, unlike for conductivity, the even and odd parts do not correspond to the symmetric and anti-symmetric parts of the effective field tensor. Thus a separate linear response tensor has to be assigned to each part,
 \begin{align}
 {\bf B}_\text{eff}^\text{even} &= \chi^\text{even}_B \mathbf{E},\\
 {\bf B}_\text{eff}^\text{odd} &= \chi^\text{odd}_B \mathbf{E}.
 \end{align}
 The same parsing can also be done for the torque. We note that the odd part of the torque corresponds to the even part of the effective field and \emph{vice versa}. Noticeably, the odd and even parts have very different properties and correspond to different contributions of the Kubo formula: the intraband formula, Eq. \eqref{eq:intra}, corresponds to the even field, whereas the interband formula, Eq. \eqref{eq:inter}, corresponds to the odd field. Similar separation can be done for the full Kubo-Bastin formula \cite{Freimuth2014a}. Furthermore, such a separation is also commonly done for experimental measurements of SOT (see Sections \ref{MML} and \ref{s:DW}). Since the following applies equally to $\chi_B$ and $\chi_S$ we denote the tensor simply by $\chi$. Following the Neumann's principle, the tensors $\chi$ have to be invariant under all symmetry operations of the crystal. The two parts transform differently for symmetry operations that contain time-reversal symmetry. For a symmetry operation represented by a matrix $\cal R$ \cite{Zelezny2017},
 \begin{align}
 \chi^\text{even} &= \rm{det}({\cal R}){\cal R}\chi^\text{even} {\cal R}^{-1},\label{eq:chi_even_trans}\\
 \chi^\text{odd} &= \pm\rm{det}({\cal R}){\cal R}\chi^\text{odd} {\cal R}^{-1}, \label{eq:chi_odd_trans}
 \end{align}
 where $\pm$ refers to a symmetry operation with and without time-reversal, respectively, and $\rm{det}({\cal R})$ is the determinant of ${\cal R}$. By considering all the symmetry operations in the magnetic point group of the given crystal, the general form of the response tensors is found from these equations. It is also possible to treat the whole tensor together without separating it into the even and odd parts, although then some information about the structure of the torque is lost. See \cite{Wimmer2016} for a table of total $\chi_T$ tensors for all the magnetic point groups. 
 
 In systems with more than one magnetic atom in the unit cell, such as antiferromagnets, it is furthermore useful to study the symmetry of SOT on each magnetic site. Then Eqs. \eqref{eq:chi_even_trans}, \eqref{eq:chi_odd_trans} are modified as follows \cite{Zelezny2017},
 \begin{align}
 \chi^\text{even}_{a'} &= \rm{det}({\cal R}){\cal R}\chi^\text{even}_{{\it a}} {\cal R}^{-1},\label{eq:chia_even_transa}\\
 \chi^\text{odd}_{a'} &= \pm\rm{det}({\cal R}){\cal R}\chi^\text{odd}_{{\it a}} {\cal R}^{-1}, \label{eq:chia_odd_transa}
 \end{align}
 where $a$ denotes a given site and $a'$ is the site to which site $a$ transforms under symmetry operation $\cal R$. In this case it is necessary to consider the full magnetic space group and atomic positions of magnetic moments. The symmetry of $\chi_a$ is determined by symmetry operations that leave site $a$ invariant (such symmetry operations form the so-called site symmetry group), whereas the symmetry operations that transform $a$ to a different site $a'$ relate tensor $\chi_a$ to tensor $\chi_{a'}$.

 A key conclusion that can be made from Eqs. \eqref{eq:chi_even_trans}, \eqref{eq:chi_odd_trans} is that there can be no net SOT (or iSGE) if the system has inversion symmetry. However, from Eqs. \eqref{eq:chia_even_transa}, \eqref{eq:chia_odd_transa} we see that even in a system with inversion symmetry there can still be a local SOT if the inversion symmetry is broken locally, i.e., there can be SOT on site $a$, if there is no inversion symmetry which would leave site $a$ invariant.

 To understand the dependence of SOT on the direction of magnetic moments, it is helpful to expand the SOT in the direction of magnetic moments. For a collinear magnetic material,
\begin{align}
 \chi_{ij}({\bf n}) = \chi_{ij}^{(0)} + \chi_{ij,k}^{(1)} n_k + \chi_{ij,kl}^{(2)} n_k n_l + \dots,
 \label{eq:chi_expansion}
\end{align}
where ${\bf n}$ is the magnetic order parameter (the magnetization direction in ferromagnets, or the N\'eel order parameter in antiferromagnets). The even terms in the expansion correspond to the even effective field and conversely the odd terms correspond to the odd field. The symmetry of the $n$-independent expansion tensors in Eq. \eqref{eq:chi_expansion} is determined by the symmetry group of the \emph{nonmagnetic} system. For a global SOT in a ferromagnet or a local SOT in a bipartite antiferromagnet the following transformation rule is found for the expansion tensors,
 \begin{align}
 \chi_{ij,mn\dots}^{(\nu)} = \det({\cal R})^{\nu-1}{\cal R}_{ik}{\cal R}_{jl}^{-T} {\cal R}_{mo}^{-T} {\cal R}_{np}^{-T}\dots \chi_{kl,op\dots}^{(\nu)},\label{eq:trans_expansion}
\end{align}
For the global case, the nonmagnetic point group has to be used, whereas for the local case, the nonmagnetic site symmetry group has to be used instead. Since there are only 21 nonmagnetic point groups with broken inversion symmetry, it is feasible to calculate all allowed leading terms of the expansion \eqref{eq:chi_expansion}. This was done for the zeroth, first, and some second order terms in Refs.~\cite{Ciccarelli2016,Zelezny2017}. The results for the zeroth and first order terms are given in Table \ref{table:point_groups1}.

The lowest order even field is typically given by $\chi^{(0)}$, which corresponds to a field-like torque. In some cases such a term is, however, prohibited by symmetry and the lowest order even field is second order in magnetization. This is the case of the cubic zinc-blende or half-heusler crystal ferromagnets with space group $F\bar{4}$3m, for instance. Under strain, however, these materials exhibit an even field at the zeroth order in magnetization (in other words, a field-like torque) \cite{Chernyshov2009}. In contrast, MnSi and its parent compounds adopt the $P2_13$ space group and display a field-like torque even without strain \cite{Hals2013b}.
\begin{figure}[h!]
\begin{center}
\includegraphics[width=8cm]{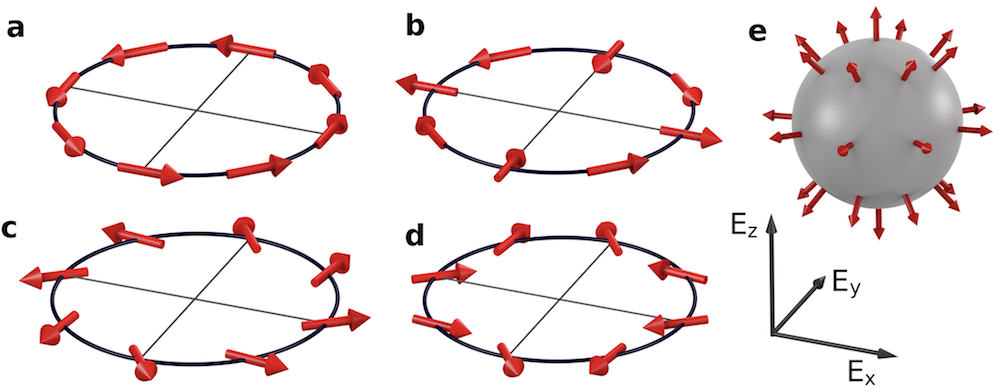}
\end{center}
\caption{(Color online) Various types of the field-like torques as a function of the electric field direction. The red arrows denote the corresponding effective field direction for (a) Rashba, (b) Dresselhaus, (c) generalized Rashba, (d) generalized Dresselhaus, and (e) Weyl coupling schemes.}
\label{fig:Rashba_Dresselhaus}
\end{figure}

More generally, the connection between the SOT field and the applied electric field can be categorized in three different types illustrated in Fig. \ref{fig:Rashba_Dresselhaus}: Rashba and Dresselhaus coupling schemes, and a coupling such that the SOT field is collinear to the electric field that we refer to as Weyl coupling [Fig. \ref{fig:Rashba_Dresselhaus}(d)]. These denominations are taken in analogy with the spin-textures in the momentum space of the Rashba, Dresselhaus and Weyl spin-orbit coupling further discussed in Subsection \ref{s:rashbatorque}. The Rashba and Dresselhaus fields are confined to a plane and only appear for electric field lying in the plane. They differ in how the effective field is changed when the electric field is rotated. In the case of standard Rashba coupling, the effective field rotates in the same direction as the electric field [Fig. \ref{fig:Rashba_Dresselhaus}(a)], whereas in the case of standard Dresselhaus coupling the effective field rotates in the opposite direction [Fig. \ref{fig:Rashba_Dresselhaus}(b)]. The generalized Rashba coupling differs from a conventional Rashba coupling in that the angle between the electric and effective field is not necessarily $90^\circ$ [Fig. \ref{fig:Rashba_Dresselhaus}(c)]. The generalized Dresselhaus coupling differs from the conventional Dresselhaus coupling in that the effective field is not necessarily parallel or perpendicular to the electric field along the crystalline axes [Fig. \ref{fig:Rashba_Dresselhaus}(d)].

As seen in Table \ref{table:point_groups1}, $\chi^{(1)}$ has always some non-zero components. These generate the lowest order odd field. It often has a damping-like character, i.e., can be written as ${\bf B}_{\rm T}\sim \mathbf{m} \times {\bm\zeta}$, where $ {\bm\zeta}$ is a vector independent of magnetization. However in some cases the first-order field does not have the damping-like form. An example of a system where no damping-like torque is allowed by symmetry is again cubic zinc-blende or half-heusler crystals. Even if the damping-like torque is allowed by symmetry there can be other first-order contributions. Magnetic dynamics induced by such torques can differ from the effect of a damping-like torque and has not been studied so far.\par
\newpage

\begingroup
\squeezetable
\begin{longtable*}{cccc}
 \hline \hline 
 \rule{0pt}{11pt} 
 Crystal system & Point group & $\chi^{(0)}$ & $\chi^{(1)}$ \\[1pt]
 \hline
 \rule{0pt}{17pt} 
 triclinic & 1 & $ \left(\begin{matrix}x_{11} & x_{12} & x_{13}\\x_{21} & x_{22} & x_{23}\\x_{31} & x_{32} & x_{33}\end{matrix}\right)
$ & $ \left(\begin{matrix}\hat{n}_{x} x_{111} + \hat{n}_{y} x_{112} + \hat{n}_{z} x_{113} & \hat{n}_{x} x_{121} + \hat{n}_{y} x_{122} + \hat{n}_{z} x_{123} & \hat{n}_{x} x_{131} + \hat{n}_{y} x_{132} + \hat{n}_{z} x_{133}\\\hat{n}_{x} x_{211} + \hat{n}_{y} x_{212} + \hat{n}_{z} x_{213} & \hat{n}_{x} x_{221} + \hat{n}_{y} x_{222} + \hat{n}_{z} x_{223} & \hat{n}_{x} x_{231} + \hat{n}_{y} x_{232} + \hat{n}_{z} x_{233}\\\hat{n}_{x} x_{311} + \hat{n}_{y} x_{312} + \hat{n}_{z} x_{313} & \hat{n}_{x} x_{321} + \hat{n}_{y} x_{322} + \hat{n}_{z} x_{323} & \hat{n}_{x} x_{331} + \hat{n}_{y} x_{332} + \hat{n}_{z} x_{333}\end{matrix}\right)
 $ \\[10pt]
monoclinic & 2 & $ \left(\begin{matrix}x_{11} & 0 & x_{13}\\0 & x_{22} & 0\\x_{31} & 0 & x_{33}\end{matrix}\right)
$ & $ \left(\begin{matrix}\hat{n}_{y} x_{1} & \hat{n}_{x} x_{13} + \hat{n}_{z} x_{12} & \hat{n}_{y} x_{3}\\\hat{n}_{x} x_{5} + \hat{n}_{z} x_{6} & \hat{n}_{y} x_{11} & \hat{n}_{x} x_{4} + \hat{n}_{z} x_{7}\\\hat{n}_{y} x_{10} & \hat{n}_{x} x_{8} + \hat{n}_{z} x_{9} & \hat{n}_{y} x_{2}\end{matrix}\right)
 $ \\[10pt]
& m & $ \left(\begin{matrix}0 & x_{12} & 0\\x_{21} & 0 & x_{23}\\0 & x_{32} & 0\end{matrix}\right)
$ & $ \left(\begin{matrix}\hat{n}_{x} x_{12} + \hat{n}_{z} x_{9} & \hat{n}_{y} x_{14} & \hat{n}_{x} x_{13} + \hat{n}_{z} x_{8}\\\hat{n}_{y} x_{3} & \hat{n}_{x} x_{11} + \hat{n}_{z} x_{10} & \hat{n}_{y} x_{4}\\\hat{n}_{x} x_{7} + \hat{n}_{z} x_{6} & \hat{n}_{y} x_{5} & \hat{n}_{x} x_{1} + \hat{n}_{z} x_{2}\end{matrix}\right)
 $ \\[10pt]
orthorhombic & 222 & $ \left(\begin{matrix}x_{11} & 0 & 0\\0 & x_{22} & 0\\0 & 0 & x_{33}\end{matrix}\right)
$ & $ \left(\begin{matrix}0 & \hat{n}_{z} x_{5} & \hat{n}_{y} x_{4}\\\hat{n}_{z} x_{1} & 0 & \hat{n}_{x} x_{6}\\\hat{n}_{y} x_{3} & \hat{n}_{x} x_{2} & 0\end{matrix}\right)
 $ \\[10pt]
& mm2 & $ \left(\begin{matrix}0 & x_{12} & 0\\x_{21} & 0 & 0\\0 & 0 & 0\end{matrix}\right)
$ & $ \left(\begin{matrix}\hat{n}_{z} x_{4} & 0 & \hat{n}_{x} x_{6}\\0 & \hat{n}_{z} x_{5} & \hat{n}_{y} x_{7}\\\hat{n}_{x} x_{3} & \hat{n}_{y} x_{2} & \hat{n}_{z} x_{1}\end{matrix}\right)
 $ \\[10pt]
tetragonal & 4 & $ \left(\begin{matrix}x_{11} & - x_{21} & 0\\x_{21} & x_{11} & 0\\0 & 0 & x_{33}\end{matrix}\right)
$ & $ \left(\begin{matrix}\hat{n}_{z} x_{6} & - \hat{n}_{z} x_{2} & \hat{n}_{x} x_{5} - \hat{n}_{y} x_{7}\\\hat{n}_{z} x_{2} & \hat{n}_{z} x_{6} & \hat{n}_{x} x_{7} + \hat{n}_{y} x_{5}\\\hat{n}_{x} x_{4} - \hat{n}_{y} x_{3} & \hat{n}_{x} x_{3} + \hat{n}_{y} x_{4} & \hat{n}_{z} x_{1}\end{matrix}\right)
 $ \\[10pt]
& -4 & $ \left(\begin{matrix}x_{11} & x_{21} & 0\\x_{21} & - x_{11} & 0\\0 & 0 & 0\end{matrix}\right)
$ & $ \left(\begin{matrix}\hat{n}_{z} x_{5} & \hat{n}_{z} x_{1} & \hat{n}_{x} x_{4} + \hat{n}_{y} x_{6}\\\hat{n}_{z} x_{1} & - \hat{n}_{z} x_{5} & \hat{n}_{x} x_{6} - \hat{n}_{y} x_{4}\\\hat{n}_{x} x_{3} + \hat{n}_{y} x_{2} & \hat{n}_{x} x_{2} - \hat{n}_{y} x_{3} & 0\end{matrix}\right)
 $ \\[10pt]
& 422 & $ \left(\begin{matrix}x_{11} & 0 & 0\\0 & x_{11} & 0\\0 & 0 & x_{33}\end{matrix}\right)
$ & $ \left(\begin{matrix}0 & - \hat{n}_{z} x_{3} & - \hat{n}_{y} x_{2}\\\hat{n}_{z} x_{3} & 0 & \hat{n}_{x} x_{2}\\- \hat{n}_{y} x_{1} & \hat{n}_{x} x_{1} & 0\end{matrix}\right)
 $ \\[10pt]
& 4mm & $ \left(\begin{matrix}0 & - x_{21} & 0\\x_{21} & 0 & 0\\0 & 0 & 0\end{matrix}\right)
$ & $ \left(\begin{matrix}\hat{n}_{z} x_{4} & 0 & \hat{n}_{x} x_{1}\\0 & \hat{n}_{z} x_{4} & \hat{n}_{y} x_{1}\\\hat{n}_{x} x_{3} & \hat{n}_{y} x_{3} & \hat{n}_{z} x_{2}\end{matrix}\right)
 $ \\[10pt]
& -42m & $ \left(\begin{matrix}x_{11} & 0 & 0\\0 & - x_{11} & 0\\0 & 0 & 0\end{matrix}\right)
$ & $ \left(\begin{matrix}0 & \hat{n}_{z} x_{3} & \hat{n}_{y} x_{2}\\\hat{n}_{z} x_{3} & 0 & \hat{n}_{x} x_{2}\\\hat{n}_{y} x_{1} & \hat{n}_{x} x_{1} & 0\end{matrix}\right)
 $ \\[10pt]
trigonal & 3 & $ \left(\begin{matrix}x_{11} & - x_{21} & 0\\x_{21} & x_{11} & 0\\0 & 0 & x_{33}\end{matrix}\right)
$ & $ \left(\begin{matrix}\hat{n}_{x} x_{7} + \hat{n}_{y} x_{2} + \hat{n}_{z} x_{8} & \hat{n}_{x} x_{2} - \hat{n}_{y} x_{7} - \hat{n}_{z} x_{3} & \hat{n}_{x} x_{6} - \hat{n}_{y} x_{9}\\\hat{n}_{x} x_{2} - \hat{n}_{y} x_{7} + \hat{n}_{z} x_{3} & - \hat{n}_{x} x_{7} - \hat{n}_{y} x_{2} + \hat{n}_{z} x_{8} & \hat{n}_{x} x_{9} + \hat{n}_{y} x_{6}\\\hat{n}_{x} x_{5} - \hat{n}_{y} x_{4} & \hat{n}_{x} x_{4} + \hat{n}_{y} x_{5} & \hat{n}_{z} x_{1}\end{matrix}\right)
 $ \\[10pt]
& 312 & $ \left(\begin{matrix}x_{11} & 0 & 0\\0 & x_{11} & 0\\0 & 0 & x_{33}\end{matrix}\right)
$ & $ \left(\begin{matrix}\hat{n}_{y} x_{3} & \hat{n}_{x} x_{3} - \hat{n}_{z} x_{4} & - \hat{n}_{y} x_{2}\\\hat{n}_{x} x_{3} + \hat{n}_{z} x_{4} & - \hat{n}_{y} x_{3} & \hat{n}_{x} x_{2}\\- \hat{n}_{y} x_{1} & \hat{n}_{x} x_{1} & 0\end{matrix}\right)
 $ \\[10pt]
& 3m1 & $ \left(\begin{matrix}0 & - x_{21} & 0\\x_{21} & 0 & 0\\0 & 0 & 0\end{matrix}\right)
$ & $ \left(\begin{matrix}\hat{n}_{y} x_{4} + \hat{n}_{z} x_{5} & \hat{n}_{x} x_{4} & \hat{n}_{x} x_{2}\\\hat{n}_{x} x_{4} & - \hat{n}_{y} x_{4} + \hat{n}_{z} x_{5} & \hat{n}_{y} x_{2}\\\hat{n}_{x} x_{3} & \hat{n}_{y} x_{3} & \hat{n}_{z} x_{1}\end{matrix}\right)
 $ \\[10pt]
hexagonal & 6 & $ \left(\begin{matrix}x_{11} & - x_{21} & 0\\x_{21} & x_{11} & 0\\0 & 0 & x_{33}\end{matrix}\right)
$ & $ \left(\begin{matrix}\hat{n}_{z} x_{6} & - \hat{n}_{z} x_{2} & \hat{n}_{x} x_{5} - \hat{n}_{y} x_{7}\\\hat{n}_{z} x_{2} & \hat{n}_{z} x_{6} & \hat{n}_{x} x_{7} + \hat{n}_{y} x_{5}\\\hat{n}_{x} x_{4} - \hat{n}_{y} x_{3} & \hat{n}_{x} x_{3} + \hat{n}_{y} x_{4} & \hat{n}_{z} x_{1}\end{matrix}\right)
 $ \\[10pt]
& -6 & $ \left(\begin{matrix}0 & 0 & 0\\0 & 0 & 0\\0 & 0 & 0\end{matrix}\right)
$ & $ \left(\begin{matrix}\hat{n}_{x} x_{1} + \hat{n}_{y} x_{2} & \hat{n}_{x} x_{2} - \hat{n}_{y} x_{1} & 0\\\hat{n}_{x} x_{2} - \hat{n}_{y} x_{1} & - \hat{n}_{x} x_{1} - \hat{n}_{y} x_{2} & 0\\0 & 0 & 0\end{matrix}\right)
 $ \\[10pt]
& 622 & $ \left(\begin{matrix}x_{11} & 0 & 0\\0 & x_{11} & 0\\0 & 0 & x_{33}\end{matrix}\right)
$ & $ \left(\begin{matrix}0 & - \hat{n}_{z} x_{3} & - \hat{n}_{y} x_{2}\\\hat{n}_{z} x_{3} & 0 & \hat{n}_{x} x_{2}\\- \hat{n}_{y} x_{1} & \hat{n}_{x} x_{1} & 0\end{matrix}\right)
 $ \\[10pt]
& 6mm & $ \left(\begin{matrix}0 & - x_{21} & 0\\x_{21} & 0 & 0\\0 & 0 & 0\end{matrix}\right)
$ & $ \left(\begin{matrix}\hat{n}_{z} x_{4} & 0 & \hat{n}_{x} x_{1}\\0 & \hat{n}_{z} x_{4} & \hat{n}_{y} x_{1}\\\hat{n}_{x} x_{3} & \hat{n}_{y} x_{3} & \hat{n}_{z} x_{2}\end{matrix}\right)
 $ \\[10pt]
& -6m2 & $ \left(\begin{matrix}0 & 0 & 0\\0 & 0 & 0\\0 & 0 & 0\end{matrix}\right)
$ & $ \left(\begin{matrix}\hat{n}_{y} x_{1} & \hat{n}_{x} x_{1} & 0\\\hat{n}_{x} x_{1} & - \hat{n}_{y} x_{1} & 0\\0 & 0 & 0\end{matrix}\right)
 $ \\[10pt]
cubic & 23 & $ \left(\begin{matrix}x_{11} & 0 & 0\\0 & x_{11} & 0\\0 & 0 & x_{11}\end{matrix}\right)
$ & $ \left(\begin{matrix}0 & \hat{n}_{z} x_{2} & \hat{n}_{y} x_{1}\\\hat{n}_{z} x_{1} & 0 & \hat{n}_{x} x_{2}\\\hat{n}_{y} x_{2} & \hat{n}_{x} x_{1} & 0\end{matrix}\right)
 $ \\[10pt]
& 432 & $ \left(\begin{matrix}x_{11} & 0 & 0\\0 & x_{11} & 0\\0 & 0 & x_{11}\end{matrix}\right)
$ & $ \left(\begin{matrix}0 & - \hat{n}_{z} x_{1} & \hat{n}_{y} x_{1}\\\hat{n}_{z} x_{1} & 0 & - \hat{n}_{x} x_{1}\\- \hat{n}_{y} x_{1} & \hat{n}_{x} x_{1} & 0\end{matrix}\right)
 $ \\[10pt]
& -43m & $ \left(\begin{matrix}0 & 0 & 0\\0 & 0 & 0\\0 & 0 & 0\end{matrix}\right)
$ & $ \left(\begin{matrix}0 & \hat{n}_{z} x_{1} & \hat{n}_{y} x_{1}\\\hat{n}_{z} x_{1} & 0 & \hat{n}_{x} x_{1}\\\hat{n}_{y} x_{1} & \hat{n}_{x} x_{1} & 0\end{matrix}\right)
 $ \\[10pt]
\hline \hline
\noalign{\vskip 1mm} 
\caption{Zeroth and first order terms in the expansion \eqref{eq:chi_expansion} for the point groups with broken inversion symmetry. The tensors $\chi^{(1)}$ have the spin-axis direction included: $\chi^{(1)}_{ij} = \chi^{(1)}_{ij,k}\hat{n}_k$. The $x$ parameters can be chosen arbitrarily for each tensor. The tensors are given in cartesian coordinate systems defined in \cite{Zelezny2017}.}
\label{table:point_groups1}
\end{longtable*}
\endgroup
\newpage

The lowest order terms frequently describe qualitative aspects of the torque both in experiments and in the theoretical calculations. The usefulness of the lowest order term is illustrated by the fact that materials with very different electronic structures but same symmetry have very similar SOTs. For instance this is the case of ferromagnetic (Ga,Mn)As and NiMnSb, or systems modeled by the 2D Rashba Hamiltonian, or antiferromagnets Mn$_2$Au and CuMnAs, discussed in Section \ref{s:antiferromagnets}. For an accurate quantitative description of the SOT, higher order terms can be important. These are not tabulated but can be produced by the publicly available code that was used for generating Table \ref{table:point_groups1} \cite{symcode}. This code can be also used to determine the full tensors $\chi_a^\text{even}$ and $\chi_a^\text{odd}$ for a given crystal. For the case of an interface with inversion symmetry breaking only (e.g., in the case of a Rashba 2D gas), one obtains
\begin{eqnarray}
&&\chi^{(0)}=x_0\left(\begin{matrix}0 & 1 & 0\\-1 & 0 & 0\\0 & 0 & 0\end{matrix}\right),\;\chi^{(1)}=x_1\left(\begin{matrix}-\hat{n}_{z} & 0 & 0\\0 & \hat{n}_{z} & 0\\\hat{n}_{x} & \hat{n}_{y} & 0\end{matrix}\right)
 \end{eqnarray}

\subsection{Spin-orbit torques due to the spin Hall effect\label{s:sheTorque}}

The SHE-SOT contribution in bilayer systems arises from the absorption of angular momentum coming from a SHE spin current generated outside the ferromagnet, e. g., in the proximate nonmagnetic metal layer \cite{Dyakonov1971b}. This is effectively the mechanism of STT where the polarizing ferromagnet in a trilayer device is replaced in this instance by the nonmagnetic metal \cite{Brataas2012b,Stiles2002}. In analogy to STT, the SHE-SOT mechanism in common metal structures is primarily damping-like in character, assuming a full absorption of the carrier spin angular momentum in the ferromagnet. Therefore in many experiments, the damping-like SOT is associated with SHE, and the extracted spin Hall angle is calculated on the basis that this is the only contribution to the damping-like SOT component. Since this is generally not the case, the spin Hall angle values extracted from these experiments should be considered only as effective phenomenological descriptions of the SOT efficiency. On the other hand, in many experiments a clear correlation between the magnitude and sign of SHE, e.g., obtained by non-local measurements \cite{Morota2011}, and damping-like SOT is observed. We do not review here the calculations of the SHE, which has been done elsewhere \cite{Sinova2015}, and focus instead on effective theoretical treatments using the spin Hall angle as a phenomenological parameter.

The SHE-SOT is present in structures where the ferromagnet is adjacent to a nonmagnetic (Pt, W, Ta, WTe$_2$, conductive Bi$_2$Se$_3$, etc.) or magnetic metal (IrMn, PtMn etc.). To model this torque, one needs to compute the spin density originating from this metal and diffusing into the ferromagnet. The simplest method is to solve the drift-diffusion equation in the presence of spin-orbit coupling and match the spin currents and accumulations at the boundary between the ferromagnet and the nonmagnetic metal using, for instance, the spin mixing conductance \cite{Haney2013b,Amin2016b,Chen2015e}. The charge and spin currents in a nonmagnetic metal with spin-orbit coupling read \cite{Dyakonov1971b,Shchelushkin2005b,Shen2014a,Pauyac2018}
\begin{eqnarray}\label{eq:jc}
&&{\bf j}_{\rm c}/\sigma_{\rm N}=-{\bm\nabla}\mu_c+\theta_{\rm sh}{\bm \nabla}\times{\bm\mu},\\\label{eq:js}
&&(2e/\hbar)\bm{\mathcal J}_{\rm s}^i/\sigma_{\rm N}=-{\bm \nabla}\mu_i-\theta_{\rm sh}{\bf e}_i\times{\bm\nabla}\mu_c-\theta_{\rm sw}{\bm\nabla}\times({\bf e}_i\times{\bm \mu}),\nonumber\\
\end{eqnarray}
where $\sigma_{\rm N}$ is the bulk conductivity and $\theta_{\rm sw}$ is the spin swapping coefficient \cite{Lifshits2009,Pauyac2018} (we comment on the spin swapping term in more detail at the end of this subsection). $\mu_c=n/e{\cal N}$ and ${\bm\mu}={\bf S}/e{\cal N}$ are the charge and spin chemical potentials, respectively, with ${\cal N}$ the density of states at the Fermi level. Formally, the drift-diffusion approach for the current-in-plane geometry is only applicable as long as the mean free path is much shorter than the layer thickness and assuming uniform spin Hall angle and conductivity in the nonmagnetic metal. This model also neglects interfacial spin-flips, or spin-memory loss \cite{Bass2007,Belashchenko2016,Dolui2017}. As discussed further below, this assumption is not always accurate. Using the spin mixing conductance, $g^{\uparrow\downarrow}$, as a boundary condition, the spin transfer arises from the absorption of the incoming transverse spin current at the interface, ${\bf T}=\bm{\mathcal J}_{\rm s}/t_{\rm F}$, $t_{\rm F}$ being the thickness of the magnet. It is composed of two components, as described in Eq. \eqref{eq:torquedef}, which read \cite{Haney2013b}
\begin{eqnarray}
&&\tau_{\rm DL}=\frac{\hbar\theta_{\rm sh}}{2e}\frac{\tilde{g}^{\uparrow\downarrow}_r+|\tilde{g}^{\uparrow\downarrow}|^2}{(1+\tilde{g}^{\uparrow\downarrow}_r)^2+\tilde{g}^{\uparrow\downarrow2}_i}\left(1-\cosh^{-1}\frac{t_{\rm N}}{\lambda_{\rm sf}}\right)\sigma_{\rm N}E,\nonumber\\\label{eq:she-ddiff1}\\
&&\tau_{\rm FL}=-\frac{\hbar\theta_{\rm sh}}{2e}\frac{\tilde{g}^{\uparrow\downarrow}_i}{(1+\tilde{g}^{\uparrow\downarrow}_r)^2+\tilde{g}^{\uparrow\downarrow2}_i}\left(1-\cosh^{-1}\frac{t_{\rm N}}{\lambda_{\rm sf}}\right)\sigma_{\rm N}E,\nonumber\\\label{eq:she-ddiff2}
\end{eqnarray}
Here, we omitted the spin swapping term in Eq. \eqref{eq:js}, and $\bm\zeta\|{\bf z}\times{\bf j}_{\rm c}$, ${\bf z}$ being normal to the interface. We also define the applied electric field $E=-\partial \mu_c/\partial x$, and the reduced mixing conductance $\tilde{g}^{\uparrow\downarrow}=g^{\uparrow\downarrow}\lambda_{\rm sf}/[\sigma_{\rm N}\tanh(t_{\rm N}/\lambda_{\rm sf})]$, while $t_{\rm N}$ and $\lambda_{\rm sf}$ are the thickness and spin relaxation length of the nonmagnetic metal, respectively. Finally, $\tilde{g}^{\uparrow\downarrow}_r$ and $\tilde{g}^{\uparrow\downarrow}_i$ refer to the real part and imaginary part of $\tilde{g}^{\uparrow\downarrow}$, respectively. In the limit of small imaginary part of the mixing conductance, $\tau_{\rm DL}\propto \eta\theta_{\rm sh}$ and $\tau_{\rm FL}\propto \eta\theta_{\rm sh} g^{\uparrow\downarrow}_i/g^{\uparrow\downarrow}_r$. The transparency coefficient $\eta=g^{\uparrow\downarrow}_r/[g^{\uparrow\downarrow}_r+\sigma_{\rm N}\tanh(t_{\rm N}/\lambda_{\rm sf})/\lambda_{\rm sf}]$ accounts for the spin current transmission through the interface, with $\eta\rightarrow 1$ when $g^{\uparrow\downarrow}_r\gg\sigma_{\rm N}/\lambda_{\rm sf}$. Equations \eqref{eq:she-ddiff1}, \eqref{eq:she-ddiff2} assume that all the spin current impinging on the interface is either reflected back to the nonmagnetic metal or absorbed by the ferromagnet. Allowing for spin-memory loss at the interface opens an additional spin dissipation channel \cite{Bass2007,Belashchenko2016,Dolui2017} and reduces the effective spin mixing conductance, leading to an {\em underestimation} of the spin Hall angle \cite{Rojas-Sanchez2014,Berger2018b}.

These expressions, although quite extensively used to interpret experimental data, must be handled with care as they disregard any corrections emerging from thickness-dependent conductivity, spin Hall angle, and spin-dependent scattering at interfaces, and assume the simplest form of the interfacial spin mixing conductance. To overcome these limitations, several methods have been employed, such as the Boltzmann transport equation where SHE is explicitly contained in the collision integral \cite{Haney2013b,Amin2016b,Engel2005}, the Kubo formula with real-space Green's function in a slab geometry \cite{Chen2017b}, the tight-binding model with random impurity potential \cite{Saidaoui2016} or transport calculations based on first principles \cite{Wang2016b,Freimuth2014a,Freimuth2015,Dolui2017}. 

Several additional features beyond the "conventional" SHE model have been identified. First, because the thickness of the nonmagnetic metal is of the same order as the mean free path (5-10 nm), the conductivity depends on the thickness of the slab \cite{Sondheimer1952}. In fact, it is well-known that spin transport in current-in-plane configuration is governed by mean free path effects, rather than by the spin diffusion length \cite{Camley1989,Zhang1993}. A direct consequence is that the spin Hall current estimated using the drift-diffusion model, Eq. \eqref{eq:she-ddiff1}, \eqref{eq:she-ddiff2}, is generally {\em overestimated} \cite{Chen2017b}. 

A second important aspect was revealed by {\em ab-initio} calculations \cite{Wang2016b,Freimuth2015}. These studies suggest that the SHE itself can be significantly enhanced close to the interface. The computed interfacial Hall angle can be an order of magnitude larger than the bulk spin Hall angle, and possibly dominate the total SHE signal. On the other hand, increasing the disorder leads to a progressive reduction of the interfacial spin Hall angle \cite{Freimuth2015}. Realistic modeling based on {\em ab-initio} simulations are further discussed in Subsection \ref{s:dftsot}.

\begin{figure}[h!]
\centering
\includegraphics[width=8cm]{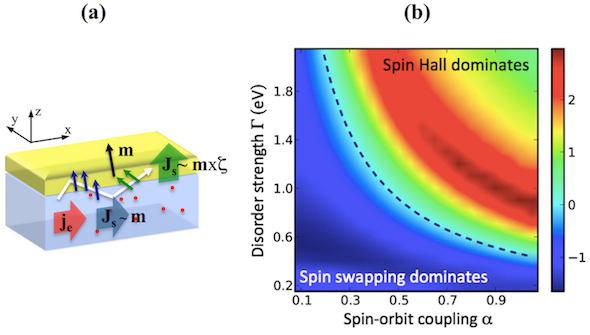}
\caption{\label{fig:swapping_torques} (Color online) Spin precession around the spin-orbit field. (a) Schematics of extrinsic spin swapping effect in a bilayer composed of a nonmagnetic metal (blue) and a ferromagnet (yellow) with magnetization {\bf m}. The spin current flowing from the ferromagnet into the nonmagnetic metal is polarized along ${\bf m}$ and precesses about the impurity-driven spin-orbit field normal to the scattering plane ${\bm\zeta}$. It produces a secondary spin current polarized along ${\bf m}\times{\bm \zeta}$ \cite{Saidaoui2016}. (b) Ratio between the magnitude of the field-like and damping-like torques $\tau_{\rm DL}/\tau_{\rm FL}$ as a function of disorder strength and spin-orbit coupling in the extrinsic spin swapping scenario. The ratio is given in logarithmic scale and the dashed line indicates $\tau_{\rm DL}=\tau_{\rm FL}$. Adapted from \citet{Saidaoui2016}.}
\end{figure}


Finally, it has been recently realized that the precession of spin currents around the spin-orbit field can substantially impact the SOT. This mechanism was originally proposed by \citet{Lifshits2009} in a different context and called {\em spin swapping}. In this mechanism, a primary spin current with spin polarization $\bm{\mathcal J}_{\rm s}\sim{\bm\sigma}$ precesses around the spin-orbit field ${\bf B}_{\rm so}$, resulting in a secondary spin current $\bm{\mathcal J}_{\rm s}\sim{\bm\sigma}\times{\bf B}_{\rm so}$. This secondary spin current can be absorbed by an adjacent ferromagnet, resulting in additional SOT components. Several flavors of this scenario have been proposed, depending on the source of the primary spin current and on the nature of the spin-orbit field. \citet{Saidaoui2016} suggested that spin-polarized electrons scattering off the ferromagnetic layer inject a primary spin current $\bm{\mathcal J}_{\rm s}\sim{\bf m}$ in the nonmagnetic metal. This primary spin current precesses around the spin-orbit field oriented normal to the scattering plane, which produces a secondary spin current $\bm{\mathcal J}_{\rm s}\sim{\bf m}\times{\bm \zeta}$ [Fig. \ref{fig:swapping_torques}(a)]. Once absorbed into the ferromagnet, this spin swapping spin current induces a field-like SOT. This mechanism only survives as long as the nonmagnetic metal thickness is comparable to the mean free path, as shown in Fig. \ref{fig:swapping_torques}(b). 


While the SHE-SOTs discussed above occur in bilayers composed of a spin current source (e.g, a heavy metal) and a spin current absorber, another configuration involving a single ferromagnet was recently investigated. By computing the spin diffusion equation in a centrosymmetric ferromagnet with spin-orbit coupling, \citet{Pauyac2018} showed how the interplay between SHE, spin swapping, and spin precession about the magnetic exchange can lead to local torques acting at the edges of the magnet. These local torques can nucleate reversed magnetic domains and initiate current-driven magnetization dynamics \cite{Wang2019}.

\subsection{Spin-orbit torques due to the inverse spin galvanic effect\label{s:rashbatorque}}

In this section we review calculations of the SOT in 2D and 3D bulk magnetic systems. Such a torque is considered to be due to the iSGE, which refers to the electrical generation of spin density when a current flows in a system lacking (bulk or interfacial) inversion symmetry. Its reciprocal effect, the SGE, is the generation of a charge current in the presence of non-equilibrium spin density (generated, e.g. by photoexcitation). Microscopically, iSGE and SGE are associated with the presence of a spin-orbit coupling that is {\em odd} in momentum $k$, due to inversion symmetry breaking \cite{Manchon2015,Winkler2003}. As a consequence, the spin texture in momentum space becomes antisymmetric in $k$ as shown in Fig. \ref{fig-Haney} for the Pt/Co interface. To keep the results tractable, theories usually consider simpler forms of odd-in-$k$ spin-orbit coupling, valid close to high symmetry points. For instance, strained zinc-blende crystals display a $k$-linear Dresselhaus spin-orbit coupling close to the $\Gamma$-point \cite{Dyakonov1986a},
\begin{equation}\label{eq:dresselk}
 \hat H_{\rm D}=\beta(\hat{\sigma}_xp_x-\hat{\sigma}_yp_y),
 \end{equation} 
 whereas interfaces display a so-called Rashba spin-orbit coupling \cite{Vasko1979,Bychkov1984}, 
\begin{equation}\label{eq:Rashbak}
 \hat H_{\rm R}=\alpha_{\rm R}\hat{\bm\sigma}\cdot({\bf p}\times{\bf z}).
 \end{equation} 
 The coefficients $\beta$ and $\alpha_{\rm R}$ are the Dresselhaus and Rashba parameters, respectively. In Weyl semimetals, the low energy bulk Hamiltonian directly connects the spin with the linear momentum, $\hat H_{\rm W}=v\hat{\bm\sigma}\cdot{\bf p}$ \cite{Weyl1929,Wan2011}. Although the spin-momentum locking scheme of real materials is in general much more complex (see Fig. \ref{fig-Haney}), these various forms of spin-orbit coupling have been widely used theoretically to study the SGE and iSGE.\par

SGE was first predicted by \citet{Ivchenko1989} and observed by \citet{Ganichev2001,Ganichev2002}. The iSGE has been predicted originally by \citet{Ivchenko1978}, followed by \citet{Aronov1989} and \citet{Edelstein1990}, and observed in non-centrosymmetric systems such as tellurium \cite{Vorobev1979}, strained semiconductors \cite{Kato2004b} and quantum wells \cite{Ganichev2004,Silov2004,Wunderlich2005,Wunderlich2004}. More recently, current-driven spin density has also been observed at the surface of transition metals \cite{Zhang2014f,Stamm2017}. In magnets lacking inversion symmetry, such as zinc-blende semiconductors \cite{Bernevig2005c,Garate2009,Hals2010}, or magnetic 2D electron gas with Rashba spin-orbit coupling \cite{Tan2007,Manchon2008b,Obata2008}, the current-driven spin density can be used to control the magnetic order parameter.\par


The iSGE-induced SOT can be derived from the dynamics of the carrier spin density {\bf S}, brought out of equilibrium by the applied electric field, in the presence of both magnetic exchange and spin-orbit interaction. For the sake of the discussion, let us consider the following model Hamiltonian
\begin{eqnarray}\label{eq:htot}
\hat{H}=\hat{H}_0+\hat{H}_{\rm ex}+\hat{H}_{\rm so},
\end{eqnarray} 
where $\hat{H}_0$ is the spin-independent part, $\hat{H}_{\rm ex}=(\Delta/2)\hat{\bm\sigma}\cdot{\bf m}$ is the $s$-$d$ exchange. The Heisenberg equation for the spin motion reads
\begin{eqnarray}
\label{s_dyn}
\frac{d{\bf S}}{dt}=\frac{\Delta}{\hbar}{\bf S}\times{\bf m}+\frac{1}{i\hbar}\langle[\hat{\bm\sigma},\hat{H}_{\rm so}]\rangle.
\end{eqnarray} 
Here $\langle\cdot\cdot\cdot\rangle$ represents quantum-mechanical averaging over the non-equilibrium carrier states and $\langle\hat{\bm\sigma}\rangle={\bf S}$. The SOT is obtained by taking the steady-state solution of Eq.~(\ref{s_dyn}) ($d{\bf S}/dt=0$) into Eq.~(\ref{eq:torque}),
\begin{equation}
{\bf T}=\frac{\Delta}{\hbar}{\bf m}\times{\bf S}=\frac{1}{i\hbar}\langle[\hat{\bm\sigma},\hat H_{\rm so}]\rangle.\label{SOT_eq}
\end{equation}
The right side of Eq.~(\ref{SOT_eq}) shows explicitly the spin-orbit coupling origin of the SOT. For discerning qualitatively distinct SOT contributions (i.e. the extrinsic versus intrinsic terms introduced in Subsection \ref{s:intrainter}), we will now use the middle expression.
 
\begin{figure}[h!]
\centering
\includegraphics[width=12cm]{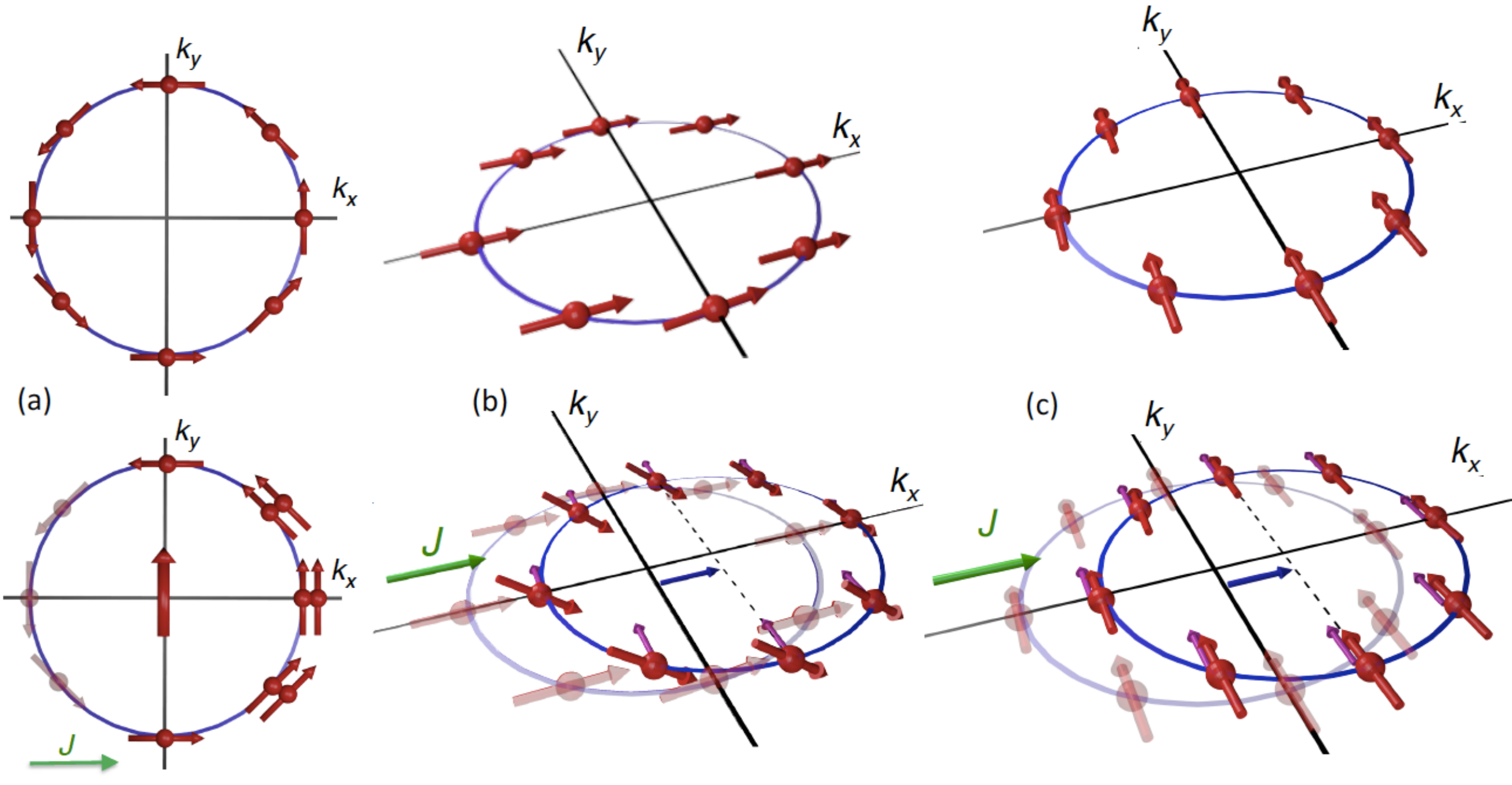}
\caption{\label{fig-iSGE-SOT} (Color online) (a) Top panel: Rashba spin-texture for one of the chiral states in equilibrium with zero net spin-density. Bottom panel: Non-equilibrium redistribution of eigenstates in an applied electric field resulting in a non-zero spin-density due to broken inversion symmetry of the spin-texture. When combined with the exchange coupling of the carrier spin-density to magnetization, this mechanism corresponds to the extrinsic (Boltzmann transport), field-like iSGE-SOT. (b) Top panel: A model equilibrium spin texture in a 2D Rashba spin-orbit coupled system with an additional time-reversal symmetry breaking exchange field of a strength much larger than the spin-orbit field. In equilibrium, all spins in this case align approximately with the $x$-direction of the exchange field (magnetization). Bottom panel: In the presence of an electrical current along the $x$-direction the
Fermi surface (circle) is displaced along the same direction. When moving in momentum space, electrons experience an additional spin-orbit field (purple arrows). In reaction to this non-equilibrium current induced field, spins tilt and generate a uniform, non-equilibrium out-of-plane spin density. (c) Top panel: Same as in (b) for $y$-direction of the exchange field. Bottom panel: Same as in (b) but now with the current induced spin-orbit field align with the exchange field, resulting in zero tilt of the carrier spins. (b) and (c) illustrate the intrinsic (Berry curvature) damping-like iSGE-SOT. Adapted from \citet{Kurebayashi2014}.}
\end{figure}

Let us first discuss the extrinsic (intraband) contribution to the spin density which, in the limit of spin-independent disorder, corresponds to the usual Boltzmann contribution. In the limit $\hat{H}_{\rm ex}\ll\hat{H}_{\rm so}$, this term is independent of the $s$-$d$ exchange \cite{Edelstein1990,Manchon2008b}. For illustration, we consider Rashba spin-orbit coupling, Eq. \eqref{eq:Rashbak}, such that the spins align perpendicular to the wavevector, ${\bm\sigma}_{\bf k}\sim {\bf z}\times{\bf k}$, as illustrated in Fig.~\ref{fig-iSGE-SOT}(a). In the absence of the electric field, $\langle {\bf k}\rangle=0$, and the equilibrium distribution of these eigenstate spin vectors adds up into a zero net spin density [Fig.~\ref{fig-iSGE-SOT}(a), top panel]. Under the applied electric field, however, the states are repopulated with a deficit/excess of left/right moving carriers with respect to the applied electric field [Fig.~\ref{fig-iSGE-SOT}(a), bottom panel]. The steady state non-equilibrium distribution is reached when balancing the carrier acceleration in the electric field with scattering against disorder, see Eq. \eqref{eq:intra}. Due to the non-centrosymmetric spin texture of the eigenstates, the non-equilibrium distribution leads to a non-zero net spin density aligned perpendicular to the electric field, ${\bf S}\sim \tau\alpha_{\rm R}\; {\bf z}\times{\bf E}$. In analogy to the Boltzmann theory of conductivity, the spin density is proportional to the momentum lifetime $\tau$ and, hence, associated with an extrinsic iSGE. Since we neglected $\hat H_{\rm ex}$ in the carrier Hamiltonian, the iSGE generated spin density ${\bf S}$ in this approximation is independent of {\bf m} and, when introduced into the middle expression of Eq.~(\ref{SOT_eq}), yields a field-like SOT, ${\bf T}_{\rm FL}\sim {\bf m}\times({\bf z}\times{\bf E})$. Incorporating the exchange field only creates small angular dependance of an otherwise constant spin density. 

 \onecolumngrid
 
 \twocolumngrid 
 
Let us now consider the intrinsic (interband) contribution, assuming the same Rashba spin-orbit coupling. Such a term is labeled intrinsic because it has a weak dependence on scattering in metallic systems. This contribution can be also derived from an intuitive picture of the Bloch dynamics of carrier spins. To do so we consider for simplicity the limit $\hat H_{\rm ex}\gg \hat H_{\rm so}$ (i.e., the opposite limit than considered above). In equilibrium, the carrier spins are then approximately aligned with the exchange field, ${\bm \sigma}_{\bf k}\approx s{\bf m}$, independent of their momentum. This is depicted in Figs.~\ref{fig-iSGE-SOT}(b,c), bottom panels, for ${\bf m}\parallel{\bf E}$ and ${\bf m}\bot{\bf E}$, respectively. The Bloch equations describe the carrier spin dynamics during their acceleration in the applied electric field, i.e., between the scattering events. Without loss of generality, we take $\mathbf{E}=E_x{\bf x}$. For $\mathbf{m}\parallel\mathbf{E}$, the equilibrium effective magnetic field acting on the carrier spins due to the exchange term is, $\mathbf{B}^{\rm eq}_{\rm eff}\approx(\Delta,0,0)$, in units of energy. During the acceleration in the applied electric field, $\frac{dp_x}{dt}=eE_x$, and the effective magnetic field acquires a time-dependent $y$-component due to $\hat H_{\rm so}$ for which $\frac{d B_{\rm eff,y}}{dt}=(\alpha_{\rm R}/\hbar)\frac{dp_x}{dt}$, as illustrated in Fig.~\ref{fig-iSGE-SOT}(b). For small tilts of the spins from equilibrium, the Bloch equation $\frac{d\mathbf{\bm\sigma}_{\bf k}}{dt}=\frac{1}{\hbar}(\mathbf{\bm\sigma}_{\bf k}\times \mathbf{B}_{\rm eff})$ yields $\sigma_{{\bf k},x}\approx s$, $\sigma_{{\bf k},y}\approx sB_{\rm eff,y}/B^{\rm eq}_{\rm eff}$, and
\begin{equation}
\sigma_{{\bf k},z}\approx -\frac{\hbar s}{(B^{\rm eq}_{\rm eff})^2}\frac{dB_{\rm eff,y}}{dt}=-\frac{s}{\Delta^2}\alpha_{\rm R} eE_x\;.
\label{s_zM_x}
\end{equation}
The non-equilibrium spin orientation of the carriers acquires a time and momentum independent $\sigma_{{\bf k},z}=\sigma_z$ component. For a general angle $\theta_{\mathbf{m-E}}$ between $\mathbf{m}$ and $\mathbf{E}$ we obtain,
\begin{equation}
\sigma_{{\bf k},z}(\mathbf{m})=\sigma_{z}(\mathbf{m})\approx \frac{s}{\Delta^2}\alpha_{\rm R} eE_x\cos\theta_{\mathbf{m-E}}\;.
\label{s_zM}
\end{equation}
The total non-equilibrium spin density, $S_z=2g_{2D}\Delta\sigma_{z}(\mathbf{m})$, is obtained by integrating $\sigma_{{\bf k},z}$ over all occupied states ($g_{2D}$ is the density of states). The non-equilibrium spin density produces an out-of-plane field which exerts a torque on the in-plane magnetization. From Eqs.~(\ref{SOT_eq}) and (\ref{s_zM}) we obtain an intrinsic damping-like SOT \cite{Kurebayashi2014},
\begin{equation}
{\bf T}_{\rm DL}=\frac{\Delta }{\hbar}(\mathbf{m}\times S_z{\bf z})\sim \mathbf{m}\times[\mathbf{m}\times(\mathbf{z}\times{\bf E})]\,.
\label{SOT_R}
\end{equation}

It is worth pointing out the analogy and differences between the intrinsic iSGE and the intrinsic SHE \cite{Murakami2003,Sinova2004}. In the SHE case where $\hat H_{\rm ex}=0$ in the paramagnet, $\mathbf{B}^{\rm eq}_{\rm eff}$ depends on the angle $\theta_{\mathbf{k}}$ of the carrier momentum with respect to $\mathbf{E}$ which implies a momentum-dependent $z$-component of the non-equilibrium spin,
\begin{equation}
\sigma_{{\bf k},z}\approx \frac{s}{\alpha_{\rm R} k^2}\alpha_{\rm R} eE_x\sin\theta_{\mathbf{k}}\;.
\label{s_zp}
\end{equation}
The same spin rotation mechanism that generates the uniform bulk spin density in the case of the intrinsic iSGE in a ferromagnet [Fig.~\ref{fig-iSGE-SOT}(b)] is responsible for the scattering-independent spin current of the SHE in a paramagnet \cite{Sinova2004}. Note that the SHE spin current yields zero spin density in the bulk while a net spin density accumulates only at the edges of the conduction channel. \par

\subsubsection{Inverse spin galvanic torque in a magnetic two-dimensional electron gas}

Because of the symmetry present in most bilayer systems considered in experiments, the Rashba spin-orbit coupling given by Eq. \eqref{eq:Rashbak} is the natural model to study, and therefore the iSGE-SOT has been alternatively called Rashba-SOT in this context. As discussed above, the Rashba torque can possess two components corresponding to the field-like and damping-like torques, see Eq. \eqref{eq:torquedef}. While the origin of the field-like torque is well understood and consistently attributed to the extrinsic intraband iSGE (Rashba-Edelstein) effect \cite{Edelstein1990}, several mechanisms contribute to the damping-like torque. The different contributions have been investigated using semiclassical Boltzmann transport equation \cite{Tan2007,Manchon2008b,Matos-Abiague2009c,Bijl2012,Kim2012b,Kim2013b,Lee2015}, or quantum-mechanical Kubo formula approaches \cite{Wang2012,Pesin2012,Wang2014,Li2015b,Qaiumzadeh2015,Freimuth2017c}.

 As discussed above, interband transitions produce an intrinsic damping-like torque in the limit of weak scattering \cite{Kurebayashi2014,Freimuth2014a}, and can be related to the Berry curvature of the electronic band structure in the mixed spin-momentum phase space \cite{Kurebayashi2014,Freimuth2014a,Lee2015,Li2015b}. As a result, one can expect "hot spots" in the band structure, i.e., points where neighboring bands get very close to each other, to give very large contribution, similarly to the case of intrinsic SHE \cite{Tanaka2008}. Notice that in the specific case of the pure 2D Rashba gas, at the first order in the Rashba parameter, vertex corrections cancel the intrinsic damping-like torque unless the momentum relaxation time is spin-dependent \cite{Qaiumzadeh2015,Freimuth2017c}, similar to the cancelation occurring for the intrinsic SHE in pure 2D Rashba gas \cite{Inoue2004}. Nevertheless, such cancellations are highly sensitive to this specific model band structure and can be considered as accidental, as discussed by \citet{Sinova2015} in the context of intrinsic SHE.
 
Extrinsic iSGE mechanisms related to spin scattering were also theoretically shown to generate a damping-like component of the SOT. In \cite{Kim2012b}, the damping-like term arises from the momentum-scattering induced spin relaxation, an effect initially proposed in metallic spin-valves and domain walls \cite{Zhang2002,Zhang2004}. In fact, when a non-equilibrium spin density ${\bf S}$ is injected into a magnet, spin relaxation generates a corrective term of the form $\sim\beta_{\rm sf}{\bf m}\times{\bf S}$, where $\beta_{\rm sf}$ is a parameter that depends on the ratio between spin precession and spin-flip scattering. In other works, this component is obtained within a quantum kinetic formalism and ascribed to spin-dependent carrier lifetimes or to a term arising from the weak-diffusion limit which in the leading order is proportional to a constant carrier lifetime \cite{Pesin2012,Wang2012b,Wang2014}. 

Finally, we stress out that the coefficients $\tau_{\rm FL,DL}$ are in principle angular dependent and display terms proportional to $\sin^{2n}\theta_{{\bf m}-{\bf E}}$, $n\in\mathbb{N}^*$. This angular dependence reflects the distortion of the band structure when changing the magnetization direction \cite{Lee2015}, as well as the anisotropic spin relaxation in the system due to D'yakonov-Perel's mechanism \cite{Ortiz2013}.

\subsubsection{Non-centrosymmetric bulk magnets}

\begin{figure}[h!]
\begin{center}
\includegraphics[width=8cm]{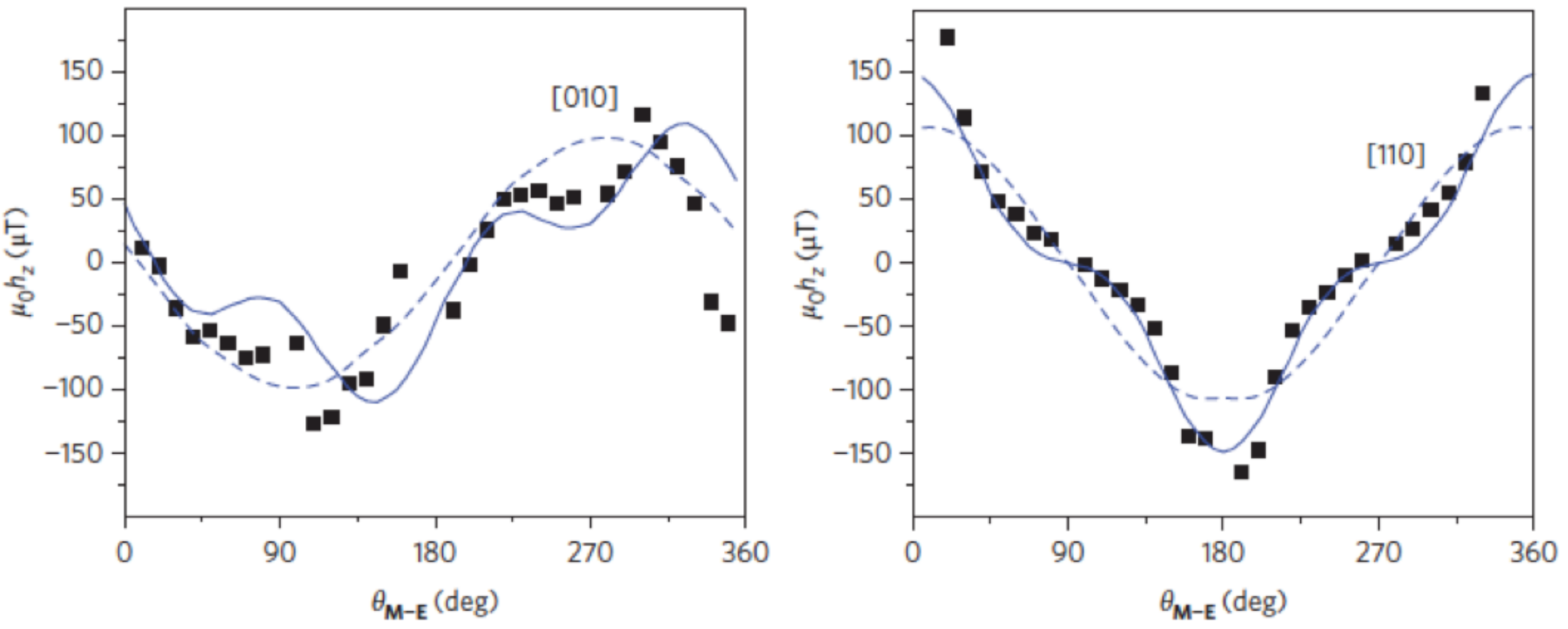}
\end{center}
\caption{(Color online) Microscopic modeling of the intrinsic SOT in bulk (Ga,Mn)As. The solid lines correspond to the numerical results and the symbols correspond to the experimental data. The dashed lines correspond to the free electron approximation. From \citet{Kurebayashi2014}.}
\label{fig:Kurebayashi}
\end{figure}

Dilute magnetic semiconductor (Ga,Mn)As has been a test-bed material for observing and exploring the bulk SOT. Hence, unlike the case of bilayer systems, all torques observed in these bulk materials arise internally with no contribution from externally injected spin currents. Current-driven torques in dilute magnetic semiconductors were first studied by \citet{Bernevig2005c}. The authors considered the Kohn-L\"uttinger Hamiltonian in the spherical approximation, augmented by a k-linear spin-orbit coupling term arising from strain of the form ${\bm \lambda}\cdot\hat{\rm{\bf J}}$, where $\hat{\rm{\bf J}}$ is the total angular momentum operator, $\lambda_x=C_4(\epsilon_{xy}k_y-\epsilon_{xz}k_z)$ and $\lambda_{y,z}$ are obtained from cyclic permutation of indices. The current-driven spin density reads
\begin{eqnarray}
{\bf S}=-\frac{e\tau}{\hbar^2}\frac{15}{4}\left(\sum_{s=\pm1}\frac{\sqrt{3n/\pi}}{(\gamma_1+2s\gamma_2)^{3/2}}\right)^{2/3}(e{\bf E}\cdot{\bm\nabla}_{\bf k}){\bm \lambda},\nonumber\\
\end{eqnarray}
where $n$ is the charge density. Because in their calculation they did not consider an exchange coupling directly, the torque induced by this iSGE is therefore a field-like torque. The intrinsic damping-like torque originating from interband transitions was later proposed by \citet{Kurebayashi2014} to interpret the experimental observation of such a damping-like torque in (Ga,Mn)As, see Fig. \ref{fig:Kurebayashi}. The theoretical investigation of SOT in dilute magnetic semiconductors was also pursued by \citet{Li2013,Li2015b}. Besides some subtleties related to the complex band structure, the numerical results obtained by these authors qualitatively confirm the general picture obtained in the context of the magnetic Dresselhaus and Rashba gas.

Apart from (Ga,Mn)As, the SOT has been studied in several other bulk systems. One of them is the ferromagnet NiMnSb \cite{Ciccarelli2016}. This half-heusler material has the same crystalline symmetry as zinc-blende (Ga,Mn)As; however, it is not a dilute-moment random alloy like (Ga,Mn)As, but a dense-moment ordered compound. Despite these differences, the SOTs found in NiMnSb are quite similar to those in (Ga,Mn)As because the torque is mostly determined by the lowest order terms in the magnetization expansion, Eq. \eqref{eq:chi_expansion}, which are the same for the two systems.

\subsubsection{Spin-orbit torques in magnetic textures\label{s:sot-texture-th}}

When itinerant electrons flow in a magnetic domain wall, their spin slightly misaligns with the local texture, resulting in STT \cite{Tatara2004,Zhang2004} (see Section \ref{s:DW} for details). When spin-orbit coupling is present, the flowing spins experience an additional precession about the spin-orbit field, resulting in enhanced spin torque \cite{Nguyen2007,Yuan2016,Stier2013,Stier2015}. An interesting consequence is the emergence of additional torque components that are proportional to the magnetization gradient \cite{Hals2013b}. Some of these contributions can be directly assigned to the presence of anomalous Hall effect and anisotropic magnetoresistance \cite{Bijl2012}, while other genuinely come from the interplay between magnetic texture and precession around the spin-orbit field. For instance, \citet{Kim2012b,Kim2013b} showed that in the presence of Rashba spin-orbit coupling, a 2D magnetic texture (i.e. a magnetic skyrmion or vortex) experiences a torque of the form ${\bf T}\propto ({\bm\nabla}\cdot{\bf m})[({\bf z}\times{\bf u})\cdot{\bm\nabla}]{\bf m}$, where ${\bf u}$ is the direction of injection. Such SOTs open interesting perspectives for the electrical manipulation of magnetic textures, discussed in Section \ref{s:DW}, but have received little attention to date.

\subsection{{\em Ab initio} modeling of spin-orbit torques in bilayer systems\label{s:dftsot}}

Although following after the studies in bulk magnets, the SOTs have been most extensively studied experimentally in NM/FM bilayer (or multilayer) structures (Section~\ref{MML}). Theories of SOT in bilayer systems based on iSGE and SHE as exposed in the previous sections present two major limitations. First, both mechanisms formally apply in very distinct situations: in the widely used diffusive model, SHE in the nonmagnetic metal is considered uniform, neglecting semiclassical size effects and possible variation of the spin Hall angle close to the interface \cite{Wang2016b,Freimuth2015}, as discussed in Subsection \ref{s:sheTorque}. In contrast, iSGE in magnetic multilayers is often modeled using the Rashba interaction, Eq. \eqref{eq:Rashbak}, which applies to 2D gases and ideally sharp interfaces only. Both approaches overlook the details of the interfacial orbital overlap, which can be quite subtle in transition metal interfaces and lead to enhanced orbital moment and related spin-orbit phenomena \cite{Blugel2007,Grytsyuk2016,Marmolejo-Tejada2017}.\par

\begin{figure}[h!]
\begin{center}
\includegraphics[width=6cm]{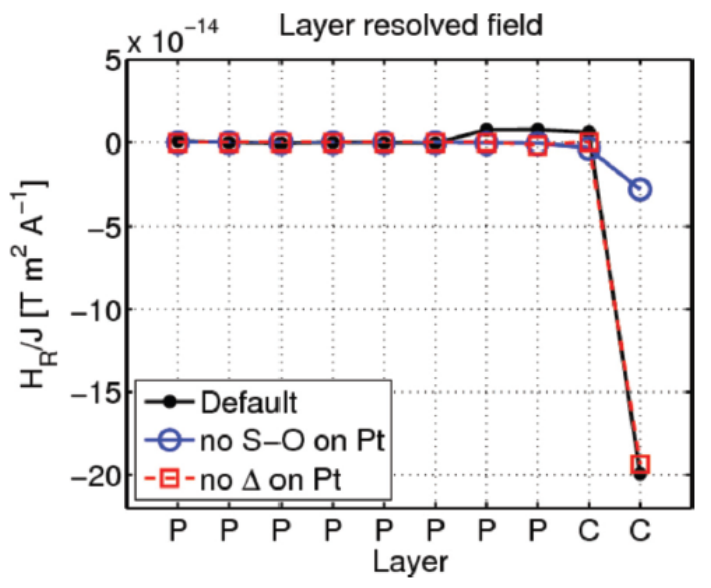}
\end{center}
\caption{(Color online) Layer-resolved field-like torque in normal Pt/Co(111) (black symbols), when turning off the induced magnetization on Pt (red symbols) and when turning off the spin-orbit coupling on Pt (blue symbols). From \citet{Haney2013a}}
\label{fig:Haney2}
\end{figure}

To overcome these issues, SOTs in Co/Pt bilayer systems have been computed within the relaxation time approximation using an {\em ab initio} density functional theory description of the whole bilayer structure \cite{Haney2013a,Freimuth2014a,Freimuth2015}. \citet{Haney2013a} focus on the extrinsic iSGE, disregarding the SHE and intrinsic contributions to iSGE. In spite of the high complexity of the band structure (see Fig. \ref{fig-Haney}), these calculations confirm the intuitive picture elaborated based on the Rashba model. In particular, they show that SOT is mostly driven by spin-orbit coupling in Pt, while the influence of induced magnetization is negligible, see Fig. \ref{fig:Haney2}. Moreover, the torque acquires a non-trivial angular dependence, and depends dramatically on the quality of the interface. Using a similar method, neglecting intrinsic contributions to both SHE and iSGE, \citet{Amin2018} computed the interfacial spin current in Co/Cu bilayers, obtaining both field-like and damping-like components, as well as an additional torque, ${\bf T}\propto {\bm\zeta}\times({\bf m}\times{\bm\zeta})$. These results are confirmed by \citet{Freimuth2018}.\par

\begin{figure}[h!]
\begin{center}
\includegraphics[width=8cm]{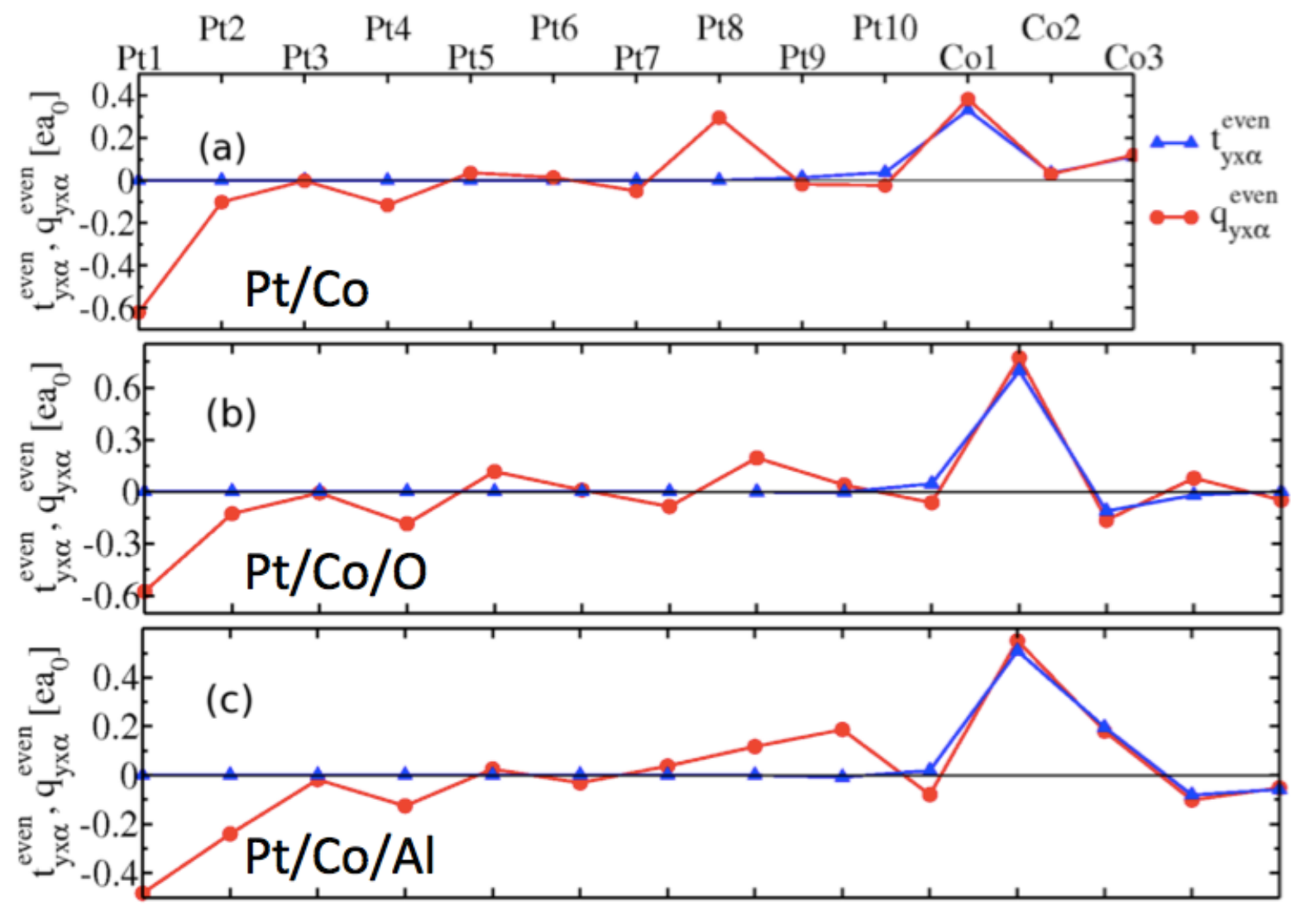}
\end{center}
\caption{(Color online) Layer-resolved damping-like torque $t^{\rm even}_{xy}$ (blue symbols) and spin current $q^{\rm even}_{xy}$ (red symbols) in (a) Pt/Co, (b) Pt/Co/O and (c) Pt/Co/Al. From \citet{Freimuth2014a}}
\label{fig:Freimuth}
\end{figure}

Alternatively, \citet{Freimuth2014a} and \citet{Geranton2015} computed the full Kubo-Bastin formula, thereby accounting for both intrinsic SHE and intrinsic iSGE. These calculations confirmed that SOTs are composed of both field-like and damping-like torques, the latter being produced by interband transitions only, see Fig. \ref{fig:Freimuth}. An interesting aspect revealed through these calculations is the high sensitivity of SOTs to interfacial engineering. In fact, the authors found that by capping the Co layer by either Al or O atoms, the damping-like torque is only slightly affected (its magnitude changes up to 50$\%$ - see Fig. \ref{fig:Freimuth}) while the field-like torque is dramatically altered and can even change its sign. In a follow-up work, \citet{Freimuth2015} reported an enhancement of SHE close to the interface, also predicted by \citet{Wang2016b} for Pt/NiFe. These studies suggest that the assumption of uniform spin Hall angle made in the diffusive approach may be valid in the strong disorder limit only. Similar results have been recently obtained by \citet{Mahfouzi2018a} in Co/Pt and Co/Pd bilayers.\par

Whereas all these studies are based on the relaxation time approximation (i.e., the effect of impurities is captured by a homogeneous broadening), \citet{Wimmer2016} calculated the torque in Pt/Fe$_x$Co$_{1-x}$/Cu superlattices using the Kubo-Bastin formula and coherent potential approximation to account for the alloy disorder. This allows for treating extrinsic scattering mechanisms (i.e., side jump and skew scattering) within the framework of density functional theory. The influence of impurities and phonon scattering on the SOT has been investigated by \citet{Geranton2016} within the Korringa-Kohn-Rostoker Green's function method. Finally, \citet{Belashchenko2019} investigated the angular dependence of the SOTs in Co/Pt bilayers using a two-terminal non-equilibrium Green's function approach with real-space Anderson disorder, uncovering a planar Hall-like contribution.

\subsection{Spin-orbit torques in antiferromagnets}\label{s:antiferromagnets}

Since the first proposal of STT in antiferromagnets \cite{Nunez2006,MacDonald2011}, several configurations have been theoretically investigated to enable large spin torque efficiency. In the course of the search for such torques it was realized that in order to efficiently manipulate the order parameter of a collinear, bipartite antiferromagnet one needs a torque that corresponds to a {\em staggered} effective magnetic field, i.e., a field with an opposite sign on the two magnetic sublattices. Such a field, unlike a homogeneous field, couples directly to the N\'eel order. The torque resulting from a staggered field has been referred to as N\'eel SOT \cite{Zelezny2014,Zelezny2017}.

The damping-like torque due to a spin current (from a SHE or a ferromagnetic polarizer) is a N\'eel torque, assuming that it has the same form on each magnetic sublattice as in ferromagnets, i.e., ${\bf T}_a = \tau_\text{DL} \mathbf{m} \times (\mathbf{m} \times {\bm\zeta})$. This form of the torque has been predicted theoretically \cite{Zelezny2014,Gomonay2012,Cheng2014c} and it was shown that it can indeed efficiently manipulate the antiferromagnetic order \cite{Gomonay2010, Gomonay2012}. A recent drift-diffusion theory confirmed that spin current injected from an adjacent ferromagnetic polarizer or induced by SHE indeed generates such a N\'eel damping-like torque \cite{Manchon2017}. An experimental indication of the presence of the SHE generated SOT in an NM/AF bilayer was reported in \citet{Reichlova2015}.

The bulk SOT can also have the N\'eel order form if the current-induced spin density has an opposite sign on the two sublattices \cite{Zelezny2014}. In a collinear antiferromagnet, the two sublattices with opposite magnetizations are connected by some symmetry operation. Typically, this is either a translation or an inversion. This symmetry operation combined with time-reversal is then a symmetry of the magnetic system which, using Eqs. \eqref{eq:chia_even_transa} and \eqref{eq:chia_odd_transa}, relates the current-induced spin densities on the two sublattices \cite{Zelezny2017}. If the sublattices are connected by translation then 
\begin{align}
 \chi^\text{even}_{B,a} &= \chi^\text{even}_{B,b},\\ 
 \chi^\text{odd}_{B,a} &= -\chi^\text{odd}_{B,b},
\end{align}
where $a,b$ denotes the two sublattices. If they are connected by inversion
\begin{align}
 \chi^\text{even}_{B,a} &= -\chi^\text{even}_{B,b},\\ 
 \chi^\text{odd}_{B,a} &= \chi^\text{odd}_{B,b}.
\end{align}
Thus in both cases there exists both a staggered component and a uniform component of the current-induced spin density and the corresponding effective field. For the magnetic dynamics only the staggered component will has an appreciable effect for the achievable magnitudes of the effective fields (a few mT). Since the component that is staggered is different in the two cases, the dynamics will differ. As discussed in Subsection \ref{s:sots-nut}, the even field typically has a field-like character, whereas the odd field is commonly damping-like. Thus in systems where magnetic sublattices are connected by translation we can expect an efficient damping-like torque, whereas in systems where the sublattices are connected by inversion a field-like torque is expected.

SOTs in antiferromagnets have been first studied in two tight-binding models \cite{Zelezny2014,Zelezny2017}: (i) the antiferromagnetic 2D Rashba gas and (ii) the bulk Mn$_2$Au. Both systems possess collinear antiferromagnetism. They together illustrate the two main types of symmetry discussed above. In the Rashba model the two sublattices are connected by translation and thus the lowest order N\'eel order SOT has a damping-like character. In the Mn$_2$Au crystal, on the other hand, the two sublattices are connected by inversion and the lowest order N\'eel torque has consequently a field-like character. Microscopic calculations based on the Kubo formula with constant relaxation time indeed show that the N\'eel SOT in the Rashba model is primarily damping-like, whereas in Mn$_2$Au it is predominantly of field-like character.

The origin of the field-like torque in the Mn$_2$Au crystal can be understood in terms of the symmetry of the nonmagnetic crystal. Without magnetism, the crystal has inversion symmetry and thus there is no net current-induced spin density. Yet, the Mn sublattices each have locally broken inversion symmetry and thus can have current-induced spin-densities that have to be precisely opposite. An intuitive explanation of this behavior is that the local inversion breaking is opposite for the two sublattices and thus the induced spin-densities are also opposite. When magnetism is added these opposite spin densities generate a staggered effective field. In the Rashba model on the other hand, the inversion breaking is the same for both sublattices and thus the field generating the field-like torque is not staggered. On the other hand, the field generating the damping-like torque is staggered, since it is proportional to the magnetic moment which is staggered in the antiferromagnet. The field-like torque in Mn$_2$Au has a Rashba-like symmetry, i.e., the effective field is on each sublattice proportional to ${\bm \zeta}$. This is because the local symmetry of the Mn sublattices is the same as that of the 2D Rashba model. Additional symmetry analysis for various types of crystalline antiferromagnets has been provided by \citet{Watanabe2018} and \citet{Zelezny2017}.

Following the calculations based on tight-binding models, the SOT was also calculated in antiferromagnets using density functional theory. Such calculations were done for Mn$_2$Au \cite{Zelezny2017} and CuMnAs \cite{Wadley2016}, which has a symmetry analogous to Mn$_2$Au. These results agree well with the tight-binding calculations in term of the magnetization and current dependence and in addition show a relatively large torque. The magnitude of the effective field is around 2 mT per 10$^7$ Acm$^{-2}$ current density for Mn$_2$Au and 3 mT per 10$^7$ Acm$^{-2}$ for CuMnAs. The switching attributed to this field-like N\'eel order torque has been observed in CuMnAs \cite{Wadley2016} and subsequently in Mn$_2$Au \cite{Bodnar2018,Meinert2018,Zhou2018b}.

\subsection{Spin-orbit torques in topological insulators\label{s:sot-ti}}

Topological insulators are a class of materials displaying intriguing properties such as insulating bulk and conductive chiral and helical surfaces \cite{Qi2011,Hasan2011,Wehling2014}. Considering the large spin-charge conversion efficiency recently reported in these systems (see Subsection \ref{MML:exp:FMTI}), they deserve special attention. The category of topological materials we are interested in are characterized by time-reversal symmetry and helical surface states: their low energy surface states are represented by a Dirac Hamiltonian of the form $\sim p_i\sigma_j$ (see Fig. \ref{fig-TI}). When electrons flow on the surface of these systems, they acquire a non-equilibrium spin density, similar to the case of the 2D Rashba gas, as demonstrated in a Bi$_2$Se$_3$ slab using {\em ab initio} calculations \cite{Chang2015b}. Since the strength of the spin-momentum coupling is quite large ($\sim4\times10^{-10}$ eV$~$m at Bi$_2$Se$_3$ surfaces, comparable to Bi/Ag surfaces, and two orders of magnitude larger than in InAlAs/InGaAs 2D gases), iSGE is expected to be very large. In addition, the absence of bulk conduction in ideal topological insulators further strongly enhances the spin-charge conversion efficiency.\par

Spin-charge conversion processes in topological insulator/insulating ferromagnet bilayers have been studied by several authors \cite{Taguchi2015,Sakai2014,Fujimoto2014,Tserkovnyak2015,Linder2014}. The low energy Hamiltonian reads $\hat{H}=v\hat{\bm\sigma}\cdot(\hat{\bf p}\times{\bf z})+\frac{\Delta}{2}\hat{\bm\sigma}\cdot{\bf m}$, where the first term models the Dirac cone and the second term is the exchange. This model applies when the Dirac states are preserved, so typically when the topological insulator surfaces are interfaced with magnetic insulators \cite{Lang2014,Li2015f,Katmis2016}. The eigenenergies read
\begin{eqnarray}
\epsilon_k^s=s\sqrt{(vk_x+\frac{\Delta}{2} m_y)^2+(vk_y-\frac{\Delta}{2} m_x)^2+\frac{\Delta^2}{4}m_z^2}.
\end{eqnarray}
When $m_z\neq0$ the surface states are gapped, whereas when $m_z=0$, the origin of the band dispersion is only shifted in the {\bf k}-plane. If the Fermi energy lies in the gap, quantum anomalous Hall effect emerges, accompanied by a quantized magnetoelectric effect, ${\bf S}=-\frac{e\hbar}{2\pi v}{\bf E}$ \cite{Qi2008,Nomura2011}. On the other hand, when the Fermi level lies above the gap, the system is metallic and the SOT possesses both field-like and damping-like components of the form \cite{Garate2010b,Ndiaye2017b}
\begin{eqnarray}
{\bf T}=\tau_{\rm FL}{\bf m}\times{\bm\zeta}+\tau_{\rm DL}m_z{\bf m}\times({\bf z}\times{\bm\zeta}),
\end{eqnarray}
where ${\bm\zeta}||{\bf z}\times{\bf E}$. While the field-like torque arises from the conventional extrinsic iSGE, the damping-like torque arising from the intrinsic interband contribution is proportional to $m_z$ and therefore {\em vanishes} when the magnetization lies in the plane of the surface, in sharp contrast with the usual damping-like torque given in Eq. (\ref{eq:torquedef}) \cite{Ndiaye2017b}. \par

The calculations discussed above are based on the 2D Dirac gas model, i.e, assuming that the transport is confined to the interface and that surface states remain intact in the presence of the proximate ferromagnetic layer. Such a model presents two major drawbacks though. First, orbital hybridization between the transition metal and the topological insulator substantially alters the surface states at Fermi energy. The presence of magnetic adatoms shifts the Dirac cone downward in energy \cite{Honolka2012,Scholz2012,Ye2012b}, and favor the presence of additional metallic bands with Rashba-like character across the Fermi level \cite{Zhang2016d,Marmolejo-Tejada2017}. 

A second limitation comes for the 3D nature of the transport. Indeed, most experiments involve topological insulators with sizable bulk conductivity, suggesting that bulk states might participate to the spin-charge conversion mechanism. Spin transport in such systems has been recently investigated using drift-diffusion model \cite{Fischer2016}, non-equilibrium Green's function technique \cite{Mahfouzi2016} and Kubo formula on a slab geometry \cite{Ghosh2017c}. The first two studies show that spin diffusion in the ferromagnet and spin-flip scattering at the interface can enhance the damping-like torque. The latter work accounts for interfacial and bulk transport on equal footing and demonstrates that a large damping-like torque is driven by the Berry-curvature of the interfacial states, whereas the SHE of the bulk states is inefficient.\par

Finally, spin-orbit charge pumping, the reciprocal effect of SOT, has also been investigated theoretically in topological insulators \cite{Ueda2012,Mahfouzi2014}, providing a charge current of the form ${\bf j}_{\rm c}=\chi_{\rm DL}m_z\frac{\partial}{\partial t}{\bf m}+\chi_{\rm FL}{\bf z}\times\frac{\partial}{\partial t}{\bf m}$~\cite{Ndiaye2017b}, where $\tau_{\rm DL, FL}=\chi_{\rm DL, FL}E$. A direct consequence of this current is the induction of an anisotropic magnetic damping on the ferromagnetic layer \cite{Yokoyama2010}. 


SOT and spin-orbit charge pumping have also been studied in various configurations involving 2D topological insulators \cite{Soleimani2017}. These studies reveal that SOT experiences a significant enhancement depending on the topological phase \cite{Mahfouzi2010,Li2016b}: the emergence of edge currents promotes a quantized charge pumping when the magnetization is perpendicular to the plane. Such investigations have been recently extended to antiferromagnetic 2D topological insulators, where time-reversal combined with a half unit cell translation is a symmetry of the system which preserves topological protection, despite the broken time-reversal symmetry of the magnetic state \cite{Ghosh2017b}.

SOTs have also been theoretically studied in magnetic 2D hexagonal lattices such as, but not limited to, graphene, silicene, germanene, stanene, transition metal dichalcogenides etc. \cite{Dyrdal2015,Li2016}. The parametric dependencies of the torque in these materials do not significantly differ from the one obtained with the Rashba model. Nonetheless, in these systems the low-energy transport occurs mostly through two independent valleys, which opens the possibility to obtain valley-dependent SOTs.

\subsection{Other spin-orbit torques\label{s:th-others}}

\subsubsection{Anisotropic magnetic tunnel junctions}

\begin{figure}[h!]
\centering
\includegraphics[width=8cm]{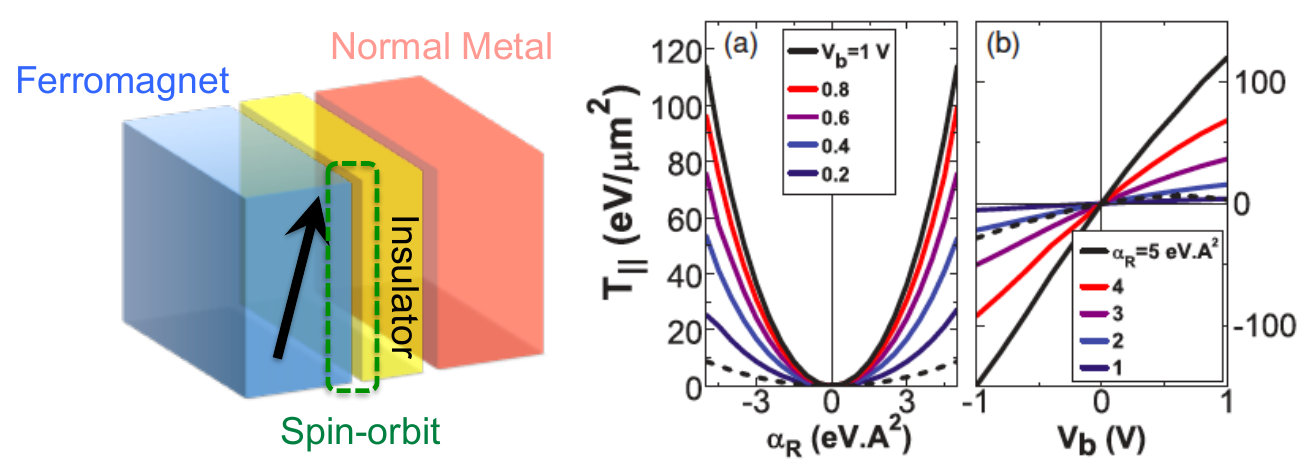}
\caption{\label{fig:fin} (Color online) Left: Schematics of a MTJ composed of a ferromagnet and a nonmagnetic metal separated by a tunnel barrier. Spin-orbit coupling is present at the interface between the ferromagnet and the tunnel barrier. Right: Rashba dependence (a) and bias dependence (b) of the damping-like torque. Adapted from \citet{Manchon2011a}}
\end{figure}

MTJs composed of a single ferromagnet with interfacial spin-orbit coupling display tunneling anisotropic magnetoresistance, i.e., a change of resistance when varying the magnetization direction \cite{Gould2004,Park2008}, see Fig. \ref{fig:fin}(a). One naturally expects that spin-polarized electrons impinging on the spin-orbit coupled interface precess about the spin-orbit field, resulting in a torque on the local magnetization \cite{Manchon2011a}. This SOT is of the form given by Eq. \eqref{eq:torquedef} with ${\bm \zeta}={\bf z}$. The field-like torque possesses an equilibrium contribution [which is nothing but the perpendicular magnetic anisotropy \cite{Manchon2011b}] and the damping-like torque is purely non-equilibrium. Both torques are linear as a function of the bias voltage, but their magnitude is quadratic in the Rashba parameter, see Fig. \ref{fig:fin}. A similar idea has been proposed by \citet{Mahfouzi2012b} by considering a topological insulator as a tunnel barrier.
 
%
%

\subsubsection{Spin-transfer torque assisted by spin-orbit coupling}
 
\begin{figure}[h!]
\centering
\includegraphics[width=8cm]{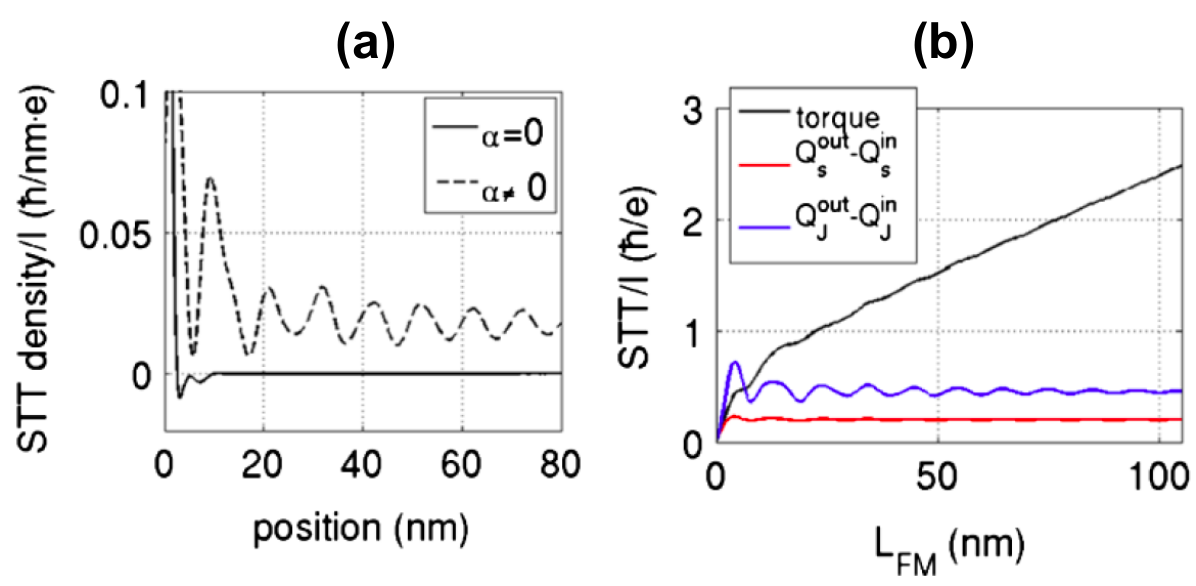}
\caption{\label{fig:haney} (Color online) (a) Spatial profile of the transverse spin density injected in the free layer of a (Ga,Mn)As-based MTJ in the presence and absence of spin-orbit coupling in the band structure; (b) Spin torque efficiency as a function of the free layer thickness. Adapted from \citet{Haney2010}.}
\end{figure}

When a spin-polarized current penetrates into a ferromagnet with spin-orbit coupling, the spin momentum precesses around an effective field that is the sum of the exchange and spin-orbit fields. This precession results in additional angular dependences of the SOT in Rashba \cite{Lee2015} and Kohn-L\"uttinger systems as discussed above \cite{Kurebayashi2014,Li2013}. Interestingly, \citet{Haney2010} showed that in a metallic spin-valve where spin-orbit coupling is present, such a precession results in an overall STT enhancement. Considering the general Hamiltonian, Eq. (\ref{eq:htot}), with $\hat{H}_{\rm so}=\xi \hat{\bf L}\cdot\hat{\bm\sigma}$, the total angular momentum $\hat{\bf J}=\hat{\bf L}+\hat{\bm\sigma}$ obeys the continuity equation
\begin{eqnarray}
\frac{d\hat{\bf J}}{dt}-{\bm\nabla}\cdot \bm{\mathcal J}_{\bf J}=-\hat{\bm\tau}_{\rm STT}-\hat{\bm\tau}_{\rm lat},
\end{eqnarray}
where $\bm{\mathcal J}_{\bf J}$ is the current density tensor for the total angular momentum, $\hat{\bm\tau}_{\rm STT}$ is the STT and $\hat{\bm\tau}_{\rm lat}=i\langle[\hat{H},\hat{\bf L}]\rangle/\hbar$ is the mechanical torque. The latter is nothing but the precession of itinerant spins about the spin-orbit field, such that the total spin torque in a spin-valve survives away from the interface, see Fig. \ref{fig:haney}(a). Due to this additional precession, the total torque extends over the whole thickness of the free layer, as displayed in Fig. \ref{fig:haney}(b). A similar effect has been identified in magnetic domain walls where spin-orbit coupling enhances spin reflection and thereby STT \cite{Nguyen2007,Yuan2016}.

\subsection{Open theoretical questions}

After a decade of theoretical progress, the key mechanisms giving rise to SOTs are now well understood at the qualitative level. Although the involved mechanisms are very different (SHE, iSGE etc.), they can all be unified using Eq. \eqref{eq:torque}, i.e. any SOT can be associated with a current-induced spin density. The various mechanisms differ in the way this spin density is generated. In the iSGE, the spin density is generated locally, whereas in the other mechanisms (SHE, spin swapping etc.), it is due to spin currents that transfer spin angular momentum from one part of the system to another. In addition, when a magnetic texture is present, the SOTs acquire new components that depend on the spatial gradient of the magnetization. The key ingredients in all these mechanisms are inversion symmetry breaking and the spin-orbit coupling. The general form of the torques can be determined using symmetry analysis.\par

Nonetheless, a number of challenges remain to be addressed on the theory side. First of all, {\em quantitative} agreement between theory and experiments is still missing. While important progress has been made using density functional theory \cite{Freimuth2014a}, most calculations adopt a simplified, unrealistic model of scattering. Including realistic disorder (structural imperfections, interfacial roughness), but also scattering from phonons and magnons will certainly improve such calculations. Moreover, understanding the interplay between bulk transport and interfacial effects, as well as the impact of interfacial orbital hybridization in magnetic multilayers will lead to the design of better SOT devices. These tasks require the development of accurate first-principles quantum transport methods \cite{Nikolic2018}.\par

Comprehensive first-principles models would be particularly useful in the context of novel materials. Among these, topological materials such as topological insulators and Weyl semimetals display exotic surface states with strong spin-orbit coupling and are regarded as promising for SOT generation. However, we lack an accurate understanding of how proximate transition metals modify these surface states \cite{Zhang2016d,Marmolejo-Tejada2017} and how bulk and surface transport cooperate to produce large SOTs \cite{Ghosh2018}. Another important class of materials exhibiting remarkable properties is the antiferromagnets. While the basic principles of SOTs and current-driven dynamics are understood, a proper description of the magnetic texture and dynamics of realistic, disordered antiferromagnets is still lacking. In addition, non-collinear antiferromagnets such as Mn$_3$X compounds, host a variety of novel phenomena, such as anomalous Hall effect (AHE) \cite{Nakatsuji2015} or "magnetic SHE" \cite{Zelezny2017b,Kimata2019}, that could be exploited in the context of SOT. These various aspects call for further theoretical endeavor.\par

Although the single-particle density functional theory provides a good description of the electronic structure of most transition metals and semiconductors, it fails in describing strongly correlated systems, such as Mott or Kondo insulators \cite{Cohen2008,Jones2015}. Utilizing more advanced many-body approaches, such as the dynamical mean field theory, might thus be necessary for accurate description of the SOT in such materials. These calculations are, however, numerically very expensive. We note also that Eq. \eqref{eq:torque} relies on a density functional theory description, and a more general expression that would be valid even in a many-body system has not been established yet.

The SOT in bilayer systems is often explained in terms of a spin current, which provides useful (but sometimes misleading) insight into the physics at stake. This concept is, however, controversial on a theoretical level. In the presence of spin-orbit coupling, the spin angular momentum is not a conserved quantity and adopting the conventional definition of the spin current tensor, $j_{{\rm s},i}^j \sim\{\hat{s}_i,\hat{v}_j\}$, can give rise to dissipationless equilibrium spin currents \cite{Rashba2003c}. For instance, in centrosymmetric crystals \citet{Shi2006} circumvent this hurdle by defining the spin current tensor $j_{{\rm s},i}^j \sim (d/dt)\{\hat{s}_i,\hat{r}_j\}$, accounting for a "torque dipole" term that ensures overall spin conservation. Until now, this new definition has not been widely adopted and the question of its applicability to heterostructures remains open. Another way to consider this problem would be to compute the current of total angular momentum \cite{An2012}. However, doing so complicates the problem because the total angular momentum of the electronic system alone is not conserved as it interacts with phonons and magnons. 

Along this line of thought, the orbital analogs of spin phenomena, such as the orbital Hall effect \cite{Tanaka2008,Go2018} and the orbital iSGE \cite{Yoda2018}, have attracted attention, but their possible connection to the SOT is yet to be explored. Unlike the SHE and the iSGE, the orbital effects exist even in the absence of spin-orbit coupling. When spin-orbit coupling is present, these effects can in principle couple to the magnetic moments and thus contribute to the SOT \cite{Go2019}. This indicates a route for the optimization of SOT via orbital engineering. The recent claim of ``maximal'' Rashba spin-splitting suggests a direction towards this end \cite{Sunko2017}.


\section{Spin-orbit torques in magnetic multilayers}
\label{MML}
This section reviews recent experimental progress in the measurement and characterization of SOT in multilayer systems. We first introduce the phenomenological description of SOT commonly used in experiments (Subsection \ref{MML:phen}) and the main techniques employed to measure SOT (Subsection \ref{MML:measurements}). Next, we present a survey of different layered materials, namely nonmagnetic metals, semiconductors, and topological insulators coupled to either ferromagnets, ferrimagnets, or antiferromagnets (Subsection \ref{MML:exp}), summarizing the most salient features of the SOT observed in these systems. Finally, we describe the SOT-induced magnetization dynamics (Subsection \ref{MML:dynamics}) and switching (Subsection \ref{MML:switching}), and conclude by highlighting examples and perspectives for the implementation of SOT in magnetic devices (Subsection \ref{MML:devices}).

\begin{figure}
	\centering
	\includegraphics[width=8.5 cm]{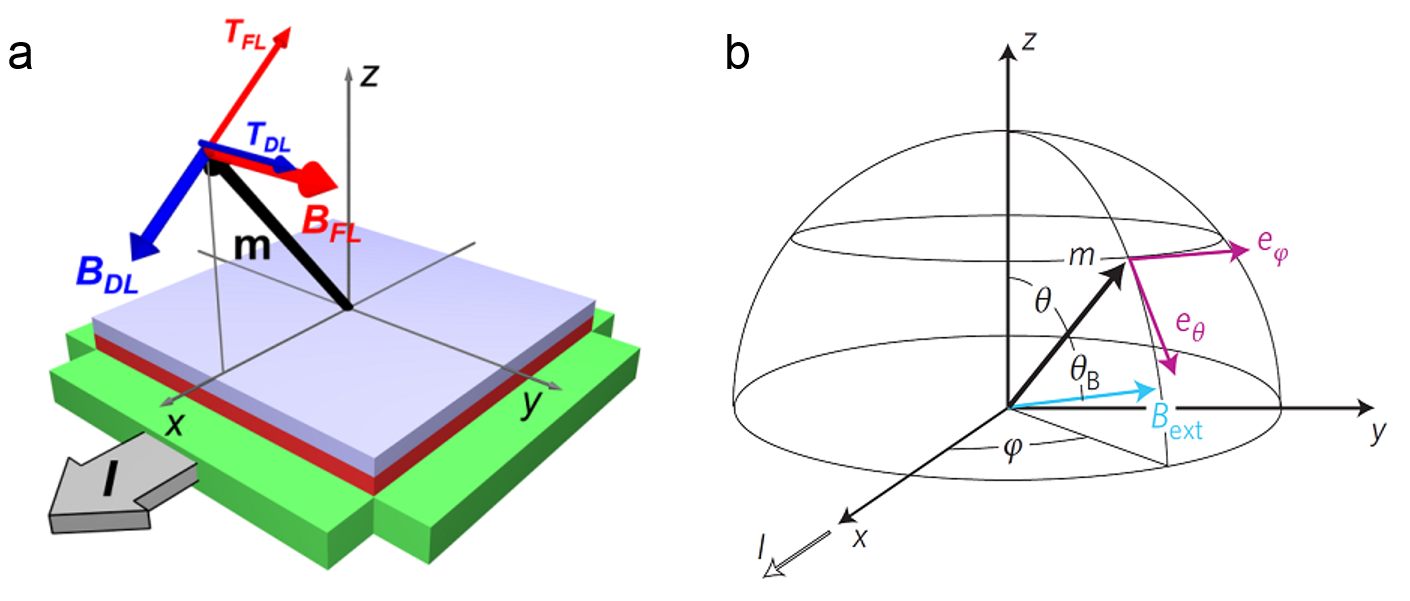}
	\caption{(Color online) (a) Spin-orbit torques and corresponding effective fields measured in Pt/Co/AlO$_x$ layers when the magnetization is tilted parallel to the current direction. (b) Schematic of the coordinate system.}
\label{Fig_SOTgeometry}
\end{figure}

\subsection{Phenomenological description} \label{MML:phen}

Current-injection in heterostructures composed of a magnetic layer adjacent to a nonmagnetic conductor with either bulk or interfacial spin-orbit coupling gives rise to a transverse spin density ${\bm\zeta} \, || \, {\bf z}\times {\bf j}_{\rm c}$ at the interface of the magnetic layer. This spin density induces both damping-like and field-like SOT components, as shown in Fig.~\ref{Fig_SOTgeometry}(a), and described by Eq.~\eqref{eq:torquedef}. For experimental purposes, it is useful to introduce two effective magnetic fields, $\mathbf{B}_{\rm DL,FL}$, which correspond to the damping-like and field-like torques and are defined by $\mathbf{T}_{\rm DL,FL} = \mathbf{M} \times \mathbf{B}_{\rm DL,FL}$. The advantage of the effective field formulation in the SOT characterization is that their action on the magnetization can be directly compared to that of a reference external field of known magnitude and direction. To the lowest order in the magnetization, for a current ${\bf j}_{\rm c}||{\bf x}$ and assuming C$_v$ symmetry (see below), Eq.~\eqref{eq:torquedef} gives
\begin{eqnarray}\label{eq:fieldEq1}
&&{\bf B}_{\rm FL}= B_{\rm FL}\, \mathbf{y},\\
&&{\bf B}_{\rm DL}= B_{\rm DL} \, \mathbf{m}\times \mathbf{y},\label{eq:fieldEq2}
\end{eqnarray}
where the field amplitudes are simply $B_{\rm FL}=\tau_{\rm FL}$ and $B_{\rm DL}=\tau_{\rm DL}$ if the torques are calculated for the unitary magnetization ${\bf m}$, as assumed in Eq.~\eqref{eq:torquedef}.\footnote{To emphasize the direction of the effective fields in perpendicularly magnetized layers, $B_{\rm FL}$ and $B_{\rm DL}$ are sometimes called "transverse field" ($H_{\rm T}$) and "longitudinal field" ($H_{\rm L}$), respectively \cite{Kim2013}.} Thus, for positive values of the SOT coefficients $\tau_{\rm FL}$ and $\tau_{\rm DL}$, ${\bf B}_{\rm FL} || \, {\bf y}$
whereas ${\bf B}_{\rm DL}$ rotates clockwise in the $xz$ plane, corresponding to ${\bf T}_{\rm DL} \, || \, - {\bf y}$.
Figure~\ref{Fig_SOTgeometry}(a) shows the orientation of the torques and effective fields for the model system Pt/Co/AlO$_x$, in which $\tau_{\rm FL}>0$ and $\tau_{\rm DL}<0$ \cite{Garello2013}. Typical values of $B_{\rm FL,DL}$ in NM/FM systems are in the range 0.1-10~mT for a current density $j_{\rm c} =10^7$~A/cm$^2$. Note also that the Oersted field due to the current flowing in the nonmagnetic layer produces an additional field $B_{\rm Oe} \approx \mu_0 j_{\rm c} t_N / 2$ antiparallel (parallel) to ${\bf y}$ for bottom (top) stacking relative to the magnetic layer. \par

In the typical NM/FM bilayer geometry shown in Fig.~\ref{Fig_SOTgeometry}(a), the SOTs are interfacial torques whose magnitude, to a first approximation, does not depend on the thickness $t_{F}$ of the ferromagnet. However, the measurable effects of the SOTs on the magnetization, namely $B_{\rm FL}$ and $B_{\rm DL}$, scale inversely with $t_{F}$ because the magnetic inertia is proportional to the volume of the ferromagnet. Keeping into account the proportional relationship between SOTs and injected current, it is thus useful to define the spin torque efficiencies
\begin{equation}
\label{eqSOT_j}
\xi^{j}_{\rm DL,FL} = \frac{2e}{\hbar} M_{\rm s} t_{F}\frac{B_{\rm DL,FL}}{j_{\rm c}},
\end{equation}
where $M_{\rm s}$ is the saturation magnetization. The parameters $\xi^{j}_{\rm DL,FL}$ represent the ratio of the effective spin current absorbed by the ferromagnet relative to the charge current injected in the nonmagnetic metal layer, and can thus be considered as effective spin Hall angles for a particular combination of nonmagnetic metal and ferromagnet. In the pure SHE-SOT picture, $\xi^{j}_{\rm DL}$ is equal to the bulk spin Hall angle of the nonmagnetic metal in the limit of a transparent interface and negligible spin memory loss. Although the SOT efficiencies are useful parameters to compare the strength of the SOT in different systems, ambiguities remain on how to estimate $j_{\rm c}$ in layered heterostructures. While some authors consider $j_{\rm c}$ to be the average current density, others apply a parallel resistor model to separate the currents flowing in the nonmagnetic metal and ferromagnetic layers. However, thickness inhomogeneities and interface scattering can significantly alter the current distribution in bilayer systems relative to such a model \cite{Chen2017b}. Even in homogeneous films, the conductivity is a strong function of the thickness \cite{Fuchs1938,Sambles1983} so that $j_{\rm c}$ changes in the bulk and interface regions of a conductor. For these reasons, the current normalization should be performed with care. Alternatively, it is possible to measure the torque efficiency per unit electric field~\cite{Nguyen2016}
\begin{equation}
\label{eqSOT_E}
\xi^{E}_{\rm DL,FL} = \frac{2e}{\hbar} M_{\rm s} t_{F}\frac{B_{\rm DL,FL}}{E} =\xi^{j}_{\rm DL,FL}/\rho,
\end{equation}
where $E=\rho j_{\rm c}$ is the electric field driving the current, which is independent of the sample thickness and can be easily adjusted in voltage-controlled experiments. Note that, in the framework of the SHE-SOT model, $\xi^{E}_{\rm DL}$ can be considered as an effective spin Hall conductivity.\par

Equations~\eqref{eq:fieldEq1} and (\ref{eq:fieldEq2}) correspond to the lowest order terms of the SOT, which are sufficient to describe many experimental results, at least on a qualitative level. On a more general level, however, higher order terms in the magnetization are allowed by symmetry. The typical polycrystalline metal bilayers have C$_{v}$ symmetry, corresponding to broken inversion symmetry along the ${\bf z}$-axis and in-plane rotational symmetry. For such systems, the torques can be decomposed into the following terms \cite{Garello2013},
\begin{eqnarray}\label{eq:angledt}
&&{\bf T}_{\rm FL}=[\tau_{\rm FL}^{\{0\}}+\sum_{n\geq1} \tau_{\rm FL}^{\{2n\}}(\sin\theta)^{2n}]{\bf m}\times{\bf y}\\
&& \qquad +{\bf m}\times({\bf z}\times{\bf m})m_x\sum_{n\geq1} \tau_{\rm FL}^{\{2n\}}(\sin\theta)^{2(n-1)},\nonumber\\\label{eq:angleft}
&&{\bf T}_{\rm DL}=\tau_{\rm DL}^{\{0\}} {\bf m}\times({\bf m}\times{\bf y}) \\
&&\qquad +m_x{\bf z}\times{\bf m}\sum_{n\geq1} \tau_{\rm DL}^{\{2n\}}(\sin\theta)^{2(n-1)},\nonumber
\end{eqnarray}
where $\theta$ is the polar angle of the magnetization defined in Fig.~\ref{Fig_SOTgeometry}(b). This formula is general and does not depend on the particular mechanism, SHE or iSGE, responsible for the spin density. In a material displaying additional symmetries, such as epitaxial films or single crystals, additional angular dependencies arise \cite{Hals2013,Wimmer2016,Zelezny2017}. This complex dependence of the SOT on the magnetization direction is best captured by writing the effective fields in spherical coordinates,
\begin{eqnarray}
\mathbf{B}_{\rm DL} = B_{\rm DL}^\theta \cos\varphi \, \mathbf{e}_{\theta} - B_{\rm DL}^\varphi \cos\theta\sin\varphi \, \mathbf{e}_{\varphi}, \label{eq:fieldEq3}\\
\mathbf{B}_{\rm FL} = -B_{\rm FL}^\theta\cos\theta\sin\varphi \, \mathbf{e}_{\theta} -B_{\rm FL}^\varphi \cos\varphi \, \mathbf{e}_{\varphi},\label{eq:fieldEq4}
\end{eqnarray}
where $\mathbf{e}_{\theta}$ and $\mathbf{e}_{\varphi}$ are the polar and azimuthal unit vectors, respectively, and $B_{DL, FL}^{\theta}$ and $B_{\rm DL, FL}^{\varphi}$ are functions of the magnetization orientation, defined by the angles $\theta$ and $\varphi$ [see Fig.~\ref{Fig_SOTgeometry}(b)]. In polycrystalline bilayers with C$_{2v}$ symmetry, the angular dependence of the polar components simplifies to a Fourier series expansion of the type $B_{\rm DL,FL}^\theta = B_{\rm DL,FL}^{\{0\}} + B_{\rm DL,FL}^{\{2\}} \sin^2\theta + B_{\rm DL,FL}^{\{4\}} \sin^4\theta + ...$. The azimuthal components, on the other hand, are found by experiments to be only weakly angle-dependent and are approximated by $B_{\rm DL,FL}^\varphi \approx B_{\rm DL,FL}^{\{0\}}$~\cite{Garello2013}.\par

\subsection{Measurement techniques} \label{MML:measurements}

Experimental measurements of SOT rely on probing the effect of the electric current on the orientation of the magnetization, e.g., by inducing resonant and nonresonant oscillations, switching, or domain wall motion. Schematically, one must first determine the magnetization angle as a function of the amplitude and phase of the applied current and, second, extract the effective magnetic fields that are responsible for the observed dynamics. In electrical and optical measurements, the magnetization dynamics is detected through changes of the transverse or longitudinal conductivity, which are mainly due to the AHE and anisotropic magnetoresistance (AMR), but include also the linear spin Hall magnetoresistance (SMR) \cite{Nakayama2013,Kim2016}, linear Rashba magnetoresistance \cite{Kobs2011,Nakayama2016}, as well as nonlinear magnetoresistance terms proportional to the current-induced spin density \cite{Avci2015a,Avci2015b,Avci2018,Olejnik2015,Yasuda2016,Yasuda2017}.
Further, current injection always results in magnetothermal effects due to the thermal gradients induced by Joule heating and asymmetric heat dissipation \cite{Avci2014b}, which affect the conductivity proportionally to $j_{\rm c}^2$. The thermal gradients that develop along ($\nabla_x T$) or perpendicular to the magnetic layer ($\nabla_z T$) contribute to the conductivity through the anomalous Nernst effect (ANE) and, to a smaller extent, through the spin Seebeck effect and the inverse spin Nernst effect. The direction of the induced voltage is $\sim {\bm\nabla}T\times {\bf m}$, which modifies both the longitudinal ($\sim m_y$) and transverse conductivities ($\sim m_x$). The relative influence of the above effects on SOT measurements depends on the system under investigation and experimental technique. The AMR, AHE, and ANE usually dominate the magnetization dependence of the conductivity and can be properly separated owing to their different symmetry and magnetic field dependence \cite{Garello2013,Avci2014b} or frequency-dependent optical response \cite{Fan2016,Montazeri2015}.
In the following, we describe the three main techniques used to characterize the SOT measurements: harmonic Hall voltage analysis, spin-torque ferromagnetic resonance (ST-FMR), and magneto-optical Kerr effect (MOKE). Less precise SOT estimates can be obtained from magnetization switching and domain wall displacements, which are discussed separately in Subsection~\ref{MML:switching} and Section \ref{s:DW}.

\begin{figure}
	\centering
	\includegraphics[width=8.5 cm]{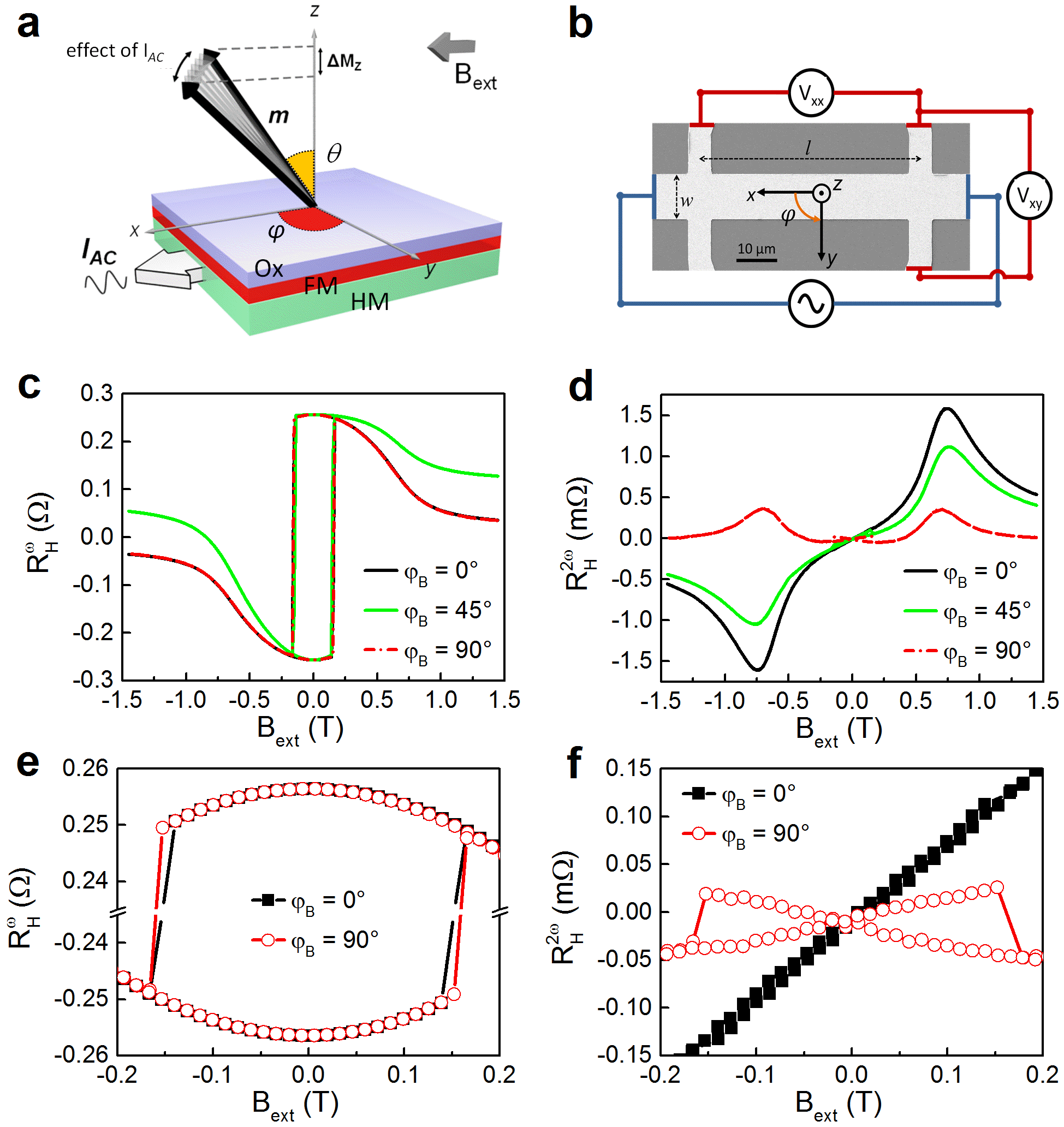}
	\caption{(Color online) (a) Schematic of the effect of an ac current on the magnetization and (b) experimental setup for harmonic Hall voltage measurements. (c) $R^{\omega}_{xy}$ and (d) $R^{2\omega}_{xy}$ of a Pt(5~nm)/Co(1~nm)/AlO$_x$ layer measured with a sinusoidal current of amplitude $j_{\rm c} = 10^7$~A/cm$^2$ and external magnetic field applied at $\varphi_B = 0^\circ, 45^\circ$, and $90^\circ$. (e,f) Close up of the curves in (c,d) showing the field range where the small angle approximation can be applied \cite{Baumgartner2018}.}
\label{Fig_HHV}
\end{figure}

\subsubsection{Harmonic Hall voltage analysis} \label{MML:HHV}
This method detects the harmonic response of the magnetization to a low frequency ac current, typically up to a few kHz. Originally, this approach was developed by assuming the simplest form of field-like torque \cite{Pi2010} and neglecting the damping-like torque and the transverse AMR (the planar Hall effect, PHE). It was soon extended to both components of the torques accounting for both the AHE and PHE \cite{Garello2013,Kim2013,Hayashi2014}, as well as for the torque angular dependence \cite{Garello2013,Qiu2014} and magnetothermal effects \cite{Avci2014b,Ghosh2017}. SOT measurements are performed by analyzing the second harmonic Hall voltage that arises due to the homodyne mixing of the ac current with the Hall resistance modulated by the oscillations of the magnetization induced by the SOTs [Fig.~\ref{Fig_HHV}(a)]. Since the magnetization dynamics is much faster than the current frequency $\omega$, the magnetization is assumed to be in quasi-static equilibrium at all times, at a position determined by the sum of the anisotropy field, external magnetic field, and current-induced fields. To first order in the current, the time-dependent Hall resistance $R_{xy}(t)$ is given by
\begin{equation}\label{eq:RH(t)}
R_{xy}({\bf B}_{\rm ext} + {\bf B}_{I}(t)) \approx R_{xy}({\bf B}_{\rm ext}) + \frac{d R_{xy}}{d{\bf B}_{I}}\cdot {\bf B}_{I} \sin(\omega t),
\end{equation}
where ${\bf B}_{\rm ext}$ is the external magnetic field and ${\bf B}_{I}= {\bf B}_{\rm DL}+{\bf B}_{\rm FL}+{\bf B}_{\rm Oe}$ is the effective current-induced field due to the sum of the damping-like and field-like SOT and the Oersted field. The Hall voltage $V_{xy}(t)= R_{xy}(t) I_0 \sin(\omega t)$ then reads
\begin{equation}\label{eq:VH(t)}
V_{xy}(t) \approx I_{0}[R^{0}_{xy} + R^{\omega}_{xy}\sin(\omega t)+R^{2\omega}_{xy}\cos(2\omega t)],
\end{equation}
where $I_0$ is the current amplitude, $R^{0}_{xy}=\frac{1}{2}\frac{d R_{xy}}{d{\bf B}_{I}}\cdot {\bf B}_{I}$, $R^{\omega}_{xy} = R_{xy}({\bf B}_{\rm ext})$, and $R^{2\omega}_{xy}=-\frac{1}{2}\frac{d R_{\omega}}{d{\bf B}_{I}}\cdot {\bf B}_{I}+R^{2\omega}_{\nabla T}$ are the zero, first, and second harmonic components of $R_{xy}$, respectively. The first harmonic term, shown in Fig.~\ref{Fig_HHV}(c) as a function of external field, is analogous to the dc Hall resistance and given by
\begin{equation}\label{eq:1stharmonic}
R^{\omega}_{xy}=R_{\rm AHE}\cos\theta + R_{\rm PHE}\sin^2\theta\sin (2\varphi),
\end{equation}
where $R_{\rm AHE}$ and $R_{\rm PHE}$ are the anomalous and planar Hall coefficients. This term serves two purposes, namely to determine the polar angle of the magnetization using Eq.~\eqref{eq:1stharmonic} when $\varphi = 0^{\circ}, 90^{\circ}$ and to measure the susceptibility of the magnetization to the magnetic field, providing self-calibration to the SOT measurement. The second harmonic term includes the SOT modulation of the Hall resistance as well as an extra contribution due to Joule heating, $R^{2\omega}_{\nabla T}$. In general, the two contributions may have a comparable magnitude and must be separated by either symmetry or magnetic field dependent measurements \cite{Avci2014b,Ghosh2017}. Assuming that $R^{2\omega}_{\nabla T}$ is negligible or has been subtracted from $R^{2\omega}_{xy}$, it is straightforward to show that
\begin{eqnarray}\label{eq:R2Reduced}
R^{2\omega}_{xy}= A_{\theta} \mathbf{B}_{I} \cdot \mathbf{e}_{\theta} + A_{\varphi} \mathbf{B}_{I} \cdot \mathbf{e}_{\varphi},
\end{eqnarray}
where $A_{\theta} = \frac{dR^{\omega}_{xy}}{dB_{\rm ext}} [I_0 \sin(\theta_B-\theta)]^{-1}$ and $ A_{\varphi} = R_{\rm PHE}\sin^{2}\theta\frac{d\sin(2\varphi)}{d\varphi} [I_0 \sin\theta_B \cos(\varphi_B-\varphi) B_{\rm ext}]^{-1}$. Here $\theta_{B}$ and $\varphi_B$ are the polar and azimuthal angles of the applied magnetic field. Equation~(\ref{eq:R2Reduced}) allows one to find the polar and azimuthal components of $B_{\rm DL}$ and $B_{\rm FL}$ as a function of the magnetization angle by measuring the dependence of $R^{2\omega}_{xy}$ on $B_{\rm ext}$. Figure~\ref{Fig_HHV}(d) shows an example of $R^{2\omega}_{xy}$ measured at $\varphi_B = 0^\circ$ and $\varphi_B = 90^\circ$. These curves are, respectively, odd and even with respect to magnetization reversal, reflecting the different symmetry of $B_{\rm DL}$ and $B_{\rm FL}$~\cite{Garello2013}. Because the damping-like torque is larger when $\mathbf{m}$ lies in the $xz$ plane, whereas the field-like torque tends to align $\mathbf{m}$ towards $y$, measurements taken at $\varphi_B = 0^\circ$ ($\varphi_B = 90^\circ$) reflect the character of the damping-like (field-like) effective fields. In general, four independent measurements at different azimuthal angles are required to determine the four effective field components in Eqs.~(\ref{eq:fieldEq3},\ref{eq:fieldEq4}).

In uniaxial and easy plane systems the number of independent measurements can be reduced to two, typically at $\varphi_B=0, \frac{\pi}{2}$ or $\varphi_B=\frac{\pi}{4}, \frac{3\pi}{4}$ \cite{Garello2013,Avci2014b}.
A further simplification is achieved using the small angle approximation, which is valid for perpendicularly magnetized samples when the magnetization deviates by at most a few degrees from the $z$-axis \cite{Kim2013,Hayashi2014}. In this case, $R^{2\omega}_{xy}$ varies linearly with the external field, as shown in Fig.~\ref{Fig_HHV}(f) and the SOTs are extracted by performing two sets of measurement at $\varphi_B=0$ and $\frac{\pi}{2}$,
\begin{eqnarray}\label{eq:SOTsmallangle}
 B_{\rm DL}=-\frac{2}{1-4r^2}(b_{x} + 2r b_{y}),\\
 B_{\rm FL}=-\frac{2}{1-4r^2}(b_{y} - 2r b_{x}),
\end{eqnarray}
where $r=R_{\rm PHE}/R_{\rm AHE}$ is the ratio between planar and anomalous Hall coefficients, $b_i=\frac{\partial R^{2\omega}_{xy}}{\partial B_{\rm ext}}/\frac{\partial^2 R^{\omega}_{xy}}{\partial B_{\rm ext}^2}$ is measured for $B_{\rm ext}\| i=x,y$, and the partial derivatives are calculated by linear fits of the curves shown in Fig.~\ref{Fig_HHV}(f). This approximation provides only the lowest order contribution to the SOTs. However, because of its simple implementation, it is widely used for characterizing the SOTs in systems with perpendicular anisotropy. Harmonic Hall voltage measurements can also be generalized to angular scans of the magnetization at constant external field, which is particularly suited for in-plane magnetized samples \cite{Avci2014b}, thus providing a versatile and sensitive method to characterize the SOTs in a variety of systems.

\subsubsection{Spin-torque ferromagnetic resonance}\label{MML:STFMR}
\begin{figure}
	\centering
	\includegraphics[width=8 cm]{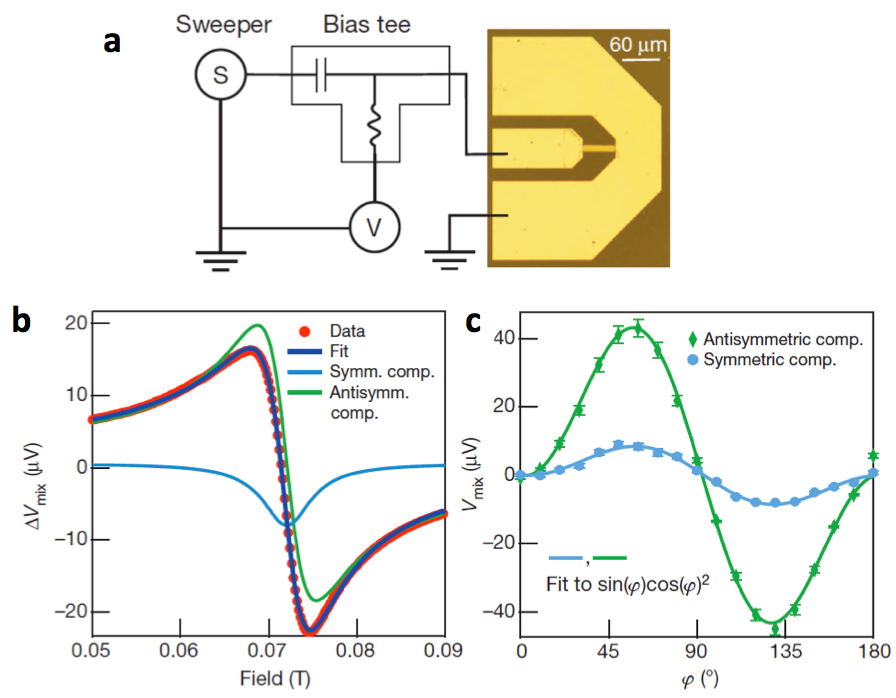}
	\caption{(Color online) (a) Schematic of the circuit used for the ST-FMR measurement and the sample
contact geometry. (b) Measured ST-FMR at room temperature with microwave
frequency $\omega/2\pi$=58 GHz for Bi$_2$Se$_3$(8 nm)/Ni$_{80}$Fe$_{20}$(16 nm). A fixed microwave power of 5 dBm is absorbed by
the device ($I_{\rm RF}=$7.7$\pm$1.1 mA) and $B$ is oriented at an angle $\varphi=\pi/4$ from
the current direction. The lines are fits to Eq. (\ref{eq:spin-fmr}) showing the
symmetric and antisymmetric resonance components. (c) Measured
dependence on the magnetic field angle $\varphi$ for the symmetric and antisymmetric
resonance components for a different sample. Adapted from \citet{Mellnik2014}.}
\label{Fig_ST-FMR}
\end{figure}

This method consists in exciting the magnetization of the ferromagnet using a radio-frequency (RF) charge current. The magnetization of the sample is excited through the spin torque and exhibits FMR when varying either the applied magnetic field or the current magnitude. This concept was initially developed in the context of MTJs \cite{Tulapurkar2005,Kubota2008,Sankey2008} and spin-valves \cite{Sankey2006} and more recently extended to the case of ultrathin magnetic bilayers \cite{Liu2011,Liu2012,Berger2018a} and bulk non-centrosymmetric magnetic semiconductors \cite{Fang2011,Kurebayashi2014}. \par

The dc voltage that develops across the sample [Fig.~\ref{Fig_ST-FMR}(a)] arises from the mixing of the RF current and the RF AMR due to the oscillating magnetization. It corresponds to the zero harmonic component in Eq. \eqref{eq:VH(t)} and here is strongly amplified due to the resonant magnetization dynamics. This rectified voltage gives information on the physical parameters of the magnetic material as well as on the nature of the torques that drive the excitation. In the context of an in-plane system with AMR driven by SOTs, the mixing voltage reads \cite{Liu2011,Reynolds2017}
\begin{eqnarray}\label{eq:spin-fmr}
&&V_{\rm mix}=-\frac{\gamma}{2}I_{\rm RF}\frac{\partial}{\partial \varphi} R\cos\varphi_B[\tau_{\rm DL}F_S(B)+\tau_{\rm FL}F_A(B)],\\\label{eq:spin-fmr-a}
&&F_S(B)=\frac{\alpha\omega^2(2B+\mu_0M_{\rm s})}{(\omega^2-\omega_0^2)^2+\alpha^2\gamma^2\omega^2(2B+\mu_0M_{\rm s})},\\\label{eq:spin-fmr-b}
&&F_A(B)=\frac{\gamma^2B(2B+\mu_0M_{\rm s})^2-\alpha\omega^2(2B+\mu_0M_{\rm s})}{(\omega^2-\omega_0^2)^2+\alpha^2\gamma^2\omega^2(2B+\mu_0M_{\rm s})},
\end{eqnarray}
where $\omega$ is the frequency of the RF current $I_{\rm RF}$ and $\omega_0=\gamma\sqrt{B(B+\mu_0M_{\rm s})}$ is the resonance frequency. The first contribution has a symmetric Lorentzian shape ($\sim F_S$) that is directly proportional to the damping-like torque, while the second has an antisymmetric shape ($\sim F_A$), providing information about the field-like torque (including the Oersted field torque). A picture of the experimental apparatus is given in Fig. \ref{Fig_ST-FMR}(a), together with the field-dependent and angular-dependent mixing voltages in Figs. \ref{Fig_ST-FMR}(b,c), respectively. This method is used extensively to probe torques in magnetic bilayers with in-plane magnetization, as well as in non-centrosymmetric bulk magnets, as explained in Section \ref{s:noncentro}. This effect is the reciprocal to spin pumping, where the field-excited precessing magnetization pumps a spin current in the adjacent nonmagnetic metal \cite{Tserkovnyak2002,Saitoh2006}.\par

Similar to other techniques, applying this method to ultrathin bilayer systems requires extreme care. First, the amplitude of the RF current generating the torques needs to be calibrated precisely using a network analyzer. Such a calibration might require thickness-dependent measurements to characterize possible size-dependent effects \cite{Nguyen2016}. Second, Eqs.~\eqref{eq:spin-fmr}, \eqref{eq:spin-fmr-b} only account for the rectification arising from AMR, but other sources such as SMR can also contribute to the mixing voltage \cite{Nakayama2013}, which should be properly accounted for \cite{Wang2016f,Zhang2016h}. Third, the phase difference between the RF current and the RF field can also have significant impact on the output signal \cite{Harder2011}. We refer the interested reader to the specialized literature for more information \cite{Harder2016}. A fourth issue is that this method assumes the simplest form of the torques, Eq. \eqref{eq:torquedef}, neglecting the angular dependence of SOTs \cite{Garello2013}.

\subsubsection{Magneto-optic Kerr effect}\label{MML:MOKE}
\begin{figure}
	\centering
	\includegraphics[width=7 cm]{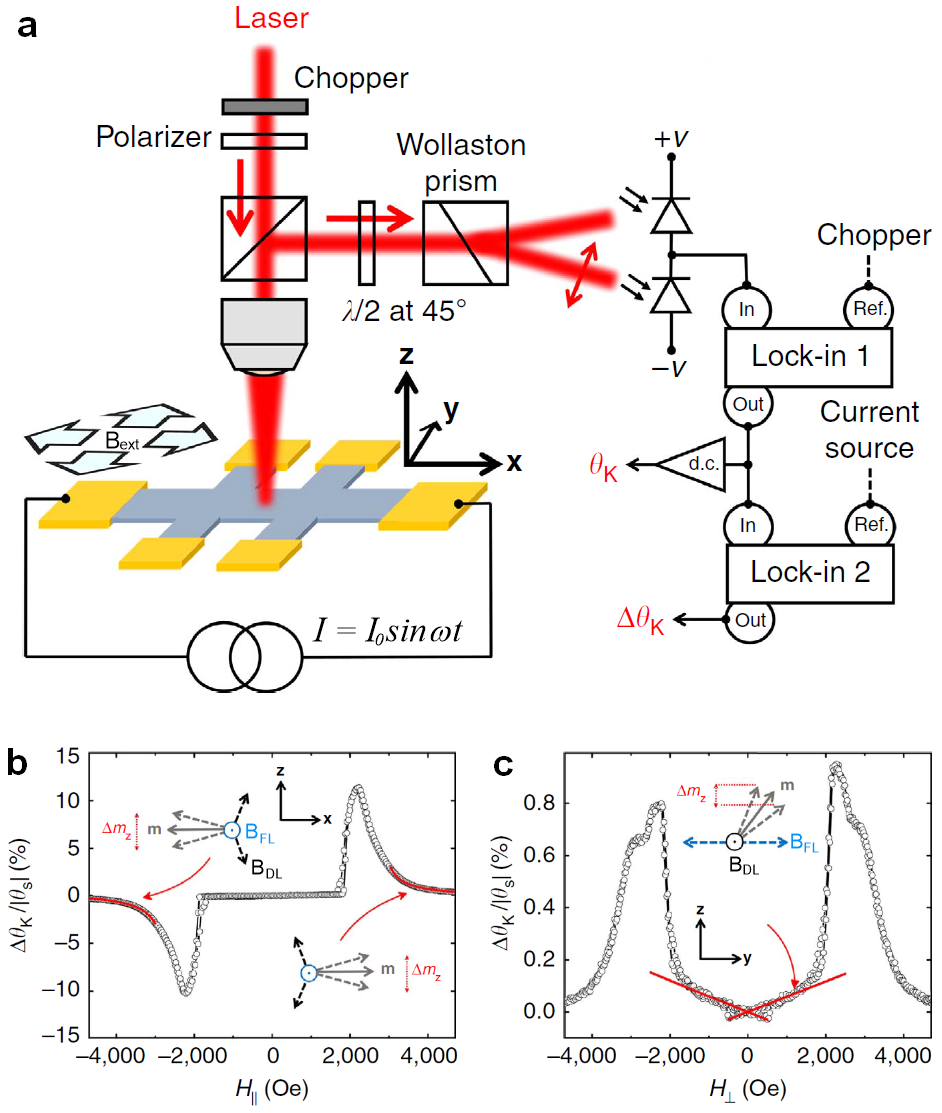}
	\caption{(Color online) (a) Schematic of a MOKE setup for SOT detection. (b,c) Differential Kerr angle $\Delta \theta_K$ measured on a Ta(5 nm)/CoFeB(1.1 nm)/MgO(2.0 nm) trilayer with perpendicular magnetic anisotropy for ${\bf B}_{\rm ext} || {\bf j}_{\rm c}$ (b) and ${\bf B}_{\rm ext} \bot {\bf j}_{\rm c}$ (c) with current amplitude $j_{\rm c} = 4.6\times 10^6$~Acm$^{-2}$. Adapted from \citet{Montazeri2015}.}
\label{Fig_MOKE}
\end{figure}
The magneto-optic Kerr effect (MOKE) allows for detecting the in-plane and out-of-plane components of the magnetization through the rotation of the light polarization upon reflection from a magnetic surface \cite{Qiu2000}. MOKE microscopy, with a wavelength-limited resolution of about 1~$\mu$m, has been used extensively to characterize SOT-induced domain nucleation and displacement~\cite{Miron2010,Emori2013,Ryu2013,Safeer2016} as well as the current-induced spin density in bare Pt and W layers \cite{Stamm2017}. MOKE-based detection schemes have been used also to estimate the SOT amplitude by measuring the oscillations of the magnetization induced by an ac current in thin metal bilayers \cite{Fan2014c}.

Vector measurements of the SOTs are based on the separate calibration of the first- and second-order magneto-optic coefficients, $f_{\bot}$ and $f_{\|}$, which parameterize the coupling of the light to the out-of-plane and in-plane magnetization, respectively \cite{Fan2016,Montazeri2015}. Such a technique measures the damping-like and field-like components of the SOT as a function of the magnetization angle via the polar and quadratic MOKE response, respectively, using only normally incident light [see Fig.~\ref{Fig_MOKE}(a)]. Similar to the Hall resistance, Eq.~\eqref{eq:RH(t)}, the Kerr rotation measured during ac current injection can be Taylor expanded as
\begin{equation}\label{eq:thetaK(t)}
\theta_{K}({\bf B}_{\rm ext} + {\bf B}_{I}(t)) \approx \theta_{K}({\bf B}_{\rm ext}) + \frac{d \theta_{K}}{d{\bf B}_{I}}\cdot {\bf B}_{I} \sin(\omega t).
\end{equation}
Here, the first term is the equilibrium Kerr angle given by $\theta_K = f_{\bot} m_{z}+f_{\|}[1/2(m_y^2-m_x^2)\sin 2\phi_p + m_{x} m_{y}\cos 2\phi_p]$, with $\phi_p$ the angle between the light polarization and ${\bf B}_{\rm ext}$, and the second term results in the differential Kerr signal $\Delta \theta_K = (d\theta_{K}/dI)I_0$ due to the current-induced fields. In analogy with the harmonic Hall voltage analysis technique, measurements of $\Delta \theta_K$ are mostly sensitive to changes of $m_z$. Thus, measurements taken with ${\bf B}_{\rm ext} || {\bf x}$ reflect the strength of the damping-like effective field,
\begin{equation}\label{eq:DeltathetaK1}
\Delta \theta_K = \frac{f_{\bot} B_{\rm DL}}{B_{\rm ext}-B_{K}} + \frac{f_{\|} \cos 2\phi_p (B_{\rm FL}+B_{\rm Oe})}{B_{\rm ext}},
\end{equation}
where $B_{K}$ is the magnetic anisotropy field and $f_{\|} \ll f_{\bot}$. Conversely, measurements taken with ${\bf B}_{\rm ext} || {\bf y}$ reflect the strength of the field-like effective field,
\begin{equation}\label{eq:DeltathetaK2}
B_{\rm FL}= \frac{-2\partial(\Delta \theta_K)/\partial B_{\rm ext}}{\partial^2 \theta_K / \partial B_{\rm ext}^2}.
\end{equation}
Figure~\ref{Fig_MOKE}(b) shows that $\Delta \theta_K$ exhibits an antisymmetric (symmetric) line shape consistent with the symmetry of $B_{\rm DL}$ ($B_{\rm FL}$) under magnetization reversal, in close analogy with $R^{2\omega}_{xy}$ [Fig.~\ref{Fig_HHV}(d)]. SOT vector measurements performed by MOKE agree well with harmonic Hall voltage \cite{Montazeri2015} and ST-FMR measurements \cite{Fan2016} and can be used to characterize the SOT in metallic as well as insulating ferromagnets. An advantage of this technique is that it is less sensitive to thermoelectric and inductive effects compared to all-electrical SOT probes, and that it offers spatial resolution comparable to the wavelength of the probing laser beam.

\subsection{Materials survey} \label{MML:exp}
\begin{table*}
  \footnotesize
  {\renewcommand{\arraystretch}{1.2}%
		\begin{tabular}{ll|llllllllll}
   \hline\hline
			Structure & & MA & Method & $B_{\rm DL}/j$ & $B_{\rm FL}/j$ & $\xi_{\rm DL}^j$ & $\xi_{\rm FL}^j$ & $\xi_{\rm DL}^E$ & $\xi_{\rm FL}^E$ \\
   \hline
   \multicolumn{12}{l}{Nonmagnetic metals} \\
   \hline
			Pt(3)/Co(0.6)/AlO$_x$(1.6) & \cite{Garello2013} & OP & HHV & -6.9 & 4 & 0.13 & -0.073& 3.5 & -2.0 \\
   Pt(3)/CoFe(0.6)/MgO(1.8)& \cite{Emori2013} & OP & HHV& -5 & 2 & 0.064&-0.024& & \\
   Ti(1)/CoFe(0.6)/Pt(5)& \cite{Fan2014c}& IP & MOKE & 3.2 & -0.3 & 0.074 & -0.008& & \\
   Pt(5)/Co(1)/MgO(2) & \cite{Nguyen2016}& OP & HHV & -4.5 & 1 & 0.11 & -0.024& 2.43 & -0.53 \\
   Pt(5)/ Ni$_{80}$Fe$_{20}$(8)/AlO$_x$(2) & \cite{Fan2016}& IP & MOKE & -0.49 & 0.71 & 0.082 & -0.12& 2.64 & -3.88 \\
   YIG(50)/Pt(4) & \cite{Montazeri2015}& IP & MOKE & 0.29 & & 0.03 & & & \\
   TmIG(8)/Pt(5) & \cite{Avci2017}& OP & HHV & 0.59 & & 0.014 & & & \\
   Ta(4)/CoFeB(1.1)/MgO(1.6) &\cite{Liu2012}& OP & HHV & 3.5 & & -0.13 & & -0.68 & \\
   Ta(3)/CoFeB(0.9)/MgO(2)&\cite{Avci2014} & OP & HHV & 3.2 & -2.1 & -0.06 & 0.04 & -0.34 & 0.22 \\
   Ta(3)/CoFeB(0.9)/MgO(2)$^{a}$ &\cite{Garello2013} & OP & HHV & 2.4 & -4.5 & -0.07 & 0.12 & -0.36 & 0.67 \\
   Ta(1.5)/CoFeB(1)/MgO(2)$^{a}$ &\cite{Kim2013} & OP & HHV & 1.35 & -4.46 & -0.03 & 0.11 & -0.14 & 0.48 \\
   Ta(2)/CoFeB(0.8)/MgO(2)$^{a}$ &\cite{Qiu2014}& OP & HHV & 4.4 & -19.4 &-0.11& 0.47 & & & & \\	
   Ta(5)/CoFeB(1.1)/MgO(2)$^{a}$ &\cite{Montazeri2015}& OP & MOKE & 2.0 & -3.3 &-0.05& 0.08 & & \\	
   W(5)/CoFeB(0.85)/Ti(1)$^{a}$ &\cite{Pai2012}& IP & ST-FMR & & & -0.33 & & & \\
   Hf(3.5)/CoFeB(1)/MgO(2)$^{a}$ &\cite{Torrejon2014}& OP & HHV & 0.8& -2.6& -0.02 & 0.06 & & \\
   Hf(3.5)/CoFeB(1.1)/MgO(2)$^{a}$ &\cite{Akyol2016}& OP & HHV & 5& & -0.17 & & & \\
   Hf(10)/CoFeB(1.1)/MgO(2)$^{a}$ &\cite{Akyol2016}& OP & HHV & -1& & 0.03 & & & \\
   Hf(1)/CoFeB(1)/MgO(2) &\cite{Ramaswamy2016}& OP & HHV & -0.24& 0.9 & 0.007 & -0.03 & & \\
   Hf(6)/CoFeB(1)/MgO(2) &\cite{Ramaswamy2016}& OP & HHV & 9& -27 & -0.28 & 0.82 & & \\
   Pd(7)/Co(0.6)/AlO$_x$(1.6) & \cite{Ghosh2017}& OP & HHV & -1.3 & 0.7 & 0.03 & -0.015& 1.0 & -0.55 \\
   \hline
   \multicolumn{12}{l}{Oxidized metals} \\
   \hline
   WO$_x$(6)/CoFeB(6)/TaN(2) &\cite{Demasius2016}& IP & ST-FMR & & & -0.49 & & & \\
   SiO$_2$/Ni$_{80}$Fe$_{20}$(8)/CuO$_x$(10) & \cite{An2016}& IP & ST-FMR & & & 0.08 & -0.08 & & \\
   Ti(1.2)/Ni$_{80}$Fe$_{20}$(1.5)/AlO$_x$(1.5) & \cite{Emori2016}& IP & ST-FMR & & 0.15 & & -0.01 & & \\
   PtO$_x$(32)/Ni$_{81}$Fe$_{19}$(5)/SiO$2$(4)  & \cite{An2018}& IP & ST-FMR & & & 0.9 & -0.2 & 8.7 & -1.8 \\
   \hline
   \multicolumn{12}{l}{Metal alloys} \\
   \hline
   CuAu(8)/Ni$_{80}$Fe$_{20}$(1.5) & \cite{Wen2017}& IP & HHV & -1.9 & 0.58 & 0.01 & -0.003 & 0.33 & -0.1 \\
   Au$_{25}$Pt$_{75}$(4)/Co(0.8)/MgO(2) & \cite{Zhu2018}& OP & HHV & -8.0 & 3.2 & 0.28 & -0.11 & 3.3 & -1.3 \\
   Ni$_{80}$Fe$_{20}$(9)/Ag(2)/Bi(4) & \cite{Jungfleisch2016b}& IP & ST-FMR & & & 0.18 & 0.14$^{b,c}$ & & \\
   \hline
   \multicolumn{12}{l}{Antiferromagnets} \\
   \hline
   IrMn(8)/ Ni$_{80}$Fe$_{20}$(4)/Al(2) &\cite{Tshitoyan2015}& IP & ST-FMR & -2.2 & -1.7 & 0.22 & 0.17 & & \\
   IrMn$_3$[001](6)/ Ni$_{80}$Fe$_{20}$(6)/TaN &\cite{Zhang2016b}& IP & ST-FMR & & & 0.20 & & & \\
   IrMn$_3$[111](6)/ Ni$_{80}$Fe$_{20}$(6)/TaN &\cite{Zhang2016b}& IP & ST-FMR & & & 0.12 & & & \\
   IrMn$_3$(5)/CoFeB(1)/MgO$^{a}$ &\cite{Wu2016b} & OP & HHV & -1.8 & 0.7 & 0.06 & -0.02 & & \\
   PtMn(8)/Co(1)/MgO(1.6) &\cite{Ou2016b} & IP & ST-FMR & & & 0.16 & -0.04 & & \\
   MgO(1.6)/Co(1)/PtMn(8) &\cite{Ou2016b} & IP & ST-FMR & & & 0.19 & $\simeq$ 0 & & \\
   \hline
   \multicolumn{10}{l}{Semiconductors and semimetals} \\
   \hline
   (Ga,Mn)As(20)/Fe(2)/Al(2) &\cite{Skinner2015} & IP & ST-FMR & -0.34$^{b}$ & 0.26$^{c,d}$ & 0.03$^{b}$ & -0.02$^{c,d}$ & &\\
   MoS$_2$(0.8)/CoFeB(3)/TaO$_x$(3) &\cite{Shao2016} & IP & HHV &$\simeq 0$ & $\simeq 0.008$ & $\simeq 0$ & -0.14 & $\simeq 0$ &-0.03 \\
   WSe$_2$(0.8)/CoFeB(3)/TaO$_x$(3) &\cite{Shao2016} & IP & HHV & $\simeq 0$ & 0.012 & $\simeq 0$ & $\simeq -0.3$ & $\simeq 0$ &-0.06\\
   WTe$_2$/Ni$_{80}$Fe$_{20}$(6)/Al(1) &\cite{MacNeill2017} & IP & ST-FMR & & & 0.04$^{b}$ & & 0.12$^{b}$ &0.09$^{c}$ \\
   \hline
   \multicolumn{10}{l}{Topological insulators} \\
   \hline
   Bi$_2$Se$_3$(8)/Ni$_{80}$Fe$_{20}$(16)/Al(2) & \cite{Mellnik2014} & IP & ST-FMR & & & 1& 1.3& 0.5 & 0.7 \\
   Bi$_2$Se$_3$(20)/CoFeB(5)/MgO(2) & \cite{Wang2015b} & IP & ST-FMR & & & 0.08$^{b}$& 0.05$^{b}$& & \\
   Bi$_2$Se$_3$(10)/Ag(8)/CoFeB(7)/MgO(2) & \cite{Shi2018} & IP & ST-FMR & 5.3 & 3.2 & 0.49 & 0.3& & \\
   (Bi,Sb)$_2$Te$_3$(8)/CoTb(8)/SiN$_x$(3) & \cite{Han2017} & OP & Coercivity & -8 & & 0.4 & & & \\
    Mn$_{0.4}$Ga$_{0.6}$(3)/Bi$_{0.9}$Sb$_{0.1}$(10)  & \cite{Khang2018} & OP & Coercivity & -2300 & & 52 & & 130 & \\
   \hline
   \multicolumn{12}{l}{$^{a}$Annealed. $^{b}$Average value. $^{c}$Sign uncertain.} \\
   \hline \hline
		\end{tabular}}
  \caption{\label{tableSOT} SOTs in magnetic multilayers. The thickness of the layers is given in nm with the topmost layer on the right. The following units are used for the effective fields and SOT efficiencies: $B_{\rm DL,FL}/j$ [mT/(10$^{11}$ A/m$^2$)], $\xi_{\rm DL,FL}^j$ [adimensional], and $\xi_{\rm DL,FL}^E$ [10$^{5}$~$(\Omega m)^{-1}$]. The sign of $B_{\rm DL}$ and $B_{\rm FL}$ is defined as in Eqs.~\eqref{eq:fieldEq3}, \eqref{eq:fieldEq4}. $\xi_{\rm DL} > 0$ corresponds to the same sign of the damping-like torque as for Pt, whereas $\xi_{\rm FL} < 0$ indicates that $B_{\rm FL}$ is opposite to the Oersted field. The magnetic anisotropy (MA) of the ferromagnetic layers is indicated as out-of-plane (OP) or in-plane (IP). The values for the OP samples are given for the magnetization lying close to the easy axis. All measurements have been carried out at room temperature. Here, HHV stands for Harmonic Hall voltage analysis.}
\end{table*}

\begin{figure}[t]
	\centering
	\includegraphics[width=6.5cm]{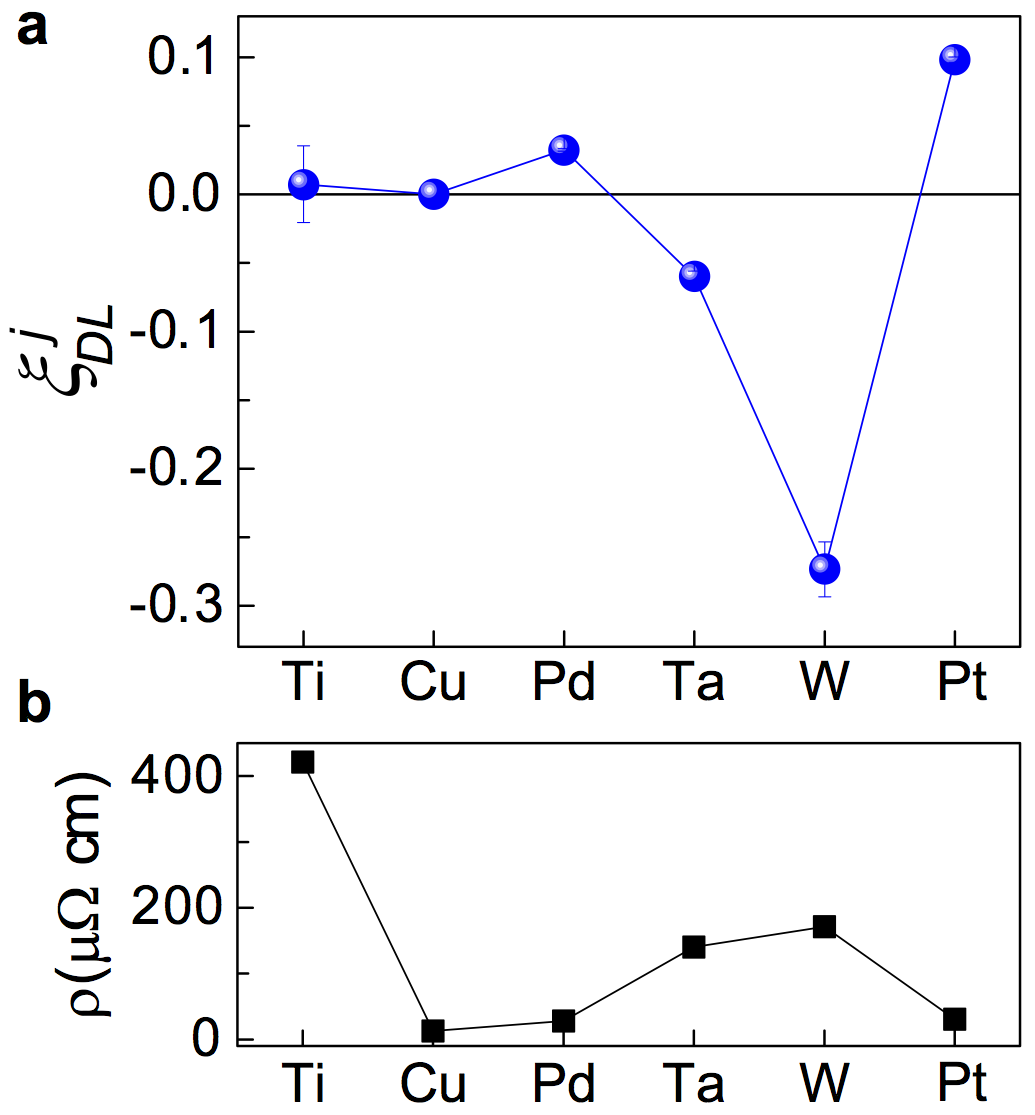}
	\caption{(Color online) \label{Fig_SOTrho}(a) Damping-like SOT efficiency in X(8 nm)/Co/AlO$_x$(2) trilayers, where X = Ti, Cu, Pd, Ta, W, Pt. The data are measured using the harmonic Hall voltage analysis method. The Co thickness is 2.5~nm except for the Pd sample where it is 0.6 nm.
(b) Room temperature resistivity of the nonmagnetic metal. Adapted from \citet{Avci2015b,Ghosh2017}.}
\end{figure}

\subsubsection{Ferromagnet/nonmagnetic metal layers} \label{MML:exp:FMNM}
The most studied SOT systems are composed of a metallic ferromagnet deposited on a nonmagnetic metal layer, often capped by an amorphous or crystalline oxide layer. These systems present strong damping-like and field-like SOTs, of the order of a few mT per $10^7$~A/cm$^2$ ($\xi^{j}\approx 0.1$, see Table~\ref{tableSOT}), are easy to fabricate, and compatible with established processing of magnetic materials for memory applications. An early experimental observation of the damping-like SOT in ferromagnetic metals was reported by \citet{Ando2008b} in a Pt/NiFe bilayer resonantly excited by an external microwave field, by measuring the change of magnetic damping upon injection of a dc current. This effect was attributed to the SHE of the Pt layer and later extended to the excitation of FMR upon injection of an RF current \cite{Liu2011}. Evidence for the field-like SOT was first reported by \citet{Miron2010} by observing that the current-induced nucleation of magnetic domains in perpendicularly magnetized Pt/Co/AlO$_x$ wires is either enhanced or quenched by applying an in-plane magnetic field at an angle of $\pm 90^\circ$ relative to the current. This effect was attributed to the action of a Rashba-like effective field and later quantitatively estimated by harmonic Hall voltage analysis measurements \cite{Pi2010,Garello2013}. \par

A major breakthrough was achieved in 2011 when bipolar magnetization switching was demonstrated in perpendicular Pt/Co/AlO$_x$ dots \cite{Miron2011b}, establishing the relevance of SOT for applications. The authors observed that the symmetry of the switching field corresponds to a damping-like torque consistent with either the SHE or the iSGE, and argued that the SHE of Pt alone could not account for the magnitude of the torque. Other experiments favored a SHE-only explanation of the switching mechanism \cite{Liu2012c}, triggering an ongoing debate on the origin of the torques (see Subsection \ref{MML:origin}). These experiments were rapidly followed by measurements of SOTs and magnetization switching in Ta/CoFeB/MgO \cite{Liu2012,Kim2013,Garello2013,Emori2013,Avci2014} and W/CoFeB/MgO layers \cite{Pai2012}, which showed that the damping-like SOT correlates with the sign of the spin-orbit coupling constant and the SHE of the nonmagnetic metal layer, whereas the field-like torque has a more erratic behavior depending on the type of ferromagnet and interface structure \cite{Pai2015}.\par

The largest SOT efficiencies are found in the 5$d$ metals, in particular for the highly resistive $\beta$-phase of W and Ta as well for fcc Pt (Fig.~\ref{Fig_SOTrho}). In metals where the spin Hall conductivity $\sigma_{\rm sh}$ is of intrinsic origin, the spin Hall angle is directly proportional to the longitudinal resistivity, given by $\theta_{\rm sh} =\sigma_{\rm sh}\rho$. Pt and Pd display a large SOT efficiency despite their moderate resistivity \cite{Nguyen2016,Ghosh2017}, which is attributed to their large intrinsic $\sigma_{\rm sh}$ and density of states at the Fermi level \cite{Freimuth2010,Freimuth2015}. Large damping-like and field-like SOT efficiencies have been reported also by replacing the nonmagnetic metal by an intermetallic antiferromagnet such as IrMn \cite{Tshitoyan2015,Oh2016,Wu2016b,Zhang2016b} and PtMn \cite{Ou2016b}, which allows for including exchange-biased systems in SOT devices (Subsection~\ref{MML:switching:zerofield}).

Enhanced efficiencies can be obtained in multilayers where the ferromagnet is sandwiched between two nonmagnetic metals with opposite spin Hall angle, giving rise to parallel damping-like torques at opposite interfaces \cite{Woo2014,Yu2016a}. In such systems, the spin current associated with the PHE in the bulk of the ferromagnet can give rise to an additional damping-like torque if the spin transfer to the nonmagnetic metals is asymmetric \cite{Safranski2019}. Results obtained on symmetric multilayers such as [Co/Pd]$_n$ \cite{Jamali2013}, on the other hand, are more controversial because of the expected compensation of the SOT from the top and bottom interfaces and the missing analysis of thermal voltages.\par

In general, significant variations of the torque efficiencies have been observed depending on multilayer composition, thickness, thermal annealing protocols, interface oxidation and dusting, as well as temperature, which we briefly describe below.

\paragraph{Thickness dependence} \label{MML:exp:FMNM:thickness}
\begin{figure}[t]
	\centering
	\includegraphics[width=7cm]{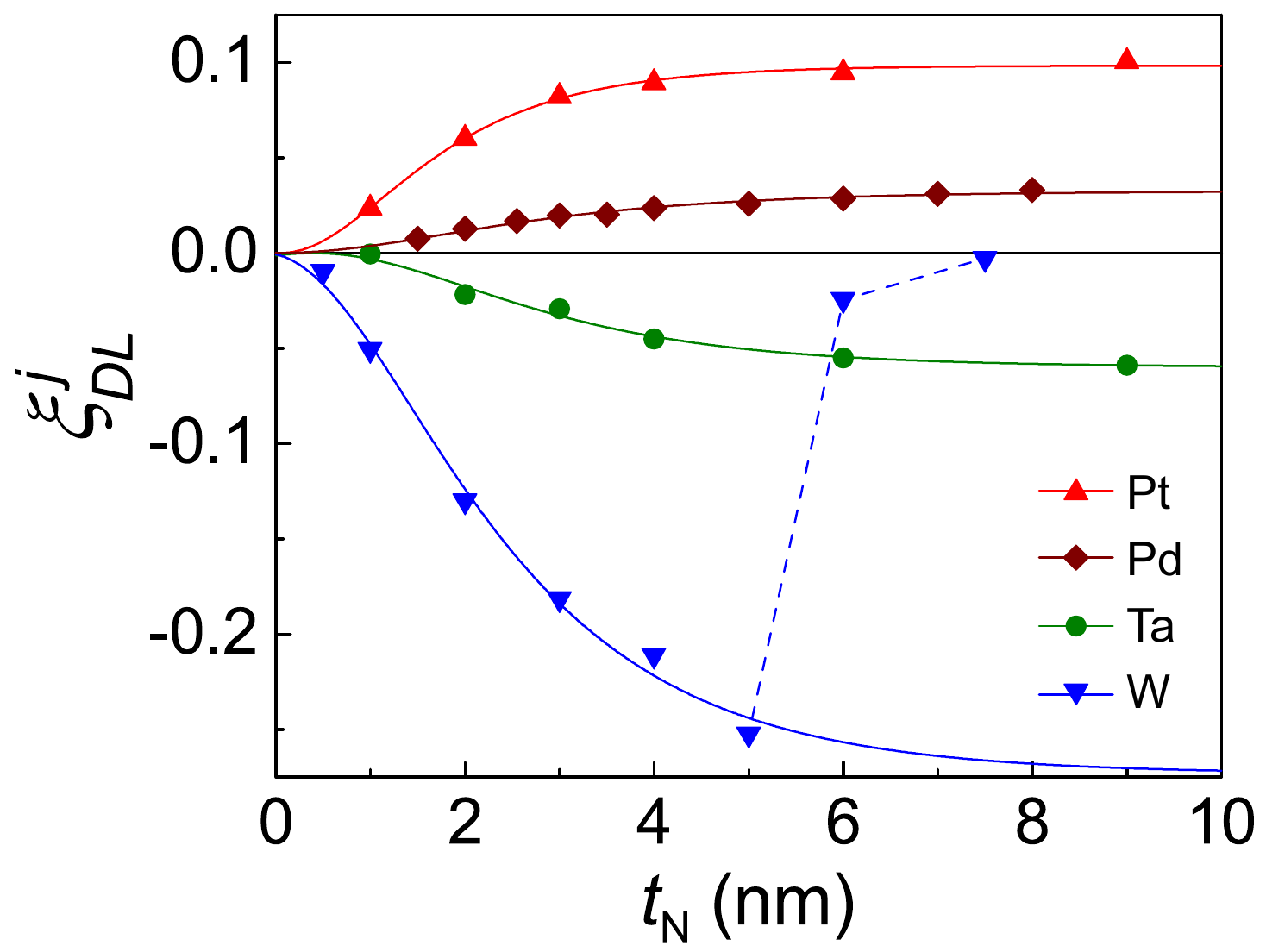}
	\caption{(Color online) Damping-like torque efficiency as a function of thickness in NM($t_N$)/Co/AlO$_x$ layers, where NM=Pt, Pd, Ta, and W \cite{GarelloUnp}. The solid lines are fit to the function $\xi^{j}_{\rm DL}[1-\mathrm{sech}(t_{N}/\lambda_{\rm sf})]$. Note that the efficiency $\xi^{j}_{\rm DL}$ of W drops abruptly between 5 and 6~nm as the crystal structure changes from the $\beta$ to the $\alpha$ phase.}\label{Fig_SOT_thickness}
\end{figure}
Assuming that the charge-spin conversion in multilayer systems occurs outside the ferromagnetic volume, one expects the SOTs to be simply inversely proportional to the ferromagnet thickness ($\sim 1/t_{F}$), as the effects of the current-induced fields are inversely proportional to the magnetic volume on which they act upon, and strongly dependent on the nonmagnetic metal thickness ($t_{N}$) as well as on interfacial properties. The influence of $t_{F}$ on the SOT has been systematically investigated in Ta/CoFeB/MgO \cite{Kim2013}, NiFe/Cu/Pt \cite{Fan2013}, Ti/CoFeB/Pt \cite{Fan2014c}, Co/Pt \cite{Skinner2014}, Pt/Co/MgO and Pt/Co$_{50}$Fe$_{50}$/MgO \cite{Pai2015}, and Pd/FePd \cite{Lee2014b}, all deposited on thermally oxidized Si. \citet{Kim2013} found that $B_{\rm FL}$ decreases strongly while $B_{\rm DL}$ remains approximately constant in Ta/CoFeB/MgO when increasing $t_{F}$ from 0.8 to 1.4~nm. \citet{Fan2014c} showed that both fields decrease when increasing $t_{F}$ from 0.7 to 6~nm, with $B_{\rm FL}$ dropping significantly faster than $1/t_{F}$. The spin torque efficiencies, $\xi^{j}_{\rm DL,FL}$, have been found to decrease in annealed Pt/Co$_{50}$Fe$_{50}$/MgO layers between 0.6 and 1~nm, but to increase in as-grown Pt/Co/MgO \cite{Pai2015}, possibly because the Co thickness has to exceed the spin absorption length (i.e., the length over which the spin current is absorbed in the ferromagnet) in order to develop the full torque or because of strain relaxation in the Pt/Co layer. Interestingly, the sign of the field-like torque is opposite in these two systems. \citet{Skinner2014} have found a sign inversion of the field-like torque in Co/Pt for a 2~nm thick Co layer, which suggests that two mechanisms with different dependence on $t_{F}$ compete to determine the total torque. 
\par

\begin{figure}[b]
	\centering
	\includegraphics[width=\columnwidth]{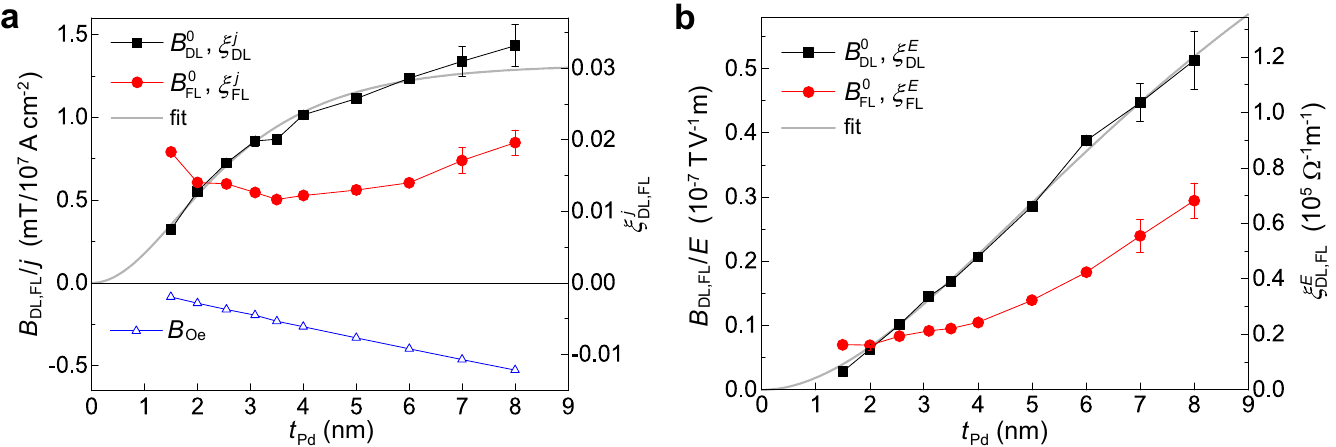}
	\caption{(Color online) SOT efficiency in Pd(t)/Co(0.6)/AlO$_x$ trilayers as a function of Pd thickness. (a) $\xi^{j}_{\rm DL,FL}$ and (b) $\xi^{E}_{\rm DL,FL}$ differ significantly from each other due to the strong decrease of the Pd resistivity with increasing thickness \cite{Ghosh2017}. The field-like torque efficiency is shown after subtraction of the Oersted field contribution $B_{\rm Oe} = \mu_0 j_{\rm Pd} t_{\rm Pd}/2$ shown by the open triangles in (a).}\label{Fig_SOT_PdCo}
\end{figure}

The dependence of the SOT on $t_{N}$ has been the focus of many studies aimed at distinguishing the bulk and interfacial nature of the torques. In the simplest theoretical models, effects coming from the interfacial Rashba interaction should be independent of $t_{N}$, whereas effects emerging from the bulk SHE should scale as $[1-\mathrm{sech}(t_{N}/\lambda_{\rm sf})]$ according to the profile of the spin density in the nonmagnetic metal layer \cite{Liu2011}. In addition, the Oersted field should increase linearly with $t_{N}$. Therefore, assuming that the overall structure (crystallinity, interface and inter-diffusion processes) is unchanged upon modifying $t_{N}$, analyzing the thickness-dependence of $\xi^{j}_{\rm DL}$ and $\xi^{j}_{\rm FL}$ should provide information about the physical origin of the torques. Figure~\ref{Fig_SOT_thickness} shows that $\xi^{j}_{\rm DL}$ of as-grown Co/AlO$_x$ layers deposited on $\beta$-Ta, $\beta$-W, and Pt increases monotonically with $t_{N}$ up to saturation, which agrees well with the SHE model assuming a spin diffusion length of the order of 1.5~nm for all metals. Such a trend is common to a variety of systems based on Ta \cite{Torrejon2014}, W \cite{Hao2015}, Pt \cite{Nguyen2016}, and Pd \cite{Ghosh2017}, suggesting that the SHE is the dominant source of the spin current causing the damping-like torque. Recent theoretical work, however, has pointed out that a similar $t_{N}$ dependence is expected for a Rashba-like damping-like torque due to interfacial spin-dependent scattering \cite{Haney2013b,Amin2016b}, so that separating the bulk and interface contributions to $\xi^{j}_{\rm DL}$ is not straightforward. Moreover, a change of sign of both $\xi^{j}_{\rm DL}$ and $\xi^{j}_{\rm FL}$ has been reported for Ta/CoFeB/MgO \cite{Kim2013,Allen2015} and Hf/CoFeB/MgO \cite{Ramaswamy2016,Akyol2016} at $t_{\rm Ta}\approx 0.5$~nm and $t_{\rm Hf}\approx 2$~nm, respectively, indicating that there are different mechanisms contributing to the torques that may compete or reinforce each other. \par

Calculations based on the drift-diffusion model of the SHE predict that the damping-like and field-like torques should have a similar dependence on $t_{N}$ and be proportional to the real and imaginary part of the spin mixing conductance of the FM/NM interface, respectively, which naturally leads to $\xi^{j}_{\rm DL} \gg \xi^{j}_{\rm FL}$ \cite{Haney2013b}.

Several reports, however, show that $\xi^{j}_{\rm FL} \gtrsim \xi^{j}_{\rm DL}$ in out-of-plane as well as in-plane magnetized layers (Table~\ref{tableSOT}) and that the dependence of $\xi^{j}_{\rm FL}$ on $t_{N}$ differs from that of $\xi^{j}_{\rm DL}$ in systems based on Ta \cite{Kim2013}, Pt \cite{Fan2014c,Nguyen2016}, and Pd \cite{Ghosh2017}, particulary at low thickness ($t_{N}<2$~nm). An example of this behavior is reported in Fig.~\ref{Fig_SOT_PdCo}(a) for a perpendicularly magnetized Pd/Co/AlO$_x$ layer, where $\xi^{j}_{\rm FL}$ clearly departs from the monotonic increase of $\xi^{j}_{\rm DL}$ as a function of $t_{\rm Pd}$. Remarkably, the thickness dependence changes when the SOT efficiency is normalized to the electric field, as in Fig.~\ref{Fig_SOT_PdCo}(b), showing that $\xi^{E}_{\rm DL}$ and $\xi^{E}_{\rm FL}$ do not saturate up to $t_{N}=8$~nm and that $\xi^{E}_{\rm FL}$ extrapolates to a finite value at $t_{\rm Pd}=0$. The difference between $\xi^{E}_{\rm DL,FL}$ and $\xi^{j}_{\rm DL,FL}$ also suggests that the thickness dependence should be analyzed with care in films when the resistivity is not homogeneous \cite{Nguyen2016,Ghosh2017}. \par

\paragraph{Interfacial tuning} \label{MML:exp:FMNM:interfacial}

The transport of charge and spin in multilayer systems is strongly affected by interface scattering and discontinuities in the electronic band structure, as is well known from early studies of the giant magnetoresistance \cite{Levy1994,Parkin1993}. Thus, significant variations of the SOTs are expected upon modification of the interfaces, even when the nonequilibrium spin density originates in the bulk of the nonmagnetic metal layer. Experimentally, it has been shown that the damping-like and field-like SOTs change dramatically upon annealing and consequent intermixing of Pt/Co/AlO$_x$ \cite{Garello2013} and Ta/CoFeB/MgO \cite{Avci2014}, as well as upon the insertion of different spacer layers between the ferromagnet and the nonmagnetic metal that is considered to be the main source of spin density \cite{Fan2013,Pai2014,Zhang2015i}.
The insertion of a light metal such as Cu has been pursued with the intention of removing the interfacial spin-orbit coupling.
\citet{Fan2013,Fan2014c} measured a field-like torque that decreases smoothly with the thickness of the Cu spacer in Pt/Cu/NiFe, indicating a nonlocal origin, but also that the $\xi^{j}_{\rm FL}/\xi^{j}_{\rm DL}$ ratio of CoFeB/Cu/Pt has a discontinuity around $t_{\rm Cu}=0.7$~nm, which points towards a modified interface effect. In fact, the insertion of a light metal, while reducing the magnetic proximity effect between the nonmagnetic metal and the ferromagnet, does not completely eliminate the interfacial spin-orbit coupling. Rather, it creates two additional interfaces on either sides of the light metal layer, with different iSGE and scattering properties. The latter effect is evident when considering that the SOTs change by as much as 50 \% for a Cu spacer thickness of the order of 1~nm \cite{Fan2014c,Nan2015,Rojas-Sanchez2014}, which is two orders of magnitude smaller than the spin diffusion length in Cu. \par

The insertion of a spacer layer can also modify the ability of the ferromagnet to absorb the incoming spin current, by modifying both the transparency \cite{Nguyen2015,Zhang2015i} and the spin memory loss at the interface \cite{Rojas-Sanchez2014,Berger2018b,Tao2018,Dolui2017}. The former accounts for the spin current backflow in the nonmagnetic metal (the larger the backflow, the smaller the spin current transmission into the ferromagnet), while the latter opens an additional spin dissipation channel at the interface (see Subsection \ref{s:sheTorque}). Both effects reduce the effective spin injection. A typical case is that of Hf, which has been shown to improve the SOT efficiency in W/Hf/CoFeB/MgO and Pt/Hf/CoFeB/MgO whilst promoting perpendicular magnetic anisotropy \cite{Pai2014} and reducing the magnetic damping \cite{Nguyen2015}. Changes in the SOT efficiency in such cases are usually interpreted in terms of an enhanced spin mixing conductance, which may also explain why the damping-like torque efficiency changes for different ferromagnets coupled to the same nonmagnetic metal, as observed, e.g., in Pt/Co/TaN ($\xi^{j}_{\rm DL}=0.11$) and Pt/NiFe/TaN ($\xi^{j}_{\rm DL}=0.05$) \cite{Zhang2015i}. Such a phenomenological parameter, however, accounts for the transmission of the bulk spin current as much as for the generation of interfacial spin currents, so that its use to estimate an asymptotic value of the bulk SHE in nonmagnetic metals can be questioned. Moreover, the spacer layer itself can be regarded as a source of spin current, as has been shown in the case of Hf \cite{Akyol2016,Ramaswamy2016}. \par

\begin{figure}[t]
	\centering
	\includegraphics[width=\columnwidth]{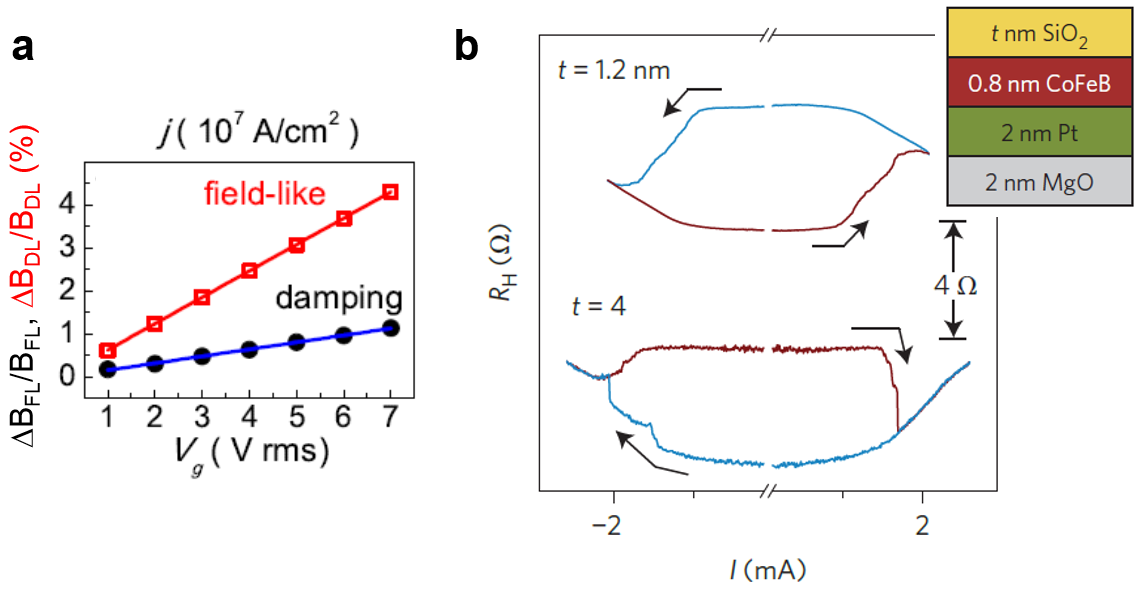}
	\caption{(Color online) (a) Effect of gate voltage on the field-like and damping-like SOT in Pt/Co/Al$_2$O$_3$ \cite{Liu2014d}. (b) Inversion of the polarity of current-induced switching for different thickness of the oxide capping layer in Pt/CoFeB/SiO$_2$ \cite{Qiu2015}.}\label{Fig_SOT_Oxidation}
\end{figure}

Another interesting aspect is the control of magnetic properties through interfacial oxidation \cite{Monso2002,Manchon2008c,Rodmacq2009} or gate voltage \cite{Weisheit2007,Maruyama2009,Shiota2012,Wang2012d,Bauer2015}. Using photoemission spectroscopy, \citet{Manchon2008c} showed that both the perpendicular magnetic anisotropy and AHE reach a maximum in Pt/Co/AlOx trilayers when the Co/AlOx interface is optimally oxidized. This effect is connected to the dependence of the interfacial magnetic anisotropy on the electron density and orbital character of the interface atoms \cite{Yang2011,Chshiev2017}. It is therefore natural to expect that other spin-orbit coupling properties, such as SOT \cite{Freimuth2014a} or DMI \cite{Belabbes2016b,Srivastava2018}, can be controlled by tuning the interfacial electron density through oxidation or by applying a gate voltage \cite{Liu2014d,Qiu2015,Emori2014b}. \citet{Miron2011b} first showed that moderate oxidation of Pt/Co/AlO$_x$ favors current-induced switching, as recently confirmed in Pt/Co/CoO$_x$ layers oxidized in air, in which up to a two-fold enhancement of the SOT efficiency was measured relative to Pt/Co/MgO \cite{Hibino2017}. On the other hand, \citet{Liu2014d} demonstrated that both field-like and damping-like torques can be modified by gating Pt/Co/Al$_2$O$_3$ multilayers [Fig.~\ref{Fig_SOT_Oxidation}(a)], obtaining an enhancement of 4\% (1\%) of the field-like (damping-like) torque for a gate voltage of about 7~V. Since the gate voltage essentially modifies the electric dipole of the Co/Al$_2$O$_3$ interface and leaves the SHE from Pt unaffected, this observation provides some indication about the origin of the SOTs in this system. \citet{Liu2014d} estimated that the SHE does not contribute to more than 20\% of the field-like torque, while the interfacial spin-orbit coupling produces about 50\% of damping-like torque. \citet{Emori2014b} carried out measurements on gated Pt/Co/GdO$_x$, showing that oxidation of the top Co interface leads to a 10-fold increase of the damping-like torque due to oxygen ion migration, which also affects the magnetic anisotropy. \citet{Qiu2015} demonstrated the spectacular impact of interfacial oxidation on SOTs in Pt/CoFeB/SiO$_2$, where the oxidation of the CoFeB/SiO$_2$ layer is varied continuously. They reported that the sign of both damping-like and field-like torques changes from positive to negative when increasing the oxidation of CoFeB [see Fig.~\ref{Fig_SOT_Oxidation}(b)]. The authors attributed this change of sign to the increase of the orbital moment of Fe and Co upon oxidation \cite{Nistor2011,Yang2011}. This results in an enhancement of the interfacial SOT at the upper CoFeB/SiO$_2$ interface that can even dominate over the SOT arising from the bottom Pt/CoFeB.

Oxidation of the bottom Pt interface in Pt/Ni$_{81}$Fe$_{19}$ bilayers also leads to drastic enhancements of both damping-like and field-like torque efficiencies, which can be controlled by the oxygen flow during sputter deposition as well as by a gate voltage \cite{An2018,An2018b}. Interestingly, the maximum SOT efficiency in this system, $\xi^{j}_{\rm DL} = 0.92$ ($\xi^{E}_{\rm DL} = 9 \times 10^{3}$~$\Omega^{-1} m^{-1}$), is reached for a fully oxidized nonconducting Pt layer. Other reports reveal an enhancement of $\xi^{j}_{\rm DL}$ from -0.14 to -0.49 upon oxidation of W in W/CoFeB/TaN \cite{Demasius2016} and the emergence of strong SOT in as-grown SiO$_x$/Co/Cu \cite{Verhagen2015} and oxidized SiO$_x$/NiFe/Cu layers \cite{An2016}, with contrasting evidence on the role played by the oxidized interfaces. These experiments show that interfacial spin-orbit coupling can produce significant field-like and damping-like torques, but also that a detailed microstructural analysis of the bulk vs interface oxidation is required to understand the role of oxygen in inducing or modifying the SOT.\par

Finally, \citet{Qiu2016} demonstrated a 3-fold enhancement of the SOT magnitude in a Pt/Co/Ni/Co multilayer by capping the system with Ru. This result is interpreted in terms of enhanced spin absorption induced by the negative spin polarization arising at the Co/Ru interface \cite{Nozaki2004} and could partly explain the very large SOT magnitude measured in synthetic antiferromagnetic domain walls \cite{Yang2015a}. Recent work on IrMn$_3$/NiFe epitaxial layers also shows that $\xi^{j}_{\rm DL}$ has a facet-dependent contribution, which arises from the different orientation of the Mn magnetic moments at different interfaces \cite{Zhang2016b}.

\paragraph{Angular dependence} \label{MML:exp:FMNM:angular}

\begin{figure}[b]
\centering
\includegraphics[width=6.5cm]{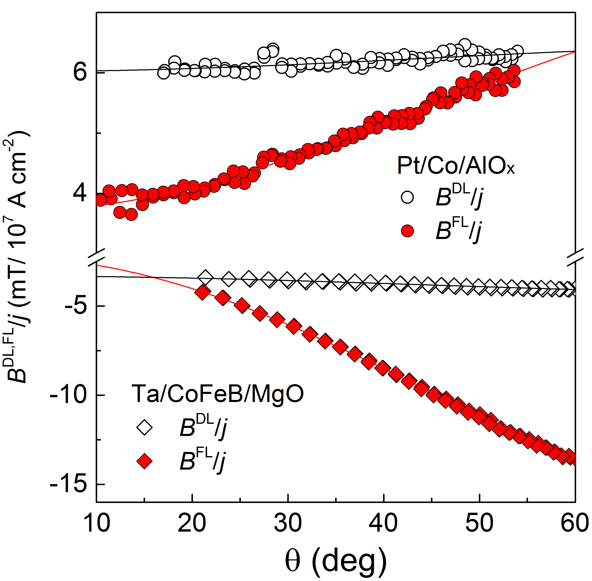}
\caption{(Color online) Angular dependence of $B_{\rm DL}$ and $B_{\rm FL}$ measured in as-grown Pt(3)/Co(0.6)/AlO$_x$ \cite{Garello2013} and Ta(3)/CoFeB(0.9)/MgO \cite{Avci2014} at room temperature. The angle $\theta$ between the magnetization and the $z$-axis is determined by anomalous Hall resistance measurements. The solid lines are fits to the function $B^{\rm DL,FL}_\theta = B^{\rm DL,FL}_0 + B^{\rm DL,FL}_2 \sin^2\theta$.
}\label{Fig_SOT_angdep}
\end{figure}

As mentioned in Subsection \ref{MML:phen}, the SOTs are anisotropic, i.e., their magnitude changes depending on the magnetization direction in a way that is more complex than described by Eq.~\eqref{eq:torquedef}. In polycrystalline systems with $C_{2v}$ symmetry, the magnitude of this anisotropy is characterized by the coefficients $\tau^{\{2n\}}_{\rm DL,FL}$ in Eq.~(\ref{eq:angledt}). As measured in Pt/Co/AlO$_x$ \cite{Garello2013}, Pt/Co/MgO \cite{Gweon2019}, Ta/CoFeB/MgO \cite{Avci2014,Qiu2014}, and Pd/Co/AlO$_x$ \cite{Ghosh2017}, the SOT anisotropy can be quite large. Figure~\ref{Fig_SOT_angdep} shows that both field-like and damping-like torques increase in absolute value when the magnetization points in-plane, which is the typical behavior observed in metal layers. The anisotropies of the field-like and damping-like components differ from each other and can reach up to a factor of 4 depending on the material and annealing conditions.\par

The angular dependence of the SOT, although quite general, provides additional clues about the physics taking place in these ultrathin layers. Different physical mechanisms can generate such an angular dependence: (i) the presence of D'yakonov-Perel relaxation \cite{Ortiz2013}, (ii) the distortion of the Fermi surface when changing the magnetization direction due to strong spin-orbit coupling \cite{Lee2015,Haney2013a}, and (iii) the angular dependence of the interfacial mixing conductance, i.e., the change of spin absorption and reflection as a function of the magnetization direction \cite{Amin2016a,Baek2018}. Additional effects related to spin scattering in the nonmagnetic metal may also be relevant, such as, e.g., spin swapping \cite{Saidaoui2016}. Interestingly, \citet{Qiu2014} reported that the angular dependence of the two torque components vanishes when decreasing the temperature, an observation that highlights the importance of scattering events in the emergence of the angular dependence of the SOTs. Finally, systems characterized by low crystalline symmetry may display additional contributions not included in Eqs.~\eqref{eq:angledt} and \eqref{eq:angleft} due to the specific symmetry of the spin and orbital textures\cite{MacNeill2017,Chen2016c}.

\paragraph{Temperature dependence} \label{MML:exp:FMNM:temperature}
\begin{figure}[t]
\centering
\includegraphics[width=8.5cm]{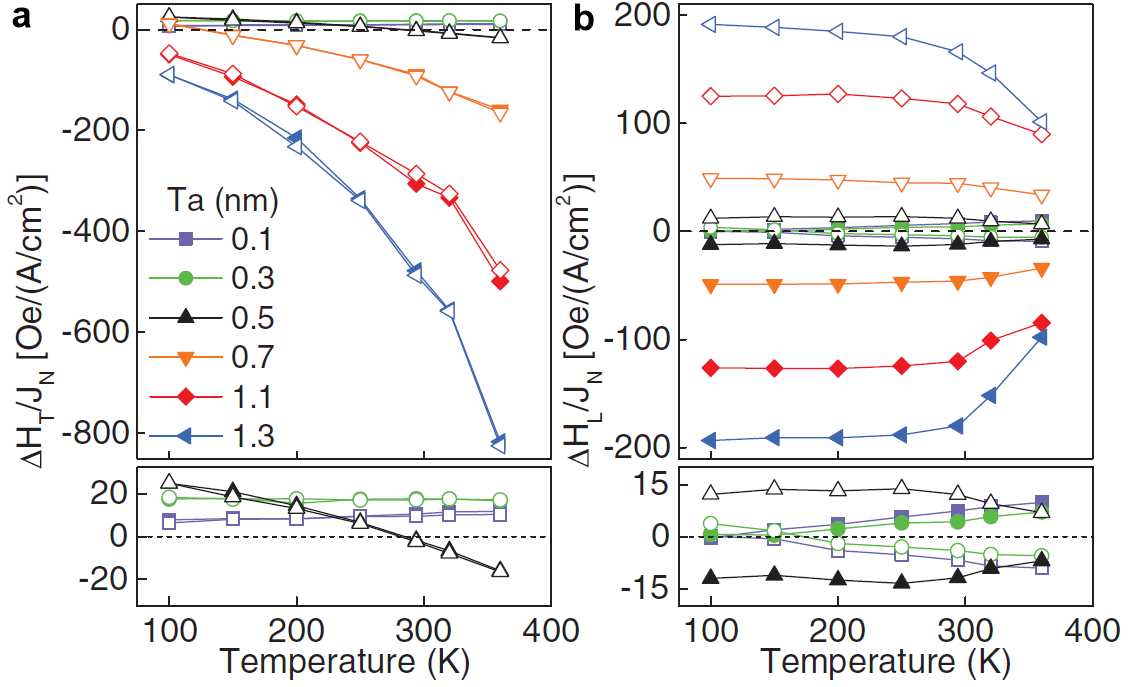}
\caption{(Color online) Temperature dependence of (a) $B_{\rm FL}/j$ ($\Delta H_{\rm T}/j$) and (b) $B_{\rm DL}/j$ ($\Delta H_{\rm L}/j$) in Ta/CoFeB(1)/MgO(2) layers with different Ta thickness \cite{Kim2014}. The bottom panels show a magnified view of the fields for the thinner Ta layers. Solid and open symbols correspond to the magnetization pointing along +${\bf z}$ and -${\bf z}$, respectively.}\label{Fig_SOT_temperature}
\end{figure}

A way to obtain information on the physics governing the SOTs is to measure their magnitude as a function of temperature. In Ta/CoFeB/MgO, \citet{Qiu2014} reported that the field-like torque decreases linearly when reducing the sample temperature, while the damping-like torque remains mostly unaffected. A qualitatively similar behavior was observed by \citet{Kim2014} in similar structures, i.e., the field-like torque decreases dramatically with the temperature, while the damping-like torque increases from 400 to 300~K and saturates at lower temperatures (Fig.~\ref{Fig_SOT_temperature}). Since the Ta resistivity is almost constant between 100 and 400~K, the relative independence of the damping-like torque on temperature is consistent with the damping-like torque being driven by the intrinsic SHE of Ta. In contrast, the strong decrease of the field-like torque suggests that scattering events involving phonons and magnons (usually stronger at disordered interfaces) play an important role in the emergence of this component. Studies of the temperature dependence of the SOTs in Pt-based structures, on the other hand, show that the field-like and damping-like SOTs are both approximately constant with temperature in as-grown Pt/Co/MgO, whereas both increase with temperature in annealed Pt/CoFeB/MgO \cite{Pai2015}. In the latter case, the field-like torque shows a much stronger change compared to the damping-like torque and even changes sign, from parallel to antiparallel to the Oersted field, at around 125~K.

In bilayers including disordered alloys of nonmagnetic metals such as Cu$_{x}$Au$_{1-x}$, where extrinsic effects dominate, the damping-like torque decreases upon reducing the temperature, consistently with an extrinsic bulk-like SHE, whereas the field-like torque increases \citet{Wen2017}. On the other hand, alloys that present a ferromagnetic to paramagnetic transition, such as Fe$_{x}$Pt$_{1-x}$ in combination with a ferromagnet such as CoFeB, display a maximum of the damping-like torque near the Curie temperature, which is attributed to spin fluctuation enhancement of the SHE arising from the interaction between the conduction electrons and the localized magnetic moments \cite{Ou2018b}. Ab-initio calculations additionally show that the generation and absorption of spin currents in an ordered FePt alloy are extremely sensitive to the distribution of defects near the interface \cite{Geranton2016}. Overall, these studies support the view that intrinsic as well as extrinsic mechanisms contribute in different proportion to the field-like and damping-like torques.

\subsubsection{Ferrimagnet and antiferromagnet/nonmagnetic metal layers} \label{MML:exp:FIMAFM}

Ferrimagnetic films were once widely used as recording media in bubble memories \cite{Bobeck1975} and magneto-optic memories \cite{Jenkins2003}. Applications included both insulating garnets \cite{Nielsen1976} and amorphous rare-earth 3$d$ transition-metal alloys \cite{Buschow1984}. Depending on their composition, ferrimagnets can exhibit a magnetization compensation temperature ($T_{\rm M}$) where the magnetizations
of the two antiparallel coupled sublattices cancel each other and, similarly, an angular momentum compensation temperature ($T_{\rm A}$) where the total angular momentum of the two sublattices vanishes \cite{Nielsen1976,Buschow1984,Hirata2018}. The frequency of the uniform spin precession mode as well as the magnetic damping are expected to diverge at $T_{\rm A}$ \cite{Wangsness1954,Stanciu2006}, which makes these materials extremely interesting for ultrafast switching \cite{Stanciu2007,Mangin2014} as well as fast domain wall motion \cite{Kobayashi2005,Kim2017c,Caretta2018,Siddiqui2018}.

SOT-induced switching of ferrimagnets has been reported for amorphous ferrimagnetic alloys, such as Ta/TbFeCo \cite{Zhao2015}, Ta/TbCo \cite{Finley2016}, Pt/GdCo \cite{Mishra2017}, and Pt/GdFeCo \cite{Roschewsky2017}, as well as rare earth garnets such as Tm$_3$Fe$_5$O$_{12}$/Pt \cite{Avci2017,Avci2017b,Velez2019} and Tm$_3$Fe$_5$O$_{12}$/W \cite{Shao2018}. In contrast to ferromagnets, the reduced saturation magnetization of these systems allows for switching relatively thick layers, up to 30~nm \cite{Roschewsky2017}, at current densities of the order of $10^7$~A/cm$^{2}$. Moreover, because of the alternance of magnetic moments with opposite orientation on neighboring atomic sites, spin dephasing due to spin precession in metallic ferrimagnets partially cancels out, which allows a spin current to propagate for several nm inside these materials, as reported for Pt/[Co/Tb]$_{N}$ multilayers and Pt/CoTb amorphous alloys \cite{Yu2019}. These properties, combined with the bulk perpendicular magnetic anisotropy of rare-earth 3$d$ transition-metal compounds, make ferrimagnets very interesting for applications requiring relatively thick magnetic layers.

Measurements of the SOT as a function of temperature \cite{SeungHam2017,Ueda2017} and composition \cite{Finley2016,Roschewsky2017,Je2018} show that the damping-like effective field tends to diverge as $B_{\rm DL} \propto 1/M_s$ near $T_{\rm M}$, whereas $\xi^{j}_{\rm DL}$ is roughly constant across $T_{\rm M}$, as expected. In some cases, however, a disproportionate scaling between $B_{\rm DL}$ and $1/M_s$ has been observed, leading to a considerable enhancement of $\xi^{j}_{\rm DL}$ of yet unclear origin \cite{Mishra2017,Je2018,Yu2019}. As not only $M_s$ and $T_{\rm M}$, but also the magnetic anisotropy, Gilbert damping, spin-orbit scattering, and spin dephasing depend on the composition and thickness of these systems, it is not surprising that the simple $1/M_s$ scaling has no general validity. An interesting point is that, even in systems where $B_{\rm DL} \propto 1/M_s$, the threshold switching current does not decrease at $T_{\rm M}$, but rather changes smoothly as a function of composition \cite{Je2018} or thickness \cite{Yu2019}. This behavior agrees with a macrospin model based on the LLG equation for two antiferromagnetically coupled lattices, which shows that the threshold switching current scales with the effective perpendicular anisotropy \cite{Je2018}. The latter depends on the sum of effective anisotropy energy of each lattice, which does not cancel out at the compensation point.

In fully compensated bipartite antiferromagnets, simulations predict that the N\'{e}el order can be manipulated via damping-like SOT \cite{Gomonay2010} (see Subsection \ref{s:antiferromagnets}). It has been recently shown that current injection in Pt/NiO \cite{Chen2018} and Pt/NiO/Pt \cite{Moriyama2018} leads to switching of the N\'{e}el vector of up to 90~nm thick films of NiO, independently of the strain state and crystallographic orientation \cite{Baldrati2018}. Contrasting mechanisms have been proposed to explain this type of switching, based on the coherent rotation of the N\'{e}el vector \cite{Chen2018}, field-like SOT acting on uncompensated interfacial spins \cite{Moriyama2018}, as well as rotation of the N\'{e}el vector inside individual domains combined with the displacement of the domain walls driven by the damping-like SOT \cite{Baldrati2018}.

\subsubsection{Ferromagnet/semiconductor layers} \label{MML:exp:FMSC}
We now turn from purely metallic systems to ferromagnet/semiconductor bilayers, in which the semiconductor has a specific crystal structure that brings about additional symmetries on top of the one arising from interfacial inversion symmetry. For instance, in zinc-blende lattices under strain, such as GaAs, a spin accumulation can be generated via the iSGE driven by Rashba and Dresselhaus spin-orbit coupling as well as by the bulk SHE (see Sect. \ref{s:noncentro}). Differently from the commonly studied polycrystalline NM/FM samples, where the iSGE-based and the SHE-based mechanisms are indistinguishable in the lowest order terms \cite{Garello2013}, the dependence of the torques on the angle of the current relative to the high symmetry directions of the semiconductor crystal provides a direct means to disentangle the SHE and iSGE contributions. \citet{Skinner2015} proved this point by investigating the SOTs of an epitaxial Fe(2 nm)/(Ga,Mn)As(20 nm) bilayer using the ST-FMR technique. The GaAs host was doped with high enough concentration of substitutional Mn acceptors to increase the semiconductor conductivity, but low enough so that (Ga,Mn)As remains paramagnetic at room temperature. It was then shown that the field-like and damping-like torques have similar magnitude, with the first originating from the iSGE with Dresselhaus symmetry and the second from the SHE-like spin current generated inside the paramagnetic $p$-doped GaAs layer. \citet{Chen2016c}, on the other hand, showed that the SOT of epitaxial Fe films grown on non-conducting GaAs(001) originate from the interfacial iSGE and have mixed Rashba and Dresselhaus symmetry, which also leads to the emergence of an unusual crystalline magnetoresistance \cite{Hupfauer2015}. The interfacial spin-orbit interaction and SOT can further be modulated by applying a gate voltage across the Schottky barrier at the Fe/GaAs interface \cite{Chen2018c}.

\begin{figure}[t]
\centering
\includegraphics[width=1\columnwidth]{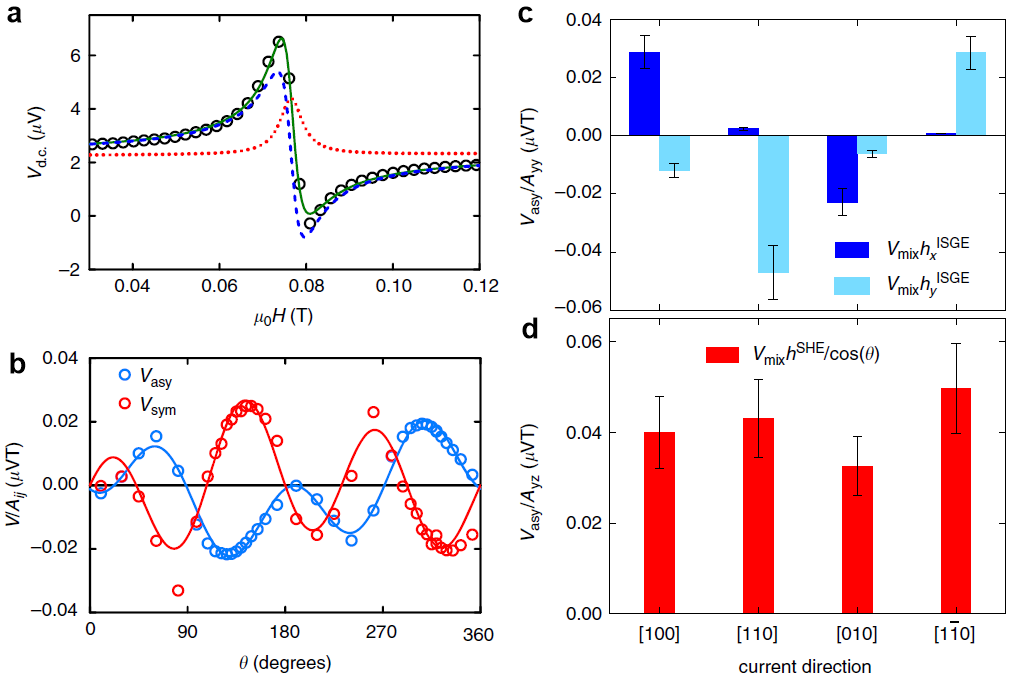}
\caption{(Color online) (a) Electrical excitation and detection of FMR induced by a 16.245 GHz
RF current in a Fe(2 nm)/(Ga,Mn)As(20 nm) bilayer. A typical ST-FMR curve (points) is shown as a function of external field. The dc voltage is fitted (solid green line) by a combination of symmetric (red dotted line) and antisymmetric (blue dashed line) Lorentzians. (b) Dependence of the fitted Lorentzian amplitudes on the in-plane magnetization angle for a device with current in the [010] direction. (c) iSGE dependence on the direction of the current. The fitted in-plane field coefficients (representing the field-like torque) for a set of devices in different crystal directions. (d) The fitted out-of-plane field coefficient (representing the damping-like torque) for the same devices. Adapted from \citet{Skinner2015}.}\label{Fig_SOT_FMSC}
\end{figure}

Evidence of strong SOTs due to the iSGE has been observed also in heterostructures involving transition metal dichalcogenides and metallic ferromagnets. Van der Waals crystals provide a unique platform for generating SOTs because they have strong spin-orbit coupling, a range of broken crystal symmetries, and can be prepared as monolayer crystalline films by exfoliation or chemical vapor deposition methods \cite{Xu2014,Manchon2015}. \citet{Shao2016} showed that the field-like torque in 1~nm CoFeB deposited on monolayer MoS$_2$ and WSe$_2$ is of the order of 0.1-0.14 mT/($10^7$ A/cm$^{2}$), independently of temperature, and is consistent with iSGE-induced spin density, whereas the damping-like torque is negligibly small. Sizable damping-like SOTs, on the other hand, have been reported for NiFe deposited on MoS$_2$ \cite{Zhang2016} and on
the Weyl semimetal WTe$_2$ \cite{MacNeill2017}. The latter case is of particular interest as the surface crystal structure of WTe$_2$ has only one mirror plane and no two-fold rotational invariance about the $c$-axis, which allows for a damping-like torque that is directed out-of-plane when the current is applied along a low-symmetry axis of the surface. Such a damping-like torque is forbidden by symmetry in NM/FM bilayers, where the direction of the incoming spin polarization is in-plane. The possibility of controlling the allowed symmetry of the damping-like SOT in multilayer samples is particularly attractive for counteracting the torque due to magnetic damping during magnetization reversal in systems with perpendicular magnetic anisotropy. Further, the current-induced spin density in two-dimensional materials is expected to be extremely sensitive to gating, thus allowing for tuning the SOT efficiency.

\subsubsection{Ferromagnet/topological insulator layers} \label{MML:exp:FMTI}

Three dimensional topological insulators are materials that have insulating bulk and conductive surface states \cite{Hasan2011,Roche2015}. The surface states are protected by time-reversal symmetry and have a Dirac-like linear dispersion characterized by spin-momentum locking (Fig.~\ref{fig-TI}), a property that makes them very attractive in the context of SOT and spintronics. Moreover, owing to hexagonal-warping of the Dirac cone \cite{Kuroda2010}, a current carried by the topological surface states can generate a nonequilibrium spin density with both in-plane and out-of-plane components, which can induce out-of-plane and in-plane torques onto an adjacent magnetic layer.

The most thoroughly investigated topological insulators to date are the bismuth and antimony chalcogenides M$_{2}$Q$_{3}$, where M = Bi, Sb and Q = Se, Te.
In intrinsic systems, the Fermi level resides in the bulk energy gap and thus only intersects the topological surface states. However, these materials are narrow gap semiconductors that are very sensitive to doping by impurities or crystalline defects, which typically shifts the Fermi level to the conduction band. Furthermore, unintentional surface doping caused by the formation of extrinsic defects or the adsorption of impurities leads to the emergence of a two-dimensional electron gas with strong Rashba-split bands that are wrapped by the topological Dirac states \cite{King2011}.
These effects have a strong influence on SOT, which has not yet been fully understood.

Topological insulator thin films are usually grown by molecular beam epitaxy \cite{He2013}. In order to favor surface transport over bulk conduction, it is necessary to minimize defects such as Se/Te vacancies, dislocations, and twin domains. This task has proven to be quite challenging, requiring careful optimization of the lattice matching with the substrate and growth kinetics \cite{Richardella2015,Bonell2017}. Bulk insulating materials can be obtained also by growing naturally compensated ternary alloys such as (Bi$_{1-x}$Sb$_{x}$)$_{2}$Te$_{3}$, which exploit the tendency of Bi$_{2}$Se$_{3}$ and Bi$_{2}$Te$_{3}$ to be $n$-type and of Sb$_{2}$Te$_{3}$ to be $p$-type \cite{Zhang2011b}.

Since topological insulators involve heavy elements and spin-momentum locked electron states, they are expected to show large charge-spin conversion and SOT efficiency when interfaced with a magnetic layer. However, three issues arise when considering these systems. First, the proximity between a ferromagnet and a topological insulator induces complex electronic hybridization effects, which go beyond the simple notion of a magnetic exchange field breaking time reversal symmetry and opening a gap in the surface states \cite{Wray2011}. Using first-principles calculations, \citet{Zhang2016d} and \citet{Marmolejo-Tejada2017} showed that charge transfer between Bi$_{2}$Se$_{3}$ and 3$d$ metal layers such as Co, Ni, and Cu, shifts the topological surface states below the Fermi energy, where hybridization with the metal bands destroys or heavily distorts the helical spin structure. Crucially for SOTs, it was found that proximity spin-orbit coupling also modifies the electronic states of the ferromagnet adjacent to Bi$_{2}$Se$_{3}$, leading to a Rashba-like spin texture \cite{Marmolejo-Tejada2017}. Thus, the properties of magnetic/topological insulators bilayers, even in the theoretical approximation of ideal materials and interfaces, cannot be extrapolated from those of the parent layers. Second, the interface chemistry between a topological insulator such as Bi$_{2}$Se$_{3}$ and typical contact metals (Pd, Ir, Cr, Co, Fe, Ni) leads to the formation of metal selenides, metallic Bi, or intermetallic alloys, which can evidently alter the properties of the pristine materials \cite{Walsh2017}. Third, because of the competition between bulk and surface conduction, which depends on temperature and extrinsic factors, it is hardly possible to determine the current distribution in magnetic/topological insulators bilayers. This uncertainty makes it difficult to identify the electronic states responsible for charge-spin conversion as well as to provide consistent estimates of the SOT efficiency in different systems.

Regardless of the role played by the topological surface states, mounting experimental evidence suggests that strong spin-momentum coupling can be achieved in these materials.
Spin-charge conversion has been reported by spin pumping for bismuth and antimony chalcogenide layers adjacent to metallic ferromagnets \cite{Deorani2014,Shiomi2014,Jamali2015,Kondou2016,Mendes2017} and insulating ferrimagnets \cite{Wang2016l,Tang2018} as well as by magnetoresistance measurements \cite{Ando2014,Li2014,Yasuda2016}. Current-induced SOTs have been demonstrated by ST-FMR in Bi$_2$Se$_3$/NiFe and Bi$_2$Se$_3$/CoFeB bilayers \cite{Mellnik2014,Wang2015b}, gate control of the torque efficiency \cite{Fan2016}, and magnetization switching \cite{Fan2014a,Yasuda2017,Wang2017d,Han2017,Khang2018}. In these experiments, the reported damping-like torque efficiency is widely distributed from 0.01 to 2 for Bi$_2$Se$_3$, reaching $\sim 50$ in Bi$_2$Sb$_3$/MnGa \cite{Khang2018} and even larger values in (Bi$_{0.5}$Sb$_{0.5}$)$_2$Te$_3$/(Cr$_{0.08}$Bi$_{0.54}$Sb$_{0.38}$)$_2$Te$_3$ \cite{Fan2014a}. In the latter case, however, the SOT analysis is complicated by nonlinear Hall effects \cite{Yasuda2017}.

Demonstrations of room temperature SOT-driven switching with threshold currents that are about one order of magnitude smaller compared to NM/FM bilayers are particularly interesting in view of possible applications. \citet{Wang2017d} reported switching of in-plane magnetized Bi$_2$Se$_3$/NiFe with a critical current of $\sim 6\times10^5$ A/cm$^2$, while \citet{Han2017} demonstrated switching of perpendicularly magnetized Bi$_2$Se$_3$/CoTb at $\sim 3\times10^6$ A/cm$^2$ and \citet{Khang2018} obtained a similar switching threshold for the high coercivity system Bi$_2$Sb$_3$/MnGa. Whereas all these studies were performed on topological insulators grown by molecular beam epitaxy, \citet{Mahendra2018} used sputtering to grow Bi$_2$Se$_3$/Ta/CoFeB/Gd/CoFeB heterostructures with perpendicular anisotropy promoted by the 0.5~nm thick Ta layer. Due to its polycrystalline nature, the Bi$_2$Se$_3$ layer is highly resistive, one order of magnitude larger than \cite{Mellnik2014}, thereby enabling the current to flow mostly through the interface and in the ferromagnetic layer, which enhances the SOT efficiency. However, the role, if any, of the topological surface states in these sputtered layers remains to be proven, together with the stoichiometric profile of the Bi$_2$Se$_3$ films. Other strategies to improve the SOT efficiency in these systems rely on the use of spacer layers, such as Ag, which favor the formation of Rashba-split bands with strong spin-momentum coupling \cite{Shi2018}, the creation of Rashba-Dirac coupled systems \cite{Eremeev2015}, and the search for novel topological materials \cite{Rojas-Sanchez2016,Manna2018}.
\par

\subsubsection{Two-dimensional alloys and oxide interfaces} \label{MML:exp:SA}

Spin pumping measurements performed on heterostructures consisting of a ferromagnetic layer and an interface alloy with strong Rashba-like spin-orbit coupling, such as Ag/Bi \cite{Rojas-Sanchez2013b}, Cu/Bi \cite{Isasa2016}, and Cu/Bi$_2$O$_3$ \cite{Karube2016}, have revealed large spin-to-charge conversion efficiencies due to the SGE. This effect converts a nonequilibrium {\em spin density} {\bf S} into an interfacial 2D charge current $\tilde{j}_{\rm c}$ \cite{Ivchenko1978}. Owing to the interfacial nature of $\tilde{j}_{\rm c}$, the spin-to-charge conversion is given by the inverse Rashba-Edelstein "length" 
\begin{equation}
\lambda_{\rm IREE} =\frac{\hbar}{2e}\frac{\tilde{j}_{\rm c}}{j_{\rm s}},
\end{equation}
where $j_{\rm s}$ is the spin current density pumped by the ferromagnet and associated with the spin density {\bf S}, expressed in $(\hbar/2e)$A/m$^{2}$, and $\tilde{j}_{\rm c}$ is measured in A/m. In the framework of the Rashba model, it can be shown that $\lambda_{\rm IREE} = \alpha_{\rm R}\tau/\hbar$, where $\alpha_{\rm R}$ is the Rashba coupling strength and $\tau$ the momentum relaxation time at the Rashba-split Fermi surface \cite{Gambardella2011,Shen2014a}. Typical values of $\lambda_{\rm IREE}$ range from 0.1-0.3
~nm in NiFe/Ag/Bi \cite{Rojas-Sanchez2013b,Zhang2015a} to -0.6~nm in NiFe/Cu/Bi$_2$O$_3$ \cite{Karube2016}.

Comparison with the inverse SHE in NM/FM bilayers is achieved by converting the effective spin Hall angle (or the SOT efficiency) into $\lambda_{\rm IREE}$ by taking
\begin{equation}
\lambda_{\rm IREE}=\theta_{\rm sh}\lambda_{\rm sf}\tanh(t_I/2\lambda_{\rm sf}),
\end{equation}
where $t_I$ is the "thickness" of the interface layer in which the spin-to-charge conversion takes place \cite{Rojas-Sanchez2013b,Rojas-Sanchez2016b}. The maximum attainable length is therefore $\lambda_{\rm IREE}^{\rm max}=\theta_{\rm sh}\lambda_{\rm sf}$ for $t_I\gg\lambda_{\rm sf}$. For values of $\theta_{\rm sh}$ between 0.1 and 0.3, and $\lambda_{\rm sf}=1.5 - 2$~nm as typical of Pt, Ta, and W, one obtains $\lambda_{\rm IREE}=0.15 - 0.6$~nm, which is comparable to $\lambda_{\rm IREE}$ of the Ag/Bi interface.

Current injection in such systems results in sizable damping-like and field-like SOT due to the iSGE, as demonstrated by ST-FMR for NiFe/Ag/Bi \cite{Jungfleisch2016b} as well as for the oxidized heavy metal interfaces described in Sect.~\ref{MML:exp:FMNM:interfacial} \cite{Fujiwara2013a,An2016,An2018,An2018b,Demasius2016}. In this situation, a 2D charge current produces a 3D non-equilibrium spin density, and the relation between the effective spin Hall angle and the Rashba-Edelstein length reads \cite{Laczkowski2017}
 \begin{equation}
\frac{1}{\lambda_{\rm REE}}=\frac{2e}{\hbar}\frac{j_{\rm s}}{\tilde{j}_{\rm c}}=\frac{\theta_{\rm sh}}{t_I}\tanh(t_I/2\lambda_{\rm sf}).
\end{equation}
Therefore, the maximum charge-to-spin conversion is $\left(\frac{1}{\lambda_{\rm REE}}\right)^{\rm max}=\frac{\theta_{\rm sh}}{2\lambda_{\rm sf}}$ when $t_I\ll\lambda_{\rm sf}$. In other words, within this picture, a figure of merit of spin-to-charge conversion is $\theta_{\rm sh}\lambda_{\rm sf}$, while for charge-to-spin conversion it is $\theta_{\rm sh}/\lambda_{\rm sf}$.

Prominent spin-charge interconversion effects are observed also in 2D electron gases confined at the interface between two insulating oxides, such as LaAlO$_3$ and SrTiO$_3$ \cite{Ohtomo2004}. These systems host a variety of unusual electronic phases \cite{Zubko2011} as well as tunable carrier density and Rashba spin-orbit interaction \cite{Caviglia2010}. Even in the absence of heavy metals, the large interfacial electric fields and long electron relaxation time result in extraordinarily large $\lambda_{\rm IREE}$, which can be further modulated by electric gating \cite{Lesne2016,Song2017}. Spin pumping experiments on SrTiO$_3$/LaAlO$_3$/NiFe reveal that $\lambda_{\rm IREE}$ changes from 2 to -6~nm as the Fermi level is raised through the crystal-field split interface states, namely from a single low-lying band with $d_{\rm xy}$ character to the higher-lying heavier $d_{\rm xz,yz}$ bands, where $\alpha_{\rm R}$ is largest \cite{Lesne2016,Seibold2017}. The observation of strong SOT in SrTiO$_3$/LaAlO$_3$/CoFeB at room temperature \cite{Wang2017e} shows that the spin current generated by the iSGE at the oxide interface can be effectively absorbed by a magnetic layer deposited on a few nm thick LaAlO$_3$, likely via inelastic electron tunneling promoted by defect states in the oxide layer. Interfaces between complex oxides thus represent a notable alternative to heavy metal systems for the generation of SOT, offering additional tools to tune $\lambda_{\rm REE}$ by controlling the interplay of crystal field and spin-orbit effects in multifunctional heterostructures.

\subsubsection{Metallic spin-valves} \label{MML:exp:spinvalves}
Recent theoretical \cite{Taniguchi2015c,Freimuth2017b} and experimental works \cite{Humphries2017,Bose2018,Baek2018} pointed out the possibility to induce SOT in all-metallic spin-valves by in-plane current injection. These structures, similar to those employed for generating STT (Fig.~\ref{fig-SOT-STT}), are FM$_{\rm ref}$/spacer/FM$_{\rm free}$ trilayers where FM$_{\rm ref}$ is a fixed reference ferromagnet with magnetization along ${\bf p}$, FM$_{\rm free}$ is the ferromagnet on which the SOT is measured, and the spacer is a light metal (e.g., Cu or Ti) such that no or little SHE or iSGE are expected from it. According to \citet{Taniguchi2015c}, a spin current polarized along ${\bm \zeta}\parallel {\bf p}$ is generated by either the AHE or PHE in the bulk of the reference layer and absorbed by the free layer, giving rise to both damping-like and field-like SOTs according to Eq.~\eqref{eq:torquedef}. However, \citet{Baek2018} and \citet{Humphries2017} pointed out that spin filtering caused by spin-orbit scattering and spin-orbit precession experienced by electrons flowing at the interface between magnetic and nonmagnetic layers give rise to spin currents polarized along ${\bm \zeta}\parallel {\bf z}\times{\bf j}_{\rm c}$ and ${\bm \zeta}\parallel {\bf p}\times({\bf z}\times{\bf j}_{\rm c})$, respectively, which are potentially stronger than the bulk spin currents generated by the AHE and PHE. These predictions have been verified in trilayers with both in-plane \cite{Bose2018,Baek2018} and out-of-plane ${\bf p}$ \cite{Humphries2017}, for which the SOT symmetry is consistent with the latter mechanisms and allows also for field-free switching of the free layer \cite{Baek2018}.

\subsubsection{Established features and open questions} \label{MML:origin}
The complexity and interplay of the different charge-spin conversion mechanisms outlined in Section \ref{s:6} underpins an ongoing debate on the origin of SOTs and on strategies to improve their efficiency. Below, we summarize the most important findings drawn from experimental investigations of metallic layers:

\begin{itemize}[noitemsep,wide=0pt, leftmargin=\dimexpr\labelwidth + 2\labelsep\relax]

\item In most NM/FM systems, the sign of the damping-like torque is consistent with that of the SHE of the bulk nonmagnetic metal. Additionally, nonmagnetic metal elements with strong SHE present large damping-like torques. The magnitude and the sign of the damping-like torque can be modified by changing the oxidation state or the capping layer of the ferromagnet interface that is not in contact with the nonmagnetic metal. Significant damping-like torques have been reported also for ferromagnetic layers adjacent to metal alloys and oxide layers with a strong iSGE.

\item The field-like torque is of the same order of magnitude as the damping-like torque. The sign and magnitude of the field-like torque are not consistent with the predictions of the drift-diffusion model based on the bulk SHE.

\item The damping-like and field-like torques typically increase with the thickness of the nonmagnetic metal layer and saturate after a few nm. The dependence of the two torques on the nonmagnetic metal thickness is not the same.

\item The temperature dependence of the field-like and damping-like torques is different, indicating the distinct role of electron scattering by magnons or phonons.

\item Extrinsic effects related to both interface and bulk electron scattering are significant and can give rise to both damping-like and field-like torques. The SOTs are typically large in high resistivity metals and correlate with the presence of strong SMR in NM/FM bilayers and crystalline AMR in ferromagnet/semiconductor layers.

\item The angular dependence of the torques shows that interfacial spin-orbit coupling, either through D'yakonov-Perel relaxation, Fermi surface distortion or anisotropic mixing conductance, plays a relevant role.

\item The insertion of a nonmagnetic light metal spacer between the ferromagnet and a nonmagnetic metal reduces magnetic proximity effects in the nonmagnetic metal, but creates additional interfaces that can contribute to the generation of spin currents. Both the damping-like and field-like torques change upon the insertion of nonmagnetic and magnetic spacers.

\item 2D materials, topological insulators, and oxide heterostructures provide large SOTs when interfaced with magnetic layers, consistently with the iSGE arising from spin-momentum locked interface states. Additional contributions to the SOT may result from the SHE in systems with residual bulk conductivity. The symmetry of the SOTs generated by single crystal layers is determined by the current injection direction relative to the crystal axes.

\item Both damping-like and field-like torques can be controlled through interface engineering, such as gate voltage, oxidation, or capping layer, which offers an efficient way to improve charge-spin conversion in NM/FM as well as 2D systems.

\end{itemize}

SOT measurements in multilayer systems are often interpreted assuming either the SHE-SOT model or the Rashba-type iSGE. Such approaches are appealing because of their simplicity, but neglect important aspects of the generation of SOT. The one-dimensional drift-diffusion theory based on the bulk SHE (Subsection \ref{s:sheTorque}) is the most commonly employed model to relate the torque amplitude to the spin Hall conductivity of the nonmagnetic metal. Such a model includes the probability of spin transmission at the interface through the spin mixing conductance parameter, but neglects the interface-generated spin density by either the iSGE or spin-dependent electron scattering as well as the spin memory loss. Another major limitation of this model is that it assumes constant parameters $\sigma_N$, $\lambda_{\rm sf}$, and $\theta_{\rm sh}$ throughout the nonmagnetic metal layer, which is unjustified on both theoretical and experimental grounds. When the thickness of the nonmagnetic metal is comparable to the electronic mean free path (of the order of the grain size or $t_N$, i.e., a few nm in sputtered samples), semiclassical size effects become important and govern the current distribution in NM/FM bilayers \cite{Camley1989,Zhang1993}. Neglecting these effects can lead to a wrong estimation of the SOT efficiency and $\lambda_{\rm sf}$ \cite{Chen2017b}. Moreover, ab-initio calculations have shown that the intrinsic spin Hall conductivity can strongly vary close to the interface, leading to an enhancement of the SOT \cite{Wang2016b,Freimuth2015}.

On the other hand, most SOT models based on interfacial Rashba spin-orbit coupling assume a static spin polarization localized at a sharp interface between the nonmagnetic metal (or the oxide) and the ferromagnet. Considering the complexity of the real ultrathin magnetic multilayers involving complex orbital hybridization, disordered interfaces, and spin-dependent semiclassical size effects, it is quite unclear how these two models (bulk SHE and interfacial Rashba-like iSGE) apply to real systems. Spin pumping experiments at Bi surfaces have been interpreted as evidence for either an interface-enhanced SHE \cite{Hou2012} or the iSGE \cite{Rojas-Sanchez2013b}. Angle-resolved photoemission studies, on the other hand, provide evidence that the iSGE is not a pure 2D effect in metallic thin films: the presence of magnetic exchange \cite{Krupin2005}, out-of-plane spin polarization \cite{Takayama2011}, spin-momentum locked quantum well states in the ferromagnet \cite{Moras2015}, and topologically protected surface states \cite{Thonig2016,Marmolejo-Tejada2017} significantly alters the Rashba effect at metallic interfaces compared to model semiconducting heterostructures.

Extrinsic effects such as impurity and interface scattering induce additional spin currents that propagate through or away the magnetic layer and are polarized in directions different from the standard SHE and Rashba models, calling for a generalization of the spin current sources and spin mixing conductance \cite{Amin2016a,Amin2016b,Chen2015c,Baek2018,Humphries2017,Saidaoui2016}. For example, electron scattering from an interface with spins parallel and antiparallel to the local spin-orbit field have different reflection and transmission probabilities, leading to a net spin current polarized parallel to the ${\bf y} = {\bf z}\times {\bf j}_{\rm c}$ direction, identical to that of the spin current due to the SHE. Additionally, if the electrons carry a net spin polarization along ${\bf p}$, as in a ferromagnet, precession about the spin-orbit field results in a transverse spin current with polarization parallel to the ${\bf p} \times {\bf y}$ direction \cite{Baek2018}. The interplay between all these effects makes it questionable to draw a clear separation between the SHE and iSGE in metallic structures, even when considering idealized theoretical models of these heterostructures.\par

Finally, the typical SOT bilayers are usually only a few nanometer thick. The phenomenological notion of an interface between bulk regions, as well as the interpretation in terms of bulk SHE, appear unjustified based on both theoretical and practical grounds. A full quantum-mechanical treatment of the SOT in realistic three-dimensional structures including disorder is therefore essential to reach consistency between experiments and theory.

\subsection{Magnetization dynamics} \label{MML:dynamics}

\begin{figure}[t]
\centering
\includegraphics[width=8cm]{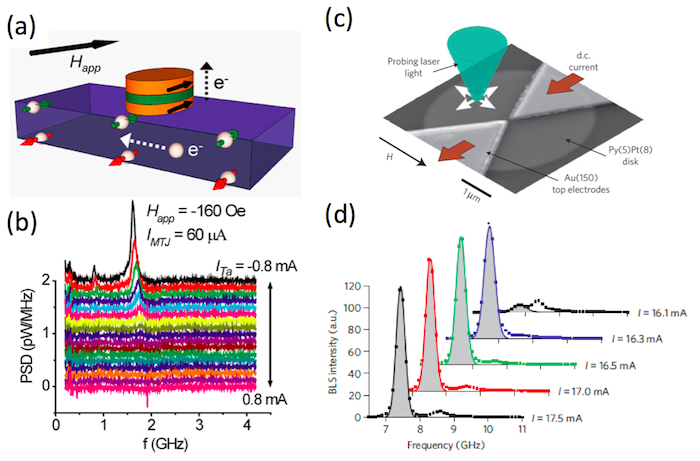}
\caption{(Color online) (a,c) Two configurations for the SOT-induced nano-oscillator and (b,d) their corresponding excitation spectrum. (a,b) Nanopillar deposited on top of a nonmagnetic metal \cite{Liu2012b}, and (c,d) Local injection into an extended ferromagnet \cite{Demidov2012}.}\label{fig:stno}
\end{figure}

It is well known that, due to the STT, a spin-polarized electric current injected into a ferromagnetic layer through a nanocontact leads to the emission of spin waves \cite{Berger1996,Tsoi1998}. This effect provides a way to realize tunable spin-torque nano-oscillators, which can serve as active microwave components in integrated circuits \cite{Tsoi2000,Kiselev2003,Rippard2004,Demidov2010,Madami2011}.
The discovery of SOT has led to new paradigms to control the high-frequency magnetization dynamics by means of dc and ac currents \cite{Demidov2017}. In contrast to STT, SOTs allow for the compensation of magnetic damping and the generation of spin waves in spatially extended regions of a magnetic material. Moreover, since the spin and charge currents follow separate paths, the electrical current does not need to flow through the active magnetic layer, allowing for the excitation of both conducting and insulating magnetic materials. SOTs thus enable efficient and flexible device geometries for the generation and amplification of magnetic oscillations as well as for the propagation and manipulation of coherent spin waves, opening entirely new perspectives in the field of magnonics \cite{Chumak2015}.

To a first approximation, the effects of the SOTs on the magnetization dynamics are described by Eqs.~\eqref{eq:LLG} and \eqref{eq:torquedef}. Consequently, one expects that the field-like torque shifts the frequency spectrum of the magnetic layer, similar to an applied magnetic field, and that the damping-like torque changes the magnitude of the magnetic damping. An early demonstration of the SOT-induced modification of magnetic damping was reported
by \citet{Ando2008b} in resonantly excited Pt/NiFe, in which the width of the FMR line decreased or increased depending on the sign of the dc current injected in the bilayer. This work evidenced variations of the damping constant $\alpha$ by an amount
\begin{equation}\label{eq:alpha}
\Delta\alpha = \frac{\gamma}{2\pi f M_s t_F}\frac{\hbar}{2e}\sin\psi \, \xi^{j}_{DL} \, j_{c},
\end{equation}
where $f$ is the resonance frequency of the magnetic layer and $\psi$ is the angle between the current and the precessional axis of the magnetization. Later measurements showed how the electrical control of magnetic damping can be used to enhance the spin wave propagation length in microwave guides \cite{Demidov2014b,An2014}. The variations of $\Delta\alpha$ reported in the literature range from a few percent to complete compensation of the damping, which eventually results in the onset of steady-state auto-oscillations of the magnetization \cite{Demidov2012,Liu2012b}. It shall be noted, however, that the simple linear relationship between damping and current exemplified by Eq.~\eqref{eq:alpha} is observed only at low currents, whereas nonlinear phenomena and magnetic fluctuations lead to a more complex behavior as the damping compensation is approached \cite{Demidov2017}.

A limiting factor for achieving self-sustained oscillations is the degeneracy of spin wave modes. If the sample is large, a significant amount of modes compete with each other to absorb the excitations induced by the SOT. In such a case, the degeneracy is high and only thermal excitations can be electrically controlled rather than current-driven coherent oscillations. Achieving self-sustained magnetic oscillations requires to lift the degeneracy by reducing the size of the sample and thereby lowering the excitation threshold and excitation bandwidth. Current-driven magnetic oscillations where thus reported in a 3-terminal CoFeB/MgO/CoFeB MTJ fabricated on top of a large Ta buffer layer [see Fig. \ref{fig:stno}(a)] \cite{Liu2012b}. Due to the reduced size of the nano pillar ($\sim 50\times80$ nm$^2$), current-driven oscillations were detected electrically through the MTJ [see Fig. \ref{fig:stno}(b)] and independent control of the excitation via the currents injected into the Ta layer and through the MTJ was achieved. More recently, \citet{Duan2014,Duan2014b} achieved SOT-driven spin wave damping control and auto-oscillations in long and narrow nanowires ($\sim1.6\mu$m$\times190$nm), where both bulk modes and edge modes were identified.\par
\begin{figure}[b]
\centering
\includegraphics[width=6cm]{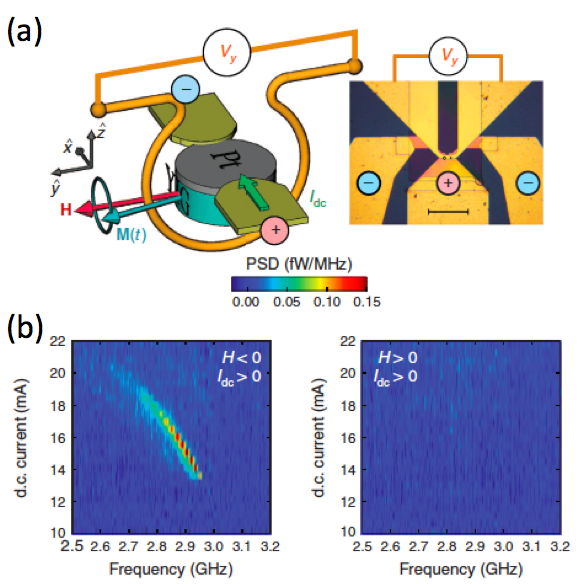}
\caption{(Color online) (a) Sketch of the measurement configuration and microscopy image of a device with two connected microdiscs (underneath the circles). The bias field $\mu_0H$ is oriented transversely to the dc current $I_{\rm dc}$ flowing in Pt. The inductive voltage $V_y$ produced in the antenna by the precession of the YIG magnetization $M(t)$ is amplified and monitored by a spectrum analyzer. (b) Power spectral density (PSD) maps measured on a 4 mm YIG/Pt disc at fixed $|\mu_0H|$=47 mT and variable $I_{\rm dc}$. The two panels correspond to two different polarities of $\mu_0H$. An auto-oscillation signal is detected above a threshold current of $\pm$13mA if $\mu_0H\cdot I_{\rm d}<$0, in agreement with the symmetry of the torque. Adapted from \citet{Collet2016}.}\label{fig:stno-YIG}
\end{figure}

Another successful configuration was realized by \citet{Demidov2012} by locally injecting a spin current in an extended ferromagnetic layer [see Fig. \ref{fig:stno}(c)]. The local injection creates a spin wave bullet, i.e., a spin wave packet localized in space through non-linear energy losses \cite{Slavin2005}. This self-localization enables the selection of a small number of spin wave modes that reveal themselves in the coherent auto-oscillation. 
The microwave spectrum of such a nano-oscillator presents features similar to "traditional" STT point-contact oscillators \cite{Bonetti2010,Slavin2005}, namely a spin wave "bullet" and a propagating spin wave mode \cite{Liu2013}. These devices, similar to STT point-contact oscillators, are characterized by a strong nonlinearity, which enables their efficient synchronization to external RF signals over a broad frequency range \cite{Demidov2014} as well as the synchronization of different oscillators placed next to each other at distances of up to several microns \cite{Awad2017}.
\par

A unique feature of SOTs is that they provide inter-conversion between the spin and charge currents in an electrical conductor and the magnon currents in a magnetic insulator. In an influential experiment, \citet{Kajiwara2010} proposed that SOTs can convert a dc electric current flowing in a Pt wire deposited on a YIG film into a spin wave propagating through the YIG film, which can then be detected by a Pt electrode at a different location using spin pumping. In this experiment, which remains controversial, SOT needs to be large enough to compensate the damping of the fundamental FMR spin wave mode. This is particularly difficult to achieve in ultralow magnetic damping materials, as SOTs excite a broad range of modes. In YIG, \citet{Xiao2012} argued that surface spin waves are preferentially excited compared to bulk spin waves, which renders the observation of current-driven auto-oscillations very sensitive to both the size of the YIG layer and to the quality of the interface with Pt. \citet{Hamadeh2014} showed that the magnetic losses of spin wave modes in micron-sized YIG(20nm)/Pt(8nm) discs can be reduced or enhanced depending on the polarity and intensity of the dc current flowing through Pt, reaching complete compensation of the damping of the fundamental mode for a current density of $3\times 10^7$~A cm$^{-2}$, and eventually inducing coherent SOT-induced auto-oscillations \cite{Collet2016}, see Fig. \ref{fig:stno-YIG}. By using Bi-substituted YIG films with perpendicular magnetic anisotropy, it is further possible to prevent the self-localization of the magnetization oscillations, thus leading to the propagation of coherent magnons into an extended magnetic insulator film \cite{Evelt2018b}.\par

An alternative strategy to circumvent the hurdles posed by the generation of coherent spin waves is to utilize thermal magnons, in the same spirit as in spin caloritronics experiments \cite{Uchida2010,Bauer2012}. In this case, SOTs excite a broad range of magnons with characteristic frequencies much higher than the fundamental FMR mode ($\sim k_{\rm B}T$ rather than a few GHz) and able to transmit information over long distances \cite{Zhang2012d,Bender2012}. Using a non-local setup consisting of two parallel Pt electrodes deposited on an extended YIG film, \citet{Cornelissen2015} and \citet{Goennenwein2015} demonstrated SOT-driven injection, transmission, and detection of thermal magnons over distances up to 40 $\mu$m, with a crossover between a linear transport regime dominated by thermal exchange magnons at low current and non-linear transport regime dominated by subthermal magnetostatic magnons at high current \cite{Thiery2018}. These studies have recently been extended to antiferromagnets, such as $\alpha$-Fe$_2$O$_3$, where antiferromagnetic magnons can carry spin information over a few tens of microns \cite{Lebrun2018}.

Overall, the SOT approach is very attractive for controlling the magnetization dynamics of a broad class of ferromagnetic and antiferromagnetic materials, including both conducting and insulating systems. Because no electric current is required to flow between the magnetic layer and the spin-orbit coupled electrodes, low-damping magnetic dielectric materials can be used as the carriers of magnetic information over large distances. Moreover, SOTs can be applied to arbitrarily large areas of a magnetic film, unlike STT, which is limited to pillar-shaped nanostructures, allowing for spin wave amplification through the compensation of damping. SOTs can thus be utilized for the generation of propagating spin waves, the enhancement of their propagation range, and their detection in a single integrated nanomagnonic device, with the perspective of implementing coherent-wave computing and information processing.

\subsection{Magnetization switching} \label{MML:switching}
\begin{figure*}[t!]
\centering
\includegraphics[width=15cm]{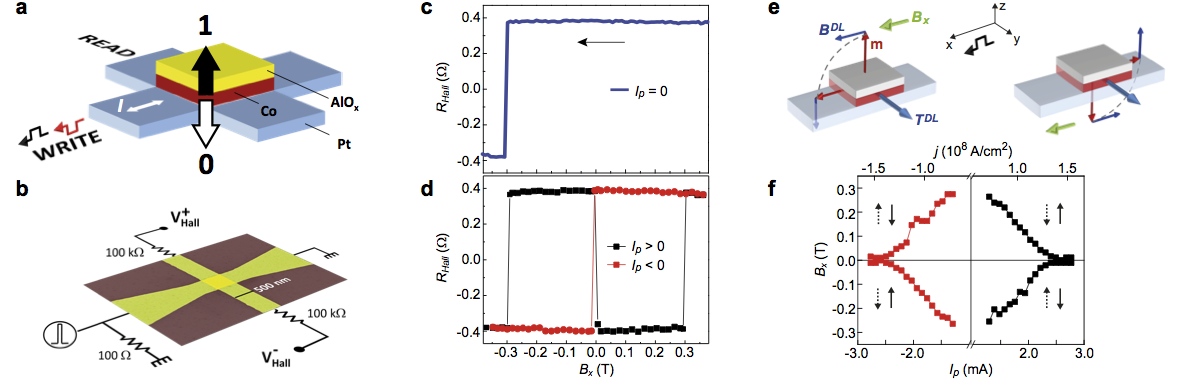}
\caption{(Color online) (a) Schematic of a Co(0.6nm)/AlO$_x$(2nm) dot patterned on top of a 3 nm thick Pt Hall cross. Black and white arrows indicate the equilibrium magnetization states of the Co layer. (b) Detection scheme and scanning electron micrograph of the sample. (c) $m_z$ measured by the anomalous Hall resistance during a downward sweep of the external field $B_x$ applied parallel to the current direction. The field has a $2^{\circ}$ out-of-plane tilt to unambiguously define the residual $z$ component. (d) The same measurement recorded after the injection of positive (black squares) and negative (red circles) current pulses of amplitude $I_p = 2.58$~mA, showing bipolar switching of $m_z$. (e) Macrospin model showing the stable (right) and unstable (left) magnetic configurations depending on the sign of $B_{\rm DL}$ relative to $B_x$. (f) Switching diagram: the dots show the minimum in-plane field at which switching becomes deterministic as a function of the injected current. Dashed (solid) arrows indicate the magnetization direction before (after) switching. Adapted from \citet{Miron2011}.}\label{Fig_SOT_switching}
\end{figure*}
The realization of current-driven magnetization switching has been a major milestone in the progress towards SOT devices. \citet{Miron2011b} and \citet{Liu2012,Liu2012c} demonstrated that, in the presence of a constant in-plane magnetic field, the magnetization direction of a perpendicularly magnetized ultrathin trilayer (Pt/Co/AlO$_x$ and Ta/CoFeB/MgO) can be reversibly switched by injecting bipolar current pulses at current densities of the order of $10^7- 10^8$ A/cm$^2$ (see Fig. \ref{Fig_SOT_switching}). This observation was soon confirmed by several groups using different magnetic stacks and heavy metal substrates \cite{Avci2012,Avci2014,Emori2013,Pai2012,Yu2014b}, as well as antiferromagnets \cite{Fukami2016a,Oh2016,Wadley2016}, magnetic insulators \cite{Avci2017,Li2016c}, and topological insulators \cite{Han2017,Wang2017d,Mahendra2018}.
Notably, earlier investigations of Pt/Co/Pt and Pt/Co also reported an effect of very small current densities on the low temperature coercivity of Co, albeit mainly attributed to Joule heating \cite{Lin2006,Xie2008,Riss2010}.

The switching of a perpendicularly magnetized layer can be qualitatively explained by considering the combined action of the damping-like torque and in-plane field $B_x$ in a simple macrospin picture, as shown in Fig.~\ref{Fig_SOT_switching}(e). In the Pt/Co/AlO$_x$ stack, a positive current pulse induces an effective field $B_{\rm DL}$, such that the magnetization can rotate from up to down if $B_{\rm DL}$ is initially parallel to $B_x$, but cannot rotate from down to up if $B_{\rm DL}$ is antiparallel to $B_x$. When the current polarity is reversed, the sense of rotation changes, such that bipolar switching is achieved by either current or in-plane field reversal, as shown in Fig.~\ref{Fig_SOT_switching}(f). More generally, the transferred angular momentum is transverse to both the current direction and the normal to the plane, which alone cannot ensure reversible magnetization switching between the +${\bf z}$ and -${\bf z}$ directions. Hence, the damping-like torque must be supplemented by the in-plane field $B_x$ that breaks the symmetry along the current direction and determines the outcome of the switching process. In the macrospin approximation, the threshold switching current is given by \cite{Lee2013d}
\begin{equation}\label{eq:jsw_op}
j_{sw,\bot}=\frac{2e}{\hbar}\frac{M_s t_F}{\xi^j_{DL}}\left(\frac{B_{K,\bot}}{2}-\frac{B_{x}}{\sqrt{2}} \right),
\end{equation}
where $B_{K,\bot}$ is the perpendicular anisotropy field. In-plane magnetized samples, on the other hand, switch at zero external field as long as the magnetization has a nonzero component in the ${\bf y}$ direction, which can be induced by shape anisotropy \cite{Liu2012,Fukami2016b}. In this case, the threshold current has the same form as that of the conventional
STT switching for free and fixed layers with in-plane magnetization \cite{Sun2000}, and is given by \cite{Lee2013d}
\begin{equation}\label{eq:jsw_ip}
j_{sw,||}=\alpha \frac{2e}{\hbar}\frac{M_s t_F}{\xi^j_{DL}}\left(B_{K,||}+\frac{B_{d}}{2} \right),
\end{equation}
where $B_{K,||}$ is the in-plane anisotropy field and $B_{d}$ the demagnetizing field. Equations \eqref{eq:jsw_op} and \eqref{eq:jsw_ip} exemplify the relationship between the power required for switching, the thermal stability of a magnet (determined by $B_K$) and $\xi^j_{DL}$. However, the actual mechanism of SOT switching is more complex than coherent magnetization reversal under the action of the damping-like torque alone.\par

\begin{figure} [t]
	\centering
	\includegraphics[width=7cm]{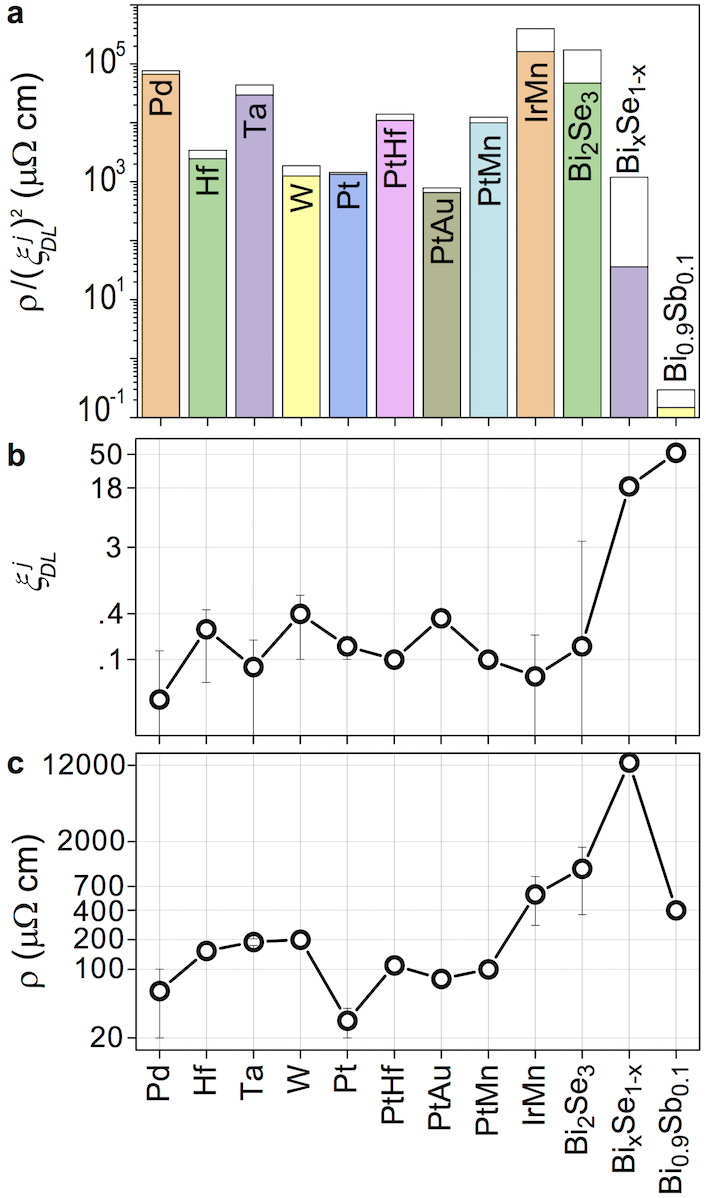}
	\caption{(Color online) (a) Switching efficiency $\rho/(\xi^j_{DL})^2$ (solid bars) calculated using the SOT efficiency (b) and resistivity (c) of different nonmagnetic layers. The values are taken from Table~\ref{tableSOT}, with error bars representing experimental spreads, when available. The open bars represent the switching efficiency calculated for a NM/FM stack including a fictitious CoFeB layer according to Eq.~\eqref{eq:Psw}.}\label{Fig_SOT_efficiency}
\end{figure}
In realistic systems, $j_{\rm sw}$ depends on the factors appearing in Eqs.~\eqref{eq:jsw_op}, \eqref{eq:jsw_ip} as well as on the DMI, domain pinning field, device geometry, size, temperature, and duration of the current pulses. The temperature, which is determined by the current distribution in the bilayer as well as by the thermal conductivity of the different materials in the stack, plays a major role, both in activating the switching as well as in changing critical parameters such as $M_s$, $B_K$, $\alpha$, and $\xi^j_{DL}$ during switching. These factors vary significantly from experiment to experiment, so that a comparative estimate of the switching efficiency for different material systems can be highly misleading. Nonetheless, an approximate figure of merit for the switching efficiency can be calculated by taking the threshold power density $P_{\rm sw}=j_{\rm sw}^2\rho \propto \rho/(\xi^j_{DL})^2$ estimated using the macrospin approximation, which is independent of the device size, temperature, and pulse length. Figure~\ref{Fig_SOT_efficiency}(a) presents a comparison of $\rho/(\xi^j_{DL})^2$ for different nonmagnetic material systems (solid bars), based on the values of $(\xi^j_{DL})^2$ and $\rho$ reported in (b) and (c), respectively. Within the confines of such a comparison, Pt and W emerge as the best heavy metal elements, whereas topological insulators offer the largest gains in efficiency. It shall be noted, however, that the current distribution in NM/FM bilayers can significantly alter the efficiency, especially if the resistivity of the ferromagnet is much smaller than that of the nonmagnetic material. By using a simple parallel resistor model, the threshold power density can be estimated as
\begin{equation}\label{eq:Psw}
P_{sw}\propto \left(\frac{\rho_N t_F}{\rho_F t_N} +1 \right) \frac{\rho_N}{(\xi^j_{DL})^2}.
\end{equation}
The open bars in Fig.~\ref{Fig_SOT_efficiency}(a) show the change of the efficiency calculated using Eq.~\eqref{eq:Psw} for a bilayer with $t_N=4$~nm, $t_F=1$~nm, and $\rho_F=100$~$\mu\Omega$cm, as typical, e.g., of CoFeB.

\begin{figure*} [t]
	\centering
	\includegraphics[width=16.8cm]{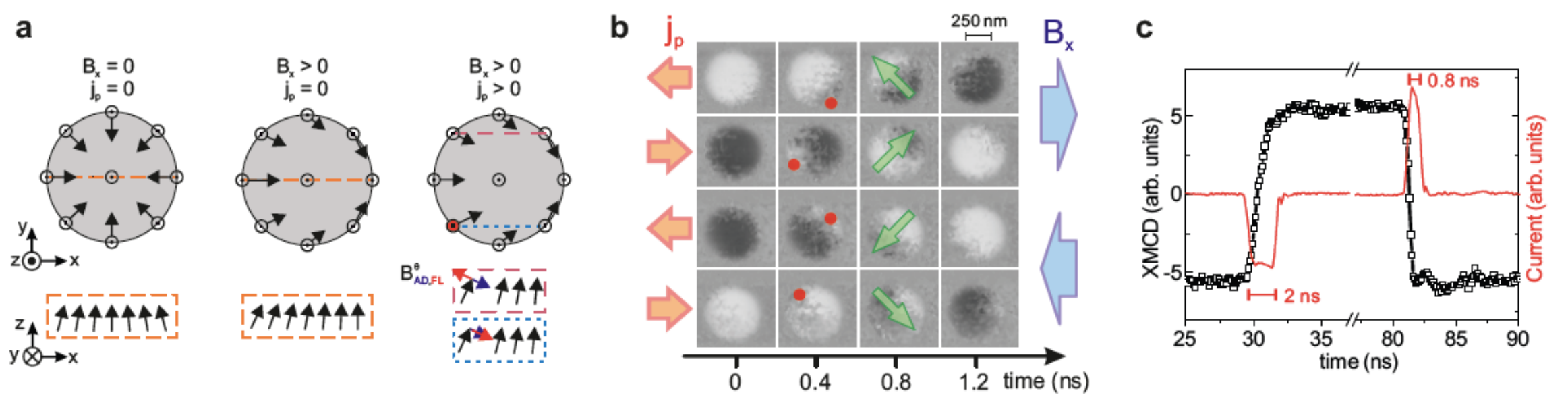}
	\caption{(Color online) (a) Schematics of the tilting of the magnetization at the edges of a Pt/Co/AlO$_x$ dot due to the DMI (left), DMI and external field $B_x$ (middle), DMI, $B_x$, and current (right). The polar components of the damping-like and field-like effective fields add up at the nucleation point. (b) Snapshots of the reversal process of a circular dot for different combinations of current and field measured by time-resolved scanning x-ray transmission microscopy. The red dot and green arrows indicate the nucleation point and the domain wall propagation direction, respectively. The pulse duration is 2~ns.
(c) Time trace of the average out-of-plane magnetization (black squares) during current injection (red line). The amplitude of the first (second) pulse is $j_p = 3.1 \times 10^8$ ($4.4 \times 10^8$) A/cm$^2$; $B_x = 0.11$~T. Adapted from \citet{Baumgartner2017}.}\label{Fig_DW_switching}
\end{figure*}

\subsubsection{Switching mechanism} \label{MML:switching:mechanism}

Although the macrospin model reproduces qualitatively the stability phase diagram of rather extended films \cite{Liu2012c}, magnetization switching in structures larger than the width of a domain wall ($\gtrsim 10$~nm) occurs by nucleation and expansion of magnetic domains. The magnetization reversal process is thus closely related to the SOT-driven dynamics of N\'{e}el-type domain wall in the presence of DMI (see Section \ref{s:DW}). Different switching models have been proposed based on micromagnetic simulations \cite{Finocchio2013,Perez2014,Mikuszeit2015,Martinez2015} and spatially-resolved MOKE measurements \cite{Emori2013,Ryu2013,Yu2014b,Safeer2016}. In such models, the domain nucleation is either random and thermally-assisted \cite{Finocchio2013,Perez2014,Lee2014} or determined by the combined action of DMI, external field, and edge effects \cite{Mikuszeit2015,Martinez2015,Pizzini2014}, followed by domain wall propagation across the magnetic layer driven by the damping-like torque.\par

Time-resolved x-ray microscopy measurements of circular shaped Pt/Co/AlO$_x$ dots \cite{Baumgartner2017} eventually confirmed the edge nucleation models, further showing that the nucleation point is deterministic and alternates between the four quadrants of a dot depending on the sign of the magnetization, $B_x$, DMI, damping-like and field-like torque, as illustrated in Fig.~\ref{Fig_DW_switching}. These measurements also showed that switching is achieved within the duration of the current pulse with an incubation time below the time resolution of the experiment ($\approx 100$~ps) and fast propagation of a tilted domain wall across the dot \cite{Baumgartner2018} with domain wall velocities of the order of 400 m/s. As the switching unfolds along a reproducible and deterministic path, the timing and the extent of magnetization reversal can be reliably controlled by the amplitude and duration of the current pulses \cite{Baumgartner2017}. Measurements performed by time-resolved MOKE on larger dots with a thinner Co layer, on the other hand, show significant after-pulse magnetic relaxation \cite{Decker2017}, which is ascribed to long-lasting heating effects and weaker magnetic anisotropy compared to \citet{Baumgartner2017}. After-pulse relaxation has been observed also in Ta/CoFeB/MgO dots for current pulses exceeding 2~ns, attributed to domain wall reflection at the sample edges that is favored by the lower DMI and Gilbert damping of Ta/CoFeB/MgO \cite{Yoon2017}. These different results reveal how the reversal path is determined by the balance between damping-like and field-like torques, DMI, magnetic anisotropy, and temperature. For samples matching the width of the current line, the Oersted field can also facilitate or hinder the reversal \cite{Baumgartner2017,Aradhya2016}. In all cases, however, SOT switching is bipolar and robust with respect to multiple cycling events as well as to the presence of defects.

\subsubsection{Switching speed} \label{MML:switching:speed}
\begin{figure}[b!]
\centering
\includegraphics[width=\columnwidth]{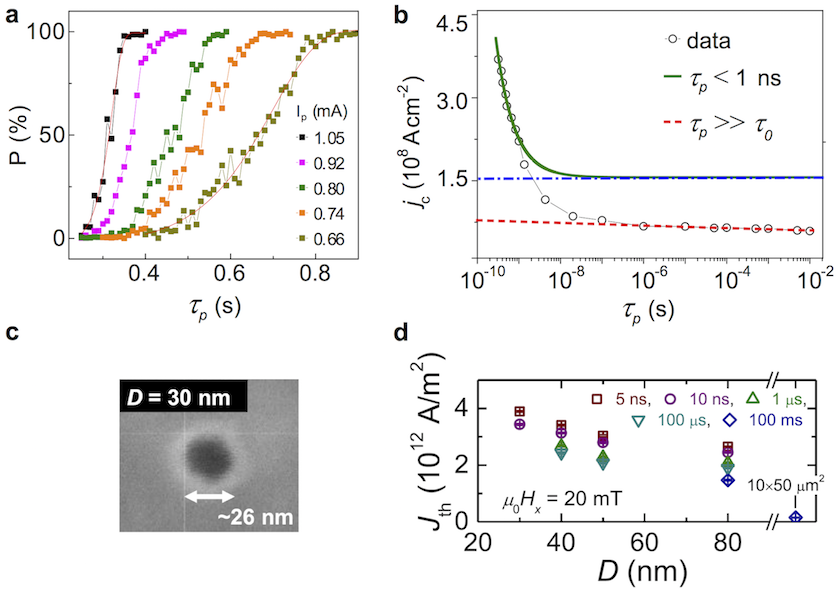}
\caption{(Color online) (a) Switching probability $P$ of a square Pt(3nm)/Co(0.6nm)/AlO$_x$ dot with a lateral size of 90~nm as a function of the current pulse duration $\tau_p$ at fixed in-plane field $B_{x}=91$~mT. (b) Critical current density as a function of pulse duration defined at $P=90$~\%. The green solid line is a fit to the data in the short-time regime ($\tau_{p} < 1$~ns), the red dashed line is a fit in the thermally activated regime ($\tau_{p}\geq 1$~$\mu$s). The blue dash-dotted line represents the intrinsic critical current $j_{c0}$. Adapted from \citet{Garello2014}. (c) Scanning electron microscope image of a Ta(5nm)/CoFeB(1.2)/MgO dot with a nominal diameter $D$ of 30 nm. (d) Device diameter dependence of the critical current density at various $\tau_p$. Adapted from \citet{Zhang2015h}.}\label{Fig_SOT_fast}
\end{figure}

One of the most attractive features of SOT switching is the timescale of magnetization reversal. Because the switching speed scales with the lateral dimensions of the sample, and the domain wall velocity can attain up to 750 m/s \cite{Miron2011b,Yang2015a}, the reversal time can be reduced to well below 1~ns in dots of 100~nm size \cite{Garello2014}. Figure~\ref{Fig_SOT_fast}(a) shows that the switching probability of perpendicularly magnetized Pt/Co/AlO$_x$ dots has a narrow distribution as a function of pulse length $\tau_p$, which decreases to below 100~ps as the current density increases. In this study, a switching probability of 100\% was demonstrated down to $\tau_p=180$~ps, consistently with reversal due to domain nucleation and propagation. The critical switching current $j_{\rm sw}$ is characterized by a long and a short time scale regime, shown in Fig.~\ref{Fig_SOT_fast}(b), similar to STT-induced switching in metallic spin valves \cite{Liu2014e}. $j_{\rm sw}$ depends weakly on $\tau_{p}$ above 10~ns, as expected for a thermally-activated reversal process \cite{Bedau2010}, and scales linearly with $\tau_{p}^{-1}$ below about 1~ns, as expected in the intrinsic regime where the reversal time is inversely proportional to the transferred angular momentum.\par

Studies of how $j_{\rm sw}$ scales as a function of dot size have been performed for Ta and W/CoFeB/MgO dots \cite{Zhang2015h,Zhang2018d}. $j_{\rm sw}$ was found to increase by one order of magnitude going from micrometer-sized Ta/CoFeB/MgO stripes to 80~nm dots, and to remain approximately constant upon further reduction of the dot size down to 30 nm [Fig.~\ref{Fig_SOT_fast}(c,d)]. This behavior was interpreted as a signature of incipient monodomain behavior, even though no precessional switching was observed, contrary to the prediction of macrospin models \cite{Lee2013d,Park2014}. An additional feature that makes SOT switching very attractive for applications is that the incubation time required to start the process appears to be negligible \cite{Garello2014}. The SOT geometry, in which ${\bf T}_{\rm DL}$ can be made orthogonal to the quiescent magnetization, implies that the magnetization reacts immediately to the current, contrary to STT-induced switching, in which ${\bf T}_{\rm DL}$ is initially zero for collinear magnetic layers until thermal fluctuations induce a misalignment of the free layer magnetization that is sufficient to trigger the reversal, leading to ns-long random delays \cite{Hahn2016,Devolder2008}.\par

\begin{figure*}[t]
\centering
\includegraphics[width=14cm]{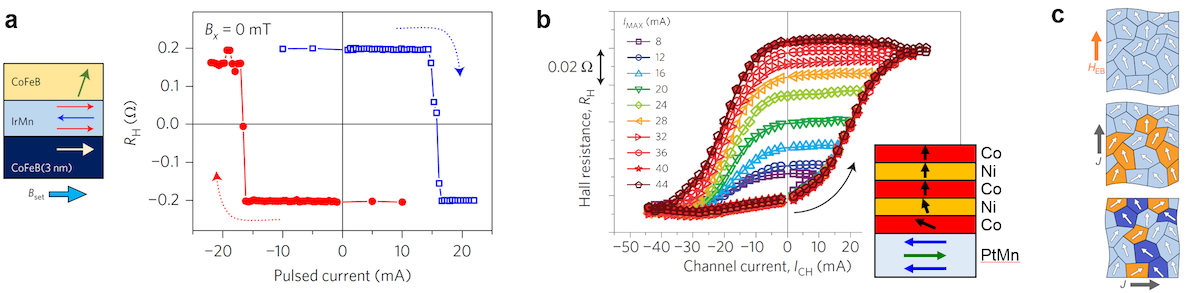}
\caption{(Color online) (a) Left: schematic of the in-plane effective field induced by exchange bias. The colored arrows in the layers indicate the direction of the magnetic moments. Field-free switching of Ta(5 nm)/CoFeB(3 nm)/IrMn(3 nm)/CoFeB/MgO sample as a function of current \cite{Oh2016}. (b) Magnetization loops of a [Co(0.3)/Ni(0.6)]$_2$/Co(0.3) multilayer on PtMn(8nm) measured after the application of current pulses of increasing amplitude up to the maximum specified in the legend. The black arrow indicates the position from which the measurement starts after initialization by a negative pulse. Adapted from \citet{Fukami2016a}. (c) Model representing the uncompensated spin direction in each grain of the antiferromagnet at the interface with the ferromagnet. Top: situation after field-cooling showing an average exchange bias field (orange arrow). A current pulse along the exchange bias direction (middle) or perpendicular to it (bottom) switches the regions of the ferromagnetic layer coupled to only one type of antiferromagnetic domain. Switched regions are indicated in orange and blocked regions are indicated in dark blue. Adapted from \citet{Brink2016}.}\label{Fig_SOT_exbias}
\end{figure*}
\subsubsection{Zero field switching} \label{MML:switching:zerofield}
A critical issue for perpendicularly magnetized layers is the need to apply an external field $B_x$ to uniquely define the switching polarity, as shown in Fig.~\ref{Fig_SOT_switching}. Although $B_x$ by itself cannot switch the magnetization because it is orthogonal to the easy axis, fields ranging from 1 to 100~mT are typically required to achieve deterministic reversal, depending on the current density as well as on the magnetic anisotropy of the layers \cite{Avci2014}. Several approaches have been demonstrated to solve this issue by substituting $B_x$ with a real or effective field embedded into a device. The first working concept by \citet{Miron2011} was to deposit two 50~nm thick CoFe layers on either side of the magnetic dot, providing a dipolar in-plane field parallel to the current. This solution, however, is not practical for device integration because it limits the scalability of a matrix of such dots or MTJs.\par

\citet{Lau2016} have shown that it is possible to embed an in-plane magnetized CoFe layer directly into the stack, and provide an effective $B_x$ on the perpendicular CoFe free layer via interlayer exchange coupling mediated by nonmagnetic Ru or Pt spacers. Such an approach allows for varying the sign of $B_x$ upon changing the spacer thickness, but may not be easily integrated into standard MTJ architectures. A more straightforward approach relies on the stray field projected by an in-plane magnetized layer placed on top of the free layer/barrier/reference layer stack \cite{Zhao2017}, provided that such a field does not reduce the TMR. Alternatively, the ferromagnet can be deposited directly on top of a few nm-thick antiferromagnet like IrMn or PtMn \cite{Brink2016,Fukami2016a,Oh2016}. The antiferromagnetic layer provides an in-plane exchange bias field but also the source of the spin density, which enables the switching of perpendicular ferromagnetic layers
in zero field at current densities of the order of $3\times 10^{7}$~A/cm$^2$. The switching process in ferromagnetic/antiferromagnetic systems takes place in a step-wise manner, as schematized in Fig.~\ref{Fig_SOT_exbias}(c), depending on the microstructure of the antiferromagnetic layer and the local direction of the exchange bias field \cite{Brink2016,Fukami2016a,Kurenkov2017}. This behavior can be also exploited to introduce analogue memristive properties into three-terminal MTJ devices \cite{Fukami2016a}. \par

Another practical solution consists in employing a magnetic spin-valve composed of a bottom reference ferromagnet and a top recording free layer in the current-in-plane configuration (see Sect.~\ref{MML:exp:spinvalves}). If the magnetization of the bottom ferromagnet points along the current direction, the spin current resulting from spin-orbit interfacial scattering has a component ${\bm \zeta}\parallel {\bf p}\times({\bf z}\times{\bf j}_{\rm c})\equiv {\bf z}$ polarized along the easy axis of the free layer, which has been show to induce field-free switching in CoFeB/NiFe/Ti/CoFeB/MgO spin-valves \cite{Baek2018}. The combination of STT and SOT in perpendicular MTJs also leads to field-free magnetization
switching of the free layer, which is particularly promising for applications \cite{Wang2018a}.

Finally, an elegant approach to this problem is to introduce lateral symmetry breaking in the magnetic structure. Thickness gradients of the oxide and ferromagnetic layers have been shown to induce an out-of-plane field-like torque \cite{Yu2014,Yu2014b} or a tilted anisotropy \cite{You2015,Torrejon2015}, both conducive to zero field switching, whereas asymmetric patterning of the magnetic and conductive layers has been used to control the switching polarity via nonreciprocal domain wall propagation \cite{Safeer2016}. Recently, artificial nanomagnets consisting of adjacent out-of-plane and in-plane magnetized regions coupled by the DMI have also been shown to exhibit field-free switching, with interesting implications to cascade linear and planar arrays of nanomagnets \cite{Luo2019}.

\subsection{Memory and logic devices \label{MML:devices}}

\begin{figure}[t]
\centering
\includegraphics[width=1\columnwidth]{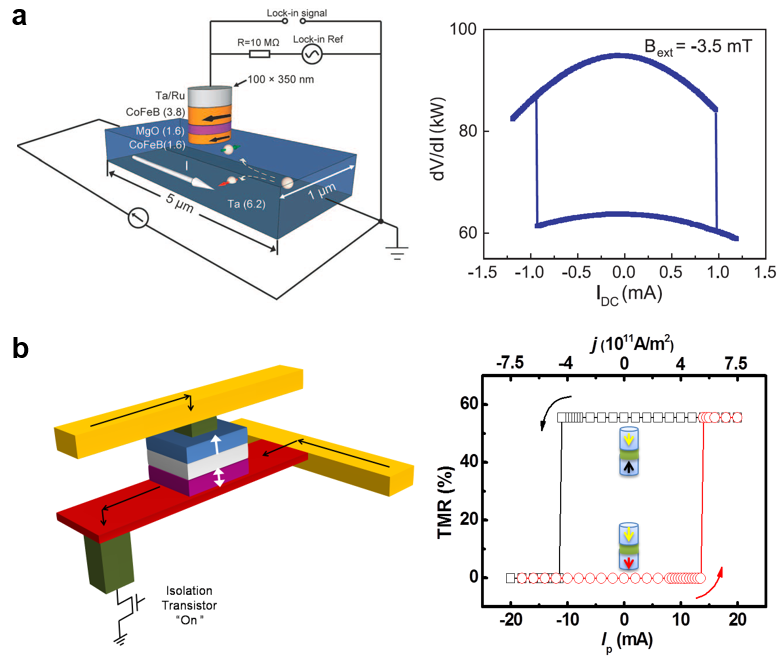}
\caption{(Color online) (a) Left: SOT-induced switching for an in-plane magnetized nanomagnet at room temperature: schematic of the three-terminal MTJ device and the circuit used in the measurements. Right: TMR of the device as a function of the applied dc current. An in-plane external field of 3.5 mT is applied to set the device at the center of the minor loop, although this is not required for switching the in-plane magnetized free layer. Adapted from \citet{Liu2012}. (b) Left: Schematic of a three-terminal
MTJ with perpendicular magnetization. Right: TMR as a function of current amplitude $I_p$ injected in the Ta electrode using 50 ns long pulses under an in-plane magnetic field of 40 mT. Adapted from \citet{Cubukcu2014}.}\label{Fig_SOT_MTJ}
\end{figure}

SOT-operated devices can find application in memory as well as logic architectures where current-induced switching is required to control the magnetization of one or several magnetic elements \cite{Lee2016d}. MTJs with in-plane \cite{Liu2012,Pai2012,Yamanouchi2013} and perpendicular \cite{Cubukcu2014,Garello2018} magnetization provided the first demonstration of three-terminal devices in which the write operation is performed by SOTs (Fig.~\ref{Fig_SOT_MTJ}). 

MTJs constitute the building blocks of MRAMs, where the bit state is encoded in the high (low) TMR corresponding to antiparallel (parallel) alignment of the magnetization of the free and reference layers. The ever increasing need for faster data storage and retrieval has placed MRAMs in a prime position to replace or complement CMOS-based memory technologies, owing to the intrinsic nonvolatility, low write energy, low standby power, as well as superior endurance and resistance to radiation of MTJ bit cells compared to semiconductor memories \cite{Apalkov2016,Hanyu2016}. State-of-the-art MRAMs incorporate STT as the writing mechanism \cite{Kent2015}. STT brings great advantages in terms of scalability and integration with peripheral electronics, since the critical switching current scales with the area of the free layer and requires only two terminals to perform the read and write operations [Fig.~\ref{Fig_SOT_MTJ}(a)]. However, as the write and read currents flow along the same path through the oxide tunnel barrier, a compromise between conflicting requirements must be achieved, namely a thin barrier for low current switching and a thick barrier for high TMR. Moreover, because the STT reversal process is thermally activated, a large overdrive current is required for fast switching, which can damage the tunnel barrier, while the finite probability to not switch at high currents and to switch at low current leads to write error rates that are larger than desired \cite{Oh2009}. 

Three-terminal MTJ devices based on SOT offer critical advantages in this respect, as the free layer can be switched without passing a current through the oxide and reference layers [Fig.~\ref{Fig_SOT_MTJ}(b)]. The separation of the read and write current paths in the MTJ allows for optimal tuning of the barrier independently of the write process and increases the endurance of the MTJ. Moreover, the deterministic character of SOT switching enables sub-ns reversal of perpendicular MTJs \cite{Cubukcu2018} and low error rates in in-plane MTJs down to 2~ns long current pulses \cite{Aradhya2016}. The three-terminal configuration of MTJs operated by SOT, on the other hand, implies a larger footprint of the bit cell compared to two-terminal MTJs. As the actual size of the bit cell depends on the size and number of the transistors required to control data flow, the area penalty depends on the particular cell architecture and may not be so large. Importantly, the three-terminal configuration also allows for voltage control of the magnetic anisotropy of the free layer, which enables write speed acceleration \cite{Yoda2017}, lower current thresholds, as well as selective SOT switching of several MTJs sharing a single write line \cite{Kato2018}.
The analysis of SOT-MRAMs at the circuit- and architecture-level \cite{Oboril2015,Prenat2016} reveals that this technology can be advantageously introduced in the data cache of processors, offering a strong reduction of the power consumption relative to volatile memories, comparable performances to STT-MRAMs and significant gains in terms of reliability and speed. \par

SOTs hold great promise also for driving magnetic cellular automata \cite{Cowburn2000}, domain wall logic \cite{Allwood2005}, and MTJ-based logic devices \cite{Yao2012,Guo2014}.
In the first two types of devices, SOTs offer unique features such as the clocking of nanomagnetic logic arrays by in-plane current injection \cite{Bhowmik2014} and the efficient manipulation of domain walls \cite{Yang2015a,Safeer2016}. In hybrid CMOS/magnetic devices based on MTJs, SOTs can perform similar functions as STT \cite{Yao2012,Guo2014}, but also enable novel architectures. Recent proposals include MTJ devices that exploit gate-voltage-modulated SOT switching for the parallel initialization of programmable logic arrays \cite{Lee2016d}, four terminal devices that allow for direct cascading at high operation gain and low switching power \cite{Kang2016}, and nonvolatile flip-flops for power gating \cite{Jabeur2014,Kwon2014,Hanyu2016}. More futuristic ideas concern "probabilistic spin logic" systems in which SOTs are used to control the stochastic switching of thermally activated nanomagnets \citet{Camsari2017} as well as neuromorphic computing architectures \cite{Locatelli2013,Sengupta2015,Borders2018}. Other unconventional memory and logic architectures can be envisaged based on purely planar structures. In such a case, the SOTs would provide the writing mechanism while the reading operation can be performed by the AHE \cite{Moritz2008} or the unidirectional SMR \cite{Avci2015a,Avci2018,Olejnik2015}.

A critical issue in this wide range of applications is the dynamic power consumption relative to the thermal stability factor of nanomagnets, $\Delta = B_{K}M_{s}V_{F}/2k_{B}T$, where $V_{F}$ is the volume of the ferromagnet. In perpendicularly magnetized structures with $\Delta \gtrsim 500$, the critical current density ranges from $10^7$ to a few times $10^8$~A/cm$^2$ depending on the switching speed [Fig. \ref{Fig_SOT_fast}(b)]. However, because the critical current scales with the lateral cross-section of a device, the switching of a 50~nm wide dot is predicted to require less than 200~$\mu$A and a write energy smaller than 100~fJ at 1.5 ns \cite{Cubukcu2018}, which is close to the best results obtained so far for perpendicular STT-MRAM devices. \par

Very promising figures of merit in this context have been obtained for in-plane CoFeB layers with $\Delta \gtrsim 35$ by dusting the W/CoFeB interface with Hf, which allows for critical current densities of the order of $5\times 10^6$~A/cm$^2$ at 2 ~ns \cite{Shi2017}. The power dissipated in the current lines is also a matter of concern, as some of the most efficient NM/FM combinations are based on the high-resistive phase of W and Ta \cite{Liu2012,Pai2012} (Fig.~\ref{Fig_SOTrho}). The search for novel SOT materials is thus focusing on systems that combine large charge-spin conversion efficiency with low resistivity or whose magnetic properties can be strongly modulated by a gate voltage.
While there are still margins of improvement, SOT devices already offer an unprecedented variety of applications and compatibility with different classes of materials, which extends the range of spintronics well beyond the prototypical spin-valve and MTJ structures of the past two decades.


\section{Spin-orbit torques in noncentrosymmetric magnets\label{s:noncentro}}

SHE and iSGE are known as distinct but companion phenomena from their initial observations in nonmagnetic semiconductor structures \cite{Silov2004,Kato2004b,Kato2004d,Ganichev2004b,Wunderlich2004,Wunderlich2005,Ivchenko2008,Belkov2008}. As discussed in the previous section, both iSGE and SHE have been utilized for electrically generating SOTs in metallic magnetic multilayers. The primary focus of the present section is to discuss the experiments performed on bulk non-centrosymmetric magnets, including dilute magnetic semiconductors \cite{Chernyshov2009,Endo2010,Fang2011,Kurebayashi2014}, magnetic half-heusler compounds \cite{Ciccarelli2016} and antiferromagnets \cite{Wadley2016,Bodnar2018,Meinert2018,Zhou2018b}. This type of systems is particularly interesting as SHE is absent (there is no adjacent nonmagnetic metal), so that the observed SOTs are solely attributed to iSGE.\par

In analogy to the galvanic (voltaic) cell, the term spin galvanic effect (SGE) was coined for a phenomenon in which an externally induced non-equilibrium spin density generates an electrical current (voltage) \cite{Ganichev2002}. Inversely the iSGE, sometimes also called the Rashba-Edelstein effect, then refers to an externally applied electrical current that generates a spin density \cite{Ivchenko1978,Ivchenko1989,Aronov1989,Edelstein1990,Malshukov2002,Inoue2003}. The theory of iSGE was discussed in details in Subsection \ref{s:rashbatorque}. We start in Subsection~\ref{iSGE-SOT-nonmagnetic} with initial observations of the iSGE in nonmagnetic GaAs structures and continue in Subsection~\ref{iSGE-SOT-ferromagnetic} by discussing the iSGE induced SOTs in bulk ferromagnets, namely in the low Curie temperature, dilute-moment semiconductor (Ga,Mn)As, and in the high Curie temperature, dense-moment metal NiMnSb. The physics of staggered iSGE spin densities in locally non-centrosymmetric lattices and corresponding N\'eel SOTs is reviewed in Subsection~\ref{iSGE-SOT-antiferromagnet} based on studies in antiferromagnetic CuMnAs and Mn$_2$Au. We conclude in Subsection~\ref{iSGE-SOT-magnonic} by discussing the SGE and spin-orbit-driven magnonic charge pumping phenomena that are reciprocal to the iSGE and SOT, respectively.

\subsection{nonmagnetic GaAs structures}
\label{iSGE-SOT-nonmagnetic}

\begin{figure}[h!]
\centering
\includegraphics[width=8cm]{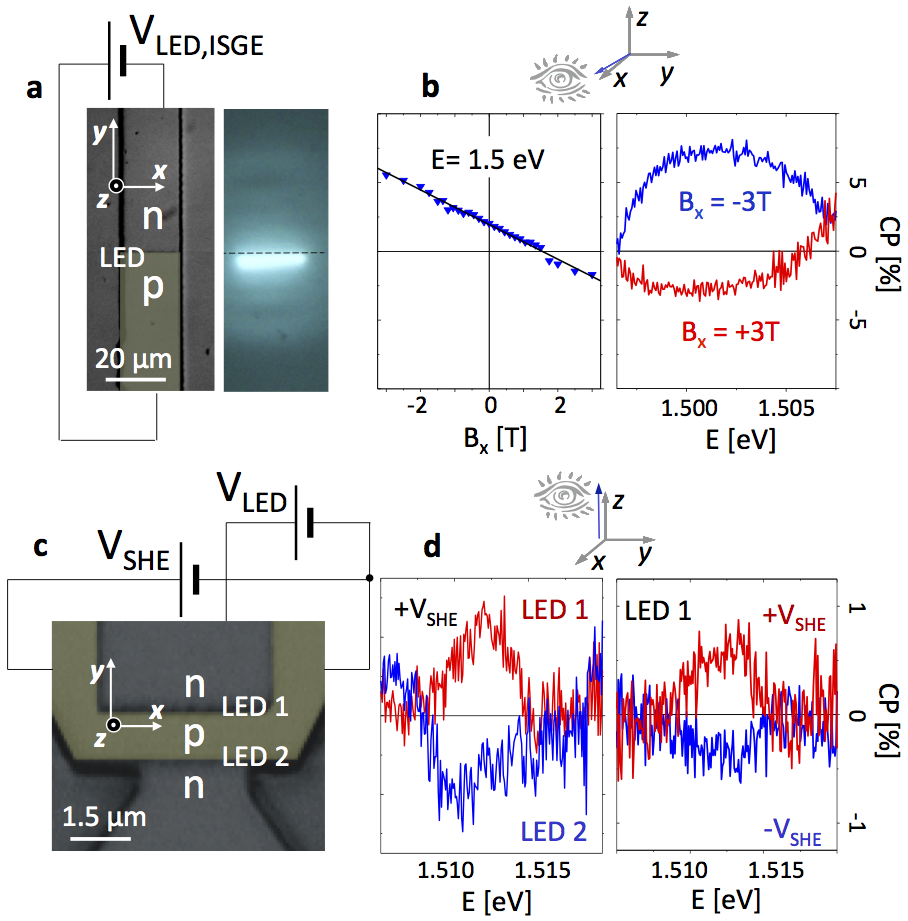}
\caption{(Color online) (a) Electron micrograph of the device and an optical image of the emitted light in the experimental detection of the iSGE by circularly-polarized electro-luminescence. The net in-plane spin polarization is detected by placing the LED across the transport channel and using an in-plane observation angle. (b) Right: spectral dependence of the circular polarization of the emitted light. Left: the dependence of the circular polarization on the external in-plane magnetic field. (c,d) Experimental detection of the SHE by two LEDs placed along the edges of the conduction channel and using an out-of-plane observation angle. Adapted from \citet{Wunderlich2004,Wunderlich2005}.}
\label{fig_iSGE-SOT-nonmagnetic}
\end{figure}

Initial observations of the iSGE were made in parallel with the initial SHE experiments, in both cases in semiconductors and employing optical detection methods \cite{Silov2004,Kato2004b,Kato2004d,Ganichev2004b,Wunderlich2004,Wunderlich2005,Ivchenko2008,Belkov2008}. In cite{Wunderlich2004,Wunderlich2005}, iSGE and SHE were detected in the same asymmetrically confined hole gas in a AlGaAs/GaAs semiconductor heterostructure. The experiments are shown in Fig.~\ref{fig_iSGE-SOT-nonmagnetic}. The current-induced spin density was measured by detecting the circularly polarized electroluminescence from a built-in planar {\em p-n} light emitting diode (LED). Since in this semiconductor heterostructure the iSGE has the Rashba symmetry and the corresponding in-plane polarization (perpendicular to the applied electric field) is uniform, the LED was placed across the hole transport channel and an in-plane observation angle was used [see Fig.~\ref{fig_iSGE-SOT-nonmagnetic}(a)]. The measured non-zero circular polarization at zero magnetic field [see Fig.~\ref{fig_iSGE-SOT-nonmagnetic}(b)] is then a signature of the iSGE spin density of current-carrying holes that radiatively recombined with electrons at the detection LED. For comparison, the SHE experiment is displayed in Figs.~\ref{fig_iSGE-SOT-nonmagnetic}(c,d). Here opposite out-of-plane spin densities accumulate only at the edges and, correspondingly, the detecting LEDs are fabricated along the edges of the transport channel and the emitted light observation angle is out-of-plane. 

The remarkable strength of these relativistic phenomena was already recognized in the initial experiments performed in the strongly spin-orbit coupled GaAs valence band. The effective iSGE fields inferred from Fig.~\ref{fig_iSGE-SOT-nonmagnetic}(b) are in Teslas. In other words, the $\sim1-10$~\% spin polarization was achieved in the microchip at a $\sim100$~$\mu$A current, compared to a $\sim100$~A superconducting magnet that would generate the same degree of spin density in the semiconductor via an external magnetic field. Using Maxwell's equations physics one needs $10^6\times$ larger equipment with $10^6\times$ larger current than using Dirac equation physics in the iSGE (SHE) microchips to achieve the same polarization in the nominally nonmagnetic system. 

When the current is switched off these large spin densities immediately vanish, which makes the iSGE and SHE phenomena in nonmagnetic crystals impractical for spintronic memory applications. However, shortly after their initial discovery, it was realized theoretically \cite{Bernevig2005c,Manchon2008b,Garate2009,Zelezny2014} and subsequently verified in experiments \cite{Chernyshov2009,Ciccarelli2016,Wadley2016}, that iSGE represents uniquely efficient means for electrical writing of information when the non-equilibrium, spin-orbit-induced charge polarizations are exchange-coupled to ferromagnetic or antiferromagnetic moments. These are discussed in the following subsections.

\subsection{Bulk ferromagnetic (Ga,Mn)As and NiMnSb}
\label{iSGE-SOT-ferromagnetic}

\begin{figure}[h!]
\centering
\includegraphics[width=8cm]{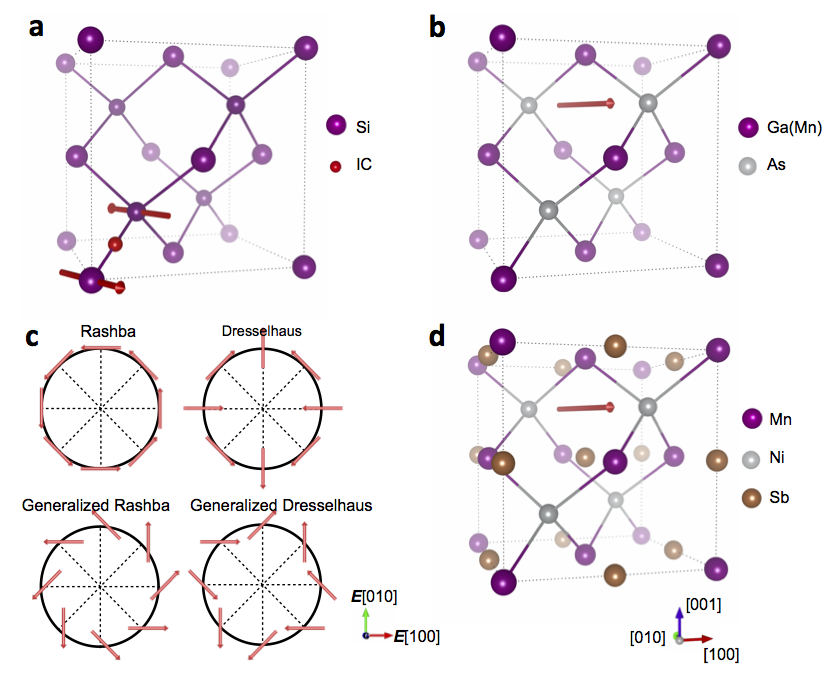}
\caption{(Color online) (a) Cartoon representation of opposite iSGE spin densities generated at the locally non-centrosymmetric inversion-partner lattice cites of the Si lattice. (b) Cartoon representation of a net uniform iSGE spin density generated over a non-centrosymmteric unit cell of a zinc-blende GaAs lattice. Exchange coupling between the iSGE spin density of carriers and equilibrium dilute ferromagnetic moments on Mn atoms results in the SOT. (c) Different symmetries of iSGE spin density as a function of the electric field direction corresponding to different non-centrosymmetric crystal point groups. (d) Same as (b) for a room-temperature dense-moment ferromagnet NiMnSb. Adapted from \citet{Ciccarelli2016}.}
\label{fig_iSGE-SOT-symmetry}
\end{figure}

One can picture iSGE based on simple symmetry rules. Fig. \ref{fig_iSGE-SOT-symmetry} represents the iSGE spin densities in three selected systems: (i) Si diamond lattice, (ii) GaAs zinc-blende crystal and (iii) NiMnSb non-centrosymmetric magnet. A priori, since Si diamond-lattice possesses inversion symmetry, iSGE vanishes globally at the level of the unit cell. But due to the {\em local} inversion symmetry breaking, iSGE generates two spin densities, ${\bf S}_A=-{\bf S}_B$, pointing in the opposite direction on the two non-centrosymmetric, inversion-partner sites of the Si diamond-lattice unit cell, as shown in Fig.~\ref{fig_iSGE-SOT-symmetry}(a) and discussed theoretically in \citet{Ciccarelli2016} and references therein. This staggered-symmetry spin density induced by the iSGE can generate an efficient SOT in collinear antiferromagnets as further discussed in Subsection~\ref{iSGE-SOT-antiferromagnet}. \par

On the other hand, the zinc-blende lattice of GaAs [or (Ga,Mn)As] and of the closely related half-heusler lattice of NiMnSb are examples of crystals that lack an inversion center in the unit cell. This can result in a non-zero net spin density, illustrated in Figs.~\ref{fig_iSGE-SOT-symmetry}(b,d), that generates an efficient SOT in ferromagnets, provided that the iSGE-induced spin density is exchange coupled to the ferromagnetic moments. As discussed earlier in detail in Subsection \ref{s:rashbatorque}, depending on the crystal symmetry, the iSGE can be composed of three distinct terms: generalized Rashba and Dresselhaus terms, shown in Fig.~\ref{fig_iSGE-SOT-symmetry}(c), and a term describing a response collinear to the electric field. 

\begin{figure}[h!]
\centering
\includegraphics[width=8cm]{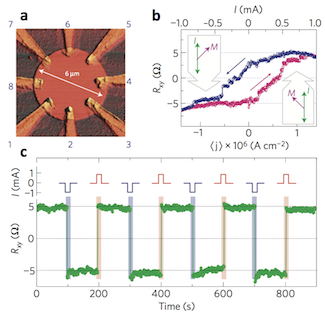}
\caption{(Color online) (a) An atomic force micrograph of the sample used to detect the SOT in GaMnAs. (b) $R_{xy}$ shows hysteresis as a function of the current for a fixed external magnetic field $H=6$~mT applied at an angle $\phi_H=72^\circ$. (c) The magnetization switches between the $[010]$ and $[\bar{1}00]$ directions when alternating $\pm1$~mA current pulses are applied. The pulses have 100~ms duration and are shown schematically above the data curve. Adapted from \citet{Chernyshov2009}.}
\label{fig_iSGE-SOT-GaMnAs-1}
\end{figure}

The experimental discovery of the iSGE-induced SOT was reported in a (Ga,Mn)As sample whose image is shown Fig.~\ref{fig_iSGE-SOT-GaMnAs-1}(a) \cite{Chernyshov2009,Endo2010}. The experiment demonstrated not only the presence of the iSGE effective field of the expected Dresselhaus symmetry for the strained (Ga,Mn)As epilayer, but also demonstrated that iSGE was sufficiently strong to reversibly switch the direction of magnetization. Data in Fig.~\ref{fig_iSGE-SOT-GaMnAs-1}(b) were taken at external magnetic field magnitude and angle fixed close to the switching point between the [010] and [100] easy-axes. The measured transverse AMR, used for the electrical readout, forms a hysteresis loop as the writing iSGE current is swept between $\pm1$~mA. The loop corresponds to the electrical switching between the [010] and [100] easy-axes. Here 100~ms current pulses of a 1~mA amplitude and alternating polarity were sufficient to permanently rotate the direction of magnetization, as highlighted in Fig.~\ref{fig_iSGE-SOT-GaMnAs-1}(c).

\begin{figure}[h!]
\centering
\includegraphics[width=8cm]{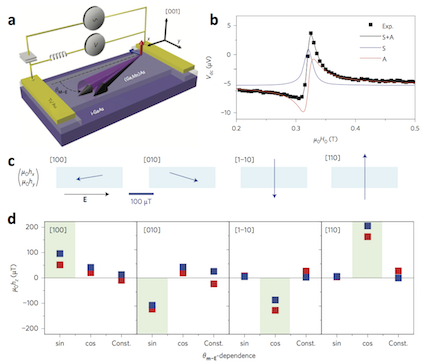}
\caption{(Color online) Schematic of the (Ga,Mn)As sample, measurement set-up and magnetization precession. The injected microwave current drives FMR, which is detected via the dc voltage $V_{dc}$ across the microbar. $\theta_{{\bf m}-{\bf E}}$ is the angle of the static magnetization direction measured from the current flow direction. Arrows represent in-plane (blue) and out-of-plane (red) components of the instantaneous non-equilibrium iSGE spin density induced by the microwave current that drives the magnetization. (b) A typical ST-FMR signal driven by an alternating current at 11~GHz and measured by $V_{dc}$ as a function of external magnetic field. Data were fitted by a combination of symmetric (S) and antisymmetric (A) Lorentzian functions. (c) Direction and magnitude of the in-plane spin-orbit field (blue arrows) within the microbars (light blue rectangles). The direction of the electric field is represented by ${\bf E}$. (d) Coefficients of the $\cos\theta_{{\bf m}-{\bf E}}$ and $\sin\theta_{{\bf m}-{\bf E}}$ fits to the angle dependence of the out-of-plane SOT field for the sample set. In this out-of-plane data, two samples are shown in each microbar direction and are distinguished by blue and red square data points. The symmetries expected for the damping-like SOT, on the basis of the theoretical model for the Dresselhaus spin-orbit Hamiltonian [see Eq. \eqref{eq:dresselk}], are shown by light green shading. Adapted from \citet{Kurebayashi2014}.}
\label{fig_iSGE-SOT-GaMnAs-2}
\end{figure}

A detailed analysis of the magnitude and symmetry of iSGE effective fields in (Ga,Mn)As was performed by employing an all-electrical ST-FMR technique, sketched in Fig.~\ref{fig_iSGE-SOT-GaMnAs-2}(a) \cite{Fang2011,Kurebayashi2014} and presented in Subsection \ref{MML:STFMR}. Here an electric current oscillating at microwave frequencies is used to create an oscillating effective SOT field in the magnetic material being probed, which makes it possible to characterize individual nanoscale samples with uniform magnetization profiles \cite{Fang2011}. For detection, a frequency mixing effect based on the AMR was used. When magnetization precession is driven, there is a time-dependent change $\Delta R(t)$ in longitudinal resistance from the equilibrium value $R$ (owing to the AMR). The resistance oscillates with the same frequency as the microwave current, thus causing frequency mixing, and a directly measurable dc voltage $V_{dc}$ is generated. This voltage provides a probe of the amplitude and phase of magnetization precession with respect to the microwave current.

The FMR vector magnetometry on the driving SOT fields revealed a dominant Dresselhaus and a weaker Rashba contribution [Fig.~\ref{fig_iSGE-SOT-GaMnAs-2}(a)] \cite{Fang2011}. By separating the symmetric and antisymmetric parts of the mixing $V_{dc}$ signal [Fig.~\ref{fig_iSGE-SOT-GaMnAs-2}(b)], it was possible to identify both the field-like and the damping-like SOT components \cite{Kurebayashi2014}. It was shown that the damping-like SOT plays a comparably important role in driving the magnetization dynamics in (Ga,Mn)As as the field-like SOT [Figs.~\ref{fig_iSGE-SOT-GaMnAs-2}(c,d)].

\begin{figure}[h!]
\centering
\includegraphics[width=8cm]{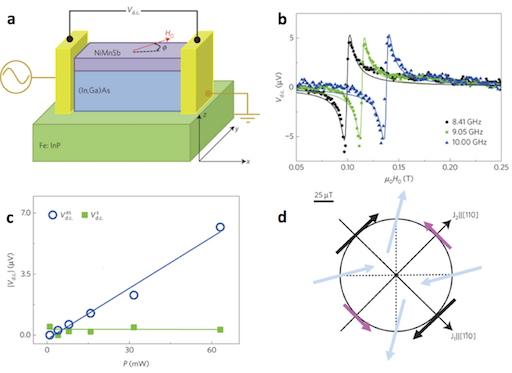}
\caption{(Color online) (a) Schematic of the NiMnSb epilayer sample and measuring set-up. A microwave current is passed in the bar and excites ST-FMR. By measuring the longitudinal dc voltage, the magnitude of the spin-orbit driving field is deduced. (b) The rectified voltage showing FMR for different frequencies of the microwave current. The Lorentzians are well fitted by an antisymmetric line-shape (continuous line) at all frequencies. (c) Power dependence of the symmetric and antisymmetric components of the rectified voltage. (d) Polar plot illustrating the direction of the spin-orbit field for current flowing along different crystal directions of NiMnSb. Adapted from \citet{Ciccarelli2016}}.
\label{fig_iSGE-SOT-NiMnSb}
\end{figure}

The FMR technique was also employed in the study of the iSGE-induced SOT in the room-temperature, dense-moment metal ferromagnet NiMnSb, as shown in Fig.~\ref{fig_iSGE-SOT-NiMnSb}. In agreement with the symmetry expectations for the strained half-heusler lattice of the NiMnSb epilayer, and in agreement with the results in the directly related zinc-blende lattice of (Ga,Mn)As, the observed field-like component has a dominant Dresselhaus symmetry [Fig.~\ref{fig_iSGE-SOT-NiMnSb}(d)]. Unlike (Ga,Mn)As, the damping-like SOT was not identified in NiMnSb [Fig.~\ref{fig_iSGE-SOT-NiMnSb}(b,c)]. This is likely due to the higher conductivity of metallic NiMnSb. While the extrinsic field-like SOT scales with the conductivity, the intrinsic contribution to the damping-like SOT is scattering-independent to lowest order (see Subsection \ref{s:rashbatorque}), implying that the higher conductivity of the NiMnSb metal might favor the field-like SOT. 

\subsection{Collinear antiferromagnets}\label{iSGE-SOT-antiferromagnet}

Compensated two-spin-sublattice antiferromagnets have north poles of half of the microscopic atomic moments pointing in one direction and the other half in the opposite direction. This makes the uniform external magnetic field inefficient for switching magnetic moments in antiferromagnets. The complete absence of electromagnets or reference permanent magnets in the SOT scheme for writing ferromagnetic memory bits, discussed above, has served as the key for introducing the physical concept for the efficient control of magnetic moments in antiferromagnets \cite{Zelezny2014}. 

Two distinct scenarios have been considered for the SOT on antiferromagnetic spin sublattices $A/B$, $\frac{\partial}{\partial t}{\bf m}_{A/B}\sim {\bf m}_{A/B}\times {\bf B}^{\rm eff}_{A/B}$ \cite{Zelezny2014}. One in which the crystal is globally non-centrosymmetric. Here an example is the half-heusler antiferromagnet CuMnSb \cite{Forster1968} or any thin-film antiferromagnet with structural inversion asymmetry. The efficient torque in this case is the damping-like SOT which, assuming e.g. Rashba spin-orbit coupling, is driven by an effective field ${\bf B}^{\rm eff}_{A/B}\sim({\bf E}\times{\bf z})\times{\bf m}_{A/B}$. Here ${\bf B}^{\rm eff}_{A/B}$ is staggered due to the opposite magnetizations on the two spin sublattices of the antiferromagnet, ${\bf m}_{A}=-{\bf m}_{B}$. The field-like SOT in these globally non-centrosymmetric crystals in not efficient for antiferromagnets since the effective field driving the field-like torque, ${\bf B}^{\rm eff}_{A/B}\sim{\bf E}\times{\bf z}$, is not staggered.

In Fig.~\ref{fig_iSGE-SOT-symmetry}(a) we illustrated that in crystals with two inversion-partner lattice sites in the unit cell, the iSGE can generate a staggered spin density. This leads to the second scenario in which the field-like component of the SOT is efficient in an antiferromagnet whose magnetic spin sublattices $A/B$ coincide with the two inversion-partner crystal-sublattices. In this case the effective field driving the field-like SOT has the staggered form (again assuming the Rashba symmetry): ${\bf B}^{\rm eff}_{A}\sim{\bf E}\times{\bf z}$ and ${\bf B}^{\rm eff}_{B}\sim-{\bf E}\times{\bf z}$. Mn$_2$Au and CuMnAs are examples of high N\'eel temperature antiferromagnetic crystals in which this scenario applies \cite{Zelezny2014,Wadley2016}.

\begin{figure}[h!]
\centering
\includegraphics[width=8cm]{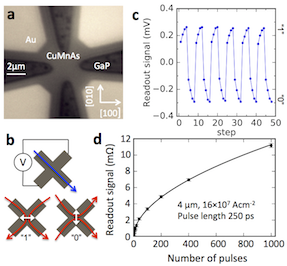}
\caption{(Color online) (a) Optical microscopy image of the device containing Au contact pads and the antiferromagnet CuMnAs cross-shape cell on the GaP substrate. (b) Top: The readout current (blue arrow) and transverse voltage detection geometry. Bottom: Write pulse current lines (red arrows) labeled "1" (left) and "0" (right) and the corresponding preferred antiferromagnetic moment orientations (white double-arrows). (c) Readout signals after repeated four write pulses with current lines along the [100] direction ("0") followed by four pulses with current lines along the [010] direction ("1"). (d) Readout signal as a function of the number of pulses in the train of pulses for the individual pulse length of 250~ps. Adapted from \citet{Olejnik2017}.}
\label{fig_iSGE-SOT-antiferro-1}
\end{figure}

Figure~\ref{fig_iSGE-SOT-antiferro-1} illustrates the experimental realization of electrical switching by the staggered SOT field in a memory bit cell fabricated from a single-crystal epitaxial film of a CuMnAs antiferromagnet \cite{Wadley2016,Olejnik2017}. Writing current pulses are sent through the four contacts of the bit-cell to generate current lines in the central region of the cross along one of two orthogonal axes, representing "0" and "1" [Fig.~\ref{fig_iSGE-SOT-antiferro-1}(b)]. The writing current pulses give preference to domains with antiferromagnetic moments aligned perpendicular to the current lines (Rashba-like symmetry). Electrical readout is performed by running the probe current along one of the arms of the cross and by measuring the antiferromagnetic transverse AMR across the other arm [Fig.~\ref{fig_iSGE-SOT-antiferro-1}(b)]. The write/read functionality of the CuMnAs memory cells was verified to be not significantly perturbed in a superconducting magnet generating a magnetic field as strong as 12~T \cite{Wadley2016}. This highlights the efficiency of the staggered SOT fields whose inferred magnitude allowing to switch the antiferromagnetic moments is only in the mT range.\par

The bit-cell write/read signals can be sent at ambient conditions by placing the CuMnAs chip on a standard printed circuit board connected to a personal computer via a 5~V USB interface \cite{Olejnik2017}. Fig.~\ref{fig_iSGE-SOT-antiferro-1}(c) shows an example of data obtained from this proof-of-concept antiferromagnetic memory device. Apart from demonstrating the application potential of antiferromagnets in spintronics thanks to the SOT, it also illustrates a deterministic multi-level switching of the antiferromagnetic bit cell. Here successive $\sim\mu$s writing pulses along one of the current path directions produce reproducible step-like changes in the memory readout signal. A photoemission electron microscopy study of CuMnAs has associated the multi-level electrical switching signal with the antiferromagnetic moment reorientations within multiple domains \cite{Grzybowski2016}. 

The observation of SOT-driven switching has been recently extended to Mn$_2$Au, where a large AMR ratio up to 6 \% is obtained \cite{Bodnar2018,Meinert2018,Zhou2018b}. The general switching features are quite similar to the ones observed in CuMnAs, revealing the multidomain magnetic structure of the system. Upon increasing the applied current, the N\'eel order of the different magnetic domains is progressively reoriented under thermal activation \cite{Meinert2018}, in sharp contrast with the fast switching obtained in NM/FM bilayers driven by domain wall nucleation/propagation (see Subsection \ref{MML:switching}). This progressive switching seems to be a specific property of antiferromagnetic materials, as it was also reported in the case of field-free switching in AF/FM metallic bilayers \cite{Fukami2016a,Brink2016,Oh2016} (see Subsection \ref{MML:switching:zerofield}). In agreement with the multi-domain picture, experiments in the AF/FM bilayers showed that the number of intermediate levels decreases with the decreasing size of the device and finally evolves into a binary mode below a certain threshold \cite{Kurenko2017}.

The multi-level nature of antiferromagnetic bit cells opens the possibility for combining memory, logic and neuromorphic functionalities (e.g., pulse-counter) within the cell \cite{Olejnik2017}. Another unique merit of antiferromagnets is the THz scale of the internal spin dynamics which in combination with the SOT physics opens the door to ultra-fast switching schemes. Fig.~\ref{fig_iSGE-SOT-antiferro-1}(d) shows initial results of experiments in this direction demonstrating a deterministic memory-counter functionality in a multi-level CuMnAs memory cell for $\sim$ 1000 pulses with individual pulse-length scaled down to 250~ps. In these experiments, current pulses were delivered via wire-bonded contacts for which pulse-length $\sim$ 100 ps is at the limit achievable with common current-pulse setups. \par

Subsequently, reversible switching with analogous characteristics was demonstrated using 1 ps long writing pulses \cite{Olejnik2018}. A non-contact technique was employed for generating the ultra-short current pulses in the antiferromagnetic memory cell via THz electromagnetic transients to overcome the above limit of common contact current-pulse setups. Remarkably, the writing energy did not increase when down-scaling the pulse-length from ns to ps. This is in striking contrast to ferromagnetic STT \cite{Bedau2010} or SOT \cite{Garello2014} memories in which the theoretically extrapolated writing energy at ps would increase by three orders of magnitude compared to the state-of-the-art ns-switching devices. While readily achievable in antiferromagnets, the ps range remains elusive for ferromagnets because, in frequency terms, it far exceeds the GHz-scale of the FMR in typical ferromagnets.

All the above SOT experiments on antiferromagnets employed 90$^{\rm o}$ switching of the N\'eel vector. 180$^{\rm o}$ switching has been also recently demonstrated in CuMnAs by alternating the sign of the writing current. The readout of the reversed N\'eel vector memory states was performed electrically using a second-order magnetoresistance whose presence relies on the broken time-reversal and space-inversion symmetries in the antiferromagnetic crystal of CuMnAs \cite{Godinho2018}. Microscopically, the mechanism of this second-order magnetoresistance in CuMnAs was ascribed to a transient tilt of the N\'eel vector due to the SOT combined with the AMR.

\subsection{Antiferromagnetic topological Dirac fermions\label{iSGE-SOT-Dirac}}

\begin{figure}[h!]
\centering
\includegraphics[width=8cm]{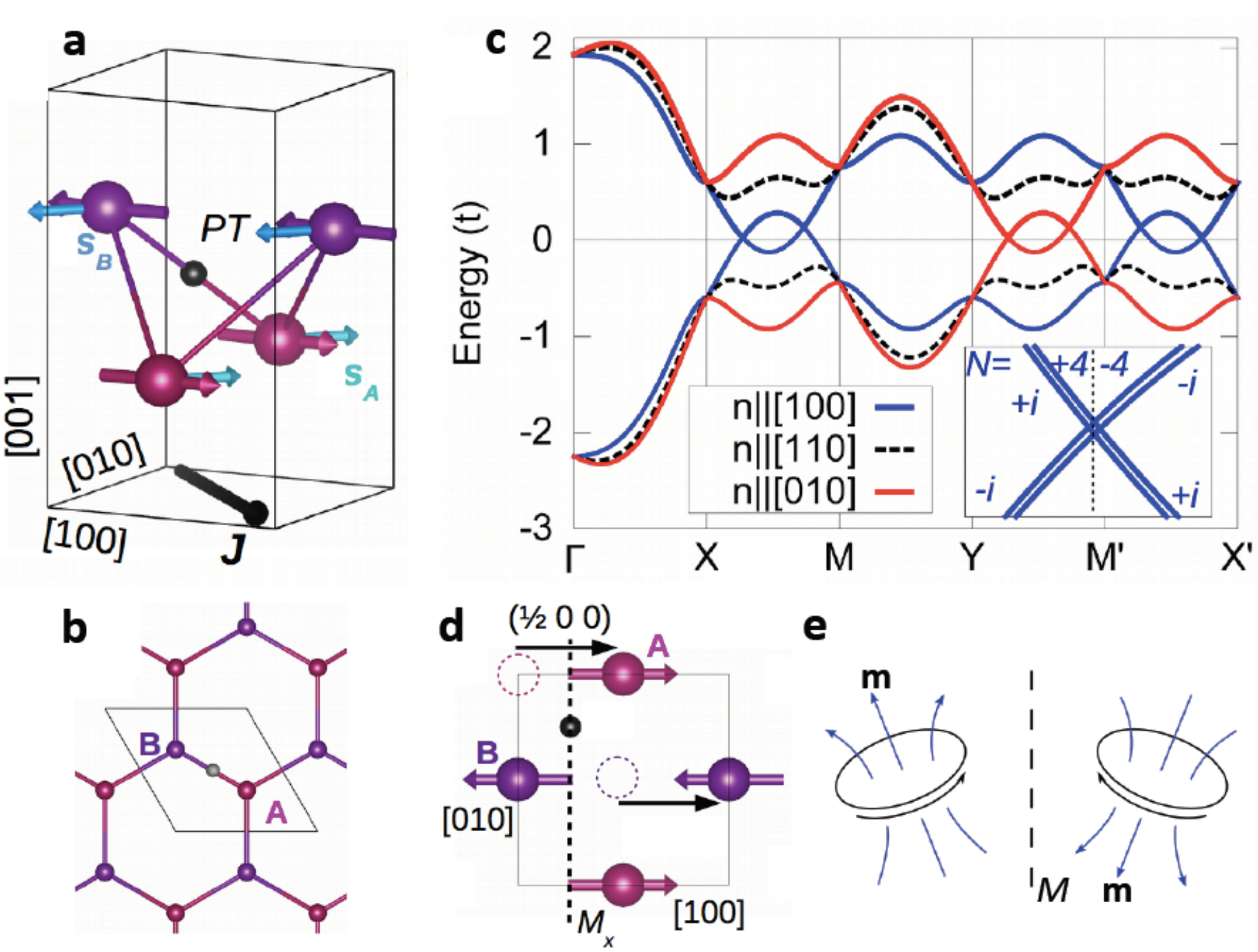}
\caption{(Color online) (a) Mn antiferromagnetic spin sublattices of CuMnAs denoted by purple and pink balls with thick arrows. The antiferromagnet order breaks time-reversal symmetry ($\cal{T}$) and space-inversion symmetry ($\cal{P}$), however, the combined $\cal{P}\cal{T}$ symmetry is preserved. Staggered current-induced spin density on the sublattices $A$ and $B$ is denoted by cyan and blue arrows. (b) Graphene crystal with two C-sites per unit cell in analogy with the Mn-sites in CuMnAs. (c) Band dispersion of the minimal antiferromagnet model based on CuMnAs illustrating the control of the Dirac points by the direction of the N\'eel vector {\bf n}. Topological indices of the Dirac point are shown in the inset (for the sake of clarity the degenerate bands are slightly shifted). (d) Top view of the model quasi-2D-antiferromagnetic lattice of CuMnAs highlighting the non-symmorphic glide mirror plane symmetry, combining mirror plane (${\cal M}_x$) reflection with a half-unit-cell translation along the $x$-axis. (e) An axial vector {\bf m} under mirror (${\cal M}$) reflection. Adapted from \citet{Smejkal2017}.}
\label{fig_iSGE-SOT-antiferro-2}\end{figure}

Recently, a new concept has been theoretically proposed. It follows from the observation that the staggered SOT fields can co-exist with topological Dirac fermions in the band structure of antiferromagnets because of the serendipitous overlap of the key symmetry requirements \cite{Smejkal2017}. Therefore, one can use SOT to reorient the N\'eel vector in antiferromagnets in order to control such topological Dirac fermions. This is illustrated in Fig.~\ref{fig_iSGE-SOT-antiferro-2}(a,b) on examples of the CuMnAs where the SOT switching was experimentally verified, as mentioned above, and of the graphene lattice representing the Dirac systems \cite{CastroNeto2009b}: (i) The two-Mn-site primitive cell of CuMnAs favors band crossings in analogy with the two-C-site graphene lattice. (ii) In the paramagnetic phase, CuMnAs has time reversal ($\cal{T}$) and space inversion ($\cal{P}$) symmetries. It guarantees that each band is doubly-degenerate forming a Kramer's pair, in analogy to graphene. In the antiferromagnetic phase, this degeneracy is not lifted because the combined $\cal{PT}$ symmetry is preserved, although the $\cal{T}$ symmetry and the $\cal{P}$ symmetry are individually broken \cite{Herring1966,Chen2014,Tang2016,Smejkal2017}. (iii) The combined $\cal{PT}$ symmetry is just another way of expressing that the two antiferromagnetic spin sublattices conincide with the two inversion-partner crystal-sublattices. As explained above, this condition leads to an efficient field-like SOT in bipartite antiferromagnets.

An additional crystal symmetry is needed to mediate the dependence of Dirac quasiparticles on the N\'eel vector orientation [Fig.~\ref{fig_iSGE-SOT-antiferro-2}(c)]. In graphene there is no symmetry that protects the four-fold degeneracy of Dirac crossings of two Kramer's pair bands in the presence of spin-orbit coupling \cite{Kane2005}. In CuMnAs, on the other hand, the Dirac crossings are protected by a non-symmorphic, glide mirror plane symmetry \cite{Young2015}, $\mathcal{G}_{x}=\left\lbrace {\cal M}_{x} \vert \frac{1}{2} 0 0 \right\rbrace$, as long as the N\'eel vector is aligned with the [100] axis. $\mathcal{G}_{x}$ combines the mirror symmetry ${\cal M}_{x}$ along the (100)-plane with the half-primitive cell translation along the [100] axis [Fig.~\ref{fig_iSGE-SOT-antiferro-2}(d)]. Due to the mirror-reflection behavior of the axial vectors of magnetic moments [Fig.~\ref{fig_iSGE-SOT-antiferro-2}(e)], the $\mathcal{G}_{x}$ symmetry, and thus also the Dirac crossing protection, is broken when the antiferromagnetic moments are reoriented into a general crystal direction by the SOT.

\subsection{Magnonic charge pumping in (Ga,Mn)As}
\label{iSGE-SOT-magnonic}

We conclude this section by briefly discussing the SGE, which is a reciprocal phenomenon to the iSGE, and its counterpart in magnets termed the magnonic charge pumping \cite{Ciccarelli2014}. The latter, in turn, is a reciprocal phenomenon to the SOT. Following theoretical predictions \cite{Ivchenko1978,Ivchenko1989,Aronov1989,Edelstein1990,Malshukov2002,Inoue2003}, the SGE was initially observed in an asymmetrically confined GaAs quantum well \cite{Ganichev2002}. The key signature of the SGE is the electrical current induced by a non-equilibrium, but uniform, polarization of electron spins. In the non-equilibrium steady-state, the spin-up and spin-down sub-bands have different populations, induced in \citet{Ganichev2002}'s experiment by a circularly polarized light excitation. Simultaneously, the two sub-bands for spin-up and spin-down electrons are shifted in momentum space due to the inversion asymmetry of the semiconductor structure which leads to an inherent asymmetry in the spin-flip scattering events between the two sub-bands. This results in the flow of the electrical current.

\begin{figure}[h!]
\centering
\includegraphics[width=8cm]{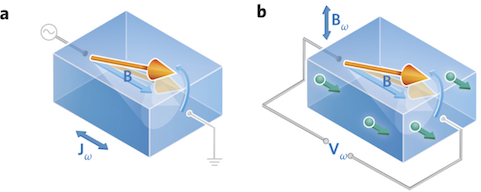}
\caption{(Color online) (a) A charge current through (Ga,Mn)As results in a non-equilibrium spin polarization of the carriers, which exchange-couples to the magnetization and exerts the SOT. An alternating current generates a time-varying torque, which drives magnetic precession resonantly when a magnetic field is applied. (b) The reciprocal effect of (a) termed the magnonic charge pumping. From \citet{Ciccarelli2014}.}
\label{fig_iSGE-SOT-magnonic}
\end{figure}

The Onsager reciprocity relations imply that there is also a reciprocal phenomenon of the iSGE induced SOT in which electrical signal due to the SGE is generated from magnetization precession in a uniform, spin-orbit coupled magnetic system with broken space inversion symmetry (see Fig.~\ref{fig_iSGE-SOT-magnonic}) \cite{Hals2010,Kim2012,Tatara2013}. In this reciprocal SOT effect no secondary spin-charge conversion element is required and, as for the SOT, (Ga,Mn)As with broken inversion symmetry in its bulk crystal structure and strongly spin-orbit coupled holes represents a favorable model system to explore this phenomenon. The effect was observed in (Ga,Mn)As and termed the magnonic charge pumping \cite{Ciccarelli2014}. This effect is physically similar to the SGE (or alternatively called inverse Rashba-Edelstein effect) observed at Bi/Ag(111) \cite{Rojas-Sanchez2013b} or topological insulators surfaces \cite{Shiomi2014}.

\subsection{Established features and open questions}
SOTs in bulk non-centrosymmetric crystals are relatively well understood since here the iSGE mechanism is not complemented by the SHE, and ST-FMR experiments in external magnetic fields can provide a quantitatively accurate vector analysis of the SOT fields. Among the remaining open questions is what material parameters control the relative strengths of field-like and damping-like SOTs. Regarding the crystal symmetries of the SOT, Rashba and Dresselhaus spin-orbit fields have been already identified whereas an experimental evidence of the Weyl symmetry SOT [Fig. \ref{fig:Rashba_Dresselhaus}(e)] is yet to be demonstrated. 

Compared to ferromagnets, the research of SOTs in antiferromagnetic crystals is still in its infancy. The insensitivity to external magnetic fields makes the experimental calibration of the staggered SOT field strength difficult to perform. In experiments performed to date, the current-induced switching shows clear signatures of a heat-assisted mechanism \cite{Olejnik2018,Bodnar2018,Meinert2018}. On the one hand, this is favorable for lowering the effective magnetic anisotropy barrier but, on the other hand, it may limit the accessible writing frequency and current amplitudes, and further complicates the experimental determination of the strength of the SOT fields. Therefore, only the experimental switching current amplitudes have been reported in antiferromagnets so far. In CuMnAs structures, the switching current densities are in the 10$^8$ A/cm$^{2}$ range for $\sim$ ns long writing pulses which is comparable to the common ferromagnetic Co/Pt SOT devices \cite{Olejnik2017,Garello2014}.

SOT electrical writing speeds, defined as the inverse of the writing pulse length, in the CuMnAs antiferromagnet have been experimentally demonstrated to reach the THz range, which far exceeds the SOT writing speeds in ferromagnets. Antiferromagnets also offer the possibility to combinate the SOT with topological Dirac fermions, which are prohibited by symmetry in ferromagnets. An indication of the SOT-induced opening and closing of the Dirac crossing has been already reported in an experimental and theoretical study of the AMR in Mn$_2$Au \cite{Bodnar2018}. The ultimate strength of this topological AMR, however, has been predicted for purely semimetallic antiferromagnets in which the Dirac points are at the Fermi level and no other trivial bands are crossing the Fermi energy \cite{Smejkal2017}. If experimentally demonstrated, it would have important implications not only for the basic research of topological phenomena in magnetic systems but may also provide the desired large magnetoresistance allowing for a better scalability of the readout signals in antiferromagnetic spintronic devices.

\section{Spin-orbit torques and non-uniform magnetic textures\label{s:DW}}

The electrical manipulation of magnetic textures using SOTs opens stimulating perspectives for applications. In Section~\ref{MML}, we already mentioned that domain wall nucleation and propagation play an important role in the context of SOT-driven switching. In addition, intentional and well-controlled domain wall manipulation constitutes the basis of alternative, domain wall-based racetrack memories \cite{Parkin2008,Parkin2015} and logic concepts \cite{Allwood2005}. In this context, a major breakthrough has been the recent realization and control of individual metastable skyrmions at room temperatures \cite{Fert2017,Jiang2017c}, which show promising potential for such applications \cite{Fert2013,Zhang2015d}. Nonetheless, evaluating SOTs in magnetic textures poses a specific challenge compared to the magnetically uniform thin films discussed in Section~\ref{MML}. While SOTs induce a rotation of the magnetization that can be 'simply' recorded through magnetometry (AMR, AHE or MOKE), in magnetic textures one can only evaluate the {\em global} impact of the SOTs through the texture motion and deformation. This feature transforms the magnetometry issue to a magnetic microscopy issue. The present section addresses SOT-driven domain wall and skyrmion motion and dynamics in detail.\par

Starting with a phenomenological description of the influence of current-induced torques on domain wall motion in Subsection \ref{s:DW-dyna}, we then discuss its experimental observation in in-plane and perpendicularly magnetized domain walls in Subsections \ref{s:DW-XY} and \ref{s:DW-Z}, respectively. Recent progress achieved on ferrimagnetic and antiferromagnetic systems is presented in Subsection \ref{s:AF-DW}. 
Note that the role of domain wall nucleation and propagation in SOT driven switching has been discussed in Subsection \ref{MML:switching}.

\subsection{Domain wall dynamics under current}
\label{s:DW-dyna}

The dynamics of magnetic textures is governed by the LLG equation, Eq. \eqref{eq:LLG}, at the basis 
of the continuous theory of magnetic structures, called micromagnetics \cite{Hubert1998}.
In this framework, the local magnetization vector is written 
${\bf M}\left( {\bf r},t\right)=M_\mathrm{s} {\bf m}\left( {\bf r},t\right)$,
where the spontaneous magnetization modulus $M_\mathrm{s}$ depends on temperature, whereas the unit vector
${\bf m}$ specifies its local orientation as a function of space and time. The torques induced by
an electrical current on the magnetic texture are of two forms. On the one hand, the STT is generally written, in its local version,
as the sum of so-called adiabatic and non-adiabatic terms \cite{Beach2008}
\begin{equation}
\label{eq:DW-STT}
(\gamma/M_{\rm s}){\bf T}_\mathrm{STT} = -\left( {\bf u} \cdot {\bf \nabla} \right) {\bf m} +
\beta{\bf m} \times \left[\left( {\bf u} \cdot {\bf \nabla} \right) {\bf m} \right],
\end{equation}
where the velocity ${\bf u}$ is proportional to the electrical current density in the magnetic material,
its spin polarization etc., and where $\beta$ is the non-adiabaticity factor (no dimensions).
This torque is proportional to the gradient of magnetization along the current direction and thus vanishes in the domains.
On the other hand, the SOT is expressed by Eq. \eqref{eq:torquedef}. It does not depend on the gradient of the magnetization at the lowest order, hence acts also on the magnetization within the domains [for higher order expansion, see \citet{Bijl2012}]. Note that in general when a current is applied to a magnet/metal bilayer, it flows both into the magnet, leading to STT, and into the metal, leading to SOT in the magnet as well as to an Oersted field. We thus need to study the effect of these three torques on domain walls.\par

\begin{figure}
\includegraphics[width=8cm]{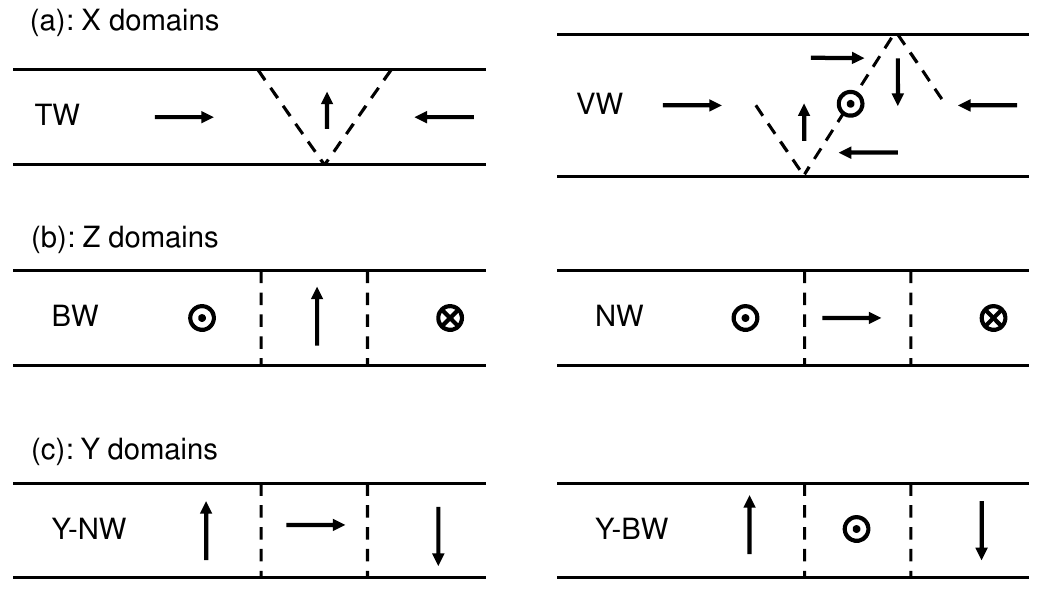}
\caption{Schematic of domain wall structures in nanostrips relevant for SOT studies.
(a) When the magnetic easy axis is along the nanostrip ($x$ axis), typically for small magnetic anisotropy, 
magnetostatics leads to two basic structures, the transverse wall at small width and thickness (left), and the 
vortex wall at larger lateral dimensions (right) \cite{McMichael1997,Nakatani2005}. Note that these domain walls have a non-zero magnetostatic charge.
(b) When the easy axis is perpendicular ($z$ axis), typically for strong interface anisotropy, magnetostatics
favors the Bloch wall (left) but the interfacial DMI can favor the N\'eel wall \cite{Heide2008,Thiaville2012}, and fix its chirality (right). A `bulk' DMI would favor the Bloch wall, and fix its chirality. (c) The last case of a transverse easy axis ($y$ axis) is rare.
The associated walls, known for a long time \cite{Hubert1998}, are the N{\'{e}}el wall (Y-N\'eel wall) at small thickness (left),
and the Bloch wall (Y-Bloch wall) at larger thickness (right). In the absence of DMI, the domain wall magnetization is uncorrelated to the
magnetization in the domains, so that domain or domain wall magnetization arrows can be reversed with no 
change of energy.
The dashed lines outline the shape of the domain walls.}
\label{fig:DW-structures}
\end{figure}

A qualitative analysis of these current-induced torques is instructive. For this we consider, for each torque term ${\bf T}$, the effective field ${\bf B}_{\bf T}$ obtained by writing 
${\bf T}={\bf M}\times{\bf B}_{\bf T}$, the evaluation being performed at the center of the domain wall. From the solved form of Eq.~(\ref{eq:LLG}), i.e., with $\frac{\partial {\bf m}}{\partial t}$ only on the left-hand side, one sees that the magnetization dynamics is driven by the total effective field ${\bf B}_{\bf M} - {\bf B}_{\bf T}$ [where the minus sign is consistent with Eq. \eqref{eq:LLG}], with on the one hand a precession around it driven by the gyromagnetic ratio $\gamma$, and on the other hand a relaxation towards it driven by the damping parameter $\alpha$. To analyze the impact of current-induced torques on the domain wall motion, we need to know the types of magnetic domain walls in samples
where large current pulses can be applied (typically $10^{11}$~A/m$^2$). In order to promote large current densities while avoiding excessive sample Joule heating, these samples have the shape of nanostrips, with a width $w$ of about a few hundreds of nanometers, and a thickness $h$ of the order of a few nanometers (the thickness being generally thinner for interfacial SOT). As shown in Fig.~\ref{fig:DW-structures}, a limited number of domain wall structures has to be considered, according to
the magnetic anisotropy of the sample.

\subsubsection{Steady domain wall dynamics}
\label{s:DW-steady}

In order to get a {\em steady-state} current-induced domain wall motion (CIDM) under a torque ${\bf T}$, one needs the
effective field ${\bf B_{\bf T}}$ to be directed along the domains' magnetization (a subtlety exists for the vortex wall, as the magnetization is not uniform outside the vortex core, see below). This prescription derives from the analogy with the case of an applied magnetic field. It expresses a domain wall motion controlled by damping: if the effective field favors one domain, this domain steadily grows and hence the domain wall moves.

The STT torques, Eq. \eqref{eq:DW-STT}, depend on the magnetization gradient along the current direction, i.e., 
$\sim({\bf u}\cdot{\bm\nabla}){\bf m}$ ($\equiv\partial{\bf m}/\partial x$ with the axes convention defined in Fig.~\ref{fig:DW-structures}). From Fig.~\ref{fig:DW-structures} we see that at the domain wall center this
derivative is along the magnetization of the domain on the right of the domain wall
(an exception to this rule is afforded by the vortex wall, where the magnetization streamlines are reoriented
by 90$^\circ$ through the vortex structure). By construction, the effective field ${\bf B}_{\bf T}$ associated with the adiabatic STT is orthogonal to the domains' magnetization, so that it cannot lead to steady domain wall motion.
On the other hand, the effective field associated with the non-adiabatic STT lies along the domain
magnetization.
This explains qualitatively the rule for steady STT driven domain wall motion, given by the velocity formula
${\bf v} = \left( \beta / \alpha \right) {\bf u}$ in which domain walls move
along the carriers for positive current polarization (majority spin polarization of the current)
and positive $\beta$ factor \cite{Thiaville2005,Zhang2004}.
The same conclusions are reached for the vortex wall case, by considering the surrounding of the vortex core
instead of the domains.

We now perform the same analysis for the SOTs. The effective field associated with the field-like SOT reads ${\bf B}_{\rm FL}=-(\tau_{\rm FL}/M_{\rm s}) {\bm\zeta}$ with ${\bm\zeta}||{\bf y}$ for a current along $x$, considering the Rashba symmetry of the spin-orbit coupling.
This field is oriented like the main part of the Oersted field (as $w \gg h$ the $y$ component
of the stray field dominates the $z$ component).
The results for the various domain wall structures are summarized in Table~\ref{tab:DW-HT}, a generalization of those 
of \citet{Khvalkovskiy2013}: apart from the obvious case of
$y$ easy axis \cite{Obata2008}, no steady domain wall motion is expected.
On the other hand, for the damping-like SOT with
${\bf B}_{\rm DL}=-(\tau_{\rm DL}/M_{\rm s}) {\bf m} \times {\bm\zeta} $,
only the N\'eel wall for the $z$ easy axis is expected to be set in steady motion.

\subsubsection{Precessional domain wall dynamics}
\label{s:DW-precess}

Another characteristic regime of domain wall motion is called {\em precessional}, meaning that the domain wall magnetization is rotating in a given direction around 
the domains' magnetization. Following very general arguments initially due to \citet{Slonczewski1972} according to which the domain wall position
and the angle of the domain wall magnetization are coupled variables in the Hamilton sense, a continuously precessing
domain wall magnetization induces an overall domain wall motion.

The simplest known case of precession occurs under a large enough field applied along the domains' magnetization, the field being larger 
than the so-called Walker field.
In that case, this precession-induced velocity opposes that due to the applied field, hence the term
of Walker breakdown stressing that domain wall velocity drops above the Walker field.
The Walker threshold occurs because the domain wall structure deformation by domain wall magnetization rotation around the applied field can be
counter-balanced by internal energies (anisotropy, demagnetizing field, DMI etc.), up to a certain limit.
The same breakdown is therefore also expected when the effective field ${\bf B}_{\bf T}$ is along the domains' magnetization
and large enough [with the subtleties that for STT, the velocity increases above the threshold when $\beta < \alpha$
\cite{Thiaville2005,Zhang2004}, while for damping-like SOT the threshold is never reached \cite{Thiaville2012}].

Domain wall magnetization rotation also occurs by relaxation towards the current-induced effective fields ${\bf B}_{\bf T}$.
If these fields are below the `breakdown' threshold, a domain wall position shift will appear as a result of the 
domain wall structure transformation when current is applied.
When current goes back to zero, and provided the sample is perfect, the opposite domain wall position shift
will however occur as the domain wall recovers its initial structure. Note that several devices based on an anticipated 
stick-slip domain wall motion under application of dissymmetric pulses with short rise-time and long fall-time have been 
proposed, based on this phenomenon.
A partial list of cases with domain wall shift was presented in \citet{Khvalkovskiy2013}. 
The full list is given in Table~\ref{tab:DW-HT}.
When the effective field related to a current-induced torque is large enough, the domain wall structure goes to its image where the domain wall magnetization has been reversed.
Whether this process continues or not depends on the power to which the domain wall magnetization enters the expression
of the effective field ${\bf B}_{\bf T}$.
If this power is odd, the opposite field will act on the opposite domain wall magnetization, leading to indefinite
precession of domain wall magnetization and hence to long-term precessional domain wall motion.
If the power is even, however, indefinite precession will not occur and only a domain wall position shift will occur.
These cases are also indicated in Table~\ref{tab:DW-HT}. The table shows that field-like SOT (and Oersted field) can only drive domain walls in the $y$-easy axis situation, see Y-domain walls in Fig. \ref{fig:DW-structures}. 

\begin{table}
\begin{tabular}{c||cc|cc}
 DW & STT ad. & STT na. & FL SOT/ & DL SOT \\
  &   &   & Oersted &  \\ \hline \hline 
TW & N & Y  & N  & N \\
  & odd & even & even & null \\ \hline
VW & N & Y  & N  & N \\
  & odd & even & even & odd \\ \hline
BW & N & Y  & N  & N \\
  & odd & even & even & null \\ \hline
NW & N & Y  & N  & Y \\
  & odd & even & even & odd  \\ \hline
Y-NW & N & Y  & Y  & N \\
  & odd & even & even & odd \\ \hline
Y-BW & N & Y  & Y  & N \\
  & odd & even & even & odd \\ \hline \hline
\end{tabular}
\caption{\label{tab:DW-HT} 
Characteristics of the effective field ${\bf B}_{\bf T}$ associated with the
current-induced torques, evaluated at the center of the domain wall types shown in Fig.~\ref{fig:DW-structures}.
For each case, the first line indicates (Y/N) whether or not this effective field drives the domain wall into steady
motion.
The second line indicates (null/odd/even) if this field is zero and, when
it is not, if it is even or odd with respect to the domain wall magnetization, the case "odd" leading to
long-term precessional domain motion.}
\end{table}

With this analysis in mind, we turn in the next subsections to each situation, reviewing the experimental
reports existing on the subject.

\subsection{In-plane magnetized samples}
\label{s:DW-XY}

\subsubsection{Soft samples (X domains)}
\label{sec:ip}

These samples have been the workhorse of the initial studies of the STT, leading to the
definition of the adiabatic and non-adiabatic STT terms.
As Table~\ref{tab:DW-HT} shows, such samples are generally not adequate to test the SOT.
The vortex wall is a special case in this picture, being a composite object that can easily
deform by lateral motion of the vortex core, inducing a displacement of the whole
wall along the nanostrip [see e.g., \cite{Beach2008,Tretiakov2008,Clarke2008}].
As a result, under adiabatic STT for example, the vortex core displaces laterally (along $y$),
leading to a longitudinal domain wall displacement (along $x$). The effect is however transient as the core eventually stops
or disappears at the nanostrip edge, transforming the vortex wall into a transverse wall.
The same effect is expected under SOT.

Micromagnetic studies of the effect of disorder on CIDM by STT show that disorder induces, on top of the expected current threshold for domain wall motion, a modification of the linear regime (change of slope, offset), as well as a suppression of velocity breakdown \cite{Nakatani2003,Thiaville2009,Thiaville2005}.
The modification of the linear regime may be partly understood by introducting a larger effective damping constant for a magnetic texture (such as a domain wall) moving in a disordered medium \cite{Min2010}.

Up till now, only two studies have considered X domains with adjacent heavy metal layers. An early study on Pt/NiFe \cite{Vanhaverbeke2008} investigated the influence of the current direction on the domain wall polarity (i.e. the direction of the domain wall's transverse magnetization). Another more recent study addressed thermal effects in Ta/NiFe/Pt \cite{Torrejon2012}. Moreover, typical thicknesses of the ferromagnetic film were 10~nm, so that the effect of the interfacial torques is strongly reduced. Note that the Oersted field effect was directly observed in the case of a bilayer sample \cite{Uhlir2011} by time-resolved photoelectron emission microscopy using x-ray magnetic circular dichroism, a technique that could be used to measure the field-like SOT {\em in situ}. Simulations have shown that field-like SOT modifies the STT driven dynamics \cite{Seo2012}.

Trilayer samples, typically Co/Cu/NiFe where easier domain wall motion and higher velocities have been observed, are a special case that could not be understood in the frame of STT plus Oersted fields.
It was thus proposed that perpendicular spin currents may play some role \cite{Pizzini2009,Uhlir2010}.
\citet{Khvalkovskiy2009} performed a numerical exploration of the effect of various forms of SOT on both transverse wall and vortex wall, taking ${\bm \zeta} = {\bf x}$ and ${\bm \zeta}={\bf z}$, i.e. the two cases that are not considered 
in standard SOT configuration [the latter case was investigated in \cite{Khvalkovskiy2013} for transverse wall].
The results show that indeed in some cases domain wall sustained motion is expected (field-like SOT for ${\bm\zeta}={\bf x}$, damping-like SOT for ${\bm\zeta}={\bf z}$ for a vortex wall), but their relation to the experimental situation is unclear.
Another family of bilayer samples are the synthetic antiferromagnets. In CoFe/Ru/CoFe, a very low threshold for CIDM has been measured \cite{Lepadatu2017}, and attributed to the intrinsic dynamics of antiferromagnetically-coupled transverse walls, driven by non-adiabatic STT (see Subsection \ref{s:AF-DW}).

\subsubsection{Anisotropic samples with Y domains}

In the case of Y-domain walls (see Fig. \ref{fig:DW-structures}) the field-like SOT is directly active \cite{Obata2008}.
Such samples require an in-plane anisotropy that is stronger than the magnetostatic energy cost.
This has been realized by growing epitaxial layers on single-crystal substrates.
One example is (Ga,Mn)As grown on (001) GaAs \cite{Thevenard2017}, where structures with
X domains and Y domains were compared, on 50~nm thick layers so that bulk SOT would be active.
Large current-induced effects were observed, that strongly differed in the two cases, but no 
simple and global understanding of the observed effects could be found.

Another way to obtain such structures is to use large magnetostriction materials, as growth-induced
stress is relaxed at the edges of a nanostrip, modifying the anisotropy locally.
As a result, transverse Y domains were observed in Ni$_{80}$Pd$_{20}$ films \cite{Chauleau2011}.
No study of CIDM could however be realized on such samples,
as the Curie temperature was rapidly reached.

\subsection{Perpendicularly magnetized samples}
\label{s:DW-Z}

A numerical micromagnetic study \cite{Fukami2008} demonstrated the interest of perpendicularly magnetized samples
for CIDM: as the sample thickness is reduced, the energy cost of a N{\'{e}}el wall relative to the Bloch wall
decreases linearly. Thus, the Walker breakdown field also decreases linearly, as well as the current threshold for domain wall motion under
the adiabatic STT. In addition, microscopic STT theories predicted that the non-adiabatic torque might be larger in narrow domain walls \cite{Tatara2004,Waintal2004,Bohlens2010,Akosa2015}. 
Experimentally, studies first focused on the influence of the electric current on the domain wall depinning 
\cite{Ravelosona2005,Boulle2008,Burrowes2010}. 
The results seemed encouraging, but there were only few systems exhibiting CIDM without 
the assistance of external field. 
One of these systems are the Co/Ni multilayers where the predictions of the adiabatic STT
model were most clearly evidenced \cite{Koyama2011}: (i) the existence of an intrinsic critical current that 
depends on the geometric structure of the domain wall rather than the extrinsic pinning; 
(ii) the independence of the critical current on a perpendicular magnetic field.

\subsubsection{Demonstrations of spin-orbit torques in current-induced domain wall motion}

\begin{figure}[h!]
\centering
\includegraphics[width=8cm]{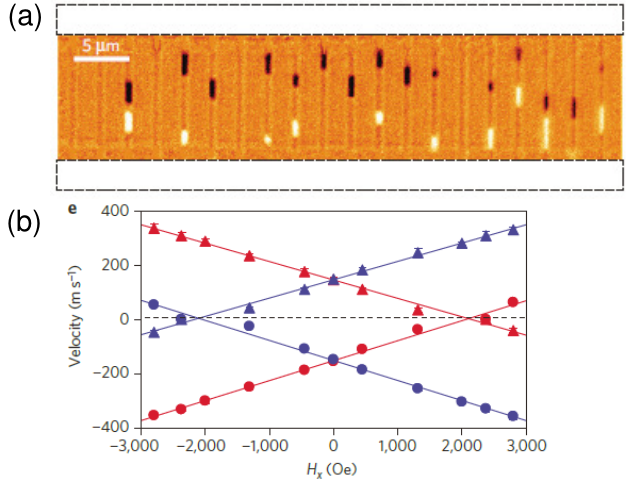}
\caption{ (a) Differential Kerr microscopy imaging of domain wall displacements (stripes of black or white contrast) 
in an array of Pt/Co 0.6~nm/AlOx 500~nm wide nanostrips, after 20 current pulses ($J=1.2 \times 10^{12}$~A/m$^2$,
3~ns duration) \cite{Miron2011}. 
(b) Observation of chiral effects: the velocity of up/down and down/up domain walls (blue and red) is the same, but 
becomes different when an in-plane field is applied [sample Pt/CoNiCo/TaN, current density 
$J=1.5 \times 10^{12}$~A/m$^2$, either positive (triangles) or negative (circles)].
Within the DMI-SOT model, the DMI field strength is indicated by the value of the crossing field, where the domain wall 
velocity changes sign \cite{Ryu2013}.}
\label{fig:DW-PtCo}
\end{figure}

Among the materials with perpendicular anisotropy, the Pt/Co/AlOx trilayers in particular have attracted a 
lot of interest. 
The domain wall motion was found to be significantly faster \cite{Moore2008,Miron2011,Baumgartner2017} compared to the previous 
observations in NiFe or Co/Ni films [Fig. \ref{fig:DW-PtCo}(a)]. 
Besides the practical importance of fast domain wall motion, the physical parameter determining this improvement was 
the structural inversion asymmetry \cite{Miron2009}. 
Indeed, while Pt/Co/AlOx supports fast CIDM, magnetically similar Pt/Co/Pt symmetric 
layers do not exhibit any CIDM at all \cite{Miron2009,Cormier2010}. 
These first observations were initially analyzed within the framework of the STT model, including the influence 
of the field-like SOT, which was discovered at the same time. 
It was proposed that the broken symmetry could accelerate the spin flip rate and enhance the non-adiabatic torque, the field-like SOT 
stabilizing the Bloch wall structure to prevent the Walker breakdown \cite{Miron2011}. This idea has been pursued by several theoretical investigations of the ability of field-like SOT to delay \cite{Ryu2012b} or suppress the Walker breakdown \cite{Stier2014,Linder2013b,Risinggard2017}.

At that stage, there was still a major discrepancy between the STT model and the experiment: the domain walls move in 
the direction of the electric current and not along that of the electron flow \cite{Moore2009b}. 
This intriguing observation motivated several theoretical studies, which found that the combination of 
STT and SOT could in certain cases produce backwards motion \cite{Boulle2014,Kim2012b}. 
However these scenarii were not robust: the backward motion was only obtained for certain values of the physical 
parameters and only in a certain range of current density. 
In parallel, it was observed that nearly symmetric Pt/Co/Pt samples exhibit CIDM if an 
external in-plane field is applied parallel to the current, sufficiently large to convert Bloch walls to N\'eel walls \cite{Haazen2013}
[the N\'eel wall structure under in-plane field was later confirmed by anisotropic magnetoresistance
measurements \cite{Franken2014}].
The damping-like SOT mechanism was shown to be compatible with all observations, especially (i) the reversal of the domain wall motion
upon locating the thicker Pt layer below or above the Co layer; (ii) the reversal
of the domain wall motion upon change of sign of the in-plane field and (iii) the fact that two successive domain walls always move in opposite directions.
The latter point is of crucial importance: all N\'eel walls, having the same magnetization, feel the same damping-like SOT and are hence
displaced in opposite directions, like under an easy-axis ($z$ here) field.

In this context, a breakthrough was the micromagnetic study \cite{Thiaville2012} of the dynamics of N\'eel walls under 
magnetic field and damping-like SOT, in the case where such walls are stabilized by the interfacial DMI.
The DMI \cite{Dzyaloshinskii1957,Moriya1960} is an antisymmetric exchange interaction that is allowed when the
medium does not have inversion symmetry. The general form of the DMI energy density reads $W_{\rm DMI}=D_{ij}{\bf e}_i\cdot\left({\bf m}\times\frac{\partial {\bf m}}{\partial j}\right)$, where the coefficient $D_{ij}$ possesses the same symmetries as the SOT response function, $\chi_{ij}$, discussed in Subsection \ref{s:sots-nut} \cite{Freimuth2014}. Hence, the generalization of SOT symmetries suggested by Fig. \ref{fig:Rashba_Dresselhaus} also applies to DMI. In an isotropic bulk material without inversion symmetry (like a heap of screws), to the lowest order in gradient expansion, DMI in continuous micromagnetic form is expressed by an energy density \cite{Bogdanov1989}
\begin{equation}
W_{\rm 3D}=D_{\rm 3D}{\bf m}\cdot({\bm\nabla}\times{\bf m}).
\label{eq:dmi-3d}
\end{equation}
Such an interaction favors helicoidal magnetization rotations of a given handedness. Referring to Fig.~\ref{fig:DW-structures}, this form of DMI stabilizes chiral Bloch walls or Y-Bloch walls.\par

On the other hand, at the interface between two dissimilar materials where inversion symmetry is structurally 
broken \cite{Fert1990}, assuming the highest symmetry ($C_{\infty v}$) and considering the lowest order in spatial gradient, one obtains \cite{Bogdanov1989,Heide2008} 
\begin{equation}
W_{\rm 2D}=D_{\rm 2D}{\bf m}\cdot[({\bf z}\times{\bm\nabla})\times{\bf m}].
\label{eq:dmi-2d}
\end{equation}
This interaction, called interfacial DMI, favors cycloidal magnetization rotations of a given handedness. Again referring to Fig.~\ref{fig:DW-structures}, this form of DMI stabilizes chiral N\'eel walls (but none of the Y-N\'eel walls).
The immediate consequence is that chiral N\'eel walls move under damping-like SOT without any in-plane field, with successive
walls moving in the same direction as their domain wall magnetizations are opposite.
Such a motion, already obtained with STT, is required for domain wall racetrack applications \cite{Parkin2008}.
Another notable feature of the domain wall dynamics under DMI and damping-like SOT is that the relative sign of domain wall velocity with 
respect to that of the current is given by the product of the sign of the damping-like SOT and the sign of the DMI.

Interfacial DMI was already evidenced in magnetic atomic monolayers or bilayers by spin-polarized
scanning tunneling microscopy that revealed magnetization cycloids of fixed handedness 
\cite{Bode2007,Meckler2009}. However, these were situations of very large DMI so that the uniform magnetic
state was destabilized.
For the Pt/Co/AlOx case, direct proof that domain walls are chiral N\'eel walls was obtained by NV-center magnetic microscopy
\cite{Tetienne2015}, and by x-ray magnetic circular dichroism \cite{Boulle2016}.
In addition, spin-polarized low energy electron microscopy has shown the change of domain wall structure
from chiral N\'eel wall to achiral Bloch wall as a function of the thickness of the magnetic layer \cite{Chen2013c,Chen2013d}, confirming the interfacial DMI description.

The prediction of \citet{Thiaville2012} was immediately backed by two experimental papers \cite{Emori2013,Ryu2013}.
As the sign of the SHE (hence of the damping-like SOT) was known from other measurements, the direction of domain wall motion
under current could be related to the sign of DMI [Fig. \ref{fig:DW-PtCo}(b)].
This sign was later obtained by several other techniques, so that presently estimates of interfacial DMI
for a fair number of NM/FM interfaces exist.
In this picture, the Pt/Co interface stands out with one of the largest interfacial DMI constant
$D_\mathrm{s} \approx -1.7$~pJ/m \cite{Belmeguenai2015}.
One of the techniques for determining the DMI consists in applying an additional in-plane field in order to compensate the DMI effective field on the domain wall. At this compensation, the domain wall velocity crosses zero \cite{Emori2013} [for an example see Fig.~\ref{fig:DW-PtCo}(b)].


\subsubsection{Domain wall motion under spin-orbit torque}

We now describe in more detail the dynamics of domain walls under SOTs and DMI. Once the torques are known and quantified, the study of their impact on domain wall motion should ultimately be performed by numerical micromagnetic simulations, for the sample parameters and geometrical dimensions.
For the physical understanding, however, simplified and as analytical as possible models are helpful.
The simplest model was exposed in Subsection~\ref{s:DW-dyna}.
The next level of complexity is addressed by the so-called $q-\Phi$ model, that describes a 1D domain wall dynamics for an assumed domain wall profile described by only two variables, namely the domain wall position $q$ and the angle $\Phi$ of the domain wall magnetization within the plane
orthogonal to the easy axis \cite{Slonczewski1972,Schryer1974}.
For SOT-driven domain wall motion assisted by DMI, the model was established by \citet{Thiaville2012}, and further developed
to incorporate in-plane fields \cite{Emori2013} and STT \cite{Torrejon2014}.
At a higher level of complexity, a numerical micromagnetic calculation is performed assuming 1D structure and dynamics, i.e., the magnetization depends only on the $x$ coordinate, the magnetostatic effects being computed for the nanostrip width $w$ and thickness $h$. Finally, for ultrathin films the full model consists of 2D numerical micromagnetics.

\begin{figure}
\includegraphics[width=8cm]{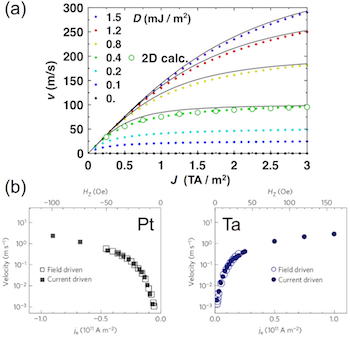}
\caption{Velocity of domain walls in ultrathin Pt/Co/oxide films with DMI, under current.
(a) Micromagnetic 1D calculations (points) of domain wall velocity {\em versus} current density, for various values of
effective DMI in a 0.6~nm Co film \cite{Thiaville2012}, considering only damping-like SOT.
Curves show the $q-\Phi$ model results for comparison.
(b) Measured domain wall velocity under current for Pt/CoFe/MgO and Ta/CoFe/MgO \cite{Emori2013}.
The CoFe film has 80-20 atomic composition and is 0.6 nm thick.
Note the log scale for velocities, and the opposite current signs for the two heavy metal layers.}
\label{fig:DW-vJ}
\end{figure}

Figure~\ref{fig:DW-vJ}(a) shows the predicted velocity {\em versus} current curves, $v(J)$, in the case of pure damping-like SOT and for various values of the effective DMI energy density, $D=D_s/h$. 
The domain wall velocity initially rises linearly with current, following a slope that does not depend on DMI and is given,
for DMI dominating the magnetostatic energy associated to a N\'eel wall and using the notation of Eq.~(\ref{eq:torquedef}),
by 
\begin{equation}
v=-\gamma\frac{\pi \Delta_{\rm W}}{2 \alpha M_{\rm s}}\tau_{\rm DL}.
\end{equation}
Here, $\Delta_{\rm W}$ is the micromagnetic domain wall width parameter. Upon further increase of the current density, the velocity saturates towards a plateau determined by the DMI strength, $v_D =\gamma \pi D / (2M_\mathrm{s})$ (derived in the same limit).
The velocity saturation is physically explained by the progressive rotation of the domain wall magnetization from N{\'{e}}el 
to Bloch around the effective field ${\bf B}_{\rm DL}$ associated with the damping-like SOT. This rotation leads to a reduction of the damping-like SOT on the domain wall, as the torque vanishes for a Bloch wall. This behavior is in good overall agreement with experiments [see Fig.~\ref{fig:DW-vJ}(b)]. With intrinsic curvature and no Walker breakdown, the velocity versus current behavior, $v(J)$, is markedly different from that expected for STT.

When DMI is not much larger than the magnetostatic energy associated to the N\'eel wall, the situation is more complex
to analyze, as the velocity $v_D$ decreases and becomes comparable to that induced by STT.
Moreover, for the $q-\Phi$ model, the analytical expressions become much more complex.
The analysis of the competition of DMI {\em versus} domain wall magnetostatics, together with that of damping-like SOT {\em versus} STT, was
performed by \cite{Torrejon2014} in the case of HM/CoFeB/MgO for HM=Hf, Ta, TaN, W i.e., the beginning of the 
5$d$ series, using the $q-\Phi$ model to analyze the experiments.
This showed that the determination of the DMI by the `crossing field' technique is strongly affected
by the STT when DMI is not large.

\subsubsection{Two-dimensional effects in current-induced domain wall motion}

\begin{figure}[h!]
\centering
\includegraphics[width=8cm]{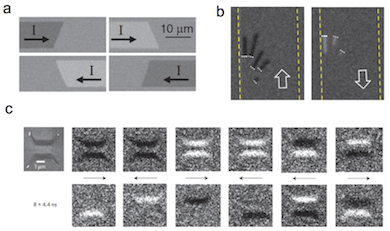}
\caption{\label{fig:Fig44-7} (a) Kerr imaging of domain wall tilting produced by current injection \cite{Ryu2012}. (b) Kerr images of current induced domain wall motion in a non-collinear geometry. (c) Magnetization reversal controlled by geometry \cite{Safeer2016}.}
\end{figure}

Unlike in-plane magnetized nanowires, where domain walls behave as quasi-1D objects, in perpendicular samples domain walls act more like 2D membranes. One of the first observations on the influence of the 2D character on the CIDM in materials with broken inversion symmetry was the occurrence of a domain wall tilt. When domain walls are displaced by sufficiently long current pulses, their end position is no longer perpendicular to the wire (at the energy minimum), but tilted at a certain angle \cite{Ryu2012,Baumgartner2018b} [see Fig. \ref{fig:Fig44-7}(a)]. The fact that this tilt is visible at rest is a proof that domain wall pinning exists.
\citet{Boulle2013} proposed that this tilting arises from the competition between the damping-like SOT and the DMI. Because the DMI energy prefers that the domain wall magnetization is perpendicular to the domain wall, the damping-like SOT acting on the domain wall magnetization modifies the domain wall angle [see also \citet{Martinez2014}]. A direct consequence of this current-driven tilting is an additional deformation of the $v(J)$ curve at large current density. For damping-like SOT only, this gives rise to a velocity increase close to the threshold for domain stability given by $\tau_{\rm DL~max}= \gamma B_{K}^\mathrm{eff}/2$ ($B_{K}^\mathrm{eff}$ being the effective perpendicular anisotropy that incorporates the demagnetizing field).\par

It was recently shown that the 2D character of CIDM can be exploited for domain wall manipulation. Since the domain wall magnetization is either aligned (Bloch wall) or perpendicular (N\'eel wall) to the domain wall direction, the control of the domain wall tilt allows for setting the SOT efficiency by modifying the angle between the electric current and the magnetization of the domain wall. Using this approach, \citet{Safeer2016} have shown that the current induced domain wall motion in the non-collinear geometry exhibits surprising features [see Fig. \ref{fig:Fig44-7}(b)]. Namely, depending on their polarity (up/down or down/up), the domain walls move faster for a certain sign of the electric current. This phenomenon links the polarity of the domain walls with their direction of motion. Therefore, by controlling the shape of a magnetic layer, one can control its magnetization reversal [Fig. \ref{fig:Fig44-7}(c)].

\subsubsection{Domain wall motion under combined spin transfer and spin-orbit torques}

A detailed study of CIDM in Pt/(Co/Ni)$_N$/Co/MgO as a function of magnetic layer thickness
(by varying the repetition number $N$) was
realized by \citet{Ueda2014}.
The (Co/Ni)$_N$ multilayer system is interesting because since magnetic anisotropy arises from the
internal Co/Ni interfaces, the total thickness of the multilayer can be changed while keeping 
the same magnetic anisotropy, which is not possible for a single Co layer.
The domain walls were observed to move along the electron flow for large thicknesses ($N > 6$), but in the opposite
direction at small thicknesses ($N < 3$).
From the application of additional easy-axis field, it was concluded that CIDM at large thickness is due
to the adiabatic STT, but that the torque on the domain wall is like a bias field for low thickness.
Applying then, in addition to current, in-plane fields in both orientations (longitudinal $x$, and 
transverse $y$), the crossing field effect [see Fig.~\ref{fig:DW-PtCo}(b)] was observed in the longitudinal 
case, in accord with the damping-like SOT in the presence of DMI.
The transverse field was observed to linearly modify the velocity of both up/down and down/up domain walls, in
the same way.
This is also consistent with the DMI and damping-like SOT mechanism, as the magnetizations of two consecutive
chiral N\'eel walls precess under the respective fields ${\bf B}_{\rm DL}$ of the damping-like SOT towards the same $y$ direction.
Thus, for not too large $y$ fields, one polarity increases this rotation and hence decreases the domain wall 
velocities, whereas the other polarity decreases this rotation and increases the velocities.
From the symmetry of the effects, the authors concluded that the field-like SOT effect was negligible.
Direct measurements of the two components of the SOT confirmed the reduced value of the field-like SOT.
This work clearly evidences the transition from bulk to interfacial CIDM and can serve as a guide for further studies of this physics.
For example, the absence of domain wall motion for $3 \le N \le 6$ was interpreted by the fact that the interfacial
DMI from the bottom Pt layer was raising too much the Walker field, so that the domain wall motion by adiabatic
STT could not be reached for the applied currents. The same mechanism applies for the combination of adiabatic STT and field-like SOT, showing that these two mechanisms
of CIDM can act in opposition. A similar transition from SOT to STT driven domain wall motion has been observed in (Co/Tb)$_N$ multilayers \cite{Bang2016}.\par

In another study in the same (Co/Ni)$_N$ system, the structure was designed such that SOT acted as a perturbation with respect to STT \cite{Yoshimura2014}.
The sample was medium-thick ($N=4$) and the structure was nominally symmetric with Pt and Ta on both
sides, with the same thicknesses. The domain wall motion, driven by STT, was modified by applying in-plane fields, both along the current ($x$) or transverse ($y$). As expected for adiabatic STT, the motion was suppressed by large in-plane fields, as these fields block the precession of the domain wall moment.
The surprise was that the domain wall motion windows were not centered at zero field, with the $x$-field offset
reversing sign between up-down and down-up domain walls.
This could be qualitatively interpreted by (i) a precession dissymmetry under in-plane field that leads 
to different residence times for N\'eel walls of opposite chiralities, and (ii) a non-compensated damping-like SOT due to a measured imbalance in the conduction of the top and bottom Pt layers. On the other hand, the independence of the $y$ field offset on the domain wall type (up/down or down/up)
is consistent with an effect of Oersted field and/or field-like SOT. This work, more generally, proposes a way to experimentally test the presence of the SOT and of the
Oersted field, as any in-plane field affects the precession of the domain wall moment triggered by STT. Here we refer also to the numerical work by \citet{Martinez2012} on the STT plus field-like SOT case, for various values on non-adiabaticity, and the micromagnetic simulations analysis by \citet{Martinez2013} and \citet{Boulle2014} of 
experimental results for Pt/Co/AlOx in terms of STT plus SOT.

\subsubsection{Motion of magnetic skyrmions under spin-orbit torques}
\label{sec:Sk}

Magnetic skyrmions with non-zero spin winding number are compact magnetic textures with a non-trivial topology, so that they cannot be removed by a continuous transformation, in the continuum limit. Although there are still arguments about the precise meaning of this terminology, we adopt here the definition agreed on by a large panel of authors \cite{Hellman2017}. There is currently an increasing interest in the electrical manipulation of such objects as they could serve as fundamental building blocks for data storage and logic devices \cite{Fert2013,Tomasello2014,Zhang2015d}. 
Skyrmions have, in addition to topology and compared to the magnetic bubbles extensively investigated
in the past \cite{Malozemoff1979}, a fixed chirality which is an important asset for SOT as can be inferred from the preceding considerations.

\begin{figure}
\includegraphics[width=8cm]{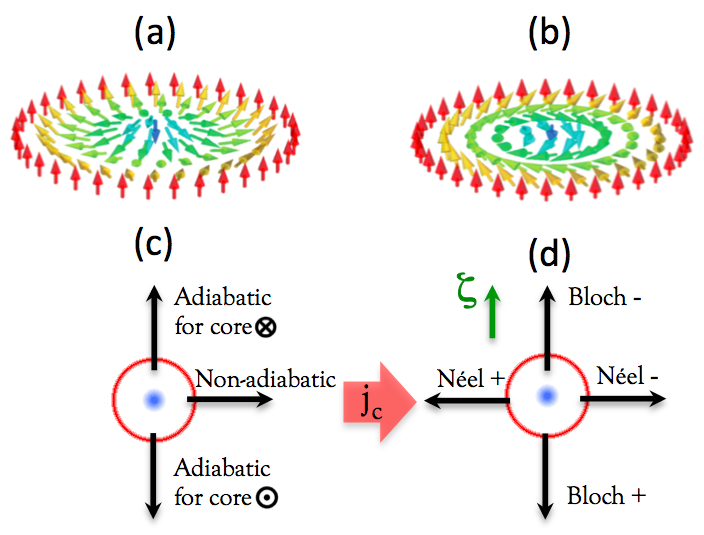}
\caption{(a) N\'eel- and (b) Bloch-skyrmions (both with negative polarity), adapted from \citet{Fert2017}. (c) Schematic of the forces (black arrow) applied to a skyrmion (circle) by a current density $j_{\rm c}$, in the case of adiabatic and non-adiabatic STT. The forces are independent on the type of the skyrmion and only depend on its core polarity. (d) Schematic of the forces (black arrows) applied to a skyrmion (circle) by a damping-like SOT with a spin polarization along ${\bm\zeta}$, as indicated by the green arrow. The forces depend on whether the skyrmion is Bloch or N\'eel.}
\label{fig:DW-Sk}
\end{figure}

A physically appealing way to understand how the various characteristics of a skyrmion affect
its response to current-induced torques is offered by Thiele's equation derived from the LLG equation to handle the steady-state motion of rigid textures \cite{Thiele1973}.
Thiele's equation has been generalized to include STT \cite{Thiaville2005} and, more recently, also SOT
\cite{Sampaio2013}.
It reads,%
\begin{equation}
{\bf G} \times \left( {\bf v}-{\bf u} \right)-
\overline{\overline{D}} \left( \alpha {\bf v} - \beta {\bf u} \right)
+ {\bf F}_\mathrm{SOT} + {\bf F} = {\bf 0},
\label{eq:DW-Thiele}
\end{equation}
where ${\bf v}$ is the in-plane velocity of the skyrmion center, ${\bf u}$ is the spin-drift velocity, $\beta$ is the
non-adiabaticity parameter related to STT [see Eq. \eqref{eq:DW-STT}], ${\bf G}$ is the so-called gyrovector, 
$\overline{\overline{D}}$ is the dissipation tensor introduced by Thiele (from which the damping
coefficient $\alpha$ was factored out when generalizing to STT), ${\bf F}$ is the other force applied to the
skyrmion (e.g., pinning), and finally ${\bf F}_\mathrm{SOT}$ is the force that SOTs apply to the skyrmion.
Topology appears in the gyrovector
${\bf G}= \left(M_\mathrm{s} h / \gamma \right) 4 \pi N_\mathrm{Sk} {\bf z}$,
that is along the film normal and proportional to the topological (or skyrmion) number
$N_\mathrm{Sk}$.
The latter is simply, for a compact texture, $N_\mathrm{Sk}=S p$ with $p$ the polarity of the
magnetization of the skyrmion center (+1 for $+{\bf z}$) and $S$ the winding number of the magnetization
(+1 for the simple skyrmions).
The dissipation tensor, diagonal for high-symmetry textures, is related to the size of the
skyrmion [see e.g. \citet{Hrabec2017}].
The force ${\bf F}$ is non-zero for example when a confining potential exists, or a small
$z$ field gradient. The STT forces on a skyrmion are illustrated in Fig.~\ref{fig:DW-Sk}(c).\par

The force from the SOT is computed by projecting the SOT on the skyrmion displacement (the procedure
by which Thiele's equation is constructed), as a volume integral for each component 
\begin{eqnarray}
F_{i\mathrm{,SOT}}=-\int d{\cal V}{{\bf B}_\mathrm{SOT} \cdot \frac{\partial}{\partial i} {\bf M}}= \tau_{\rm DL} {\bm\zeta} \cdot 
\int d{\cal V}{{\bf m} \times \frac{\partial}{\partial i} {\bf m}}.\nonumber\\
\label{eq:DW-Skforce}
\end{eqnarray}
The field-like SOT gives no contribution to the force as it acts like a constant in-plane field.
As for the damping-like SOT contribution, remembering that ${\bm\zeta} \parallel {\bf y}$ for current along $x$,
Eq.~(\ref{eq:DW-Skforce}) amounts to one term of the DMI energy density.
For the $x$ component of the force (along the current), it is the part of the interfacial DMI that involves the $x$ gradients [Eq.~(\ref{eq:dmi-2d})].
For the $y$ component of the force (transverse to the current), it is the part of the bulk DMI that involves the $y$ gradients [Eq.~(\ref{eq:dmi-3d})]. Thus, the damping-like SOT force on a skyrmion depends on its chirality and of its type (i.e.
Bloch or N{\'{e}}el), see Fig.~\ref{fig:DW-Sk}(d).
Because of the gyrotropic term, both STTs and damping-like SOT drive the skyrmion, at some angle
between the $x$ and $y$ axes, so that skyrmion motion under current alone does not allow to infer its internal structure. Nevertheless, conventional magnetic bubbles, whose lowest energy state is an achiral Bloch skyrmion because of the
absence of significant DMI, would be sorted according to their Bloch chirality by damping-like SOT. This is in contrast to skyrmions that have a definite chirality (fixed by DMI) and should all follow the same trajectory.

Metastable magnetic skyrmions have been recently obtained at room temperature in transition metal multilayers \cite{Jiang2015,Woo2016,Boulle2016,Moreau-Luchaire2016,Chen2015b,Pollard2017,Hrabec2017}. Many experiments have revealed skyrmions motion under current, either along or against the direction of the electron flow, in most cases in agreement with the DMI and damping-like SOT sign \cite{Jiang2015,Woo2016,Jiang2017b,Litzius2017,Hrabec2017,Yu2016c}. The influence of disorder and thermal fluctuations on the driven motion of single skyrmions have been investigated either using particle-simulations \cite{Lin2013c} or by micromagnetic modeling \cite{Sampaio2013}, demonstrating that skyrmions have the tendency to avoid point-like defects \cite{Iwasaki2013b,Iwasaki2013a}. Remarkably, the angle of the gyrotropic deflection (sometimes also called skyrmion Hall angle) observed in experiments is poorly reproduced by Thiele's equation. This discrepancy may be due to disorder effects, e.g. sliding along grain boundaries, as suggested by recent simulations \cite{Kim2017b,Legrand2017,Reichhardt2016,Salimath2018}. Moreover, the skyrmion dynamic deformation leads to an influence of the field-like SOT on the deflection angle \cite{Litzius2017}. Finally, the influence of the gradient of the $z$-component of the Oersted field deserves further investigation \cite{Hrabec2017}. Altogether, skyrmions appear as favorable objects to be controlled by either STT or SOT since their velocity reach that of magnetic domain walls in the same structures. Further exploration of their robustness and scalability is currently on-going \cite{Bernand-Mantel2018,Buttner2018}.

\subsubsection{Impact of disorder}

In sputtered ultrathin magnetic multilayers, disorder is so strong that CIDM only occurs at large (field or current) drive. At low drive, domain wall motion consists of thermally assisted hopping between pinning sites. This regime of domain wall motion is called creep, or depinning, depending on field magnitude and type of domain wall pinning\cite{Metaxas2007,Kim2009,Gorchon2014}. Whereas field-driven and STT-driven creep seemed to be well understood \cite{Chauve2000,Jeudy2016,DuttaGupta2016}, the situation has changed with the introduction of SOT and DMI. Several experiments of CIDM have shown that the creep regime of domain wall motion deserves further study in order to be fully understood \cite{Lavrijsen2012,Lavrijsen2015,Vanatka2015}. For instance, it was recently proposed that structural inversion asymmetry could be responsible for a chiral dissipation mechanism affecting the domain wall dynamics, called chiral damping \cite{Jue2016a,Akosa2016}. More recently it has been shown that, as the domain wall energy becomes orientation-dependent under in-plane field \cite{Pellegren2017} - an effect reinforced by DMI and strikingly evidenced by specific domain shapes \cite{Lau2016b} -, the simple creep model with uniform domain wall tension fails. Thus, the analysis of creep motion under in-plane field has to be thoroughly re-examined.

In contrast, a thorough study of the influence of disorder on CIDM in the flow or precessional regimes is still missing. Similarly to in-plane materials (see Sec.~\ref{sec:ip}), one may expect a modification of the threshold values for domain wall motion, a change of slope and/or offset of velocity in the linear regime, and suppression of velocity breakdown. Regarding the apparent offset in velocity, a simple model considering the statistical average of the inverse velocities was shown to reproduce the feature \cite{Jue2016b,Feldtkeller1968}. The suppression of velocity breakdown has been observed and numerically reproduced for field-driven motion in the presence of interfacial DMI \cite{Pham2016,Jue2016b,Yoshimura2016,Ajejas2017}. In fact, simulations reveal that this effect occurs even in the absence of disorder, as soon as the domain wall has a sufficient length to depart from 1D behavior. The novelty introduced by STT, SOT, and DMI is that the domain wall drive depends on the local domain wall orientation, which modifies the overall energetics and dynamics. In the creep regime under CIDM and field, these angle dependences are exemplified by the formation of triangular-shape domains pinned at nucleation sites \cite{Moon2013b,Moon2018}.

\subsection{Antiferromagnetic and ferrimagnetic systems\label{s:AF-DW}}

The search for extremely high domain wall velocity has recently brought perpendicularly magnetized synthetic antiferromagnet strips, such as (Co/Ni)/Ru/(Co/Ni), to the forefront. In such systems, \citet{Yang2015a} reported SOT-driven domain wall velocities as fast as 750 m/s and explained the results by the enhanced Walker breakdown threshold. Indeed, in the presence of Ru-mediated interlayer exchange coupling (RKKY) the azimuthal angles of the two antiparallel domain walls stabilize each other such that the domain wall propagates in the flow regime over a larger range of driving current densities [see also \cite{Lepadatu2017}]. In a recent work \citet{Qiu2016} observed that the presence of the Ru spacer layer may affect the spin current, leading to different SOTs for such trilayers.

Similar ideas resulted in the proposition of antiferromagnetic skyrmions, either in the bilayer form or in bulk antiferromagnets, which display no skyrmion Hall effect and could also reach very high velocity in the latter case \cite{Zhang2015c,Barker2016,Tomasello2017}. Bulk antiferromagnets are particularly interesting for their ability to support THz dynamics, and it was recently proposed that antiferromagnetic domain walls driven by SOT could reach extremely high velocities, while displaying a Lorentz contraction when reaching the spin wave group velocity \cite{Shiino2016,Gomonay2016}. The investigation of antiferromagnetic spintronics is still at its infancy though \cite{Jungwirth2016,Baltz2018}, and alternative materials are being explored. From this perspective, ferrimagnets such as FeGdCo offer an appealing platform due to the tunability of their compensation point. For instance, recent studies have demonstrated large enhancement of field-driven domain wall velocity and SOT efficiencies close to the angular momentum compensation point \cite{Kim2017c,Mishra2017}.


\section{Perspectives\label{s:7}}

SOTs offer a powerful and versatile tool to manipulate and excite magnetic order parameters, and efficiently control magnetic domain walls and skyrmions. A particularly attractive feature of these torques is their ability to excite any type of magnetic materials, ranging from metals to semiconductors and insulators, in both ferromagnetic and antiferromagnetic configurations. This versatility has led to groundbreaking accomplishments that could not be achieved with STT: the switching of single layer ferromagnets, ferrimagnets, and antiferromagnets, as well as the excitation of spin waves and auto-oscillations in planar and vertical device geometries. 

The discovery of topological materials as spin sources has opened appealing avenues for the realization of very large charge-to-spin conversion and low critical switching current. Topological insulators, Dirac semimetals, Weyl semimetals, Kondo topological insulators as well as 2D materials (bismuth chalcogenides, graphene and its siblings, transition metal dichalogenides, transition metal trihalides etc.) present a unique opportunity for the exploitation of exotic spin-charge conversion mechanisms and chiral spin textures.\par

A number of questions remain open, which will have an impact on future developments and materials design.\par

(i) Whereas the basic mechanisms behind SOT seem to be understood, a robust and systematic quantitative agreement between theory and experiment is still lacking. In particular, understanding the interplay between interfacial, bulk, but also orbital contributions to SOT, DMI and chiral damping in magnetic multilayers will indicate how to improve their efficiency.

(ii) Besides the two "flagship mechanisms" that control the current-induced dynamics (iSGE, SHE), novel phenomena have been identified recently: spin swapping \cite{Saidaoui2016}, interfacial spin currents \cite{Amin2018}, chiral damping \cite{Jue2016a} etc. What is the actual magnitude of these effects and how do they influence the magnetization dynamics? How can they be best harvested to enhance the operability of SOT devices?

(iii) The electrical control of magnetic domain walls and skyrmions substantially benefits from SOTs. Nevertheless, their behavior in the presence of disorder, and particularly the creep and depinning regimes, need to be better understood. How can these regimes provide information about the nature of the chirality (dissipation and energy)? Several novel torques have been predicted in these textures \cite{Bijl2012}, but not observed yet. In addition, topological currents have been proposed to enhance the mobility of both ferromagnetic and antiferromagnetic skyrmions \cite{Akosa2018,Abbout2018}, which calls for experimental verification.

(iv) Antiferromagnets bear outstanding promises due to the zero net magnetization and their inherent THz dynamics \cite{Jungwirth2016,Baltz2018,Jungwirth2018}. However, to date, only a few antiferromagnets have been electrically manipulated (CuMnAs, Mn$_2$Au and NiO) \cite{Wadley2016,Bodnar2018,Chen2018,Moriyama2018}. The next frontier is to extend these observations to more materials, including non-collinear antiferromagnets. The latter present the advantage of displaying AHE as well as MOKE response, enabling for the electrical and optical detection of their order parameter's orientation \cite{Nakatsuji2015}. A natural development direction will be to extend these ideas to frustrated magnets that support exotic magnetic behaviors \cite{Balents2010}.

(v) Finally, the search for most efficient sources of SOTs raises the question about the nature of spin-orbit effects in the presence of very large spin-orbit coupling. How do concepts such as spin currents, SHE, iSGE, DMI and magnetic damping evolve when the spin-orbit interaction is comparable to or larger than the crystal field? Is there a limit to the amount of angular momentum that can be transferred to the magnetic system, and if so, how can it be determined? What materials combination produces the largest torque? Similar questions can be asked when electronic correlations are important, such as in (Ce,Ca)MnO$_3$, Yb$_2$Ti$_2$O$_7$, SmB$_6$ etc.

Besides these important challenges, what makes SOTs truly attractive is their potential for efficient device operation. In a nutshell, SOTs can do everything STT can, with the crucial advantage of decoupling the injection and detection paths. This unique feature allows for the excitation and switching of large magnetic surface areas ($>\mu$m$^2$), but also the electrical control of magnetic insulators and antiferromagnets, which traditional STT cannot achieve. Its implementation does not only enhance the performance of devices (speed, power consumption) such as SOT-MRAMs, nano-oscillators or magnetic racetrack data storage devices, but it also opens thrilling perspectives beyond conventional spintronics components \cite{Sato2018}. For instance, SOT-driven memristors have been developed to be used as synapses for artificial neural networks \cite{Lequeux2016,Borders2018}, while SHE-SOT can be exploited to build stochastic parity-bits for invertible logic \cite{Camsari2017}. Finally, SOTs could be used to manipulate and explore more exotic magnetic excitations such as the ones emerging in spin liquids \cite{Balents2010}, i.e., spinons, magnetic monopoles, anyons, or even Majorana fermions.

\section*{List of abbreviations}

\noindent{\bf 1D, 2D, 3D}: One-, two- and three-dimensional 




\noindent {\bf AHE}: Anomalous Hall effect

\noindent {\bf AMR}: Anisotropic magnetoresistance

\noindent {\bf ANE}: Anomalous Nernst effect


\noindent {\bf CIDM}: Current-induced domain wall motion


\noindent {\bf DMI}: Dzyaloshinskii-Moriya interaction








\noindent {\bf FMR}: Ferromagnetic resonance



\noindent {\bf iSGE}: Inverse spin galvanic effect

\noindent {\bf LED}: Light emitting diode

\noindent {\bf LLG}: Landau-Lifshitz-Gilbert (equation)



\noindent {\bf MOKE}: Magneto-optical Kerr effect

\noindent {\bf MRAM} Magnetic random access memory

\noindent{\bf NM/FM}: Nonmagnetic metal/ferromagnet

\noindent{\bf NM/AF}: Nonmagnetic metal/antiferromagnet





\noindent {\bf RF}: Radio frequency

\noindent {\bf RKKY}: Ruderman-Kittel-Kasuya-Yosida (interaction)

\noindent {\bf SGE}: Spin galvanic effect

\noindent {\bf SHE}: Spin Hall effect

\noindent {\bf SMR}: Spin Hall magnetoresistance

\noindent {\bf SOT}: Spin-orbit torque





\noindent {\bf ST-FMR}: Spin torque ferromagnetic resonance

\noindent {\bf STT}: Spin transfer torque

\noindent {\bf TMR}: Tunnelling magnetoresistance




\section*{Acknowledgements}
A.M. was supported by the King Abdullah University of Science and Technology (KAUST). T. J. acknowledges support from the EU FET Open RIA Grant No. 766566, the Ministry of Education of the Czech Republic Grant No. LM2015087 and LNSM-LNSpin, and the Grant Agency of the Czech Republic Grant No. 14-37427G. J. S. acknowledges the Alexander von Humboldt Foundation, EU FET Open Grant No. 766566, EU ERC Synergy Grant No. 610115, and the Transregional Collaborative Research Center (SFB/TRR) 173 SPIN+X. K.G. and P.G. acknowledge stimulating discussions with C.O. Avci and financial support by the Swiss National Science Foundation (Grants No. 200021-153404 and 200020\_172775) and the European Commission under the Seventh Framework Program (spOt project, Grant No. 318144). A.T. acknowledges support by the Agence Nationale de la Recherche, project ANR-17-CE24-0025 (TopSky). J. \v{Z} acknowledges the Grant Agency of the Czech Republic grant no. 19-18623Y and support from the Institute of Physics of the Czech Academy of Sciences and the Max Planck Society through the Max Planck Partner Group programme.

\bibliography{Biblio_final}


\end{document}